\documentclass[prd,aps,nofootinbib,onecolumn,preprintnumbers,12pt]{revtex4-1}

\usepackage{amssymb}
\usepackage{amsmath}
\usepackage{bm}
\usepackage{booktabs}
\usepackage{braket}
\usepackage{breakurl}
\usepackage{caption}
\usepackage{comment}
\usepackage{color}
\usepackage{epsfig}
\usepackage{graphicx}
\usepackage{hyperref}
\usepackage{cleveref}   
\usepackage{lineno}
\usepackage{lipsum}
\usepackage{microtype}
\usepackage{multirow}
\usepackage{rotating}
\usepackage{slashed}
\usepackage{subfigure}
\usepackage{times}
\usepackage{xspace}

\allowdisplaybreaks[4]

\definecolor{nicered}{rgb}{0.7,0.1,0.1}
\definecolor{nicegreen}{rgb}{0.1,0.5,0.1}

\hypersetup{colorlinks,citecolor=nicegreen,linkcolor=nicered}

\begin{document}

\title{\texorpdfstring{$CP$}{CP} violation in two-body hadronic \texorpdfstring{$\Lambda_b$}{Lb} decays in the PQCD approach}

\author{ Jia-Jie Han$^1$, 
	Ji-Xin Yu$^1$~\footnote{Corresponding author, Email: yujx18@lzu.edu.cn},
	Ya Li$^2$~\footnote{Corresponding author, Email: liyakelly@163.com},
	Hsiang-nan Li$^3$~\footnote{Corresponding author, Email: hnli@phys.sinica.edu.tw},
	Jian-Peng Wang$^1$~\footnote{Corresponding author, Email: wangjp20@lzu.edu.cn},
	Zhen-Jun Xiao$^2$~\footnote{Corresponding author, Email: xiaozhenjun@njnu.edu.cn},
	Fu-Sheng Yu$^1$~\footnote{Corresponding author, Email: yufsh@lzu.edu.cn} }

\affiliation{
	$^1$Frontiers Science Center for Rare Isotopes, and School of Nuclear Science and Technology, Lanzhou University, Lanzhou 730000,   People’s Republic of China \\
	$^2$Department of Physics and Institute of Theoretical Physics, Nanjing Normal University, Nanjing 210023, People’s Republic of China\\
	$^3$Institute of Physics, Academia Sinica, Taipei, Taiwan 115, Republic of China
}

\begin{abstract}
	
We systematically investigate the $CP$-averaged branching ratios and $CP$ violations (CPVs) for the two-body hadronic decays $\Lambda_b\to ph$, where $h$ runs through the  mesons $\pi^-$, $\rho^-$, $a_1^-(1260)$, $K^-$, $K^{\ast -}$, $K_1^-(1270)$ and $K_1^-(1400)$, in the perturbative QCD approach to order  $\alpha_s^2$ in the strong coupling. Various topological amplitudes are obtained by incorporating subleading-twist hadron distribution amplitudes, which exhibit reasonable hierarchical patterns, sizable strong phases, and non-negligible higher-power corrections. The predicted direct CPVs in $\Lambda_b\to p\pi^-,pK^-$, different from those in similar $B$ meson decays, are as small as the current data. The low CPV in $\Lambda_b\to p\pi^-$ results from the cancellation between the $S$- and $P$-wave CPVs, while the one in $\Lambda_b\to pK^-$ is determined by the tiny $S$-wave CPV. However, individual partial-wave CPVs can exceed $10\%$, consistent with direct CPVs in $B$ meson decays. The CPVs in the $\Lambda_b\to pK_1^-(1270),pK_1^-(1400)$ channels are relatively larger. In particular, CPVs above $20\%$ appear in the up-down asymmetries associated with the final-state angular distributions of $\Lambda_b\to pK_1^-(1270),pK_1^-(1400)$, followed by the secondary $K_1\to K\pi\pi$ decays.These observables offer promising prospects for firmly establishing baryon CPVs. The decay asymmetry parameters of $\Lambda_b\to ph$ are also predicted for future experimental confrontations.
	
\end{abstract}

\maketitle

\tableofcontents

\section{Introduction}\label{sec:introduction}

The explanation of the asymmetry between matter and antimatter in the Universe requires $CP$ violations (CPVs), where $C$ and $P$ represent the charge-conjugation and parity, respectively. 
The standard model (SM) of particle physics provides the Cabibbo–Kobayashi–Maskawa (CKM) mechanism for generating CPVs, which is, however, too weak to account for the excess of matter. Additional sources of CPVs are thus needed, and their exploration has marked a long-term effort.
CPVs have been observed in $K$~\cite{Christenson:1964fg}, $B$~\cite{BaBar:2001ags,Belle:2001zzw} and $D$~\cite{LHCb:2019hro} meson decays, and consistent with the SM predictions.
However, they have not yet been firmly established in baryonic processes.
Given that visible matter in the Universe is mainly composed of baryons, it is natural to search for CPVs in baryon systems. 

Baryons cannot undergo mixing owing to baryon number conservation, and exhibit only direct CPVs in their decays. 
Experimental investigations of baryon CPVs have been extensively pursued by the BESIII, Belle and LHCb Collaborations~\cite{BESIII:2021ypr,BESIII:2018cnd,Belle:2022uod,LHCb:2017hwf,LHCb:2016yco,LHCb:2018fly,LHCb:2018fpt,LHCb:2019oke,LHCb:2019jyj,LHCb:2024yzj,LHCb:2024iis}.
The measured direct CPVs \cite{LHCb:2024iis}
\begin{equation}\label{eq:AcpExp}
	\begin{split}
		A_{CP}^{\rm dir}(\Lambda_b^0\to p\pi^-)&=(0.2\pm 0.8\pm 0.4)\%,\\
		A_{CP}^{\rm dir}(\Lambda_b^0\to pK^-)&=(-1.1\pm 0.7\pm 0.4)\%,
	\end{split}
\end{equation}
are compatible with null asymmetries within the precision of $1\%$. 
That is, the CPVs in $\Lambda_b$ baryon decays are much lower than those in similar {\it B} meson decays, although both are induced by the $b\to u\bar{u}q$ transitions, $q=d,s$.
Recently, the LHCb reported the evidence of the CPV in $\Lambda_b^0\to \Lambda^0K^+K^-$ \cite{LHCb:2024yzj} and the first observation of the baryon CPV in the $\Lambda^0_b \to pK^-\pi^+\pi^-$ mode~\cite{LHCb:2025ray}, which motivates timely related theoretical studies~\cite{Yu:2025ekh}.

Baryon processes, engaging more complicated dynamics than meson ones, impose a number of  phenomenological challenges~\cite{Han:2024kgz}.
The non-zero spin of a baryon prompts at least two partial waves, such as the $S$- and $P$-wave amplitudes in  $\Lambda_b\to p\pi^-, pK^-$, while there is only a single partial wave in $B$ meson decays into two pseudoscalar mesons. 
The polarizations of initial and final states in a baryon decay enable the construction of more observables out of angular distributions of decay products~\cite{LHCb:2016yco,Wang:2024qff,Korner:1992wi,Zhao:2024ren,Geng:2021sxe}, which shed valuable insight on involved QCD dynamics. 
A baryon contains three valence quarks, demanding at least two hard gluons to propagate momentum transfer in its exclusive decay.
More gluon exchanges may enhance higher-power contributions and modify the behavior of power expansions~\cite{Wang:2011uv,Han:2022srw}. 
Besides, there also exist additional topological diagrams, including color-commensurate $W$-emission and $W$-exchange diagrams, that serve as abundant sources
of strong phases required for direct CPVs. 

Several theoretical approaches have been proposed for analyzing two-body hadronic $B$ meson decays: the QCD
factorization (QCDF)~\cite{Beneke:1999br,Beneke:2000ry}, the soft-collinear-effective theory (SCET)~\cite{Bauer:2000yr,Bauer:2001yt,Bauer:2002nz} and the perturbative QCD (PQCD) factorization~\cite{Keum:2000wi,Lu:2000em,Keum:2000ph}. 
The QCDF and SCET are based on the collinear factorization theorem, in which $B$ meson transition form factors develop endpoint singularities if they were computed perturbatively. 
The PQCD is based on the $k_T$ factorization theorem, in which the endpoint contribution is absorbed into a transverse-momentum-dependent distribution amplitude (DA) or resummed into a Sudakov factor.
The factorizable emission, nonfactorizable emission, $W$-exchange and annihilation diagrams, being free of the endpoint singularities, are then computable.
CPVs of two-body hadronic $B$ meson decays have been successfully predicted in the PQCD \cite{Keum:2000wi,Lu:2000em,Keum:2000ph}.
However, rigorous and systematic formalisms for heavy baryon decays are still unavailable. 
The generalized factorization assumption was employed to estimate branching ratios of numerous bottom-baryon decays~\cite{Hsiao:2014mua,Hsiao:2017tif,Geng:2021nkl}. Final-state rescattering mechanism was recently developed to study CP violation of $\Lambda_b$ decays~\cite{Wang:2024oyi,Duan:2024zjv}.
As to QCD-inspired methods, the QCDF was applied to $\Lambda_b$ baryon decays under the diquark approximation \cite{Zhu:2016bra,Zhu:2018jet}, and the $\Lambda_b\to p\pi,pK$ branching ratios were evaluated in the PQCD~\cite{Lu:2009cm}, but with the results being a few times smaller than data. 

The $\Lambda_b\to p$ transition form factors with reasonable subleading-twist hadron DAs were reproduced in the PQCD recently, and the outcomes agree with those from lattice QCD and other nonperturbative methods \cite{Han:2022srw}.
The two-body charmed hadronic $\Lambda_b$ decays are also investigated in the PQCD after including high-twist DAs \cite{Zhang:2022iun,Rui:2023fiz,Rui:2022sdc}.
Those studies indicate that  exclusive heavy baryon decays can be examined reliably in this framework.
We will perform a comprehensive analysis on the two-body hadronic decays $\Lambda_b \to ph$, where $h$ runs through the pseudoscalar ($P$) mesons $\pi,K$, the vector ($V$) mesons $\rho, K^*$ and the axil-vector ($A$) mesons $a_1(1260), K_1(1270),K_1(1400)$.
A decay amplitude is expressed as a convolution of the hard kernel $T_H$ with the Sudakov factor $e^{-S}$ and the nonperturbative DAs $\Psi_{\Lambda_b}$, $\Psi_{p}$ and $\Psi_{h}$~\cite{Han:2024kgz},
\begin{equation}
	\mathcal{A}=\int_{0}^{1}[dx] \int[d^2b] \Psi_{\Lambda_b} T_H \Psi_{p} \Psi_{h} e^{-S},\label{2}
\end{equation}
where $x$ are the quark momentum fractions, $b$ are conjugate to the quark transverse momenta $k_T$, and $S$ represents the Sudakov exponent. In principle, the threshold resummation factor $S_t$ arising from the all-order organization of the large logarithms $\ln^2 x$ should be included. 
Since $S_t$ has not been derived for heavy baryon decays, we set $S_t=1$ conservatively as in our previous work~\cite{Han:2022srw}. 
For further details of the PQCD formalism and its applications to heavy baryon decays, refer to Refs.~\cite{Lu:2009cm,Han:2022srw}.

It will be demonstrated that CPVs in individual partial waves of $\Lambda_b$ decays can be as large as those in $B$ meson decays, greater than $10\%$, but the cancellation between partial waves leaves no significant deviation from zero.
The sign of a partial-wave CPV depends on the location of the weak vertex in the penguin diagram that provides the dominant source of strong phases. 
We will illustrate such cancellation in a full QCD calculation for the $\Lambda_b\to p\pi^-$ decay. 
The partial-wave CPVs in the $\Lambda_b\to p\rho^-, pK^{\ast -}, pa_1^-(1260), pK_1^-(1270)$ modes also reach 10$\%$, but cancel each other, turning in small net values. 
Stimulated by the above observation, we propose to search for the CPVs in the up-down asymmetries associated with the angular distributions of the $\Lambda_b\to pa_1(1260)$, $pK_1(1270), pK_1(1400)$ decay products. They are found to be of the order of $20\%$, granting an extraordinary opportunity to establish CPVs in the baryon sector. 
Our work unveils the distinct dynamics responsible for CPVs in bottom baryon and meson decays, and suggests to detect the former through partial-wave related CPV observables in angular distributions.

The remainder of this paper is arranged as follows. 
In Sec.~\ref{sec:framework} we present the effective Hamiltonian and classify the topologies of Feynman  diagrams relevant to the considered channels. We specify the inputs for the factorization formula in Eq.~(\ref{2}), i.e., DAs of various twists for the $\Lambda_b$ baryon,  the proton and the mesons. The numerical algorithm for handling integrable singularities present in the factorization formulas is elaborated.
The outcomes for the branching ratios and CPVs, and their phenomenological implications are discussed in Sec.~\ref{sec:results}, which is divided into three subsections focusing on decays into pseudoscalar, vector and axial-vector mesons.
Sec.~\ref{sec:summary} summarizes our findings and sketches future perspectives.
Appendix~\ref{app:aux-function} contains the auxiliary functions as the hard kernels under the Fourier transformation from the $k_T$ space to the $b$ space.
Appendix~\ref{app:decay-amplitudes} gathers the  $\Lambda_b\to ph$ decay amplitudes for the Feynman diagrams. Only one Feynman diagram is picked in each topology for simplicity.
Appendix~\ref{app:appendix-numerical-results} displays the numerical results for the  $\Lambda_b\to pV,pA$ amplitudes.

\section{Framework}\label{sec:framework}

\subsection{Effective Hamiltonian}

The weak effective Hamiltonian for the two-body hadronic $\Lambda_b$ baryon decays is given by
\begin{equation}
	\mathcal{H}_{W}=\frac{G_F}{\sqrt{2}}\left\{ V_{ub}V_{uq}^\ast\left[ C_1(\mu)O_1^u(\mu) + C_2(\mu)O_2^u(\mu) \right] - V_{tb}V_{tq}^\ast\left[ \sum_{i=3}^{10}C_i(\mu)O_i(\mu) \right]\right\},
\end{equation}
with the Fermi constants $G_F=1.166\times 10^{-5}$ GeV$^{-2}$ and the light quarks $q=d,s$. The local four-quark operators $O_i$ read
\begin{alignat}{2}
	O_1^u=&(\bar{u}_\alpha b_\beta)_{V-A} (\bar{q}_\beta u_\alpha)_{V-A},&\qquad O_2^u=&(\bar{u}_\alpha b_\alpha)_{V-A} (\bar{q}_\beta u_\beta)_{V-A},\nonumber\\
	O_3=&(\bar{q}_\alpha b_\alpha)_{V-A}\sum_{q^\prime}(\bar{q}_\beta^\prime q_\beta^\prime)_{V-A},& O_4=&(\bar{q}_\beta b_\alpha)_{V-A}\sum_{q^\prime}(\bar{q}_\alpha^\prime q_\beta^\prime)_{V-A},\nonumber\\
	O_5=&(\bar{q}_\alpha b_\alpha)_{V-A}\sum_{q^\prime}(\bar{q}_\beta^\prime q_\beta^\prime)_{V+A},& O_6=&(\bar{q}_\beta b_\alpha)_{V-A}\sum_{q^\prime}(\bar{q}_\alpha^\prime q_\beta^\prime)_{V+A},\nonumber\\
	O_7=&\frac{3}{2}(\bar{q}_\alpha b_\alpha)_{V-A}\sum_{q^\prime}e_{q^\prime}(\bar{q}_\beta^\prime q_\beta^\prime)_{V+A},& O_8=&\frac{3}{2}(\bar{q}_\beta b_\alpha)_{V-A}\sum_{q^\prime}e_{q^\prime}(\bar{q}_\alpha^\prime q_\beta^\prime)_{V+A},\nonumber\\
	O_9=&\frac{3}{2}(\bar{q}_\alpha b_\alpha)_{V-A}\sum_{q^\prime}e_{q^\prime}(\bar{q}_\beta^\prime q_\beta^\prime)_{V-A},& O_{10}=&\frac{3}{2}(\bar{q}_\beta b_\alpha)_{V-A}\sum_{q^\prime}e_{q^\prime}(\bar{q}_\alpha^\prime q_\beta^\prime)_{V-A},
\end{alignat}
where $\alpha$ and $\beta$ are the color indices, the active quarks cover $q^\prime=u,d,s,c,b$, and  $(\bar{q}_\alpha q_\beta^\prime)_{V\mp A}=\bar{q}_\alpha\gamma_\mu(1\mp\gamma_5)q_\beta^\prime$ denote the left- and right-handed currents. The expressions of the scale-dependent Wilson coefficients $C_i$ can be found in Ref.~\cite{Lu:2000em,Buchalla:1995vs}, whose values at the three specific renormalization scales $\mu=m_b/2, m_b$ and $3m_b/2$,  $m_b=4.8$ GeV being the $b$ quark mass, to leading-logarithm accuracy in QCD are listed in Table~\ref{tab:scale-dependence-Wilson-Coe}. The variation of the Wilson coefficients in the scale $\mu$ constitutes one of the sources of theoretical uncertainties in our analysis.
\begin{table}[htbp]
	\centering
	\footnotesize
	\renewcommand{\arraystretch}{1.4}
	\caption{Wilson coefficients at the  renormalization scales $\mu=m_b/2, m_b$ and $3m_b/2$ with $m_b=4.8$ GeV to leading-logarithm accuracy.}
	\label{tab:scale-dependence-Wilson-Coe}
	\begin{tabular*}{160mm}{c@{\extracolsep{\fill}}cccccccccc}
		\toprule[1pt]
		\toprule[0.7pt]
		$\mu$ & $C_1$ & $C_2$ & $C_3(10^{-2})$ & $C_4( 10^{-2})$ & $C_5( 10^{-2})$ & $C_6(10^{-2})$ & $C_7( 10^{-3})$ & $C_8( 10^{-3})$ & $C_9( 10^{-3})$ & $C_{10}( 10^{-3})$ \\
		\toprule[0.7pt]
		$m_b/2$ & $-0.39$ & $1.19$ & $1.94$ & $-3.90$ & $1.15$ & $-5.20$ & $1.05$ & $0.63$ & $-9.49$ & $3.13$ \\
		$m_b$    &  $-0.27$ & $1.12$ & $1.26$ & $-2.70$ & $0.85$ & $-3.26$ & $1.09$ & $0.40$ & $-8.95$ & $2.16$ \\
		$3m_b/2$ &  $-0.22$ & $1.09$ & $0.98$ & $-2.16$ & $0.71$ & $-2.49$ & $1.10$ & $0.30$ & $-8.71$ & $1.71$ \\
		\toprule[0.7pt]
		\toprule[1pt]
	\end{tabular*}
\end{table}

\subsection{Topological diagrams}

\begin{figure}[htbp]
	\centering
	\includegraphics[width=1.0\linewidth]{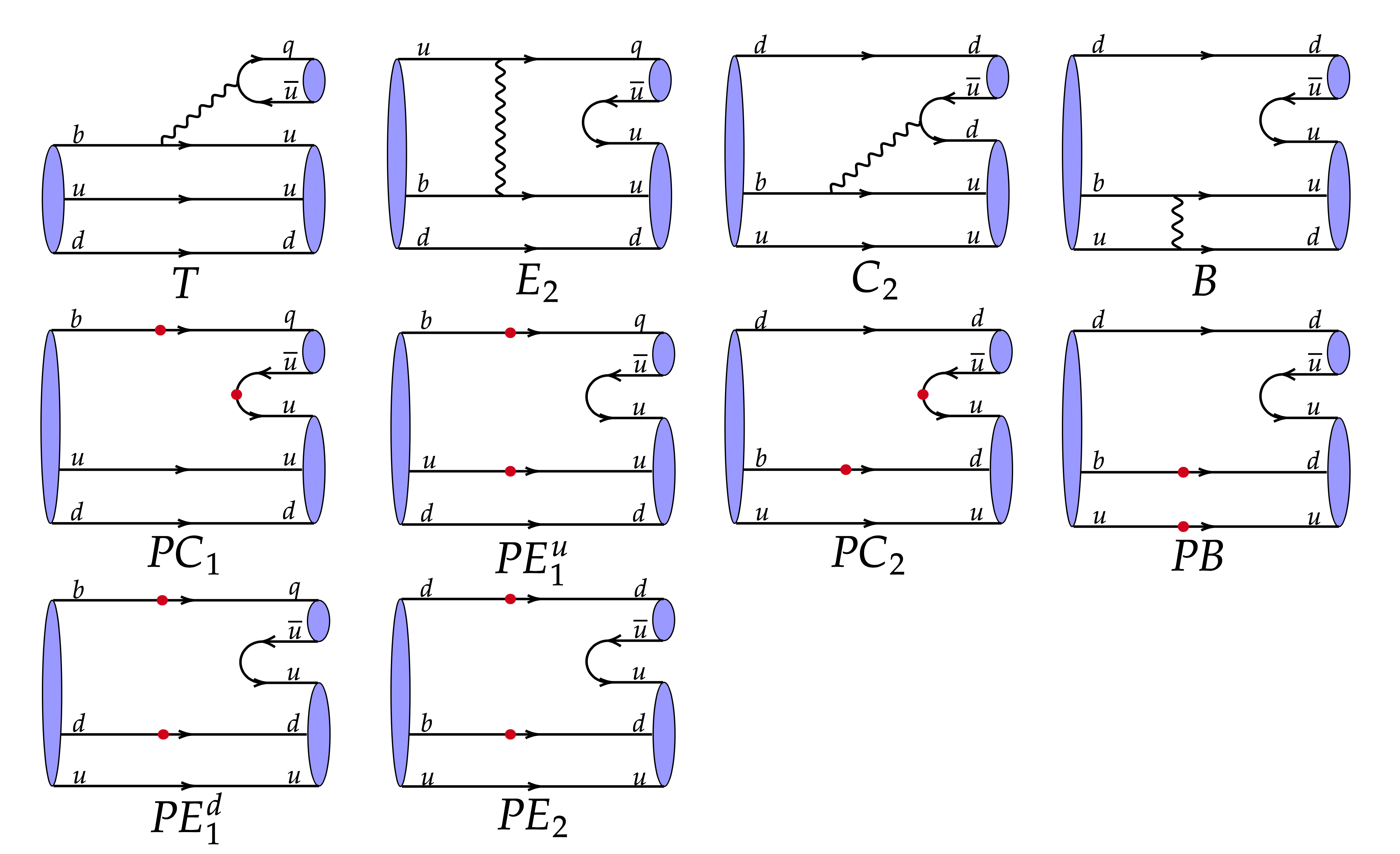}
	\caption{Topological diagrams contributing to the $\Lambda_b\to ph$ decays.}
	\label{fig:topo}
\end{figure}

The topological-diagram approach~\cite{Chau:1995gk} has been successfully applied to predictions for observables in charm hadron decays~\cite{Yu:2017zst,Jia:2024pyb}. Topological diagrams, which include full strong interaction, are categorized according to the weak vertex insertion and quark flows.
Those for the $\Lambda_b\to p h$ decays with the meson $h$ being  formed by $\bar{u}d$ or $\bar{u}s$ quarks are shown in Fig.~\ref{fig:topo}.
Four types of tree diagrams are identified~\cite{Han:2024kgz}:
\begin{itemize}
	\item $T$, the external-$W$ emission diagram, which can be separated into the factorizable piece $T^f$ and the non-factorizable piece $T^{nf}$;
	\item $C_2$, the internal-$W$ emission diagram, which is also called the color-commensurate $W$-emission diagram in some references;
	\item $E_2$, the $W$-exchange diagram, where the quark from the weak vertex goes into the meson $h$;
	\item $B$, the bow-tie $W$-exchange diagram, where the light spectator quark goes into the meson $h$.
\end{itemize}
The topologies $C_2$, $E_2$ and $B$, existing in baryon decays, have no counterparts in meson decays.

There are six types of penguin diagrams~\cite{Han:2024kgz}:
\begin{itemize}
	\item $PC_1$, $PC_2$, $PE_1^u$ and $PB$, whose quark flows are related to those of the tree diagrams $T$, $C_2$, $E_2$ and $B$, respectively, by the Feriz transformation. The $PC_1$ topology can be further classified into the factorizable piece $PC_1^f$ and the non-factorizable piece $PC_1^{nf}$;
	\item $PE_1^d$ and $PE_2$, which are related to each other through the Feriz transformation without the corresponding tree diagrams.
\end{itemize}

The $T^f$ diagram dominates the tree contributions to the $\Lambda_b\to p h$ decays as a consequence of the color transparency mechanism, and can be well estimated by the factorization hypothesis. All the other tree topologies give nonfactorizable contributions. Following the power counting rules in the SCET, we attain the hierarchy among the various tree topologies~\cite{Leibovich:2003tw}
\begin{equation}
	\frac{|C_2|}{|T|}\sim \frac{|E_2|}{|T|}\sim \frac{|B|}{|T|}\sim \mathcal{O} \left(\frac{\Lambda_{\rm QCD}}{m_b} \right)^2,
\end{equation}
with the QCD scale $\Lambda_{\rm QCD}$.

\subsection{Kinematics}
We adopt the light-cone coordinates for assigning the kinematic variables of the $\Lambda_b\to ph$ decays. 
The initial $\Lambda_b$ baryon momentum $p_{\Lambda_b}$, the proton momentum $p_p$ and the  meson momentum $q$ are chosen, in the rest frame of the $\Lambda_b$ baryon, as
\begin{equation}
	\begin{split}
		p_{\Lambda_b}=&\frac{m_{\Lambda_b}}{\sqrt{2}}(1,1,\bm{0}_T),\\
		p_p=&\frac{m_{\Lambda_b}}{\sqrt{2}}(\eta^+,\eta^-,\bm{0}_T),\\
		q=&p_{\Lambda_b}-p_p=\frac{m_{\Lambda_b}}{\sqrt{2}}(1-\eta^+,1-\eta^-,\bm{0}_T),
	\end{split}
\end{equation}
where the variables 
\begin{equation}
	\eta^{\pm}=\left[m_{\Lambda_b}^2-m_h^2+m_p^2\pm \sqrt{(m_{\Lambda_b}^2-m_h^2+m_p^2)^2-4m_{\Lambda_b}^2m_p^2}\right]/(2m_{\Lambda_b}^2),
\end{equation}
are determined by the momentum conservation with the $\Lambda_b$ baryon (proton, meson) mass $m_{\Lambda_b}$ ($m_p$, $m_h$). The longitudinal and transverse polarization vectors of a spin-1 final-state meson, 
\begin{align}
	\epsilon_L=\frac{m_h}{\sqrt{2}m_{\Lambda_b}}(\eta^+-1,1-\eta^-,\bm{0}_T),\quad\epsilon_T=(0,0,\bm{1}_T),
\end{align}
are fixed by the normalization and orthogonality conditions, which respect $\epsilon_L^2=-1$ and $\epsilon_L\cdot q=0$.

The valence quark momenta in the three involved hadrons are parameterized as
{\footnotesize
	\begin{alignat}{3}
		k_1=&(\frac{m_{\Lambda_b}}{\sqrt{2}},\frac{m_{\Lambda_b}}{\sqrt{2}}x_1,\bm{k}_{1T}),&\quad k_2=&(0,\frac{m_{\Lambda_b}}{\sqrt{2}}x_2,\bm{k}_{2T}),&\quad k_3=&(0,\frac{m_{\Lambda_b}}{\sqrt{2}}x_3,\bm{k}_{3T}),\nonumber\\
		k_1^\prime=&(\frac{m_{\Lambda_b}}{\sqrt{2}}\eta^+x_1^\prime,0,\bm{k}_{1T}^\prime),&k_2^\prime=&(\frac{m_{\Lambda_b}}{\sqrt{2}}\eta^+x_2^\prime,0,\bm{k}_{2T}^\prime),&k_3^\prime=&(\frac{m_{\Lambda_b}}{\sqrt{2}}\eta^+x_3^\prime,0,\bm{k}_{3T}^\prime),\nonumber\\
		q_1=&(0,\frac{m_{\Lambda_b}}{\sqrt{2}}y(1-\eta^-),\bm{q}_T),&q_2=&(0,\frac{m_{\Lambda_b}}{\sqrt{2}}(1-y)(1-\eta^-),-\bm{q}_T),
\end{alignat}}
where $k_{1}$ is the $b$ quark momentum, and the other momenta are designated in the Feynman diagrams in Appendix.~\ref{app:decay-amplitudes}. The momentum fractions $x_{1,2,3}^{(\prime)}$ and the transverse momenta $\bm{k}_{1T,2T,3T}^{(\prime)}$ obey
\begin{align}
	\sum_{i=1}^{3}x_i^{(\prime)}=1, \qquad \sum_{i=1}^{3}\bm{k}_{iT}^{(\prime)}=0,
\end{align}
respectively.
We stress that the quark momenta in the $\Lambda_b$ baryon should be parametrized as 
{\footnotesize\begin{equation}
		k_1=(\frac{m_{\Lambda_b}}{\sqrt{2}}(1-x_3),\frac{m_{\Lambda_b}}{\sqrt{2}}(1-x_2),\bm{k}_{1T}),\quad k_2=(0,\frac{m_{\Lambda_b}}{\sqrt{2}}x_2,\bm{k}_{2T}),\quad k_3=(\frac{m_{\Lambda_b}}{\sqrt{2}}x_3,0,\bm{k}_{3T}),
\end{equation}}
to maintain the virtuality of hard gluons as evaluating the topologies $C_2$, $B$, $PC_2$ and $PB$. This special kinematic assignment is allowed in view that light quarks in a $\Lambda_b$ baryon are soft, and their momenta can point to different directions in principle.

\subsection{Distribution amplitudes}\label{sec:LCDA}

\begin{figure}[htbp]
	\begin{minipage}{0.42\linewidth}
		\includegraphics[width=\linewidth]{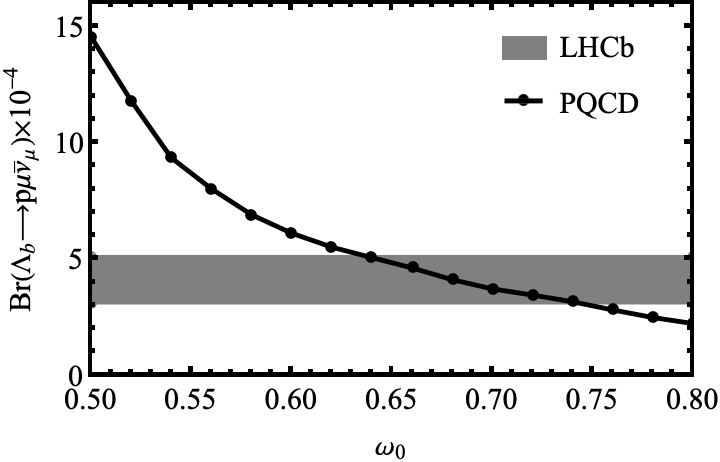}
	\end{minipage}
	\begin{minipage}{0.42\linewidth}
		\includegraphics[width=\linewidth]{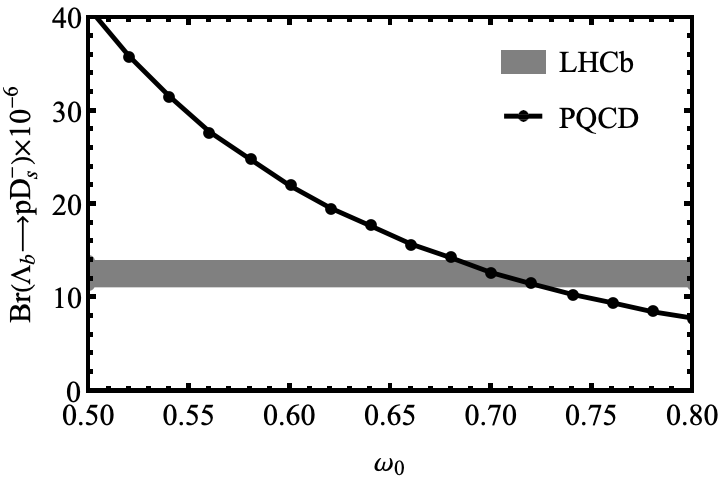}
	\end{minipage}
	\caption{ Dependencies of  $Br(\Lambda_b\to p\mu\bar{\nu}_\mu)$  (left) and $Br(\Lambda_b\to pD_s^-)$ (right) on $\omega_0$.}
	\label{fig:determine-omega0}
\end{figure}

The $\Lambda_b$ baryon DAs are defined by the matrix elements of nonlocal operators sandwiched between the vacuum and the $\Lambda_b$ baryon state. 
The general Lorentz structures of these matrix elements can be found in~\cite{Ball:2008fw,Bell:2013tfa,Wang:2015ndk,Ali:2012zza},
{\small
	\begin{equation}
		\begin{split}
			(Y_{\Lambda_b})_{\alpha\beta\gamma}(x_i,\mu)&\equiv \frac{1}{2\sqrt{2}N_c}\int \prod_{l=2}^{3} \frac{dz_l^-d\bm{z}_l}{(2\pi)^3}e^{ik_l\cdot z_l} \epsilon^{ijk}\langle 0|T[b_\alpha^i(0) u_\beta^j(z_2)d_\gamma^k(z_3)]|\Lambda_b\rangle\\
			&=\frac{1}{8\sqrt{2}N_c}\Big\{f_{\Lambda_b}^{(1)}(\mu)[M_1(x_2,x_3)\gamma_5C^T]_{\gamma\beta}+f_{\Lambda_b}^{(2)}(\mu)[M_2(x_2,x_3)\gamma_5C^T]_{\gamma\beta}\Big\}[u_{\Lambda_b}]_\alpha,
		\end{split}\label{eq:Lb-LCDA-define}
\end{equation}}
where $N_c$ is the number of colors, $i', j', k'$ are the color indices, $\alpha,\beta,\gamma$  are the Dirac indices, the normalization constants $f_{\Lambda_b}^{(1)} \approx f_{\Lambda_b}^{(2)}\equiv f_{\Lambda_b}= 0.022\pm 0.001$ GeV$^3$ are quoted from the leading-order sum-rule result~\cite{Groote:1997yr}, $C$ stands fors the charge conjugation matrix, and $u_{\Lambda_b}$ is the $\Lambda_b$ baryon spinor.
The DAs in Eq.~(\ref{eq:Lb-LCDA-define}) are decomposed into
\begin{equation}
	\begin{split}
		M_1(x_2,x_3)=&\frac{\slashed{\bar{n}}\slashed{n}}{4}\psi_3^{+-}(x_2,x_3)+\frac{\slashed{n}\slashed{\bar{n}}}{4}\psi_3^{-+}(x_2,x_3),\\
		M_2(x_2,x_3)=&\frac{\slashed{n}}{\sqrt{2}}\psi_2(x_2,x_3)+\frac{\slashed{\bar{n}}}{\sqrt{2}}\psi_4(x_2,x_3),
	\end{split}
\end{equation}
with the light-like vectors $n=(1,0,\bm{0}_T)$ and $\bar{n}=(0,1,\bm{0}_T)$.
The models with the exponential ansatz \cite{Bell:2013tfa} are employed 
\begin{equation}
	\begin{split}
		\psi_2(x_2,x_3)=& \frac{x_2x_3}{\omega_0^4}m^4_{\Lambda_b}e^{-(x_2+x_3)m_{\Lambda_b}/\omega_0},\\
		\psi_3^{+-}(x_2,x_3)=& \frac{2x_2}{\omega_0^3}m^3_{\Lambda_b}e^{-(x_2+x_3)m_{\Lambda_b}/\omega_0},\\
		\psi_3^{-+}(x_2,x_3)=& \frac{2x_3}{\omega_0^3}m^3_{\Lambda_b}e^{-(x_2+x_3)m_{\Lambda_b}/\omega_0},\\
		\psi_4(x_2,x_3)=& \frac{1}{\omega_0^2}m_{\Lambda_b}^2e^{-(x_2+x_3)m_{\Lambda_b}/\omega_0},
		\label{eq:Lb-LCDA-exp-model}
	\end{split}
\end{equation}
where the parameter $\omega_0$, with the typical value lower than $1$ GeV, measures the average energy of the two light quarks in a $\Lambda_b$ baryon. We extract $\omega_0$ from the LHCb data for the $\Lambda_b\to p\mu\bar{\nu}_\mu$ and $\Lambda_b\to pD_s^-$ branching ratios, $Br(\Lambda_b\to p\mu\bar{\nu}_\mu)=(4.1\pm 1.0)\times 10^{-4}$ and $Br(\Lambda_b\to pD_s^-)=(12.5\pm 1.3)\times 10^{-6}$. The former for a semileptonic mode depends on the $\Lambda_b\to p$ transition form factors, and the latter involves only the $T$ topology. The comparison between the data and our predictions for various $\omega_0$  sets the conservative range $\omega_0=0.7\pm 0.1$ GeV as exhibited in Fig.~\ref{fig:determine-omega0}, which is higher than the one chosen in light-cone sum rules~\cite{Wang:2015ndk,Huang:2022lfr}. The common extraction $\omega_0\approx 0.7$ GeV from the two disparate pieces of data supports the consistency of our framework.

The final-state proton DAs up to twist 6 have been defined in Ref.~\cite{Braun:2000kw} through the corresponding momentum-space projector 
{\small
	\begin{equation}\label{eq:proton DAs}
		\begin{split}
			(\overline{Y}_P)_{\alpha\beta\gamma}&(x_i^\prime,\mu)\equiv \frac{1}{2\sqrt{2}N_c}\int \prod_{l=2}^{3} \frac{dz_l^-d\bm{z}_l}{(2\pi)^3}e^{ik'_l\cdot z_l} \epsilon^{i^\prime j^\prime k^\prime}\langle p(p^\prime)|T[\bar{u}_\alpha^{i^\prime}(0) \bar{u}_\beta^{j^\prime}(z_2)\bar{d}_\gamma^{k^\prime}(z_3)]|0\rangle\\
			=&\frac{-1}{8\sqrt{2}N_c}\Big\{
			S_1 m_p C_{\beta\alpha} (\bar{N}^+ \gamma_5)_\gamma + S_2 m_p C_{\beta\alpha} (\bar{N}^- \gamma_5)_\gamma + P_1 m_p (C\gamma_5)_{\beta\alpha} \bar{N}^+_\gamma\\
			&+ P_2 m_p (C\gamma_5)_{\beta\alpha} \bar{N}^-_\gamma+ V_1 (C\slashed{P})_{\beta\alpha} (\bar{N}^+\gamma_5)_\gamma + V_2 (C\slashed{P})_{\beta\alpha} (\bar{N}^-\gamma_5)_\gamma\\
			&+ V_3 \frac{m_p}{2} (C\gamma_\perp)_{\beta\alpha}(\bar{N}^+\gamma_5\gamma^\perp)_\gamma+ V_4 \frac{m_p}{2} (C\gamma_\perp)_{\beta\alpha}(\bar{N}^-\gamma_5\gamma^\perp)_\gamma + V_5\frac{m_p^2}{2Pz} (C\slashed{z})_{\beta\alpha}(\bar{N}^+\gamma_5)_\gamma\\
			&+ V_6\frac{m_p^2}{2Pz} (C\slashed{z})_{\beta\alpha}(\bar{N}^-\gamma_5)_\gamma+ A_1 (C\gamma_5\slashed{P})_{\beta\alpha} (\bar{N}^+)_\gamma+ A_2 (C\gamma_5\slashed{P})_{\beta\alpha} (\bar{N}^-)_\gamma\\
			&+ A_3 \frac{m_p}{2} (C\gamma_5\gamma_\perp)_{\beta\alpha}(\bar{N}^+\gamma^\perp)_\gamma+ A_4 \frac{m_p}{2} (C\gamma_5\gamma_\perp)_{\beta\alpha}(\bar{N}^-\gamma^\perp)_\gamma+ A_5\frac{m_p^2}{2Pz} (C\gamma_5\slashed{z})_{\beta\alpha}(\bar{N}^+)_\gamma \\
			&+ A_6\frac{m_p^2}{2Pz} (C\gamma_5\slashed{z})_{\beta\alpha}(\bar{N}^-)_\gamma- T_1 (iC\sigma_{\perp P})_{\beta\alpha}(\bar{N}^+\gamma_5\gamma^\perp)_\gamma- T_2 (iC\sigma_{\perp P})_{\beta\alpha}(\bar{N}^-\gamma_5\gamma^\perp)_\gamma \\
			&- T_3 \frac{m_p}{Pz}(iC\sigma_{Pz})_{\beta\alpha}(\bar{N}^+\gamma_5)_\gamma- T_4 \frac{m_p}{Pz}(iC\sigma_{zP})_{\beta\alpha}(\bar{N}^-\gamma_5)_\gamma - T_5\frac{m_p^2}{2Pz}(iC\sigma_{\perp z})_{\beta\alpha}(\bar{N}^+\gamma_5\gamma^\perp)_\gamma \\
			&- T_6\frac{m_p^2}{2Pz}(iC\sigma_{\perp z})_{\beta\alpha}(\bar{N}^-\gamma_5\gamma^\perp)_\gamma+ T_7\frac{m_p}{2}(C\sigma_{\perp\perp^\prime})_{\beta\alpha}(\bar{N}^+\gamma_5\sigma^{\perp\perp^\prime})_\gamma \\
			&+ T_8\frac{m_p}{2}(C\sigma_{\perp\perp^\prime})_{\beta\alpha}(\bar{N}^-\gamma_5\sigma^{\perp\perp^\prime})_\gamma
			\Big\},
		\end{split}
\end{equation}}
with the color indices $i', j', k'$. The light-like vector $P$ is written, in terms of the proton momentum $p_p$ and the light-like vector $z$, as
\begin{equation}
	P^\mu=p_p^\mu-\frac{1}{2}z^\mu\frac{m_p^2}{Pz},
\end{equation}
with $Pz\sim p_p z\sim 1$, and the shorthand notation $\sigma_{Pz}=\sigma^{\nu\mu}P_\nu z_\mu$ has been used. Furthermore, $\bar{N}^+=\bar{N}\slashed{z}\slashed{P}/(2P z)$ and $\bar{N}^-=\bar{N}\slashed{P}\slashed{z}/(2P z)$ are introduced to represent the "large" and "small" components of the proton spinor $\bar{N}$, respectively. The symbol $\perp$ labels the projection perpendicular to $z$ or $P$, and the contraction $\gamma_\perp\gamma^\perp=\gamma^\mu g_{\mu\nu}^\perp\gamma^\nu$ is perceived with $g_{\mu\nu}^\perp=g_{\mu\nu}-(P_\mu z_\nu+z_\mu P_\nu)/Pz$.
The functions $V_i$, $A_i$, $T_i$, $S_i$ and $P_i$ are classified according to their  twists:

\begin{itemize}
	\item Twist-3 DAs\\
	\begin{align}
		V_1(x_i^\prime)=&120x_1^\prime x_2^\prime x_3^\prime[\phi_3^0+\phi_3^+(1-3x_3^\prime)],\nonumber\\
		A_1(x_i^\prime)=&120x_1^\prime x_2^\prime x_3^\prime(x_2^\prime-x_1^\prime)\phi_3^-,\nonumber\\
		T_1(x_i^\prime)=&120x_1^\prime x_2^\prime x_3^\prime[\phi_3^0+\frac{1}{2}(\phi_3^--\phi_3^+)(1-3x_3^\prime)];
	\end{align}
	\item Twist-4 DAs\\
	\begin{align}
		V_2(x_i^\prime)=&24x_1^\prime x_2^\prime[\phi_4^0+\phi_4^+(1-5x_3^\prime)],\nonumber\\
		V_3(x_i^\prime)=&12x_3^\prime[\psi_4^0(1-x_3^\prime)+\psi_4^-(x_1^{\prime 2}+x_2^{\prime 2}-x_3^\prime(1-x_3^\prime))+\psi_4^+(1-x_3^\prime-10x_1^\prime x_2^\prime)],\nonumber\\
		A_2(x_i^\prime)=&24x_1^\prime x_2^\prime(x_2^\prime-x_1^\prime)\phi_4^-,\nonumber\\
		A_3(x_i^\prime)=&12x_3^\prime(x_2^\prime-x_1^\prime)[(\psi_4^0+\psi_4^+)+\psi_4^-(1-2x_3^\prime)],\nonumber\\
		T_2(x_i^\prime)=&24x_1^\prime x_2^\prime[\xi_4^0+\xi_4^+(1-5x_3^\prime)],\nonumber\\
		T_3(x_i^\prime)=&6x_3^\prime[(\xi_4^0+\phi_4^0+\psi_4^0)(1-x_3^\prime)+(\xi_4^-+\phi_4^--\psi_4^-)(x_1^{\prime 2}+x_2^{\prime 2}-x_3^\prime(1-x_3^\prime))\nonumber\\
		&+(\xi_4^++\phi_4^++\psi_4^+)(1-x_3^\prime-10x_1^\prime x_2^\prime)],\nonumber\\
		T_7(x_i^\prime)=&6x_3^\prime[(-\xi_4^0+\phi_4^0+\psi_4^0)(1-x_3^\prime)+(-\xi_4^-+\phi_4^--\psi_4^-)(x_1^{\prime 2}+x_2^{\prime 2}-x_3^\prime(1-x_3^\prime))\nonumber\\
		&+(-\xi_4^++\phi_4^++\psi_4^+)(1-x_3^\prime-10x_1^\prime x_2^\prime)],\nonumber\\
		S_1(x_i^\prime)=&6x_3^\prime(x_2^\prime-x_1^\prime)[(\xi_4^0+\phi_4^0+\psi_4^0+\xi_4^++\phi_4^++\psi_4^+)+(\xi_4^-+\phi_4^--\psi_4^-)(1-2x_3^\prime)],\nonumber\\
		P_1(x_i^\prime)=&6x_3^\prime(x_2^\prime-x_1^\prime)[(\xi_4^0-\phi_4^0-\psi_4^0+\xi_4^+-\phi_4^+-\psi_4^+)+(\xi_4^--\phi_4^-+\psi_4^-)(1-2x_3^\prime)];
	\end{align}
	\item Twist-5 DAs\\
	\begin{align}
		V_4(x_i^\prime)=&3[\psi_5^0(1-x_3^\prime)+\psi_5^-(2x_1^\prime x_2^\prime-x_3^\prime(1-x_3^\prime))+\psi_5^+(1-x_3^\prime-2(x_1^{\prime 2}+x_2^{\prime 2}))],\nonumber\\
		V_5(x_i^\prime)=&6x_3^\prime[\phi_5^0+\phi_5^+(1-2x_3^\prime)],\nonumber\\
		A_4(x_i^\prime)=&3(x_2^\prime-x_1^\prime)[-\psi_5^0+\psi_5^-x_3^\prime+\psi_5^+(1-2x_3^\prime)],\nonumber\\
		A_5(x_i^\prime)=&6x_3^\prime(x_2^\prime-x_1^\prime)\phi_5^-,\nonumber\\
		T_4(x_i^\prime)=&\frac{3}{2}[(\xi_5^0+\psi_5^0+\phi_5^0)(1-x_3^\prime)+(\xi_5^-+\phi_5^--\psi_5^-)(2x_1^\prime x_2^\prime-x_3^\prime(1-x_3^\prime))\nonumber\\
		&+(\xi_5^++\phi_5^++\psi_5^+)(1-x_3^\prime-2(x_1^{\prime 2}+x_2^{\prime 2}))],\nonumber\\
		T_5(x_i^\prime)=&6x_3^\prime[\xi_5^0+\xi_5^+(1-2x_3^\prime)],\nonumber\\
		T_8(x_i^\prime)=&\frac{3}{2}[(\psi_5^0+\phi_5^0-\xi_5^0)(1-x_3^\prime)+(\phi_5^--\phi_5^--\xi_5^-)(2x_1^\prime x_2^\prime-x_3^\prime(1-x_3^\prime))\\
		&+(\phi_5^++\phi_5^+-\xi_5^+)(1-x_3^\prime-2(x_1^{\prime 2}+x_2^{\prime 2}))],\nonumber\\
		S_2(x_i^\prime)=&\frac{3}{2}(x_2^\prime-x_1^\prime)[-(\psi_5^0+\phi_5^0+\xi_5^0)+(\xi_5^-+\phi_5^--\psi_5^0)x_3^\prime+(\xi_5^++\phi_5^++\psi_5^0)(1-2x_3^\prime)],\nonumber\\
		P_2(x_i^\prime)=&\frac{3}{2}(x_2^\prime-x_1^\prime)[(\psi_5^0+\phi_5^0-\xi_5^0)+(\xi_5^--\phi_5^-+\psi_5^0)x_3^\prime+(\xi_5^+-\phi_5^+-\psi_5^0)(1-2x_3^\prime)];
	\end{align}
	\item Twist-6 DAs\\
	\begin{align}
		V_6(x_i^\prime)=&2[\phi_6^0+\phi_6^+(1-3x_3^\prime)],\nonumber\\
		A_6(x_i^\prime)=&2(x_2^\prime-x_1^\prime)\phi_6^-,\nonumber\\
		T_6(x_i^\prime)=&2[\phi_6^0+\frac{1}{2}(\phi_6^--\phi_6^+)(1-3x_3^\prime)].
	\end{align}
\end{itemize}
The values of $\phi_i^{\pm,0}, \psi_i^{\pm,0}$ and $\xi_i^{\pm,0}$ are connected to the eight independent parameters $f_N$, $\lambda_1$, $\lambda_2$, $V_1^d$, $A_1^u$, $f_1^d$, $f_2^d$ and $f_1^u$~\cite{Braun:2000kw}, 
\begin{align}
	\phi_3^0&=\phi_6^0=f_N,\qquad \phi_4^0=\phi_5^0=\frac{1}{2}(\lambda_1+f_N),\nonumber\\
	\xi_4^0&=\xi_5^0=\frac{1}{6}\lambda_2,\qquad \psi_4^0=\psi_5^0=\frac{1}{2}(f_N-\lambda_1),\nonumber\\
	\phi_4^-&=\frac{5}{4}\left(\lambda_1(1-2f_1^d-4f_1^u)+f_N(2A_1^u-1)\right),\nonumber\\
	\phi_4^+&=\frac{1}{4}\left(\lambda_1(3-10f_1^d)-f_N(10V_1^d-3)\right),\nonumber\\
	\psi_4^-&=-\frac{5}{4}\left(\lambda_1(2-7f_1^d+f_1^u)+f_N(A_1^u+3V_1^d-2)\right),\nonumber\\
	\psi_4^+&=-\frac{1}{4}\left(\lambda_1(-2+5f_1^d+5f_1^u)+f_N(2+5A_1^u-5V_1^d)\right),\nonumber\\
	\xi_4^-&=\frac{5}{16}\lambda_2(4-15f_2^d),\qquad \xi_4^+=\frac{1}{16}\lambda_2(4-15f_2^d),\nonumber\\
	\phi_5^-&=\frac{5}{3}\left(\lambda_1(f_1^d-f_1^u)+f_N(2A_1^u-1)\right),\nonumber\\
	\phi_5^+&=-\frac{5}{6}\left(\lambda_1(4f_1^d-1)+f_N(3+4V_1^d)\right),\nonumber\\
	\psi_5^-&=\frac{5}{3}\left(\lambda_1(f_1^d-f_1^u)+f_N(2-A_1^u-3V_1^d)\right),\nonumber\\
	\psi_5^+&=-\frac{5}{6}\left(\lambda_1(-1+2f_1^d+2f_1^u)+f_N(5+2A_1^u-2V_1^d)\right),\nonumber\\
	\xi_5^-&=-\frac{5}{4}\lambda_2f_2^d,\qquad \xi_5^+=\frac{5}{36}\lambda_2(2-9f_2^d),\nonumber\\
	\phi_6^-&=\frac{1}{2}\left(\lambda_1(1-4f_1^d-2f_1^u)+f_N(1+4A_1^u)\right),\nonumber\\
	\phi_6^+&=-\frac{1}{2}\left(\lambda_1(1-2f_1^d)+f_N(4V_1^d-1)\right).
	\label{eq:proton DA para}
\end{align}
These parameters have been derived in QCD sum rules (QCDSR)~\cite{Braun:2000kw,Braun:2006hz} and lattice QCD (LQCD)~\cite{RQCD:2019hps}, whose results are summarized in Table~\ref{tab:proton DA eight para}. They come with large uncertainties, reflecting our limited knowledge on baryon DAs. We will take the set of parameters from  QCDSR~\cite{Braun:2006hz} for the numerical investigations.

\begin{table}[htbp]
	\footnotesize
	\centering
	\caption{The eight independent parameters for Eq.~(\ref{eq:proton DA para}) at the scale $1$ GeV from QCDSR and LQCD.}
	\begin{tabular*}{160mm}{c@{\extracolsep{\fill}}cccc}
		\toprule[1pt]
		\toprule[0.7pt]
		&$f_N({\rm GeV}^2)$&$\lambda_1({\rm GeV}^2)$&$\lambda_2({\rm GeV}^2)$&$V_1^d$\\
		\toprule[0.7pt]
		QCDSR(2001)\cite{Braun:2000kw}&$(5.3\pm0.5)\times10^{-3}$&$-(2.7\pm0.9)\times10^{-2}$&$(5.1\pm1.9)\times10^{-2}$&$0.23\pm0.03$\\
		QCDSR(2006)\cite{Braun:2006hz}&$(5.0\pm0.5)\times10^{-3}$&$-(2.7\pm0.9)\times10^{-2}$&$(5.4\pm1.9)\times10^{-2}$&$0.23\pm0.03$\\
		LQCD(2019)\cite{RQCD:2019hps}&$(3.54\pm0.06)\times10^{-3}$&$-(4.49\pm0.42)\times10^{-2}$&$(9.34\pm0.48)\times10^{-2}$&$0.19\pm0.22$\\
		\toprule[0.7pt]
		&$A_1^u$&$f_1^d$&$f_2^d$&$f_1^u$\\
		\toprule[0.7pt]
		QCDSR(2001)\cite{Braun:2000kw}&$0.38\pm0.15$&$0.6\pm0.2$&$0.15\pm0.06$&$0.22\pm0.15$\\
		QCDSR(2006)\cite{Braun:2006hz}&$0.38\pm0.15$&$0.4\pm0.05$&$0.22\pm0.05$&$0.07\pm0.05$\\
		LQCD(2019)\cite{RQCD:2019hps}&$0.30\pm0.32$&...&...&...\\
		\toprule[0.7pt]
		\toprule[1pt]
	\end{tabular*}
	\label{tab:proton DA eight para}
\end{table}

The DAs for a pseudoscalar meson are expanded, up to twist 3, as~\cite{Chernyak:1983ej,Braun:1988qv,Braun:1989iv,Kurimoto:2001zj}
\begin{equation}
	\begin{split}
		\Phi_{\pi(K)}(q,y)=&\int d^4 z e^{-iyq\cdot z}\langle \pi(K)(q)|\bar{q}_\beta (z)u_\alpha(0)|0\rangle\\
		=&\frac{i}{\sqrt{2N_C}}\left[\gamma_5\slashed{q}\phi_{\pi(K)}^A(y)+m_0^{\pi(K)}\gamma_5\phi_{\pi(K)}^P(y)+m_0^{\pi(K)}\gamma_5(\slashed{v}\slashed{n}-1)\phi_{\pi(K)}^T(y)\right]_{\alpha\beta},
	\end{split}
\end{equation}
where $m_0^{\pi(K)}=m_{\pi(K)}^2/(m_u+m_{d(s)})$, $m_u$ ($m_d$, $m_s$) being the $u$ ($d$, $s$) quark mass, is the chiral scale associated with the pseudoscalar meson, and the momentum fraction $y$ is carried by the \textit{quark} in the meson. The DAs are parameterized as~\cite{Ball:2004ye,Ball:2006wn},
{\footnotesize
	\begin{equation}
		\begin{split}
			\phi^A_{\pi(K)}(y)=&\frac{f_{\pi(K)}}{2\sqrt{2N_c}}6y(1-y)\left[1+a_1^{\pi(K)}C_1^{3/2}(2y-1)+a_2^{\pi(K)}C_2^{3/2}(2y-1)+a_4^{\pi(K)}C_4^{3/2}(2y-1)\right],\\
			\phi^P_{\pi(K)}(y)=&\frac{f_{\pi(K)}}{2\sqrt{2N_c}}\bigg[1+\left(0.45-\frac{5}{2}\rho_{\pi(K)}^2\right)C_2^{1/2}(2y-1)\\
			&-3\left(-0.045+\frac{9}{20}\rho_{\pi(K)}^2(1+6a_2^{\pi(K)})\right)C_4^{1/2}(2y-1)\bigg],\\
			\phi^T_{\pi(K)}(y)=&\frac{f_{\pi(K)}}{2\sqrt{2N_c}}(1-2y)\left[1+6\left(0.0975-\frac{7}{20}\rho_{\pi(K)}^2-\frac{3}{5}\rho_{\pi(K)}^2a_2^{\pi(K)}\right)(1-10y+10y^2)\right],
		\end{split}
\end{equation}}
with the decay constant  $f_{\pi(K)}=0.13$ (0.16) GeV, the Gegenbauer polynomials $C_n^{\nu}$, $\nu=1/2,3/2$ and $n=1,2,4$, and the other parameters~\cite{Ball:2004ye,Ball:2006wn},
\begin{equation}
	\begin{split}
		a_1^\pi=0,&\qquad a_1^K=0.06,\\
		a_2^{\pi}=0.25\pm0.15,&\qquad a_2^{K}=0.25\pm0.15,\\
		a_4^\pi=-0.015,&\qquad a_4^K=0,\\
		m_\pi=0.14\;\text{GeV},&\qquad m_K=0.494\;\text{GeV},\\
		m_0^\pi=(1.4\pm0.1)\;\text{GeV},&\qquad m_0^K=(1.6\pm0.1)\;\text{GeV},\\
		\rho_\pi=\frac{m_\pi}{m_0^\pi},&\qquad \rho_K=\frac{m_K}{m_0^K}.
	\end{split}
\end{equation}

The vector meson DAs are defined by~\cite{Li:2009tx}
\begin{equation}
	\begin{split}
		\Phi_{V}^L(q,\epsilon_L^\ast,y)=&\int d^4z e^{-iyq\cdot z}\langle V(q,\epsilon_L^\ast)|\bar{q}_\beta(z)u_\alpha(0)|0\rangle\\
		=&\frac{-1}{\sqrt{2N_c}}\left[ m_V\slashed{\epsilon}_L^\ast\phi_V(y)+\slashed{\epsilon}_L^\ast\slashed{q}\phi_V^t(y)+m_V\phi_V^s(y) \right]_{\alpha\beta},
	\end{split}
\end{equation}
\begin{equation}
	\begin{split}
		\Phi_{V}^T(q,\epsilon_T^\ast,y)=&\int d^4z e^{-iyq\cdot z}\langle V(q,\epsilon_T^\ast)|\bar{q}_\beta(z)u_\alpha(0)|0\rangle\\
		=&\frac{-1}{\sqrt{2N_c}}\left[ m_V\slashed{\epsilon}_T^\ast\phi_V^v(y)+\slashed{\epsilon}_T^\ast\slashed{q}\phi_V^T(y)+m_Vi\epsilon_{\mu\nu\rho\sigma}\gamma_5\gamma^\mu\epsilon_T^{\ast\nu}v^\rho n^\sigma\phi_V^a(y) \right]_{\alpha\beta},
	\end{split}
\end{equation}
with the vector meson mass $m_V$.
The twist-2 DAs are also expanded int terms of the Gegenbauer polynomials,
\begin{align}
	\phi_V(y)=&\frac{3f_V}{\sqrt{2N_c}}y(1-y)\left[ 1+a_1^\parallel C_1^{3/2}(2y-1)+a_2^\parallel C_2^{3/2}(2y-1) \right],\nonumber\\
	\phi_V^T(y)=&\frac{3f_V^T}{\sqrt{2N_c}}y(1-y)\left[ 1+a_1^\perp C_1^{3/2}(2y-1)+a_2^\perp C_2^{3/2}(2y-1) \right].
\end{align}
The asymptotic forms are assumed for the twist-3 DAs, \cite{Ali:2007ff},
\begin{align}
	\phi_V^t(y)=&\frac{3f_V^T}{2\sqrt{2N_c}}(2y-1)^2,\nonumber\\
	\phi_V^s(y)=&\frac{3f_V^T}{2\sqrt{2N_c}}(1-2y),\nonumber\\
	\phi_V^v(y)=&\frac{3f_V}{8\sqrt{2N_c}}(1+(2y-1)^2),\nonumber\\
	\phi_V^a(y)=&\frac{3f_V}{4\sqrt{2N_c}}(1-2y).
\end{align}
The Gegenbauer moments $a_n$ in the above expansions have been obtained in QCDSR~\cite{Braun:2004vf,Ball:2005vx,Ball:2006nr,Ball:2007rt}, which, together with the decay constants, are collected in Table~\ref{tab:vector meson inputs}.

\begin{table}[htbp]
	\footnotesize
	\centering
	\caption{Inputs of the decay constants (in MeV) and the Gegenbauer moments for the vector meson DAs.}
	\begin{tabular*}{160mm}{c@{\extracolsep{\fill}}ccc}
		\toprule[1pt]
		\toprule[0.7pt]
		$f_\rho$&$f_\rho^T$&$f_{K^\ast}$&$f_{K^\ast}^T$\\
		$209\pm2$&$165\pm9$&$217\pm5$&$185\pm10$\\
		\toprule[0.7pt]
		$a_{1\rho}^\parallel$&$a_{1\rho}^\perp$&$a_{1K^\ast}^\parallel$&$a_{1K^\ast}^\perp$\\
		0&0&$0.03\pm0.02$&$0.04\pm0.03$\\
		\toprule[0.7pt]
		$a_{2\rho}^\parallel$&$a_{2\rho}^\perp$&$a_{2K^\ast}^\parallel$&$a_{2K^\ast}^\perp$\\
		$0.15\pm0.07$&$0.14\pm0.06$&$0.11\pm0.09$&$0.10\pm0.08$\\
		\toprule[0.7pt]
		\toprule[1pt]
	\end{tabular*}
	\label{tab:vector meson inputs}
\end{table}

The DAs for an axial-vector with the quantum number $J^{P}=1^+$ are introduced through \cite{Li:2009tx},
\begin{equation}
	\begin{split}
		\Phi_A^L(q,\epsilon_L^\ast,y)=&\int d^4z e^{-iyq\cdot z}\langle A(q,\epsilon_L^\ast)|\bar{q}_\beta(z)u_\alpha(0)|0\rangle\\
		=&\frac{-i}{\sqrt{2N_c}}\left[ m_A\gamma_5\slashed{\epsilon}_L^\ast \phi_A(y)+\slashed{\epsilon}_L^\ast\slashed{q}\gamma_5\phi_A^t(y)+m_A\gamma_5\phi_A^s(y) \right]_{\alpha\beta},
	\end{split}
\end{equation}
\begin{equation}
	\begin{split}
		\Phi_A^T(q,\epsilon_T^\ast,y)=&\int d^4z e^{-iyq\cdot z}\langle A(q,\epsilon_T^\ast)|\bar{q}_\beta(z)u_\alpha(0)|0\rangle\\
		&\frac{-i}{\sqrt{2N_c}}\left[ m_A\gamma_5\slashed{\epsilon}_T^\ast\phi_A^v(y)+\slashed{\epsilon}_T^\ast\slashed{q}\gamma_5\phi_A^T(y)+m_Ai\epsilon_{\mu\nu\rho\sigma}\gamma^\mu\epsilon_T^{\ast \nu}v^\rho n^\sigma\phi_A^a(y) \right]_{\alpha\beta},
	\end{split}
\end{equation}
with the axial-vector meson mass $m_A$.
The twist-2 and twist-3 DAs read
\begin{align}
	\phi_A(y)=&\frac{3f_A}{\sqrt{2N_c}}y(1-y)\left[ a_{0A}^\parallel +3a_{1A}^\parallel (2y-1)+\frac{3}{2}a_{2A}^\parallel (5(2y-1)^2-1) \right],\nonumber\\
	\phi_A^T(y)=&\frac{3f_A}{\sqrt{2N_c}}y(1-y)\left[ a_{0A}^\perp +3a_{1A}^\perp (2y-1)+\frac{3}{2}a_{2A}^\perp (5(2y-1)^2-1) \right],
\end{align}
and 
\begin{align}
	\phi_A^t(y)=&\frac{f_A}{2\sqrt{2N_c}}\left[ 3a_{0A}^\perp (2y-1)^2+\frac{3}{2}a_{1A}^\perp (2y-1)(3(2y-1)^2-1) \right],\nonumber\\
	\phi_A^s(y)=&\frac{3f_A}{2\sqrt{2N_c}}(a_{0A}^\perp-a_{1A}^\perp-2a_{0A}^\perp y+6a_{1A}^\perp y-6a_{1A}^\perp y^2),\nonumber\\
	\phi_A^v(y)=&\frac{f_A}{2\sqrt{2N_c}}\left[ \frac{3}{4}a_{0A}^\parallel(1+(2y-1)^2)+\frac{3}{2}a_{1A}^\parallel (2y-1)^3 \right],\nonumber\\
	\phi_A^a(y)=&\frac{3f_A}{4\sqrt{2N_c}}(a_{0A}^\parallel-a_{1A}^\parallel-2a_{0A}^\parallel y+6a_{1A}^\parallel y-6a_{1A}^\parallel y^2),
\end{align}
respectively. The associated decay constants and Gegenbauer moments are listed in Table~\ref{tab:axialvector meson inputs}.
\begin{table}[htbp]
	\footnotesize
	\centering
	\caption{Same as Table~\ref{tab:vector meson inputs} but for the axial-vector meson DAs.}
	\begin{tabular*}{160mm}{c@{\extracolsep{\fill}}ccc}
		\toprule[1pt]
		\toprule[0.7pt]
		$f_{a_1(1260)}$&$a_{0a_1(1260)}^\parallel$&$a_{1a_1(1260)}^\parallel$&$a_{2a_1(1260)}^\parallel$\\
		$238\pm10$&1&0&$-0.02\pm0.02$\\
		&$a_{0a_1(1260)}^\perp$&$a_{1a_1(1260)}^\perp$&$a_{2a_1(1260)}^\perp$\\
		&0&$-1.04\pm0.34$&0\\
		\toprule[0.7pt]
		$f_{K_{1A}}$&$a_{0K_{1A}}^\parallel$&$a_{1K_{1A}}^\parallel$&$a_{2K_{1A}}^\parallel$\\
		$250\pm13$&1&$0.00\pm0.26$&$-0.05\pm0.03$\\
		&$a_{0K_{1A}}^\perp$&$a_{1K_{1A}}^\perp$&$a_{2K_{1A}}^\perp$\\
		&$0.08\pm0.09$&$-1.08\pm0.48$&$0.02\pm0.20$\\
		\toprule[0.7pt]
		$f_{K_{1B}}$&$a_{0K_{1B}}^\parallel$&$a_{1K_{1B}}^\parallel$&$a_{2K_{1B}}^\parallel$\\
		$190\pm10$&$0.14\pm0.15$&$-1.95\pm0.45$&$0.02\pm0.10$\\
		&$a_{0K_{1B}}^\perp$&$a_{1K_{1B}}^\perp$&$a_{2K_{1B}}^\perp$\\
		&1&$0.17\pm0.22$&$-0.02\pm0.22$\\
		\toprule[0.7pt]
		\toprule[1pt]
	\end{tabular*}
	\label{tab:axialvector meson inputs}
\end{table}

The physical mass eigenstates $K_1(1270)$ and $K_1(1400)$ are considered to be the mixtures of the $K_{1A}(^3P_1)$ and $K_{1B}(^1P_1)$ states with the mixing angle $\theta_{K_1}$:
\begin{equation}
	\left(
	\begin{array}{c}
		|K_1(1270)\rangle\\
		|K_1(1400)\rangle
	\end{array}
	\right)
	=
	\left(
	\begin{array}{cc}
		\sin\theta_{K_1}&\cos\theta_{K_1}\\
		\cos\theta_{K_1}&-\sin\theta_{K_1}
	\end{array}
	\right)
	\left(
	\begin{array}{c}
		|K_{1A}\rangle\\
		|K_{1B}\rangle
	\end{array}
	\right),
\end{equation}
in which $\theta_{K_1}$ is chosen as the typical values $30^\circ$ and $60^\circ$ suggested in Refs.\cite{Cheng:2013cwa,Hatanaka:2008xj,Cheng:2011pb,Shi:2023kiy}.

For reader's reference, we present the explicit expressions for the relevant Gegenbauer polynomials $C_n^\nu(t)$, 
\begin{equation}
	\begin{split}
		C_1^{3/2}(t)=&3t,\\
		C_2^{1/2}(t)=&\frac{1}{2}(3t^2-1),\\
		C_2^{3/2}(t)=&\frac{3}{2}(5t^2-1),\\
		C_4^{1/2}(t)=&\frac{1}{8}(3-30t^2+35t^4),\\
		C_4^{3/2}(t)=&\frac{15}{8}(1-14t^2+21t^4),
	\end{split}
\end{equation}
and the definitions of the meson decay constants, 
\begin{equation}
	\begin{split}
		\langle P|(\bar{q}q^\prime)_{V\mp A}|0\rangle=&\pm if_Pp_\mu, \qquad\quad \langle P|(\bar{q}q^\prime)_{S\mp P}|0\rangle=\pm if_Pm_{0P},\\
		\langle V|(\bar{q}q^\prime)_{V\mp A}|0\rangle=&f_Vm_V\epsilon_\mu^\ast,  \qquad\quad \langle V|(\bar{q}q^\prime)_{S\mp P}|0\rangle=0,\\
		\langle A|(\bar{q}q^\prime)_{V\mp A}|0\rangle=&\mp if_Am_A\epsilon_\mu^\ast,  \qquad \langle A|(\bar{q}q^\prime)_{S\mp P}|0\rangle=0.
		\label{eq:matrix-vaccum-to-meson}
	\end{split}
\end{equation}

\subsection{Numerical integration}

The PQCD factorization formulas and the auxiliary functions from Fourier transformation 
in Appendices~\ref{app:aux-function} and~\ref{app:decay-amplitudes} are extremely complicated, containing $11\sim 13$ dimensional integrations. 
Computing high dimensional integrals effectively and making precise predictions for heavy baryon decays  thus remain a numerical challenge.
Various algorithms have been proposed, among which the \texttt{VEGAS+} is one of the most widely used in particle physics.
The classic \texttt{VEGAS} performs a multidimensional Monte Carlo integration based on adaptive importance sampling, originally developed by G.P. Lepage~\cite{Lepage:1977sw}. It then evolved to the \texttt{VEGAS+} by incorporating a second strategy known as adaptive stratified sampling~\cite{Lepage:2020tgj}. It is believed that the \texttt{VEGAS+} is much more efficient for handling integrands with multiple peaks compared to the \texttt{VEGAS}. See Ref.~\cite{Lepage:2020tgj} and references therein for more details of the \texttt{VEGAS+} algorithm.

\begin{figure}[htbp]
	\begin{minipage}{0.45\linewidth}
		\includegraphics[width=\linewidth]{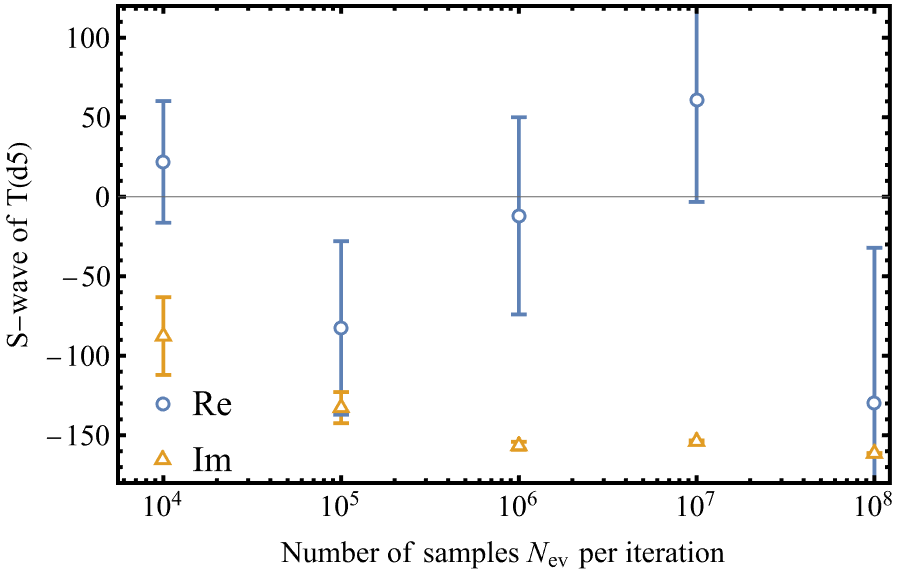}
	\end{minipage}
	\begin{minipage}{0.45\linewidth}
		\includegraphics[width=\linewidth]{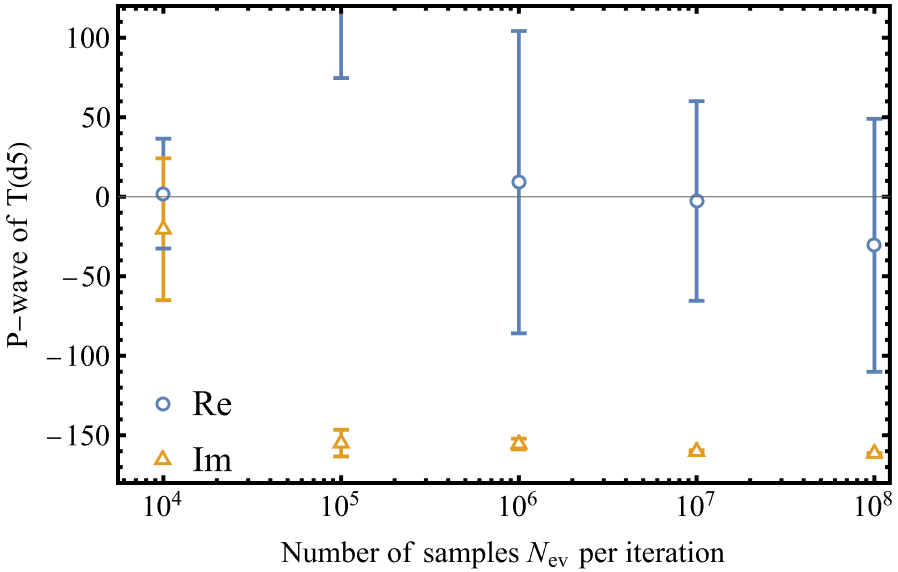}
	\end{minipage}
	\caption{(left) Numerical results of the $T(d5)$ contribution to
		the $S$-wave from the  \texttt{VEGAS+} after $15$ iterations with increasing numbers of samples; (right) same as the left panel but for the $P$-wave.}
	\label{fig:before-cut}
\end{figure}

When applying the \texttt{VEGAS+} to the integrations involved in this work, we encounter a subtlety, which can be illustrated by taking the diagram $T(d5)$ in Fig.~\ref{fig:feynmanT} as an example. This diagram gives a non-factorizable amplitude with the auxiliary function $F_4$,  
The numerical outcomes from the \texttt{VEGAS+} for the $T(d5)$ contributions after $15$ iterations are plotted in Fig.~\ref{fig:before-cut}, where the number of samples increases from $10^4$ to $10^8$. It is observed that the statistical errors in the imaginary parts decrease as $1/\sqrt{N_{ev}}$, consistent with the characteristics of a Monte Carlo algorithm. However, the real parts do not converge, meaning that the \texttt{VEGAS+} is ineffective for the real parts.

\begin{figure}[htbp]
	\begin{minipage}{0.45\linewidth}
		\includegraphics[width=\linewidth]{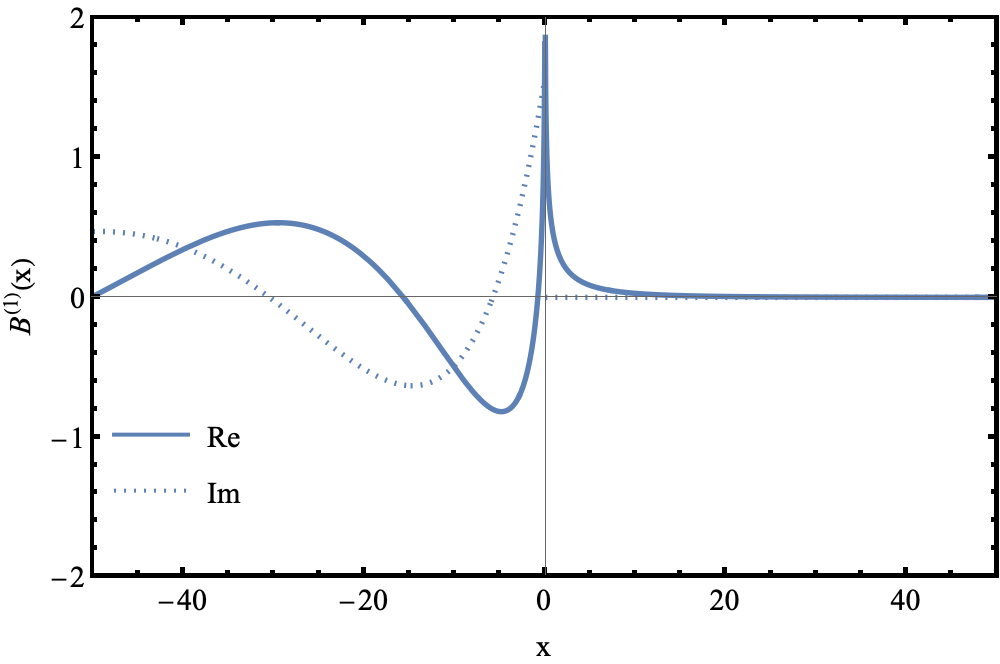}
	\end{minipage}
	\begin{minipage}{0.45\linewidth}
		\includegraphics[width=\linewidth]{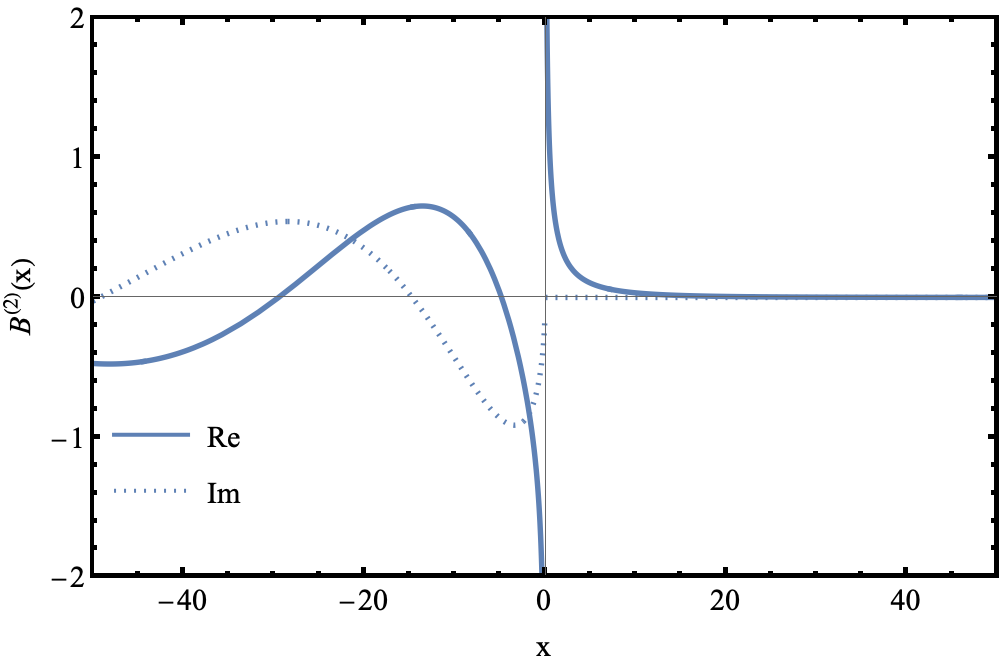}
	\end{minipage}
	\caption{The Bessel functions $B^{(1)}$ (left) and  $B^{(2)}$ (right).}
	\label{fig:bessel-function}
\end{figure}

We find that the real parts of the diagrams with the auxiliary function $F_1$ can be accurately acquired, but those depending on $F_{2,3,4}$ cannot. To elaborate the issue, we introduce the functions $B^{(1)}(x) \equiv K_0(\sqrt{x})\Theta(x) + \frac{i\pi}{2} H_0^{(1)}(\sqrt{-x})\Theta(-x)$ and $B^{(2)}(x) \equiv K_1(\sqrt{x})\Theta(x) - \frac{i\pi}{2}H_1^{(1)}(\sqrt{-x})\Theta(-x)$, where the step function $\Theta(x)$ equals $1$ when $x>0$ and $0$ otherwise. The former (latter) corresponds to the Bessel function appearing in $F_1$  ($F_{2,3,4}$). As described in Fig.~\ref{fig:bessel-function}, both $B^{(1)}(x)$ and $B^{(2)}(x)$ are singular at $x=0$ with the limits,
\begin{equation}
	\begin{split}
		\displaystyle\lim_{x\to 0^-}B^{(1)}(x)&=-\displaystyle\lim_{x\to 0^-}\text{log}(\sqrt{-x})+\frac{i\pi}{2},\\
		\displaystyle\lim_{x\to 0^+}B^{(1)}(x)&=-\displaystyle\lim_{x\to 0^+}\text{log}(\sqrt{x}),\\
		\displaystyle\lim_{x\to 0^-}B^{(2)}(x)&=\displaystyle\lim_{x\to 0^-}\frac{-1}{\sqrt{-x}},\\
		\displaystyle\lim_{x\to 0^+}B^{(2)}(x)&=\displaystyle\lim_{x\to 0^+}\frac{1}{\sqrt{x}}.
	\end{split}\label{37}
\end{equation}
The real part of $B^{(1)}(x)$ behaves like $\text{log}(\sqrt{|x|})$ on both sides of $x=0$, which can be well managed by the \texttt{VEGAS+}. However, the real part of $B^{(2)}(x)$ approaches $\pm \infty$ from the two sides of $x=0$, whose incomplete cancellation causes the numerical instability.

\begin{figure}[htbp]
	\begin{minipage}{0.45\linewidth}
		\includegraphics[width=\linewidth]{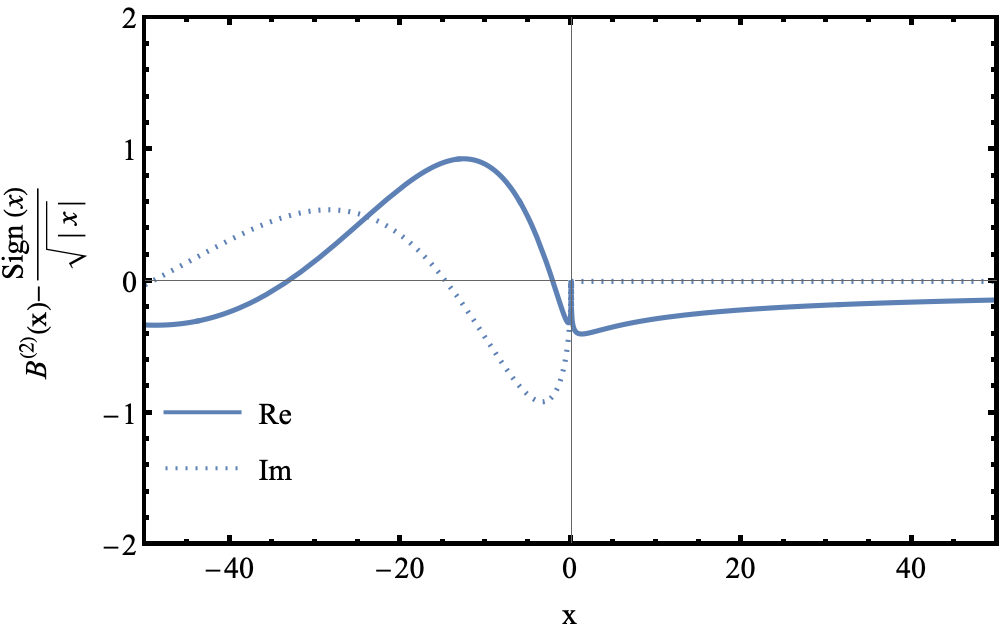}
	\end{minipage}
	\begin{minipage}{0.45\linewidth}
		\includegraphics[width=\linewidth]{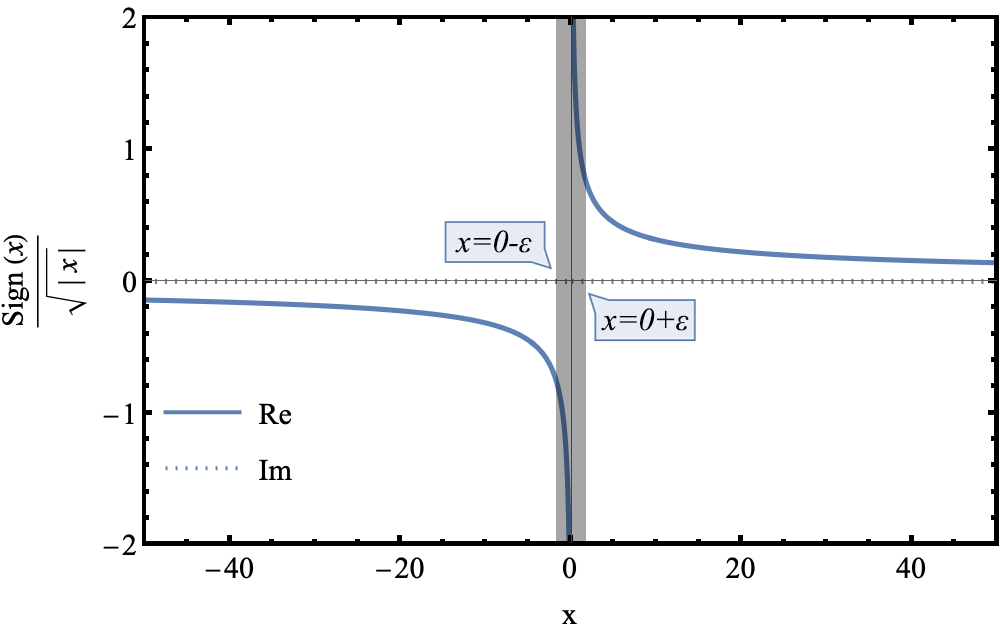}
	\end{minipage}
	\caption{The function $B^{(2)}(x)-\text{Sign}(x)/\sqrt{|x|}$ (left) ; the function $\text{Sign}(x)/\sqrt{|x|}$  (right).}
	\label{fig:B2-separate}
\end{figure}

\begin{figure}[htbp]
	\begin{minipage}{0.45\linewidth}
		\includegraphics[width=\linewidth]{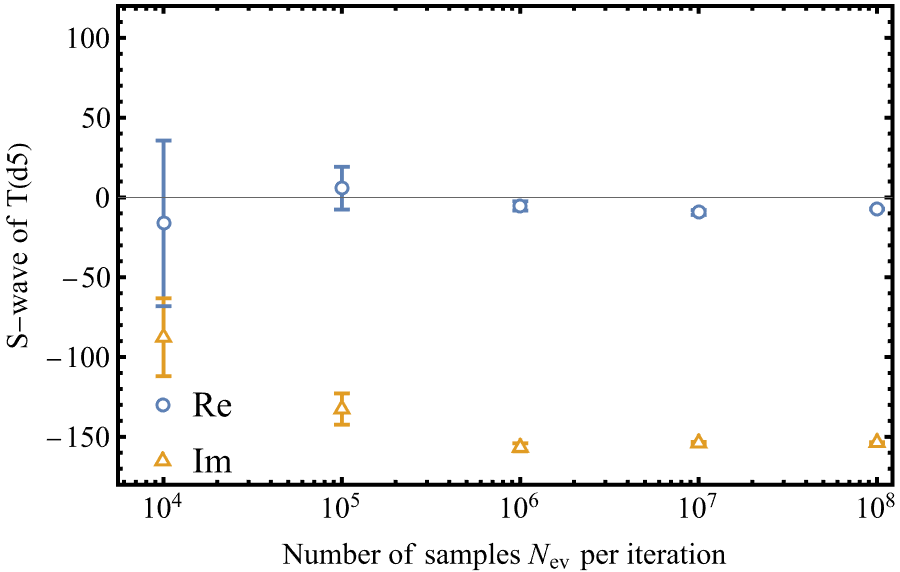}
	\end{minipage}
	\begin{minipage}{0.45\linewidth}
		\includegraphics[width=\linewidth]{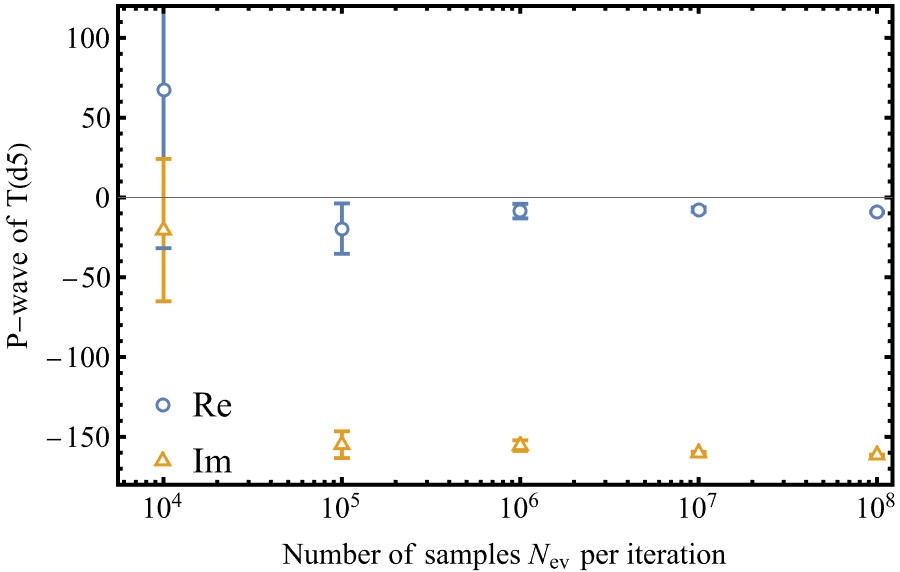}
	\end{minipage}
	\caption{Same as Fig.~\ref{fig:before-cut} but the integrals are treated under our algorithm.}
	\label{fig:after-cut}
\end{figure}

The aforementioned  divergences of $B^{(2)}$ ought to cancel, since they are from an integrable singularity as shown in Eq.~(\ref{37});
the integrand takes the same values in magnitude but with opposite signs at $x=\epsilon$ and $x=-\epsilon$, where $\epsilon$ is a tiny parameter. We then decompose the function $B^{(2)}(x)$ into two terms, $B^{(2)}(x)-\text{Sign}(x)/\sqrt{|x|}$ and $\text{Sign}(x)/\sqrt{|x|}$, as depicted in Fig.~\ref{fig:B2-separate}. 
The former is finite everywhere, whose integration is convergent and can be evaluated reliably by the \texttt{VEGAS+}.
The latter is divergent at $x=0$ but antisymmetric about the origin.
Therefore, the contributions to its integral from the intervals $(-\epsilon,0)$ and $(0,\epsilon)$ in $x$ are expected to cancel exactly and negligible.
This operation replaces the integrations involving $F_{2,3,4}$ by the sum of two pieces, one with $B^{(2)}(x)-\text{Sign}(x)/\sqrt{|x|}$ in the entire range of $x$ and the other with $\text{Sign}(x)/\sqrt{|x|}$ in the region outside $(-\epsilon, \epsilon)$.
It has been checked that the cut parameter $\epsilon$ within the range $0.0005\sim 0.005$ leads to accurate outcomes with controllable uncertainties. Here we set $\epsilon=0.001$, for which the errors in the real parts of the $T(d5)$ contributions indeed decrease as $1/\sqrt{N_{ev}}$, as indicated in Fig.~\ref{fig:after-cut}. It confirms the effectiveness of our algorithm.

\subsection{Applicability of PQCD}
The typical PQCD factorization formula for a $\Lambda_b\to ph$ decay amplitude has been quoted in Eq.~(\ref{2}). 
The Wilson coefficients in the weak effective Hamiltonian evolve from the electroweak scale $m_W$ down to the hard scale $t$ that represents the maximal virtuality of internal particles in relevant Feynman diagrams. The Wilson coefficients are combined with a second-stage renormalization-group evolution from $t$ to $1/b$, below  which infrared logarithms are absorbed into the hadron DAs \cite{Keum:2000ms}.
The hard kernels are then computed to fixed orders in the strong coupling $\alpha_s$ at the scale $t$. Hence, the applicability of the PQCD approach lies in the sufficiently large hard scale above $O(1)$ GeV. 

We have successfully reproduced the $\Lambda_b\to p$ transition form factors with physically reasonable results in our previous study \cite{Han:2022srw}. The detailed examination of the portions of the form factor $f_1(q^2=0)$ in different ranges of $\alpha_s/\pi$ (see Fig.~5 in \cite{Han:2022srw}) confirms that all contributions arise from the region with  $\alpha_s/\pi < 0.18$, implying the average hard scale around $t\sim \sqrt{\Lambda_{\rm QCD}m_{\Lambda_b}}$. This ensures the dominance of perturbative dynamics, which allows reliable evaluations of $\Lambda_b$ baryon decay amplitudes. Although next-to-leading-order (NLO) PQCD calculations have not been fully implemented, it is believed that NLO corrections are under control. We will estimate higher-order effects by varying the hard scales in the next section.
The mild impacts on our predictions for $CP$ asymmetries will further justify the applicability of the PQCD approach to $\Lambda_b$ baryon decays.

Hadron DAs are essential inputs in the PQCD framework, whose universality enables consistent applications to various modes. The precision of PQCD results is of course subject to the knowledge of these DAs. Current understanding of baryon internal structures and their associated DAs remain incomplete, which consequently introduce uncertainties in our analyses. This source of uncertainties will be also investigated in the next section by tuning the parameters involved in the hadron DAs. Unambiguous determinations of baryon DAs are necessary for improving the theoretical precision. 

The development of the PQCD formalism for heavy baryon decays is still in the preliminary stage. Several aspects need to be explored, such as the extension to higher twists (higher powers in $1/m_{\Lambda_b}$) and resummation effects of the threshold logarithms, which are important in the endpoint regions of parton momentum fractions. The assessment of the uncertainties from the above sources goes beyond the scope of the present work.

\section{Numerical Results}\label{sec:results}

\subsection{Two-body \texorpdfstring{$\Lambda_b \to ph$}{Lb->pP} decays}

We start with the invariant amplitude of the $\Lambda_b(\frac{1}{2}^+)\to p(\frac{1}{2}^+)h(0^-)$ decay~\cite{Cheng:1996cs}, $h=\pi^-$ or $K^-$, 
\begin{equation}
	\mathcal{A} = i\bar{u}_p(p_p)(S+P\gamma_5)u_{\Lambda_b}(p_{\Lambda_b}),
\end{equation}
where $u_p$ is the proton spinor, and $S$ ($P$) denotes the parity-violating $S$-wave (parity-conserving $P$-wave) amplitude.
CPVs are generated by the interference between the contributions of the {\it tree} operators $\mathcal{T}$ and the {\it penguin} operators $\mathcal{P}$. 
The $S$- and $P$-wave amplitudes are split into~\cite{Han:2024kgz}
\begin{equation}
	\begin{split}
		S &= \lambda_\mathcal{T} |S_\mathcal{T}| e^{i\delta_\mathcal{T}^S} + \lambda_\mathcal{P} |S_\mathcal{P}| e^{i\delta_\mathcal{P}^S}, \\
		P &=  \lambda_\mathcal{T} |P_\mathcal{T}| e^{i\delta_\mathcal{T}^P} + \lambda_\mathcal{P} |P_\mathcal{P}| e^{i\delta_\mathcal{P}^P}, 
	\end{split}\label{39}
\end{equation}
where $\delta$'s represent the strong phases, and $\lambda_\mathcal{T}=V_{ub}V_{ud(s)}^*$ and $\lambda_\mathcal{P}=V_{tb}V_{td(s)}^*$ are the products of the CKM matrix elements.
It is reminded that the $S$-wave amplitude flips sign under the $C$ transformation, but the $P$-wave one does not~\cite{Wang:2024qff},
\begin{equation}
	\begin{split}
		S\to \bar{S} &= -\lambda_\mathcal{T}^\ast |S_\mathcal{T}| e^{i\delta_\mathcal{T}^S} - \lambda_\mathcal{P}^\ast |S_\mathcal{P}| e^{i\delta_\mathcal{P}^S}, \\
		P\to\bar{P} &=  \lambda_\mathcal{T}^\ast |P_\mathcal{T}| e^{i\delta_\mathcal{T}^P} + \lambda_\mathcal{P}^\ast |P_\mathcal{P}| e^{i\delta_\mathcal{P}^P}. 
	\end{split}
\end{equation}

It is equivalent to formulate the decay $\Lambda_b\to ph$ in terms of the helicity amplitudes,
defined as
\begin{equation}
	H_{\lambda_p,\lambda_h}=\langle p(\lambda_p)h(\lambda_h) | H_W | \Lambda_b(\lambda_{\Lambda_b}) \rangle.
\end{equation}
The $\Lambda_b$ spin component obeys $\lambda_{\Lambda_b}=\lambda_p-\lambda_h$ with the proton helicity $\lambda_p$ and the meson spin $\lambda_h$. The helicity amplitudes $H_{\lambda_p,\lambda_h}$ are related to the partial-wave amplitudes via
\begin{align}
	H_{1/2,0}&=-iM_+S+iM_-P,\nonumber\\
	H_{-1/2,0}&=-iM_+S-iM_-P,\label{42}
\end{align}
with the coefficients $M_\pm^2\equiv (m_{\Lambda_b}\pm m_p)^2-m_h^2=2m_{\Lambda_b}(E_p\pm m_p)$, $E_p$ being the proton energy.

We calculate the contributions from all diagrams to the invariant amplitudes $S_{\mathcal{T},\mathcal{P}}$ and $P_{\mathcal{T},\mathcal{P}}$ in Eq.~(\ref{39}), including their strong phases, for the $\Lambda_b\to p\pi^-$, $pK^-$ decays in the PQCD approach. The results can be transformed into the helicity amplitudes in Eq.~(\ref{42}) straightforwardly.
It is worth mentioning that the nonfactorizable diagrams from the topologies $PC_2$ and $PE_1^u$ are evaluated for the first time. 
Our predictions for the $\Lambda_b\to p\pi^-$ ($pK^-$) amplitudes are presented in Tables~\ref{tab:pi} and \ref{tab:pi-helicity} (\ref{tab:K} and \ref{tab:K-helicity}), where only the central values are given for clarity.

The contributions to the decay amplitudes have been categorized according to topologies in the above tables. The external-$W$ emission topology $T$ is further divided into the factorizable part $T^f$ and the non-factorizable part $T^{nf}$, and the same division applies to the $PC_1$ topology. 
The topological amplitudes, excluding the CKM matrix elements, respect the hierarchy relations,  
\begin{equation}
	\begin{split}
		|T^f| \gg |E_2| \gtrsim& |PC_1^f| \gtrsim |T^{nf}| > |C_2| > |PC_2| > |B|\\
		& \gtrsim |PE_1^u| > |PE_1^d| \sim |PB| \sim |PE_2| \sim |PC_1^{nf}|\; \text{ ($S$ wave)},\\
		|T^f| \gg |T^{nf}|& \gg |E_2| > |C_2| > |B| > |PC_2| \gtrsim |PC_1^{nf}| \\
		&\gtrsim |PE_1^u| > |PC_1^{f}| > |PE_1^d| \sim |P^B| \sim |PE_2|\; \text{ ($P$ wave)},
	\end{split}
	\label{eq:hierarchy-pi}
\end{equation}	
with the explicit relative ratios,
\begin{equation}
	\begin{split}
		&\frac{|C_2|}{|T|} = 0.042,\quad \frac{|E_2|}{|T|} = 0.095, \quad  \frac{|B|}{|T|} = 0.014\; \text{ ($S$ wave)},\\
		&\frac{|C_2|}{|T|} = 0.041,\quad \frac{|E_2|}{|T|} = 0.072, \quad  \frac{|B|}{|T|} = 0.024 \text{ ($P$ wave)},\\
	\end{split}
\end{equation}
for the  $\Lambda_b\to p\pi^-$ mode. 
Likewise, we have 
\begin{equation}
	\begin{split}
		& |T^f| \gg |E_2| \gtrsim |PC_1^{f}| \gtrsim |T^{nf}| > |PE_1^u| > |PE_1^d| > |PC_1^{nf}| \; \text{ ($S$ wave)},\\
		& |T^f| \gg |T^{nf}| \gg |E_2| > |PC_1^{nf}| \gtrsim |PE_1^u| > |PC_1^{f}| > |PE_1^d|\;  \text{ ($P$ wave)},
	\end{split}
	\label{eq:hierarchy-K}
\end{equation}	
\begin{equation}
	\frac{|E_2|}{|T|} = 0.104\;\text{ ($S$ wave)},\quad \frac{|E_2|}{|T|} = 0.076\;\text{ ($P$ wave)}.
\end{equation}
for the  $\Lambda_b\to pK^-$ mode. 
It is seen that Eqs.~(\ref{eq:hierarchy-pi}) and (\ref{eq:hierarchy-K}) are consistent with those from the SCET~\cite{Leibovich:2003tw}.

The $\Lambda_b\to p\pi^-,pK^-$ decays share the same topologies $T$ and $E_2$, allowing a quantitative assessment on the SU(3)$_F$ flavor symmetry breaking effects. The ratios
\begin{equation}
	\begin{split}
		&\frac{|T^f(pK)|}{|T^f(p\pi)|}=1.22, \quad \frac{|T^{nf}(pK)|}{|T^{nf}(p\pi)|}=1.03, \quad \frac{|E_2(pK)|}{|E_2(p\pi)|}=1.33  \text{ (S wave)},\\
		&\frac{|T^f(pK)|}{|T^f(p\pi)|}=1.23, \quad \frac{|T^{nf}(pK)|}{|T^{nf}(p\pi)|}=1.28, \quad \frac{|E_2(pK)|}{|E_2(p\pi)|}=1.29  \text{ (P wave)}.
	\end{split}
\end{equation}
imply that the SU(3)$_F$ breaking effects reach  roughly $30\%$, similar to the case of $B$ meson decays~\cite{Cheng:2011qh}. 
The breaking effects primarily originate from the meson decay constants in the factorizable $T_f$ topology because of $|T^f(pK)|/|T^f(p\pi)| \approx f_K/f_\pi$, and from the meson DAs in the the other topologies.

\begin{table}[htbp]
	\centering
	\renewcommand{\arraystretch}{1.2}
	\caption{Invariant amplitudes of the $\Lambda_b\to p\pi^-$ decay classified by topologies without the CKM matrix elements.}
	\label{tab:pi}
	\begin{tabular*}{160mm}{c@{\extracolsep{\fill}}cccc|cccc}
		\toprule[1pt]
		\toprule[0.7pt]
		$\Lambda_b\to p\pi^-$ & $|S|$ & $\delta^S(^\circ)$ & Real($S$) & Imag($S$) & $|P|$ & $\delta^P(^\circ)$ & Real($P$) & Imag($P$)\\
		\toprule[0.7pt]
		$T^f$ &   707.17 &     0.00 &   707.17 &     0.00 &  1004.44 &     0.00 &  1004.44 &     0.00\\
		$T^{nf}$ &    51.72 &   -96.64 &    -5.98 &   -51.38 &   267.72 &   -97.92 &   -36.90 &  -265.17\\
		$T^f+T^{nf}$ &   703.07 &    -4.19 &   701.19 &   -51.38 &  1003.22 &   -15.33 &   967.54 &  -265.17\\
		$C_2$ &    29.37 &   154.96 &   -26.61 &    12.43 &    41.51 &   179.80 &   -41.51 &     0.14\\
		$E_2$ &    66.94 &  -145.26 &   -55.01 &   -38.14 &    72.58 &   119.94 &   -36.23 &    62.89\\
		$B$ &    10.40 &   112.64 &    -4.00 &     9.60 &    23.65 &  -122.56 &   -12.73 &   -19.93\\
		\toprule[0.7pt]
		Tree  &   619.26 &    -6.26 &   615.57 &   -67.49 &   904.75 &   -14.21 &   877.08 &  -222.06\\
		\toprule[0.7pt]
		$PC_1^f$ &    58.44 &     0.00 &    58.44 &     0.00 &     2.90 &     0.00 &     2.90 &     0.00\\
		$PC_1^{nf}$ &     1.24 &  -115.38 &    -0.53 &    -1.12 &    11.16 &   -95.25 &    -1.02 &   -11.11\\
		$PC_1^f+PC_1^{nf}$ &    57.91 &    -1.11 &    57.90 &    -1.12 &    11.27 &   -80.38 &     1.88 &   -11.11\\
		$PC_2$ &    13.36 &  -116.10 &    -5.88 &   -12.00 &    14.93 &    71.96 &     4.62 &    14.20\\
		$PE_1^u$ &     9.48 &   -87.62 &     0.39 &    -9.47 &     8.83 &   114.44 &    -3.65 &     8.04\\
		$PB$ &     1.36 &   -51.30 &     0.85 &    -1.06 &     1.55 &  -159.86 &    -1.46 &    -0.53\\
		$PE_1^d+PE_2$ &     3.87 &   -98.18 &    -0.55 &    -3.83 &     1.41 &   -12.55 &     1.37 &    -0.31\\
		\toprule[0.7pt]
		Penguin  &    59.45 &   -27.54 &    52.71 &   -27.49 &    10.65 &    74.93 &     2.77 &    10.28\\
		\toprule[0.7pt]
		\toprule[1pt]
	\end{tabular*}
\end{table}

\begin{table}[htbp]
	\centering
	\renewcommand{\arraystretch}{1.2}
	\caption{Helicity amplitudes of the $\Lambda_b\to p\pi^-$ decay classified by topologies without the CKM matrix elements.}
	\label{tab:pi-helicity}
	\begin{tabular*}{160mm}{c@{\extracolsep{\fill}}cccc|cccc}
		\toprule[1pt]
		\toprule[0.7pt]
		$\Lambda_b\to p\pi^-$ & $|H_{\frac{1}{2},0}|$ & $\delta(H_{\frac{1}{2},0})^\circ$ & Real($H_{\frac{1}{2},0}$) & Imag($H_{\frac{1}{2},0}$) & $|H_{-\frac{1}{2},0}|$ & $\delta(H_{-\frac{1}{2},0})^\circ$ & Real($H_{-\frac{1}{2},0}$) & Imag($H_{-\frac{1}{2},0}$)\\
		\toprule[0.7pt]
		$T^f$ &    65.30 &    90.00 &     0.00 &    65.30 &  9344.10 &   -90.00 &     0.00 & -9344.10\\
		$T^{nf}$ &   914.77 &    -8.40 &   904.96 &  -133.62 &  1593.23 &   172.35 & -1579.06 &   212.04\\
		$T^f+T^{nf}$ &   907.53 &    -4.32 &   904.96 &   -68.32 &  9267.57 &   -99.81 & -1579.06 & -9132.06\\
		$C_2$ &    83.29 &   -13.80 &    80.89 &   -19.87 &   378.05 &    77.44 &    82.24 &   369.00\\
		$E_2$ &   577.41 &   160.66 &  -544.83 &   191.24 &   532.44 &    85.22 &    44.34 &   530.59\\
		$B$ &   159.82 &   -12.04 &   156.30 &   -33.35 &    91.09 &   109.50 &   -30.40 &    85.87\\
		\toprule[0.7pt]
		Tree  &   601.38 &     6.66 &   597.33 &    69.70 &  8280.47 &  -100.32 & -1482.88 & -8146.61\\
		\toprule[0.7pt]
		$PC_1^f$ &   369.76 &   -90.00 &     0.00 &  -369.76 &   396.97 &   -90.00 &     0.00 &  -396.97\\
		$PC_1^{nf}$ &    44.70 &    -1.64 &    44.68 &    -1.28 &    60.01 &   172.07 &   -59.43 &     8.28\\
		$PC_1^f+PC_1^{nf}$ &   373.72 &   -83.13 &    44.68 &  -371.04 &   393.21 &   -98.69 &   -59.43 &  -388.69\\
		$PC_2$ &   157.24 &   157.48 &  -145.25 &    60.24 &    20.86 &   125.84 &   -12.21 &    16.91\\
		$PE_1^u$ &   101.73 &  -168.84 &   -99.80 &   -19.70 &    28.48 &   149.33 &   -24.50 &    14.53\\
		$PB$ &    13.17 &  -109.70 &    -4.44 &   -12.40 &     9.53 &   172.34 &    -9.45 &     1.27\\
		$PE_1^d+PE_2$ &    25.76 &   157.05 &   -23.72 &    10.04 &    26.73 &  -173.96 &   -26.58 &    -2.81\\
		\toprule[0.7pt]
		Penguin  &   403.75 &  -124.47 &  -228.52 &  -332.85 &   382.37 &  -110.22 &  -132.18 &  -358.80\\
		\toprule[0.7pt]
		\toprule[1pt]
	\end{tabular*}
\end{table}

\begin{table}[htbp]
	\centering
	\renewcommand{\arraystretch}{1.2}
	\caption{Same as Table~\ref{tab:pi} but for the $\Lambda_b\to pK^-$ decay.}
	\label{tab:K}
	\begin{tabular*}{160mm}{c@{\extracolsep{\fill}}cccc|cccc}
		\toprule[1pt]
		\toprule[0.7pt]
		$\Lambda_b\to pK^-$ & $|S|$ & $\delta^S(^\circ)$ & Real($S$) & Imag($S$) & $|P|$ & $\delta^P(^\circ)$ & Real($P$) & Imag($P$)\\
		\toprule[0.7pt]
		$T^f$ &   865.44 &     0.00 &   865.44 &     0.00 &  1230.64 &     0.00 &  1230.64 &     0.00\\
		$T^{nf}$ &    53.41 &  -102.81 &   -11.84 &   -52.08 &   343.23 &   -96.76 &   -40.43 &  -340.84\\
		$T^f+T^{nf}$ &   855.18 &    -3.49 &   853.60 &   -52.08 &  1238.05 &   -15.98 &  1190.21 &  -340.84\\
		$E_2$ &    89.06 &  -138.10 &   -66.28 &   -59.48 &    94.13 &   122.31 &   -50.31 &    79.56\\
		\toprule[0.7pt]
		Tree   &   795.18 &    -8.06 &   787.31 &  -111.55 &  1169.46 &   -12.91 &  1139.90 &  -261.28\\
		\toprule[0.7pt]
		$PC_1^f$ &    76.43 &     0.00 &    76.43 &     0.00 &     3.30 &   180.00 &    -3.30 &     0.00\\
		$PC_1^{nf}$ &     1.14 &  -134.10 &    -0.79 &    -0.82 &    13.85 &   -94.36 &    -1.05 &   -13.81\\
		$PC_1^f+PC_1^{nf}$ &    75.64 &    -0.62 &    75.64 &    -0.82 &    14.48 &  -107.50 &    -4.35 &   -13.81\\
		$PE_1^u$ &    11.80 &   -89.53 &     0.10 &   -11.80 &    11.02 &   115.62 &    -4.76 &     9.93\\
		$PE_1^d$ &     7.53 &  -101.53 &    -1.50 &    -7.38 &     2.67 &    51.53 &     1.66 &     2.09\\
		\toprule[0.7pt]
		Penguin   &    76.88 &   -15.08 &    74.23 &   -20.00 &     7.66 &  -166.53 &    -7.45 &    -1.79\\
		\toprule[0.7pt]
		\toprule[1pt]
	\end{tabular*}
\end{table}

\begin{table}[htbp]
	\centering
	\renewcommand{\arraystretch}{1.2}
	\caption{Same as Table~\ref{tab:pi-helicity} but for the $\Lambda_b\to pK^-$ decay.}
	\label{tab:K-helicity}
	\begin{tabular*}{160mm}{c@{\extracolsep{\fill}}cccc|cccc}
		\toprule[1pt]
		\toprule[0.7pt]
		$\Lambda_b\to pK^-$ & $|H_{\frac{1}{2},0}|$ & $\delta(H_{\frac{1}{2},0})^\circ$ & Real($H_{\frac{1}{2},0}$) & Imag($H_{\frac{1}{2},0}$) & $|H_{-\frac{1}{2},0}|$ & $\delta(H_{-\frac{1}{2},0})^\circ$ & Real($H_{-\frac{1}{2},0}$) & Imag($H_{-\frac{1}{2},0}$)\\
		\toprule[0.7pt]
		$T^f$ &    71.74 &    90.00 &     0.00 &    71.74 & 11397.55 &   -90.00 &     0.00 & -11397.55\\
		$T^{nf}$ &  1252.42 &    -5.08 &  1247.50 &  -110.90 &  1947.26 &   172.15 & -1929.02 &   265.90\\
		$T^f+T^{nf}$ &  1248.12 &    -1.80 &  1247.50 &   -39.17 & 11297.55 &   -99.83 & -1929.02 & -11131.65\\
		$E_2$ &   785.62 &   165.31 &  -759.92 &   199.29 &   668.39 &    91.58 &   -18.44 &   668.13\\
		\toprule[0.7pt]
		Tree   &   513.20 &    18.18 &   487.58 &   160.12 & 10643.20 &  -100.54 & -1947.46 & -10463.51\\
		\toprule[0.7pt]
		$PC_1^f$ &   515.50 &   -90.00 &     0.00 &  -515.50 &   484.74 &   -90.00 &     0.00 &  -484.74\\
		$PC_1^{nf}$ &    59.00 &     0.28 &    59.00 &     0.29 &    70.45 &   171.76 &   -69.72 &    10.10\\
		$PC_1^f+PC_1^{nf}$ &   518.58 &   -83.47 &    59.00 &  -515.22 &   479.73 &   -98.36 &   -69.72 &  -474.64\\
		$PE_1^u$ &   125.61 &  -169.53 &  -123.52 &   -22.83 &    37.72 &   145.13 &   -30.95 &    21.57\\
		$PE_1^d$ &    60.63 &   163.13 &   -58.02 &    17.60 &    38.56 &   176.88 &   -38.51 &     2.10\\
		\toprule[0.7pt]
		Penguin    &   534.68 &  -103.25 &  -122.54 &  -520.44 &   471.96 &  -107.15 &  -139.17 &  -450.97\\
		\toprule[0.7pt]
		\toprule[1pt]
	\end{tabular*}
\end{table}

A hierarchy relation exists between the helicity amplitudes $H_{-1/2,0}$ and $H_{1/2,0}$ from the factorizable topology $T^f$, as manifested in Tables.~\ref{tab:pi-helicity} and \ref{tab:K-helicity}. The proton spin tends to be anti-parallel to its momentum due to the chiral current $(V-A)$ in weak interaction, granting  the dominance of $H_{-1/2,0}$ over $H_{1/2,0}$.

Tables.~\ref{tab:pi} and \ref{tab:K} and the hierarchy relations in Eqs.~(\ref{eq:hierarchy-pi}) and (\ref{eq:hierarchy-K})  declare that the factorizable penguin contribution to the $P$-wave is much lower than to the $S$-wave, i.e., $|PC_1^f|^{\text{$P$-wave}} \ll |PC_1^f|^{\text{$S$-wave}}$. 
This feature is attributed to the current structures of the penguin operators; the matrix element for the factorizable penguin $PC_1^f$ from the $(V-A)\otimes(V-A)$ operators like $O_4$ is written, in the factorization hypothesis, as
\begin{equation}
	\begin{split}
		\langle p\pi^-|O_4|\Lambda_b\rangle =& \langle p\pi^-|(\bar{d}_\alpha b_\beta)_{V-A} (\bar{u}_\beta u_\alpha)_{V-A}|\Lambda_b\rangle \\
		= & \langle \pi^-|(\bar{d}u)_{V-A}|0\rangle \left[ \langle p|(\bar{u}b)_V|\Lambda_b\rangle - \langle p|(\bar{u}b)_A|\Lambda_b\rangle \right].
	\end{split}
\end{equation}
For the $(V-A)\otimes(V+A)$ operators like $O_6$, the corresponding matrix element is approximated by
\begin{equation}
	\begin{split}
		\langle p\pi^-|O_6|\Lambda_b\rangle =& \langle p\pi^-|(\bar{d}_\alpha b_\beta)_{V-A} (\bar{u}_\beta u_\alpha)_{V+A}|\Lambda_b\rangle \\
		= & -2 \langle \pi^-|(\bar{d}u)_{S+P}|0\rangle\langle p|(\bar{u}b)_{S-P}|\Lambda_b\rangle\\
		= & \langle \pi^-|(\bar{d}u)_{V-A}|0\rangle\Big[ \frac{2m_\pi^2}{(m_b-m_u)(m_u+m_d)}\langle p|(\bar{u}b)_V|\Lambda_b\rangle\\
		& + \frac{2m_\pi^2}{(m_b+m_u)(m_u+m_d)}\langle p|(\bar{u}b)_A|\Lambda_b\rangle \Big]\\
		= & \langle \pi^-|(\bar{d}u)_{V-A}|0\rangle \left( R_1^\pi\langle p|(\bar{u}b)_V|\Lambda_b\rangle + R_2^\pi\langle p|(\bar{u}b)_A|\Lambda_b\rangle \right),
		\label{eq:LR}
	\end{split}
\end{equation}
with the chiral factors $R_1^\pi\equiv \frac{2m_\pi^2}{(m_b-m_u)(m_u+m_d)}$ and $R_2^\pi\equiv \frac{2m_\pi^2}{(m_b+m_u)(m_u+m_d)}$. 

The $S$- and $P$-wave $PC_1^f$ amplitudes then read
\begin{equation}
	\begin{split}
		(PC_1^f)^{\text{$S$-wave}}=&-\frac{G_F}{\sqrt{2}}f_hV_{tb}V_{td}^\ast \left[\frac{C_3}{3}+C_4+ \frac{C_9}{3}+C_{10}+R_1^\pi(\frac{C_5}{3}+C_6+\frac{C_7}{3}+C_8)\right]\\
		&\Big[f_1(m_h^2)(m_{\Lambda_b}-m_p)+f_3(m_h^2)m_h^2\Big],\\
		(PC_1^f)^{\text{$P$-wave}}=&-\frac{G_F}{\sqrt{2}}f_hV_{tb}V_{td}^\ast \left[\frac{C_3}{3}+C_4+ \frac{C_9}{3}+C_{10}-R_2^\pi(\frac{C_5}{3}+C_6+\frac{C_7}{3}+C_8)\right]\\
		&\Big[g_1(m_h^2)(m_{\Lambda_b}+m_p)-g_3(m_h^2)m_h^2\Big],
		\label{eq:chiral-factor}
	\end{split}
\end{equation}
respectively, where the form factors $f_{1,3}$ and $g_{1,3}$ are defined in terms of the matrix element $\langle p|\bar{u}\gamma_\mu b |\Lambda_b\rangle = \bar{u}_p(f_1\gamma_\mu +f_2i\sigma_{\mu\nu}q^\nu + f_3q_\mu)u_{\Lambda_b}$ and $\langle p|\bar{u}\gamma_\mu\gamma_5 b |\Lambda_b\rangle = \bar{u}_p(g_1\gamma_\mu +g_2i\sigma_{\mu\nu}q^\nu + g_3q_\mu)\gamma_5u_{\Lambda_b}$. 
The chiral factors $R_{1,2}^\pi$ are of $\mathcal{O}(1)$, so the negative sign of the $R_2^\pi$ piece in the second expression of Eq.~(\ref{eq:chiral-factor}) results in destruction. 
As a consequence, the $S$-wave $PC_1^f$ is enhanced, while the $P$-wave one is suppressed.

It is observed from the amplitudes in Appendix.~\ref{app:decay-amplitudes} that specific combinations of the hadron DAs may produce opposite signs between the $S$ and $P$ waves. 
For instance, Eq.~(\ref{eq:hard-kernel-Tnf(c7)}) from the combination of the $\Lambda_b$ twist-2, proton twist-3 and meson twist-2 DAs yields 
\begin{equation}
	\begin{split}
		&M_{T_{nf}(c7)}^{(V-A)\otimes(V-A)}(\Lambda_b~\text{twist-2-proton twist-3-meson twist-2})\\
		=&8m_{\Lambda_b}^4 (x_3 - y) ( - 2 A_1 \psi_2 (-1 + x_2 + y) + 2 \psi_2 (2 T_1 + V_1) (-1 + x_2 + y))\phi_m^A\\
		&+ 8m_{\Lambda_b}^4 (x_3 - y) ( - 2 A_1 \psi_2 (-1 + x_2 + y) + 2 \psi_2 (2 T_1 + V_1) (-1 + x_2 + y))\phi_m^A\gamma_5,
	\end{split}
\end{equation}
where the $S$-wave (first) piece and the $P$-wave (second) piece have the same sign. Nevertheless, the two partial waves are opposite in sign in the combination of the $\Lambda_b$ twist-3, proton twist-3 and meson twist-2 DAs,
\begin{equation}
	\begin{split}
		&M_{T_{nf}(c7)}^{(V-A)\otimes(V-A)}(\Lambda_b~\text{twist-3+proton twist-3+meson twist-2})\\
		=&8m_{\Lambda_b}^4 (x_3 - y) ( A_1\psi_3^{-+}x_2^\prime + V_1\psi_3^{-+}x_2^\prime )\phi_m^A\\
		&- 8m_{\Lambda_b}^4 (x_3 - y) ( A_1\psi_3^{-+}x_2^\prime + V_1\psi_3^{-+}x_2^\prime )\phi_m^A\gamma_5.
	\end{split}
\end{equation}
This distinction arises from the different numbers of $\gamma_5$ for the $S$ and $P$ waves endowed by the vector and axial-vector currents in weak interaction and by the Dirac structures of the hadron DAs.

We survey all combinations of the DAs, and decide the relative signs between the $S$- and $P$-wave amplitudes as exhibited in Table~\ref{tab:signs-S-and-P}, where the symbol `$+$' (`$-$') means the same (opposite) sign.
It is interesting that the relative sign between the $S$ and $P$ waves alternates with the twists of the $\Lambda_b$ or proton DAs. This feature holds for all Feynman diagrams and all types of weak interaction vertices.
The net interference  between the $S$- and $P$-wave amplitudes for a topology is then determined by the dominant Feynman diagram and DA combination involved. This accounts for the notable fluctuation of the strong phases in Tables.~\ref{tab:pi} and \ref{tab:K}. The different strong phases impact the predictions for CPVs, which will be explored in the following subsections.
\begin{table}[htbp]
	\centering
	\renewcommand{\arraystretch}{1.1}
	\begin{tabular*}{160mm}{c@{\extracolsep{\fill}}cccc}
		\toprule[1pt]
		\toprule[0.7pt]
		meson twist-2& proton~twist-3 & proton~twist-4 & proton~twist-5 & proton~twist-6 \\
		\toprule[0.7pt]
		$\Lambda_b$ twist-2 &    + &    - &     + &    - \\
		$\Lambda_b$ twist-3 &    - &    + &  - &  + \\
		$\Lambda_b$ twist-4 &    + &    - &  + &   - \\
		\toprule[0.7pt]
		meson twist-3& proton~twist-3 & proton~twist-4 & proton~twist-5 & proton~twist-6 \\
		\toprule[0.7pt]
		$\Lambda_b$ twist-2 &     - &   + &     - & +\\
		$\Lambda_b$ twist-3 &      + &   - & + &   -\\
		$\Lambda_b$ twist-4 &     - &   + & - & +\\
		\toprule[0.7pt]
		\toprule[1pt]
	\end{tabular*}
	\caption{Relative signs between the $S$- and $P$-wave amplitudes for various twist combinations of the $\Lambda_b$ and proton DAs.}
	\label{tab:signs-S-and-P}
\end{table}


With the obtained amplitudes, we predict the various observables associated with the $\Lambda_b\to p\pi^-,pK^-$ decays as elaborated below. The outcomes are assembled in Table \ref{tab:Lb2pP-observables}, where the first three errors are from the inputs in the $\Lambda_b$, proton and meson DAs, respectively, and the fourth one stems from varying the renormalization scale of the Wilson coefficients in Table~\ref{tab:scale-dependence-Wilson-Coe}. The uncertainties from the CKM matrix elements are negligible relative to the above sources.

\begin{table}[htbp]
	\centering
	\scriptsize
	\renewcommand{\arraystretch}{1.5}
	\caption{Observables associated with the $\Lambda_b\to p\pi^-,pK^-$ decays. The percentages in the parentheses stand for the proportions of the partial-wave CPVs to the direct CPVs.}
	\begin{tabular*}{160mm}{c@{\extracolsep{\fill}}ccc}
		\toprule[1pt]
		\toprule[0.7pt]
		& $Br(\times 10^{-6})$ &  &  \\
		\toprule[0.7pt]
		$\Lambda_b\to p\pi^-$ & $3.34^{+2.53+1.33+0.10+0.47}_{-1.30-1.10-0.11-0.14}$ &  & \\
		$\Lambda_b\to pK^-$ & $2.83^{+2.17+1.17+0.49+2.19}_{-1.05-0.92-0.37-0.66}$ &  &  \\
		\toprule[0.7pt]
		& $A_{CP}^{dir}$ & $A_{CP}^S(\kappa_S)$ & $A_{CP}^P(\kappa_P)$ \\
		\toprule[0.7pt]
		$\Lambda_b\to p\pi^-$ & $0.05^{+0.00+0.00+0.00+0.02}_{-0.02-0.01-0.02-0.01}$ & $0.17^{+0.01+0.01+0.03+0.04}_{-0.04-0.04-0.07-0.04}(49\%)$ & $-0.06^{+0.01+0.03+0.02+0.00}_{-0.02-0.03-0.03-0.01}(51\%)$   \\
		$\Lambda_b\to pK^-$ & $-0.06^{+0.01+0.01+0.02+0.00}_{-0.01-0.01-0.01-0.00}$ & $-0.05^{+0.02+0.02+0.04+0.00}_{-0.02-0.01-0.03-0.00}(94\%)$ & $-0.21^{+0.07+0.23+0.29+0.04}_{-0.15-0.33-0.27-0.01}(6\%)$  \\
		\toprule[0.7pt]
		& $\alpha$ & $A_{CP}^{\alpha}$ & $\langle\alpha\rangle$  \\
		\toprule[0.7pt]
		$\Lambda_b\to p\pi^-$ & $-0.94^{+0.00+0.02+0.01+0.03}_{-0.02-0.02-0.02-0.02}$ & $0.02^{+0.00+0.01+0.00+0.01}_{-0.01-0.01-0.01-0.01}$ & $-0.96^{+0.00+0.01+0.01+0.02}_{-0.00-0.01-0.01-0.01}$ \\
		$\Lambda_b\to pK^-$ & $0.23^{+0.04+0.02+0.10+0.15}_{-0.03-0.05-0.12-0.07}$ & $0.04^{+0.02+0.02+0.01+0.01}_{-0.02-0.03-0.01-0.01}$ & $0.20^{+0.02+0.01+0.11+0.14}_{-0.02-0.02-0.12-0.06}$ \\
		\toprule[0.7pt]
		& $\beta$ & $A_{CP}^{\beta}$ & $\langle\beta\rangle$  \\
		\toprule[0.7pt]
		$\Lambda_b\to p\pi^-$ & $0.34^{+0.00+0.05+0.01+0.07}_{-0.06-0.06-0.06-0.05}$ & $0.22^{+0.00+0.00+0.03+0.07}_{-0.01-0.01-0.04-0.03}$ & $0.12^{+0.00+0.05+0.03+0.00}_{-0.05-0.05-0.04-0.02}$  \\
		$\Lambda_b\to pK^-$ & $-0.39^{+0.03+0.08+0.08+0.12}_{-0.01-0.04-0.07-0.01}$ & $-0.44^{+0.01+0.01+0.02+0.08}_{-0.00-0.00-0.01-0.04}$ & $0.05^{+0.03+0.08+0.07+0.04}_{-0.01-0.05-0.07-0.02}$  \\
		\toprule[0.7pt]
		& $\gamma$ & $A_{CP}^{\gamma}$ & $\langle\gamma\rangle$ \\
		\toprule[0.7pt]
		$\Lambda_b\to p\pi^-$ &  $0.09^{+0.02+0.04+0.04+0.04}_{-0.04-0.06-0.06-0.01}$ & $0.11^{+0.01+0.02+0.03+0.03}_{-0.02-0.03-0.04-0.02}$ & $-0.02^{+0.01+0.02+0.01+0.01}_{-0.02-0.04-0.01-0.00}$ \\
		$\Lambda_b\to pK^-$ &  $0.89^{+0.02+0.04+0.04+0.00}_{-0.01-0.02-0.05-0.01}$ & $0.02^{+0.02+0.05+0.04+0.00}_{-0.01-0.03-0.04-0.00}$ & $0.87^{+0.00+0.01+0.02+0.00}_{-0.00-0.01-0.02-0.01}$ \\
		\toprule[0.7pt]
		\toprule[1pt]
	\end{tabular*}
	\label{tab:Lb2pP-observables}
\end{table}

The decay width is written as
\begin{equation}
	\Gamma = \frac{p_c}{8\pi m_{\Lambda_b}^2}\left\{ M_+^2|S|^2 + M_-^2|P|^2 \right\} = \frac{p_c}{16\pi m_{\Lambda_b}^2}\left(|H_{1/2,0}|^2+|H_{-1/2,0}|^2\right),
\end{equation}
where 
\begin{equation}
	\begin{split}
		p_c&=\sqrt{[(m_{\Lambda_b}+m_p)^2-m_h^2][(m_{\Lambda_b}-m_p)^2-m_h^2]}/(2m_{\Lambda_b}),
	\end{split}
\end{equation}
is the momentum of either the proton or the meson in the rest frame of the $\Lambda_b$ baryon. The predicted $\Lambda_b\to p\pi^-$ and $\Lambda_b\to pK^-$ branching ratios are a bit lower than the data $Br(\Lambda_b\to p\pi^-)=(4.6\pm 0.8)\times 10^{-6}$ and $Br(\Lambda_b\to pK^-)=(5.5\pm 1.0)\times 10^{-6}$. We suspect that the discrepancy is attributed to higher-order corrections, in particular those to  the non-factorizable topologies. The factorizable contributions have been more or less fixed by the measured $\Lambda_b\to p\mu\bar{\nu}_\mu$ and $\Lambda_b\to pD_s^-$ branching ratios. 
The uncertainty arising from the Wilson coefficients for the $\Lambda_b\to pK^-$ decay is larger than for $\Lambda_b\to p\pi^-$, suggesting that higher-order contributions are more important in penguin-dominant modes.

The $\Lambda_b\to p\pi^-,pK^-$ decays share the same inputs from the $\Lambda_b$ and proton DAs, so the errors from these sources are suppressed in
the difference of the branching ratios, 
\begin{equation}
	\begin{split}
		\Delta Br(\Lambda_b\to pK^-/\pi^-)&\equiv Br(\Lambda_b\to pK^-)-Br(\Lambda_b\to p\pi^-)\\
		&= (-0.51^{+0.25+0.20+0.50+1.72}_{-0.36-0.17-0.38-0.52})\times 10^{-6}.
		\label{eq:delta-BR}
	\end{split}
\end{equation}
Though the errors from the meson DAs and the Wilson coefficients keep sizable, 
Eq.~(\ref{eq:delta-BR}) indicates that the $\Lambda_b\to pK^-$ branching ratio is higher than the $\Lambda_b\to p\pi^-$ one. 
The discrepancy between  Eq.~(\ref{eq:delta-BR}) and the LHCb measurements may be also resolved by including higher-order corrections, as hinted by the crucial effects from varying the renormalization scale of the Wilson coefficients.

According to the  definition for the direct CPV of the $\Lambda_b\to ph$ decay,
\begin{equation}
	A_{CP}^{dir}(\Lambda_b\to ph)\equiv \frac{\Gamma(\Lambda_b\to ph) - \bar{\Gamma}(\bar{\Lambda}_b\to \bar{p}\bar{h})}{\Gamma(\Lambda_b\to ph) + \bar{\Gamma}(\bar{\Lambda}_b\to \bar{p}\bar{h})},
\end{equation}
we have the explicit expression
\begin{equation}
	A_{CP}^{dir} = \frac{-2M_+^2|S_\mathcal{T}|^2 r_S \sin\Delta\phi \sin\Delta\delta_S - 2M_-^2|P_\mathcal{T}|^2 r_P \sin\Delta\phi \sin\Delta\delta_P}{M_+^2|S_\mathcal{T}|^2 (1+r_S^2+2r_S\cos\Delta\phi \cos\Delta\delta_S) + M_-^2|P_\mathcal{T}|^2 (1+r_P^2+2r_P\cos\Delta\phi \cos\Delta\delta_P)},
\end{equation}
where $r_{S}=|\lambda_\mathcal{P}S_\mathcal{P}|/|\lambda_\mathcal{T}S_\mathcal{T}|$ ($r_{P}=|\lambda_\mathcal{P}P_\mathcal{P}|/|\lambda_\mathcal{T}P_\mathcal{T}|$) is the ratio of the penguin amplitude to the tree amplitude for the $S$ ($P$) wave, the difference of the weak phases $\Delta\phi\equiv \text{arg}(\lambda_\mathcal{P})-\text{arg}(\lambda_\mathcal{T})$ is identical for both waves, and $\Delta\delta_S= \delta_\mathcal{P}^S - \delta_\mathcal{T}^S$ ($\Delta\delta_P= \delta_\mathcal{P}^P - \delta_\mathcal{T}^P$) is the difference of the strong phases for the $S$ ($P$) wave. It is trivial to construct the partial-wave CPVs,
\begin{equation}
	\begin{split}
		A_{CP}^S\equiv&\frac{|S|^2-|\bar{S}|^2}{|S|^2+|\bar{S}|^2} = \frac{-2 r_S \sin\Delta\phi \sin\Delta\delta_S }{1+r_S^2+2r_S\cos\Delta\phi \cos\Delta\delta_S},\\
		A_{CP}^P\equiv&\frac{|P|^2-|\bar{P}|^2}{|P|^2+|\bar{P}|^2} = \frac{ - 2 r_P \sin\Delta\phi \sin\Delta\delta_P}{ 1+r_P^2+2r_P\cos\Delta\phi \cos\Delta\delta_P}.   \label{eq:kappaSandkappaP}
	\end{split}
\end{equation}

The net direct CPV is  related to the partial-wave CPVs through~\cite{Han:2024kgz}
\begin{equation}
	\begin{split}
		A_{CP}^{dir}& =\frac{M_+^2(|S|^2-|\bar{S}|^2) + M_-^2(|P|^2-|\bar{P}|^2)}{M_+^2(|S|^2+|\bar{S}|^2) + M_-^2(|P|^2+|\bar{P}|^2)}\\
		& = \kappa_S A_{CP}^{S} + \kappa_P A_{CP}^{P}, 
		\label{eq:CPV=kappaS+kappaP} 
	\end{split}
\end{equation}
where the weights are given by
\begin{equation}
	\begin{split}
		\kappa_S &\equiv \frac{|S|^2}{|S|^2+\frac{M_-^2}{M_+^2}\frac{1+A_{CP}^{S}}{1+A_{CP}^{P}}|P|^2} 
		=\frac{|S|^2}{|S|^2+\kappa ~r_{CP}|P|^2},
		\\
		\kappa_P &\equiv \frac{\frac{M_-^2}{M_+^2}|P|^2}{\frac{1+A_{CP}^{P}}{1+A_{CP}^{S}}|S|^2+\frac{M_-^2}{M_+^2}|P|^2} 
		=\frac{\kappa|P|^2}{|S|^2/r_{CP}+\kappa|P|^2},
		\label{eq:presice}
	\end{split}
\end{equation}
with $\kappa\equiv M_-^2/M_+^2$ and $r_{CP}=(1+A_{CP}^{S})/(1+A_{CP}^{P})$.
It states that the total direct CPV is the weighted average of the partial-wave CPVs.

It is observed in Table~\ref{tab:Lb2pP-observables} that the partial-wave CPVs can exceed $10\%$, such as $A_{CP}^{S}(\Lambda_b\to p\pi)=0.17$ and $A_{CP}^{P}(\Lambda_b\to pK)=-0.21$, similar to the direct CPVs in $B$ meson decays. 
However, the destruction between the $S$-wave CPV ($0.17$) and the $P$-wave CPV ($-0.06$) in $\Lambda_b\to p\pi^-$ reduces the direct CPV to $0.05^{+0.02}_{-0.03}$. The central value of our prediction does not exactly match the data, but the destruction mechanism in this mode, which does not exist in the corresponding decay $B\to \pi\pi$, has been demonstrated unambiguously. 
The $S$-wave CPV in the penguin-dominated  $\Lambda_b\to pK^-$ decay is small ($-$0.05) with the tiny strong phase $\Delta \delta_S=-7^\circ$, and the sizable $P$-wave CPV is down by the proportion $\kappa_P=6\%$ because of $|P|\ll |S|$. The direct CPV $A_{CP}^{dir}(\Lambda_b\to pK)=-0.06^{+0.03}_{-0.02}$ turns out to diminish as well.

We emphasize that the aforementioned destruction mechanism was not noticed before.
The previous studies based on the generalized factorization \cite{Hsiao:2014mua,Hsiao:2017tif,Geng:2020ofy} and the QCDF \cite{Zhu:2018jet} considered only the factorizable diagrams, i.e., $T$ for the tree contributions and $PC_1$ for the penguin contributions. The associated partial-wave CPVs are always minor  owing to either the small strong phase differences or the low weights of the penguin amplitudes.
Here we point out the large relative strong phases from the $PC_2$ and $PE_1^u$ diagrams, which are first realized in the PQCD formalism. 
They generate the strong phases for the $S$ and $P$ waves with opposite signs, whose cancellation is the mechanism  responsible for the diminishing direct CPV in $\Lambda_b\to p\pi$.

\begin{figure}[htbp]
	\centering
	\begin{minipage}{0.5\linewidth}
		\includegraphics[width=\linewidth]{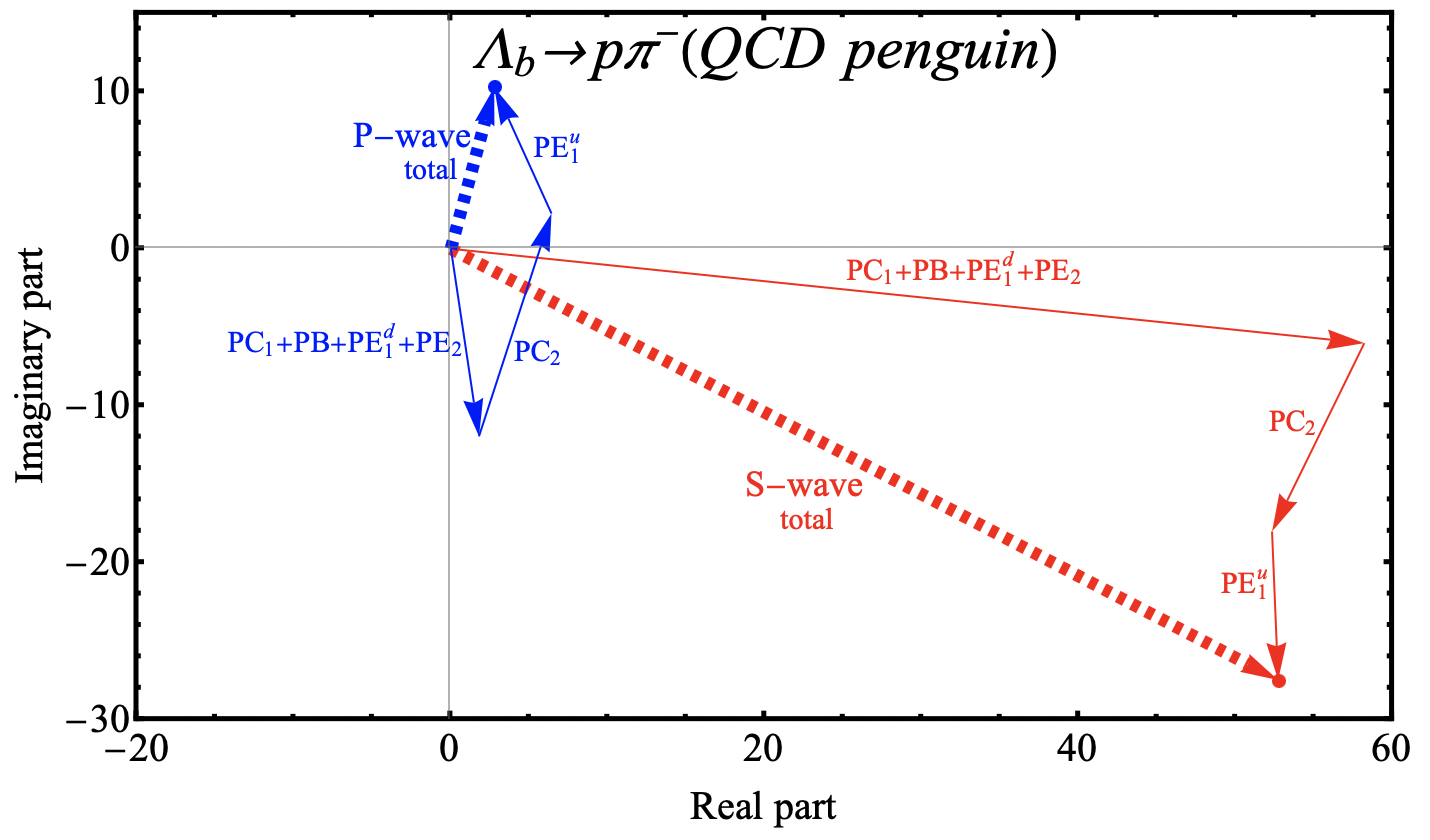}
	\end{minipage}
	
	\begin{minipage}{0.5\linewidth}
		\includegraphics[width=\linewidth]{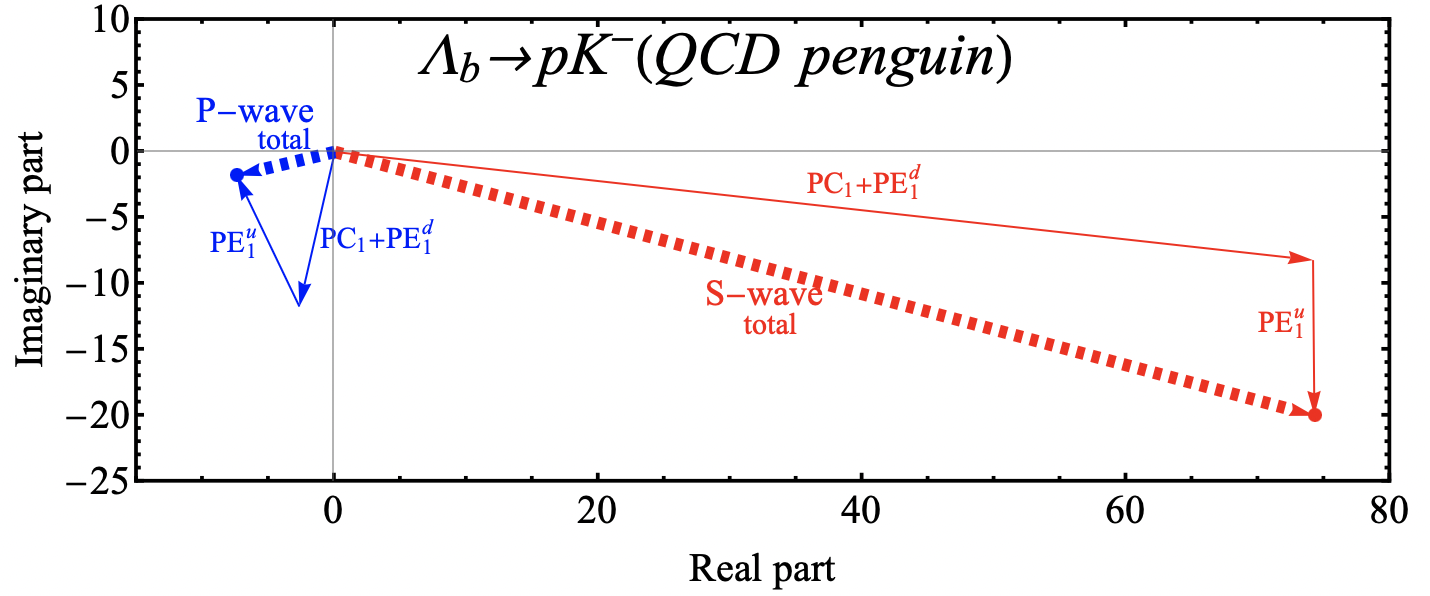}
	\end{minipage}
	\caption{QCD penguin contributions from each topological diagram to the $\Lambda_b \to p\pi^-$ decay (top) and to the $\Lambda_b \to pK^-$ decay (bottom) depicted in the complex plane. The red (blue) vectors represent the $S$-wave ($P$-wave) penguin contributions.}
	\label{fig:penguin-in-complex-plane}
\end{figure}

To visualize the cancellation between the imaginary parts or the strong phases of the partial-wave amplitudes, we display the $PC_2$, $PE_1^u$ and $PC_1$ contributions to the  $\Lambda_b\to p\pi^-$ decay in Fig. \ref{fig:penguin-in-complex-plane}.
It is evident that the combined $PC_2+PE_1^u$ constitutes the dominant source of the strong phases, which destruct between the $S$ and $P$ waves. The diagram $PC_1$ mainly contributes to the magnitude of the $S$ wave. 
The $\Lambda_b\to pK^-$ case is different as remarked before, where the diagram $PE_1^u$ also gives the strong phases with opposite signs between the $S$ and $P$ waves. However, the dominant $S$-wave CPV governs the direct CPV in this mode.

\begin{figure}[htbp]
	\centering
	\begin{minipage}{0.5\linewidth}
		\includegraphics[width=\linewidth]{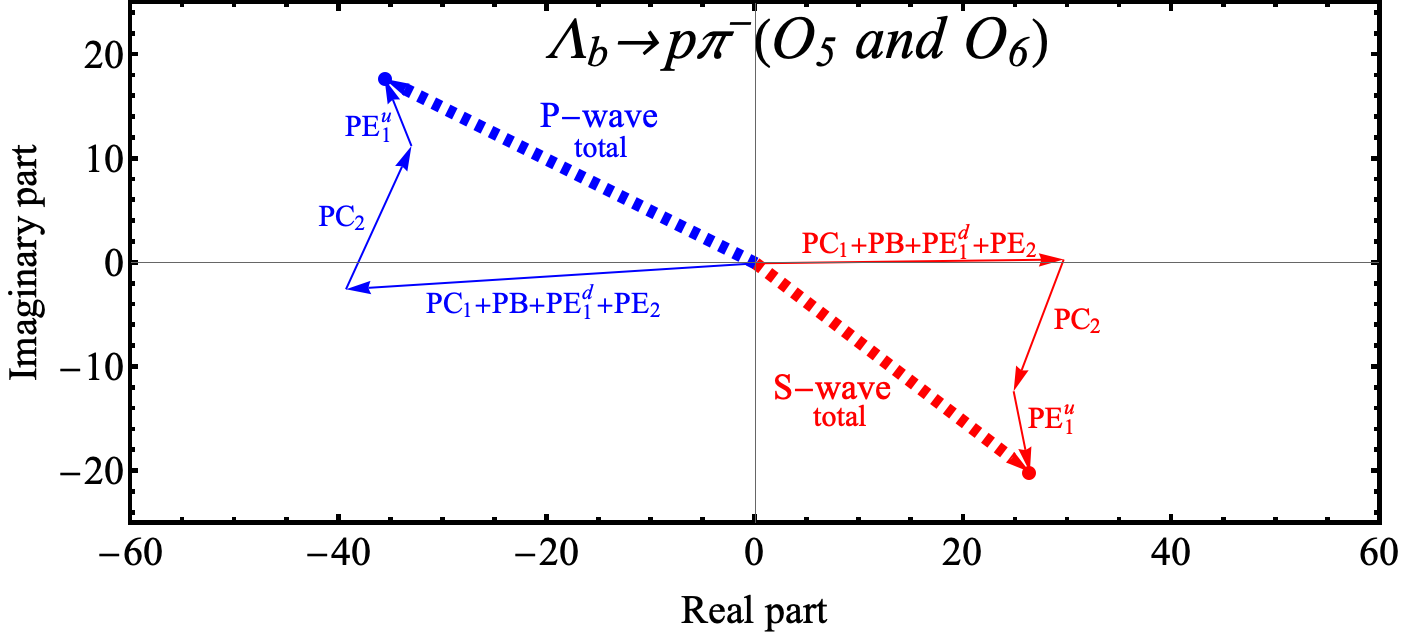}
	\end{minipage}
	
	\begin{minipage}{0.5\linewidth}
		\includegraphics[width=\linewidth]{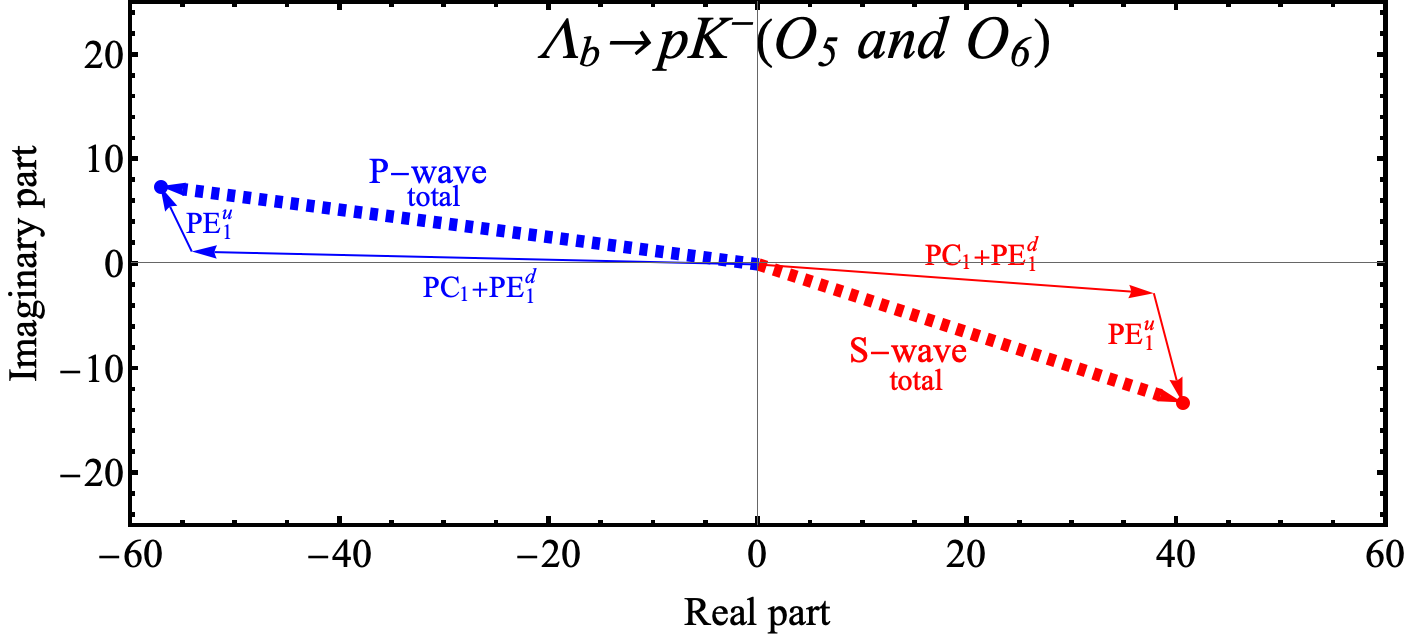}
	\end{minipage}
	\caption{$O_5$ and $O_6$ penguin contributions to the $\Lambda_b \to p\pi^-$ decay (top) and to the $\Lambda_b \to pK^-$ decay (bottom) depicted in the complex plane. The red (blue) vectors represent the $S$-wave ($P$-wave) penguin contributions.}
	\label{fig:penguin-O5O6-in-complex-plane}
\end{figure}

We explain why the \(PC_2\) and \(PE_1^u\) penguins generate the dominant strong phases with opposite signs between the $S$- and $P$-wave amplitudes.
For simplicity, we ignore the electroweak penguins and concentrate on the QCD penguins. The distinction between the contributions of \(O_{3,4}\) and \(O_{1,2}\), both being the \((V-A) \otimes (V-A)\) operators, appears only in their Wilson coefficients, such that the strong phases of the penguin amplitudes from $O_{3,4}$ are close to those of the tree amplitudes. This claim is supported by the results in Tables~\ref{tab:pi-QCDpenguin} and \ref{tab:K-QCDpenguin}. 
Taking the $\Lambda_b\to p\pi^-$ decay as an example, we have $\delta^S=-6.26^\circ$ and $-10.64^\circ$ for the $S$-wave amplitudes from the tree and $O_{3,4}$ operators, respectively, and $\delta^P=-14.21^\circ$ and $-12.75^\circ$ for the $P$-wave ones from the tree and $O_{3,4}$ operators, respectively. It is also the case for $\Lambda_b\to pK^-$.
The contributions from the \((V-A) \otimes (V+A)\) operators \(O_{5,6}\) make the substantial strong phase differences between the penguin and tree  amplitudes, which read $\delta^S=-40.61^\circ$ and $\delta^P=153.58^\circ$ for the $p\pi^-$ mode in Table~\ref{tab:pi-QCDpenguin}, and $\delta^S=-18.02^\circ$ and $\delta^P=172.56^\circ$ for the $pK^-$ mode in Table~\ref{tab:K-QCDpenguin}.
Figure~\ref{fig:penguin-O5O6-in-complex-plane}, depicting the $S$- and $P$-wave amplitudes from $O_{5,6}$, justifies our analytic argument for $S_{PC_2}\approx - P_{PC_2}$ and $S_{PE_1^u}\approx - P_{PE_1^u}$. 

\begin{table}[htbp]
	\centering\small
	\renewcommand{\arraystretch}{1.0}
	\begin{tabular*}{165mm}{|c|c@{\extracolsep{\fill}}|cccc|cccc|}
		\hline
		\hline
		\multicolumn{2}{|c|}{$\Lambda_b\to p\pi^-$} & $|S|$ & $\delta^S(^\circ)$ & Re($S$) & Im($S$) & $|P|$ & $\delta^P(^\circ)$ & Re($P$) & Im($P$)\\
		\hline
		\multirow{5}{*}{Tree}	&	$T$ &   703.07 &    -4.19 &   701.19 &   -51.38 &  1003.22 &   -15.33 &   967.54 &  -265.17\\
		& $C_2$ &    29.37 &   154.96 &   -26.61 &    12.43 &    41.51 &   179.80 &   -41.51 &     0.14\\
		& $E_2$ &    66.94 &  -145.26 &   -55.01 &   -38.14 &    72.58 &   119.94 &   -36.23 &    62.89\\
		& $B$ &    10.40 &   112.64 &    -4.00 &     9.60 &    23.65 &  -122.56 &   -12.73 &   -19.93\\
		\cline{2-10}
		& total &   619.26 &    -6.26 &   615.57 &   -67.49 &   904.75 &   -14.21 &   877.08 &  -222.06\\
		\hline\hline
		&    $PC_1$ & 59.36 & -1.62 & 59.34 & -1.68 & 13.41 & -81.85 & 1.90 & -13.27\\
		&  $PC_2$ & 13.95 & -114.84 & -5.86 & -12.66 & 13.89 & 69.92 & 4.77 & 13.05\\
		\multirow{1}{*}{QCD}   &  $PE_1^u$ & 8.52 & -93.30 & -0.49 & -8.51 & 10.23 & 112.10 & -3.85 & 9.48\\
		\multirow{1}{*}{Penguin}   & $PB$ & 1.44 & -39.65 & 1.11 & -0.92 & 1.40 & -162.08 & -1.33 & -0.43\\
		& $PE_1^d+PE_2$ & 3.98 & -101.17 & -0.77 & -3.90 & 1.35 & 7.65 & 1.34 & 0.18\\
		\cline{2-10}
		& total & 60.07 & -27.41 & 53.33 & -27.65 & 9.44 & 72.56 & 2.83 & 9.01\\
		\hline\hline
		\multirow{6}{*}{$O_{3,4}$}    &    $PC_1$ &    30.41 &    -4.26 &    30.32 &    -2.26 &    43.11 &   -16.55 &    41.32 &   -12.28\\
		&  $PC_2$ &     1.05 &   179.78 &    -1.05 &     0.00 &     1.68 &  -154.82 &    -1.52 &    -0.71\\
		&  $PE_1^u$ &     2.18 &  -159.87 &    -2.05 &    -0.75 &     3.28 &   112.54 &    -1.26 &     3.03\\
		& $PB$ &     0.36 &   103.90 &    -0.09 &     0.35 &     0.38 &  -114.38 &    -0.16 &    -0.35\\
		& $PE_1^d+PE_2$ &     2.42 &   -93.34 &    -0.14 &    -2.42 &     1.61 &    86.32 &     0.10 &     1.60\\
		\cline{2-10}
		& total &    27.47 &   -10.64 &    27.00 &    -5.07 &    39.46 &   -12.75 &    38.49 &    -8.71\\
		\hline\hline
		\multirow{6}{*}{$O_{5,6}$}    &     $PC_1$ &    29.02 &     1.15 &    29.02 &     0.58 &    39.44 &  -178.56 &   -39.42 &    -0.99\\
		&        $PC_2$ &    13.54 &  -110.82 &    -4.81 &   -12.66 &    15.13 &    65.45 &     6.29 &    13.76\\
		&        $PE_1^u$ &     7.92 &   -78.61 &     1.56 &    -7.76 &     6.95 &   111.88 &    -2.59 &     6.45\\
		&        $PB$ &     1.74 &   -46.52 &     1.20 &    -1.27 &     1.17 &  -176.00 &    -1.17 &    -0.08\\
		&        $PE_1^d+PE_2$ &     1.61 &  -113.22 &    -0.63 &    -1.48 &     1.89 &   -48.96 &     1.24 &    -1.42\\
		\cline{2-10}
		&       total &    34.69 &   -40.61 &    26.33 &   -22.58 &    39.82 &   153.58 &   -35.66 &    17.72\\
		\hline
		\hline
	\end{tabular*}
	\caption{Results of the topological amplitudes for the $\Lambda_b\to p\pi^-$ decay in units of $10^{-9}$ without the CKM matrix elements.
		The QCD penguin contributions are further broken down into those from the $O_{3,4}$ and  $O_{5,6}$ operators. 
		The magnitudes and strong phases of each topological diagram are grouped in the columns of $|S|$, $\delta^S(^\circ)$, $|P|$ and $\delta^P(^\circ)$. 
	}
	\label{tab:pi-QCDpenguin}
\end{table}

\begin{table}[htbp]
	\centering\small
	\renewcommand{\arraystretch}{1.0}
	\begin{tabular*}{165mm}{|c@{\extracolsep{\fill}}|c|cccc|cccc|}
		\hline
		\hline
		\multicolumn{2}{|c|}{$\Lambda_b\to p K^-$} & $|S|$ & $\delta^S(^\circ)$ & Re($S$) & Im($S$) & $|P|$ & $\delta^P(^\circ)$ & Re($P$) & Im($P$)\\
		\hline
		\multirow{3}{*}{Tree}	&	$T$ &   855.18 &    -3.49 &   853.60 &   -52.08 &  1238.05 &   -15.98 &  1190.21 &  -340.84\\
		& $E_2$ &  89.06 &  -138.10 &   -66.28 &   -59.48 &    94.13 &   122.31 &   -50.31 &    79.56\\
		\cline{2-10}
		& total &   795.18 &    -8.06 &   787.31 &  -111.55 &  1169.46 &   -12.91 &  1139.90 &  -261.28\\
		\hline
		\hline
		&  $PC_1$ & 76.08 & -0.81 & 76.07 & -1.07 & 15.03 & -106.59 & -4.29 & -14.40\\
		\multirow{1}{*}{QCD}  &  $PE_1^u$ & 11.88 & -88.70 & 0.27 & -11.88 & 11.41 & 114.21 & -4.68 & 10.41\\
		\multirow{1}{*}{Penguin}  & $PE_1^d$ & 7.46 & -100.12 & -1.31 & -7.34 & 2.08 & 60.60 & 1.02 & 1.81\\
		\cline{2-10}
		& total & 77.72 & -15.13 & 75.03 & -20.29 & 8.25 & -164.60 & -7.95 & -2.19\\
		\hline\hline
		\multirow{4}{*}{$O_{3,4}$}    &    $PC_1$ &    37.18 &    -4.11 &    37.09 &    -2.66 &    53.10 &   -17.42 &    50.67 &   -15.89\\
		&  $PE_1^u$ &     2.86 &  -149.71 &    -2.47 &    -1.44 &     4.48 &   111.71 &    -1.66 &     4.17\\
		& $PE_1^d$ &     3.00 &   -92.18 &    -0.11 &    -3.00 &     2.08 &    83.11 &     0.25 &     2.06\\
		\cline{2-10}
		& total &    35.23 &   -11.64 &    34.51 &    -7.11 &    50.20 &   -11.10 &    49.26 &    -9.66\\
		\hline\hline
		\multirow{4}{*}{$O_{5,6}$}    &     $PC_1$ &    39.01 &     2.33 &    38.98 &     1.59 &    54.98 &   178.45 &   -54.96 &     1.49\\
		&    $PE_1^u$ &    10.79 &   -75.30 &     2.74 &   -10.44 &     6.93 &   115.83 &    -3.02 &     6.24\\
		&        $PE_1^d$ &     4.50 &  -105.46 &    -1.20 &    -4.34 &     0.81 &   -18.16 &     0.77 &    -0.25\\
		\cline{2-10}
		&       total &    42.61 &   -18.02 &    40.52 &   -13.18 &    57.70 &   172.56 &   -57.21 &     7.47\\
		\hline
		\hline
	\end{tabular*}
	\caption{Same as Table~\ref{tab:pi} but for the $\Lambda_b\to pK^-$ decay.}
	\label{tab:K-QCDpenguin}
\end{table}

\begin{figure}[htbp]
	\centering
	\begin{minipage}{0.6\linewidth}
		\includegraphics[width=\linewidth]{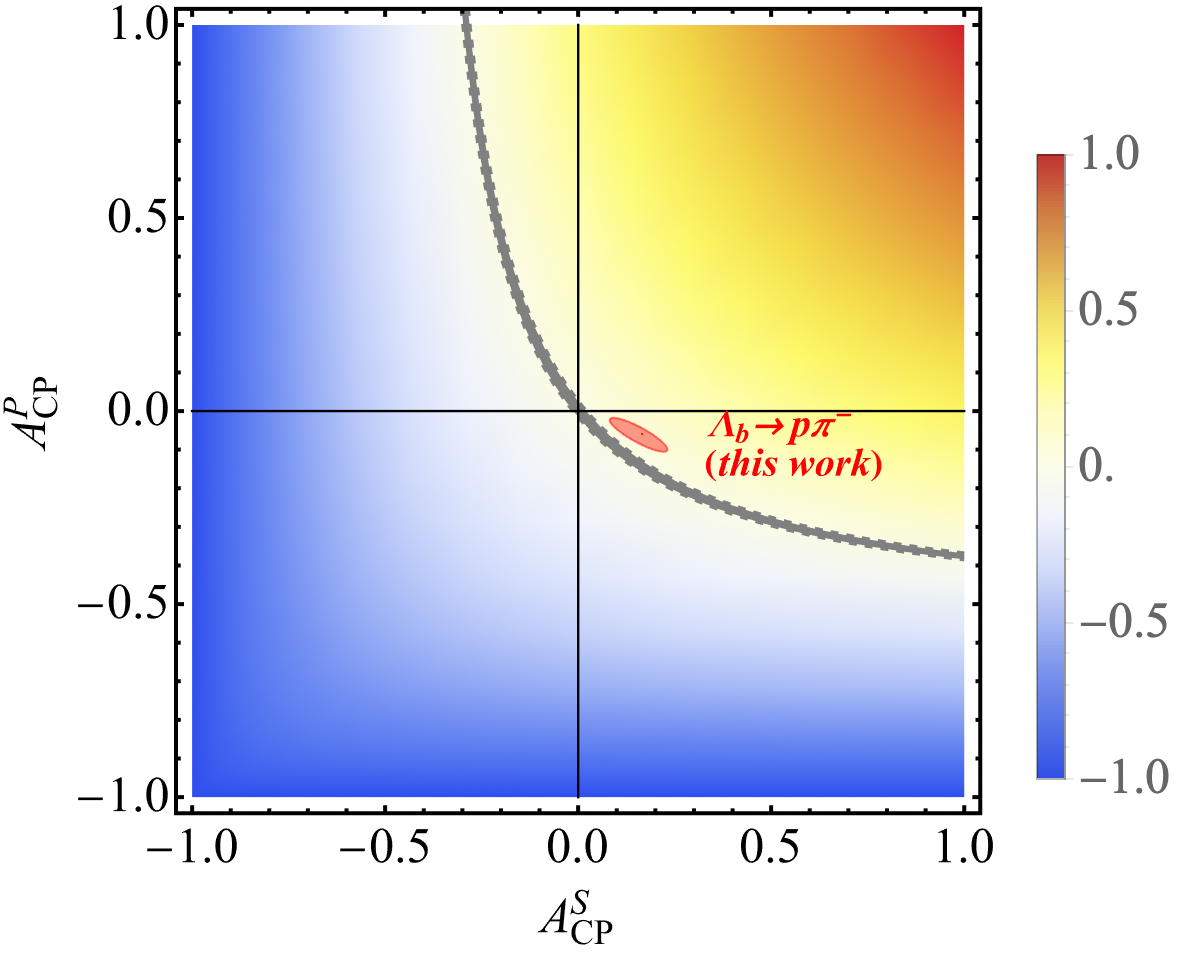}
	\end{minipage}
	\caption{Dependence of the net direct CPV on partial-wave CPVs, where the red spot corresponds to our prediction.
		The gray narrow band specifies the partial-wave CPVs allowed by the data of the direct CPV of $\Lambda_b\to p\pi^-$.}
	\label{fig:cancel-from-data}
\end{figure}

The underlying destruction between the partial-wave CPVs in the \(\Lambda_b \to p\pi^-\) decay is manifested in Figure~\ref{fig:cancel-from-data}, for the ratio of $|P|^2 / |S|^2=1.6$ obtained in this work.
It shows the dependence of the net direct CPV on partial-wave CPVs, where the red spot corresponds to our prediction. 
The gray narrow band specifies the partial-wave CPVs allowed by the data of the direct CPV.
We can perceive how the partial-wave CPVs affect the net CPV, and how close our prediction is to the data. 
It is noteworthy that the $P$ wave of the  decay $\Lambda_b\to pK^-$ reveals the highest partial-wave CPV, amounting to $21\%$, but with larger uncertainty compared to the $S$-wave CPV. 
The factorizable penguin contributions play an essential role in the $S$-wave amplitude due to the chiral factors. These contributions, being  less sensitive to the proton DAs, have smaller theoretical errors.
By contrast, the non-factorizable penguin contributions, dominating the $P$-wave amplitude, suffer significant uncertainties from the proton and kaon DAs, whch propagate into the $P$-wave CPVs.

The difference of the direct CPVs is given by
\begin{equation}
	\begin{split}
		\Delta A_{CP}^{dir}(\Lambda_b\to pK^-/\pi^-)&\equiv A_{CP}^{dir}(\Lambda_b\to pK^-)-A_{CP}^{dir}(\Lambda_b\to p\pi^-)\\
		&= -0.11^{+0.02+0.02+0.03+0.02}_{-0.01-0.00-0.01-0.02},
	\end{split}
\end{equation}
which slightly deviates from the measured value around $-0.01$. 
As mentioned in Sec.~\ref{sec:introduction} and Ref.~\cite{Han:2022srw}, the threshold resummation factor for heavy baryon decays, which smooths the endpoint behavior of hard kernels and modifies tree and penguin contributions differently, has not been taken into account in the present framework.
Implementing this factor may achieve more accurate predictions, improve the agreement with experimental data, and advance our understanding of the decay dynamics accordingly. 



We define the decay parameters for $\Lambda_b\to p\pi^-,pK^-$~\cite{Lee:1957qs},
\begin{equation}
\alpha = -\frac{2\kappa_c Re(S^\ast P)}{|S|^2+\kappa_c^2|P|^2},\quad \beta= -\frac{2\kappa_c Im(S^\ast P)}{|S|^2+\kappa_c^2|P|^2},\quad \gamma= \frac{|S|^2-\kappa_c^2|P|^2}{|S|^2+\kappa_c^2|P|^2},
\end{equation}
with the coefficient $\kappa_c=p_c/(E_p+m_p)=\sqrt{(E_p-m_p)/(E_p+m_p)}$, and their associated $CP$ asymmetries and averages,
\begin{equation}
\begin{split}
	A_{CP}^\alpha &=\frac{\alpha+\bar{\alpha}}{2},\quad A_{CP}^\beta=\frac{\beta+\bar{\beta}}{2},\quad A_{CP}^\gamma=\frac{\gamma-\bar{\gamma}}{2},\\
	\langle\alpha\rangle&=\frac{\alpha-\bar{\alpha}}{2},\quad \langle\beta\rangle=\frac{\beta-\bar{\beta}}{2},\quad \langle\gamma\rangle =\frac{\gamma+\bar{\gamma}}{2}.
\end{split}\label{63}
\end{equation}
The numerical outcomes for the above observables are supplied in \cref{tab:Lb2pP-observables}, among which the $CP$ asymmetries $A_{CP}^\beta$ are prominent for both the $\Lambda_b\to p\pi^-,pK^-$ modes. However, the information on the $\Lambda_b$ baryon and proton polarizations is needed to measure these parameters, which are experimentally inaccessible by now. 

The proton and meson masses have been retained in the kinematic variables and hard kernels in the current analysis. 
To quantify the impact of these supposed higher-power contributions, we neglect the proton and pion masses, and recalculate the $\Lambda_b\to p\pi^-$ decay amplitudes. The resultant branching ratio $3.65\times 10^{-6}$ and direct CPV $0.09$ in Table~\ref{tab:pi-ingore-final-mass} evince that the light-hadron masses produce $15\%$ and $50\%$ effects, respectively, relative to the corresponding values in Table~\ref{tab:Lb2pP-observables}. 
This test manifests the importance of power corrections in heavy baryon decays.
\begin{table}[htbp]
\centering
\renewcommand{\arraystretch}{1.2}
\caption{Same as \cref{tab:pi} but without the proton and meson masses in kinematics.}
\begin{tabular*}{165mm}{c@{\extracolsep{\fill}}cccc|cccc}
	\toprule[1pt]
	\toprule[0.7pt]
	$\Lambda_b\to p\pi^-$ & $|S|$ & $\phi(S)^\circ$ & Real($S$) & Imag($S$) & $|P|$ & $\phi(P)^\circ$ & Real($P$) & Imag($P$)\\
	\toprule[0.7pt]
	$T_f$ &   839.84 &     0.00 &   839.84 &     0.00 &   850.10 &     0.00 &   850.10 &     0.00\\
	$T_{nf}$ &    62.83 &  -100.16 &   -11.08 &   -61.84 &   240.42 &   -97.09 &   -29.69 &  -238.58\\
	$T_f+T_{nf}$ &   831.06 &    -4.27 &   828.75 &   -61.84 &   854.39 &   -16.21 &   820.41 &  -238.58\\
	$C^\prime$ &    35.65 &   162.64 &   -34.02 &    10.64 &    32.99 &  -176.98 &   -32.95 &    -1.74\\
	$E_2$ &    84.36 &  -141.20 &   -65.75 &   -52.86 &    51.84 &   128.88 &   -32.54 &    40.35\\
	$B$ &    17.54 &   104.24 &    -4.32 &    17.00 &    13.74 &  -124.50 &    -7.78 &   -11.32\\
	\toprule[0.7pt]
	Tree &   729.88 &    -6.85 &   724.66 &   -87.06 &   776.44 &   -15.79 &   747.14 &  -211.29\\
	\toprule[0.7pt]
	$P^{C_1}_f$ &    67.66 &     0.00 &    67.66 &     0.00 &     2.61 &     0.00 &     2.61 &     0.00\\
	$P^{C_1}_{nf}$ &     1.83 &  -109.17 &    -0.60 &    -1.73 &     9.87 &   -92.94 &    -0.51 &    -9.86\\
	$P^{C_1}_f+P^{C_1}_{nf}$ &    67.08 &    -1.48 &    67.06 &    -1.73 &    10.08 &   -77.93 &     2.11 &    -9.86\\
	$P^{C_2}$ &    17.24 &  -116.06 &    -7.57 &   -15.49 &    13.63 &    67.60 &     5.19 &    12.60\\
	$P^{E_1^u}$ &    11.22 &   -90.55 &    -0.11 &   -11.22 &     7.06 &   112.49 &    -2.70 &     6.52\\
	$P^B$ &     1.30 &   -42.87 &     0.96 &    -0.89 &     1.40 &  -170.77 &    -1.38 &    -0.22\\
	$P^{E_1^d}+P^{E_2}$ &     4.68 &  -101.65 &    -0.95 &    -4.59 &     1.62 &   -12.75 &     1.58 &    -0.36\\
	\toprule[0.7pt]
	Penguin &    68.39 &   -29.72 &    59.39 &   -33.91 &     9.92 &    61.02 &     4.81 &     8.68\\
	\toprule[0.7pt]
	\toprule[1pt]
\end{tabular*}
\label{tab:pi-ingore-final-mass}
\end{table}

\subsection{\texorpdfstring{Quasi-two-body $\Lambda_b \to pV$ decays}{Lb->pV}}

The $\Lambda_b \to pV$ decay amplitudes for longitudinally and transversely polarized vector mesons $V=\rho^-,K^{\ast -}$ are decomposed into
\begin{equation}
	\begin{split}
		\mathcal{A}^L=&\bar{u}_p \Big(A_1^L\gamma^\mu\gamma_5+A_2^L\frac{p_p^\mu}{m_{\Lambda_b}}\gamma_5
		+B_1^L\gamma^\mu+B_2^L\frac{p_p^\mu}{m_{\Lambda_b}} \Big) u_{\Lambda_b}\epsilon_{L \mu}^{\ast},\\
		\mathcal{A}^T=&\bar{u}_p( A_1^T\gamma^\mu\gamma_5+B_1^T\gamma^\mu ) u_{\Lambda_b}\epsilon_{T\mu}^{\ast },
	\end{split}\label{par}
\end{equation}
with the longitudinal (transverse) polarization vector $\epsilon_{L(T)}$. 
The polarization amplitudes $A_1^{L,T}$, $A_2^L$, $B_1^{L,T}$ and $B_2^L$ form the partial-wave amplitudes  
\begin{equation}
	\begin{split}
		S^L=&-A_1^L,\\
		S^T=&-A_1^T,\\
		P_1=&-\frac{p_c}{E_V}\left(\frac{m_{\Lambda_b}+m_p}{E_p+m_p}B_1^L+B_2^L\right),\\
		P_2=&\frac{p_c}{E_p+m_p}B_1^T,\\
		D=&-\frac{p_c^2}{E_V(E_p+m_p)}\left(A_1^L-A_2^L\right),
	\end{split}
\end{equation}
and the helicity amplitudes 
\begin{equation}
	\begin{split}
		H_{\frac{1}{2},1}=&- \sqrt{2}M_+A_1^T - \sqrt{2}M_-B_1^T ,\\
		H_{-\frac{1}{2},-1}=&\sqrt{2}M_+A_1^T - \sqrt{2}M_-B_1^T,\\
		H_{\frac{1}{2},0}=&\left( M_+(m_{\Lambda_b}-m_p)A_1^L-M_-p_cA_2^L+M_-(m_{\Lambda_b}+m_p)B_1^L+M_+p_cB_2^L \right)/m_V,\\
		H_{-\frac{1}{2},0}=&\left( -M_+(m_{\Lambda_b}-m_p)A_1^L+M_-p_cA_2^L+M_-(m_{\Lambda_b}+m_p)B_1^L+M_+p_cB_2^L \right)/m_V,
	\end{split}
\end{equation}
with the vector meson energy $E_V$. 


Given the numerical results in Appendix~\ref{app:appendix-numerical-results} for the above amplitudes,
the $\Lambda_b\to pV$ decay width is derived via
\begin{equation}
	\begin{split}
		\Gamma=
		&\frac{p_c}{4\pi}\frac{E_p+m_p}{m_{\Lambda_b}}\left\{ 2(|S^T|^2+|P_2|^2)+\frac{E_V^2}{m_V^2}(|S^L+D|^2+|P_1|^2) \right\}\\
		=&\frac{p_c}{16\pi m_{\Lambda_b}^2}(|H_{1/2,1}|^2+|H_{-1/2,-1}|^2+|H_{1/2,0}|^2+|H_{-1/2,0}|^2).
		\label{eq:pV-width}
	\end{split}
\end{equation}
The associated partial-wave CPVs are defined as in Eq.~(\ref{eq:kappaSandkappaP}), in terms of which the direct CPV is decomposed~\cite{Han:2024kgz},
\begin{equation}
	A_{CP}^{dir}\approx  \kappa_{S^T}A_{CP}^{S^T}  + \kappa_{P_1}A_{CP}^{P_1}+ \kappa_{P_2}A_{CP}^{P_2} + \kappa_{D+S^L}A_{CP}^{D+S^L},
	\label{eq:pV-relation-CPV-and-partialCPV}
\end{equation}
with the weights
\begin{equation}
	\begin{split}
		\kappa_{S^T}=&\frac{2|S^T|^2}{\Pi},\quad \kappa_{P_1}=\frac{E_V^2|P_1|^2}{m_V^2\Pi}, \\\kappa_{P_2}=&\frac{2|P_2|^2}{\Pi},\quad \kappa_{D+S^L}=\frac{E_V^2|D+S^L|^2}{m_V^2\Pi},\\   \Pi \equiv & 2|S^T|^2 + 2|P_2|^2 + E_V^2/m_V^2|D+S^L|^2 + E_V^2/m_V^2|P_1|^2.
	\end{split}
\end{equation}

The predictions for the branching ratios and CPVs of the $\Lambda_b\to p\rho^-,pK^{\ast -}$ decays are collected in Table~\ref{tab:Lb2pV-observables}.
Unlike $\Lambda_b\to p\pi^-,pK^-$, whose branching ratios are of the same order, the  $\Lambda_b\to p\rho^-$ branching ratio is greater than the $pK^{\ast -}$ one by a factor of $3$. 
The operators $O_{5,6,7,8}$ do not contribute to the factorizable penguin topologies as shown in Eq.~(\ref{eq:matrix-vaccum-to-meson}). 
Hence, the enhancement of the penguin amplitudes by the chiral factor  in Eq.~(\ref{eq:chiral-factor}) is absent in  the$\Lambda_b\to pK^{\ast -}$ decay. 

It is found that the $D+S^L$ and $P_1$ components dominate the direct CPVs because of the enhancement factor $E_V/m_V$.
The partial-wave CPVs of the $\Lambda_b\to pV$ decays also exceed $10\%$;
the $P_2\text{-wave}$ CPVs of $\Lambda_b\to p\rho^-$ reaches $17\%$;
the $(D+S^L)\text{-wave}$ CPV of $\Lambda_b\to pK^{*-}$ is even as large as $27\%$.
However, the direct CPVs of these modes are all minor for the similar reasons.
The relative sign of the partial-wave amplitudes in the $\Lambda_b\to pK^{*-}$ decay can be argued in the same manner. Their tiny direct CPVs  trace back to the destruction between the major $(D+S^L)$- and $P_1\text{-wave}$ CPVs, like the  $\Lambda_b\to p\pi^-$ case.
For the $\Lambda_b\to p\rho^-$ decay, the two dominant partial-wave CPVs are small, leading to a negligible direct CPV.

\begin{table}
	\centering
	\scriptsize
	\caption{Same as Table~\ref{tab:Lb2pP-observables} but for the $\Lambda_b\to p\rho^-,pK^{\ast -}$ decays.}
	\renewcommand{\arraystretch}{1.5}
	\begin{tabular*}{160mm}{c@{\extracolsep{\fill}}ccc}
		\toprule[1pt]
		\toprule[0.7pt]
		& $Br(\times 10^{-6})$ & $A_{CP}^{dir}$ & $A_{CP}^{S^T}(\kappa_{S^T})$  \\
		\toprule[0.7pt]
		$\Lambda_b\to p\rho^-$ & $9.66^{+6.23+3.23+0.21+1.89}_{-3.50-3.03-1.20-0.75}$ & $0.03^{+0.02+0.01+0.00+0.02}_{-0.02-0.03-0.03-0.02}$ & $0.01^{+0.00+0.00+0.00+0.00}_{-0.01-0.02-0.02-0.02}(7\%)$  \\
		$\Lambda_b\to pK^{\ast -}$ & $2.83^{+1.77+0.46+0.37+0.63}_{-1.29-1.23-0.53-0.66}$ & $-0.05^{+0.04+0.07+0.01+0.05}_{-0.11-0.07-0.06-0.08}$ & $-0.15^{+0.06+0.09+0.02+0.05}_{-0.00-0.04-0.05-0.00}(6\%)$  \\
		\toprule[0.7pt]
		& $A_{CP}^{S^L+D}(\kappa_{S^L+D})$ & $A_{CP}^{P_1}(\kappa_{P_1})$ & $A_{CP}^{P_2}(\kappa_{P_2})$ \\
		\toprule[0.7pt]
		$\Lambda_b\to p\rho^-$  & $0.02^{+0.03+0.04+0.02+0.05}_{-0.02-0.02-0.00-0.00}(44\%)$ & $0.03^{+0.04+0.00+0.00+0.00}_{-0.05-0.04-0.10-0.05}(45\%)$ & $0.17^{+0.00+0.00+0.01+0.03}_{-0.02-0.03-0.03-0.04}(4\%)$  \\
		$\Lambda_b\to pK^{\ast -}$  & $0.27^{+0.02+0.06+0.05+0.03}_{-0.17-0.11-0.02-0.18}(33\%)$ & $-0.23^{+0.05+0.07+0.02+0.05}_{-0.11-0.11-0.09-0.03}(55\%)$ & $-0.14^{+0.01+0.00+0.02+0.01}_{-0.04-0.09-0.02-0.03}(6\%)$  \\
		\toprule[0.7pt]
		& $\alpha$ & $A_{CP}^\alpha$ & $\langle\alpha\rangle$  \\
		\toprule[0.7pt]
		$\Lambda_b\to p\rho^-$ & $-0.83^{+0.02+0.01+0.00+0.00}_{-0.02-0.05-0.04-0.01}$ & $-0.01^{+0.01+0.01+0.01+0.00}_{-0.00-0.00-0.01-0.00}$ & $-0.83^{+0.01+0.01+0.01+0.00}_{-0.02-0.05-0.04-0.01}$  \\
		$\Lambda_b\to pK^{\ast -}$ & $-1.00^{+0.01+0.01+0.00+0.01}_{-0.00-0.00-0.00-0.00}$ &  $-0.00^{+0.00+0.00+0.00+0.00}_{-0.00-0.00-0.00-0.00}$ & $-1.00^{+0.00+0.01+0.00+0.00}_{-0.00-0.00-0.00-0.00}$  \\
		\toprule[0.7pt]
		& $\beta$ & $A_{CP}^\beta$ & $\langle\beta\rangle$  \\
		\toprule[0.7pt]
		$\Lambda_b\to p\rho^-$ & $-0.98^{+0.05+0.07+0.05+0.06}_{-0.00-0.00-0.00-0.00}$ &  $0.00^{+0.01+0.02+0.01+0.02}_{-0.00-0.00-0.00-0.00}$ & $-0.99^{+0.04+0.05+0.04+0.04}_{-0.00-0.00-0.00-0.00}$ \\
		$\Lambda_b\to pK^{\ast -}$ & $-0.90^{+0.07+0.17+0.11+0.00}_{-0.03-0.03-0.00-0.03}$ &  $-0.02^{+0.04+0.06+0.04+0.01}_{-0.00-0.04-0.00-0.00}$& $-0.88^{+0.06+0.11+0.08+0.00}_{-0.03-0.06-0.00-0.04}$ \\
		\toprule[0.7pt]
		& $\gamma$ & $A_{CP}^\gamma$ & $\langle\gamma\rangle$  \\
		\toprule[0.7pt]
		$\Lambda_b\to p\rho^-$ & $-0.11^{+0.01+0.01+0.01+0.01}_{-0.01-0.01-0.02-0.00}$ &  $-0.01^{+0.00+0.00+0.00+0.00}_{-0.00-0.00-0.00-0.00}$  & $-0.10^{+0.01+0.01+0.01+0.00}_{-0.01-0.01-0.02-0.00}$  \\
		$\Lambda_b\to pK^{\ast -}$ & $-0.12^{+0.01+0.00+0.02+0.00}_{-0.06-0.05-0.03-0.05}$ &  $0.02^{+0.01+0.03+0.01+0.01}_{-0.02-0.02-0.01-0.01}$& $-0.14^{+0.01+0.01+0.02+0.00}_{-0.04-0.07-0.04-0.04}$  \\
		\toprule[0.7pt]
		& $\Lambda$ & $A_{CP}^\Lambda$ & $\langle\Lambda\rangle$  \\
		\toprule[0.7pt]
		$\Lambda_b\to p\rho^-$ & $-0.96^{+0.05+0.06+0.04+0.05}_{-0.00-0.00-0.00-0.00}$  &  $0.00^{+0.01+0.02+0.01+0.02}_{-0.00-0.00-0.00-0.00}$  & $-0.97^{+0.04+0.04+0.03+0.04}_{-0.00-0.00-0.00-0.00}$  \\
		$\Lambda_b\to pK^{\ast -}$ & $-0.91^{+0.06+0.15+0.09+0.00}_{-0.02-0.02-0.00-0.03}$  &  $-0.01^{+0.03+0.06+0.03+0.01}_{-0.00-0.03-0.00-0.00}$& $-0.90^{+0.05+0.09+0.07+0.00}_{-0.03-0.05-0.01-0.03}$ \\
		\toprule[0.7pt]
		& $\mathcal{J}$ & $A_{CP}^\mathcal{J}$ & $\langle\mathcal{J}\rangle$  \\
		\toprule[0.7pt]
		$\Lambda_b\to p\rho^-$ & $1.66^{+0.04+0.04+0.02+0.02}_{-0.03-0.03-0.05-0.00}$  &  $-0.01^{+0.01+0.01+0.01+0.00}_{-0.01-0.01-0.01-0.00}$ & $1.67^{+0.03+0.04+0.02+0.02}_{-0.05-0.03-0.05-0.00}$  \\
		$\Lambda_b\to pK^{\ast -}$ & $1.67^{+0.02+0.00+0.04+0.00}_{-0.14-0.12-0.08-0.12}$  &  $0.04^{+0.02+0.05+0.02+0.01}_{-0.06-0.04-0.02-0.03}$ & $1.63^{+0.01+0.03+0.04+0.00}_{-0.08-0.15-0.09-0.09}$ \\
		\toprule[0.7pt]
		\toprule[1pt]
	\end{tabular*}
	\label{tab:Lb2pV-observables}
\end{table}

The decay asymmetry parameters are expressed, in terms of the helicity amplitudes, as
\begin{equation}
	\begin{split}
		\alpha&=\frac{|H_{1/2,1}|^2-|H_{-1/2,-1}|^2}{|H_{1/2,1}|^2+|H_{-1/2,-1}|^2},\quad \beta=\frac{|H_{1/2,0}|^2-|H_{-1/2,0}|^2}{|H_{1/2,0}|^2+|H_{-1/2,0}|^2},\quad \gamma=\frac{|H_{1/2,1}|^2-|H_{-1/2,-1}|^2}{|H_{1/2,0}|^2+|H_{-1/2,0}|^2},\\
		\Lambda&=\frac{|H_{1/2,1}|^2-|H_{-1/2,-1}|^2+|H_{1/2,0}|^2-|H_{-1/2,0}|^2}{|H_{1/2,1}|^2+|H_{-1/2,-1}|^2+|H_{1/2,0}|^2+|H_{-1/2,0}|^2},
	\end{split}
\end{equation}
which satisfy $\Lambda=(\beta+\alpha\gamma)/(1+\gamma)$. The associated CPVs and averages are defined as in Eq.~(\ref{63}) and by
\begin{equation}
	\begin{split}
		A_{CP}^\Lambda = \frac{\Lambda+\bar{\Lambda}}{2}, \quad \langle\Lambda\rangle= \frac{\Lambda-\bar{\Lambda}}{2}.
	\end{split}
\end{equation}
The results for the above quantities in Table~\ref{tab:Lb2pV-observables} clearly imply that the CPVs of the decay asymmetry parameters are all consistent with zero. 

\begin{figure}
	\centering
	\includegraphics[width=0.4\linewidth]{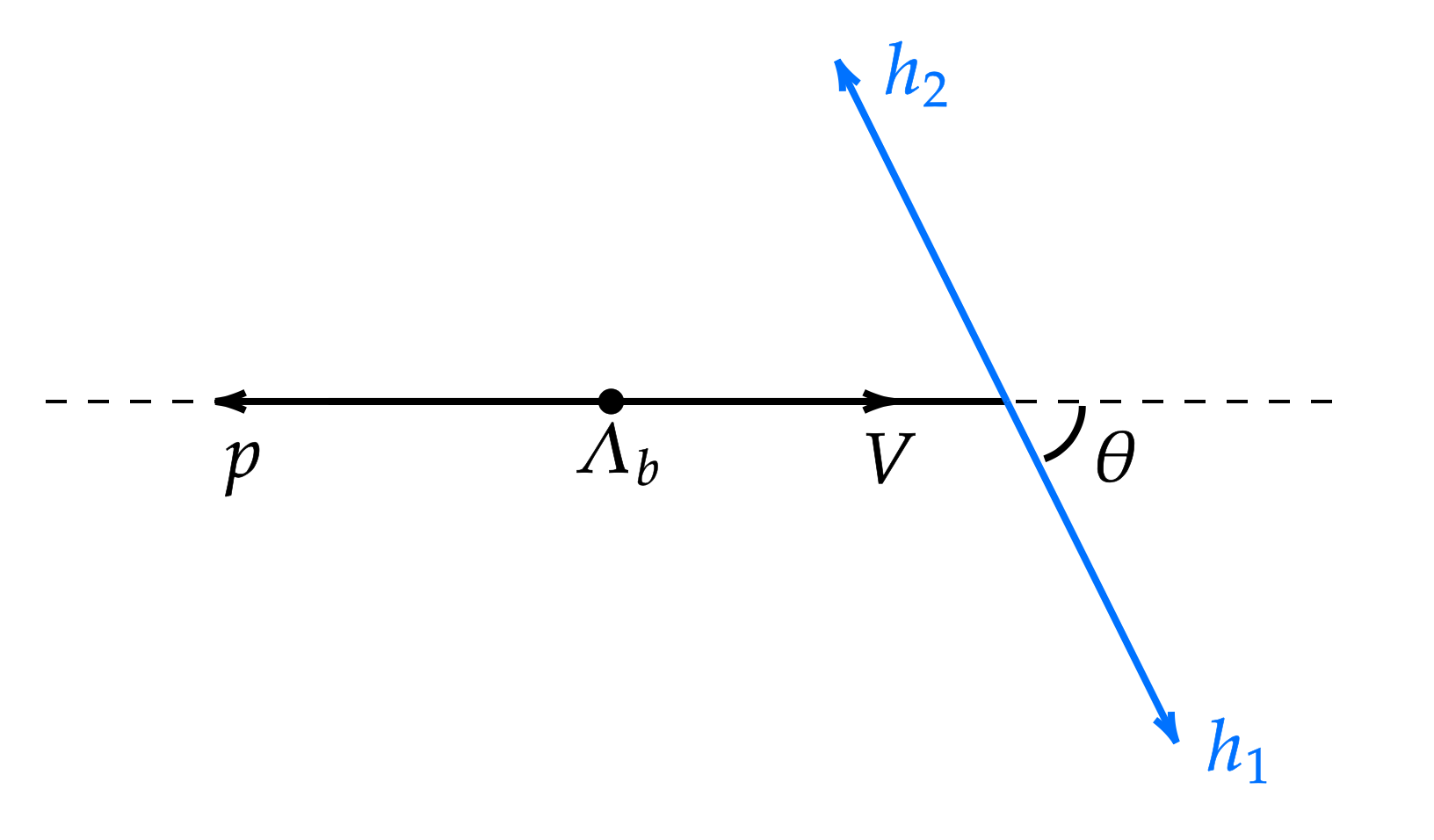}
	\caption{Configuration of the $\Lambda_b\to pV\to ph_1h_2$ decay, where $\theta$ is the angle between $h_1$ in the $V$ rest frame and the $V$ momentum in the $\Lambda_b$ rest frame.}
	\label{fig:angle-distri-pv}
\end{figure}

Since a vector meson decays strongly with a finite width, 
what we worked on are actually the three-body decays $\Lambda_b\to p(\rho^- \to)\pi^-\pi^0,p(K^{\ast -}\to)K^-\pi^0$.
Motivated by the sizable partial-wave CPVs, we investigate the potential observables for identifying CPVs involved in the angle distributions for three-body decays. 
The angle distribution of $\Lambda_b\to pV\to ph_1h_2$ in Fig.~\ref{fig:angle-distri-pv} is formulated as
\begin{equation}
	\begin{split}
		\frac{d\Gamma}{d \cos\theta}&\propto |H_{\frac{1}{2},0}|^2 + |H_{-\frac{1}{2},0}|^2 + |H_{-\frac{1}{2},-1}|^2 + |H_{\frac{1}{2},1}|^2\\
		&\hspace {0.5cm}+ (2|H_{\frac{1}{2},0}|^2 + 2|H_{-\frac{1}{2},0}|^2 - |H_{-\frac{1}{2},-1}|^2 - |H_{\frac{1}{2},1}|^2)P_2\\
		&\propto 1 + \mathcal{J} \cdot P_2,
	\end{split}
\end{equation}
where the observable
\begin{equation}
	\mathcal{J}\equiv \frac{2|H_{\frac{1}{2},0}|^2+2|H_{-\frac{1}{2},0}|^2-|H_{-\frac{1}{2},-1}|^2-|H_{\frac{1}{2},1}|^2}{|H_{\frac{1}{2},0}|^2+|H_{-\frac{1}{2},0}|^2+|H_{-\frac{1}{2},-1}|^2+|H_{\frac{1}{2},1}|^2},
\end{equation}
with the Legendre polynomial $P_2=(3\cos^2\theta -1)/2$ can be extracted from data fitting. The CPV and average for $\mathcal{J}$ are written as
\begin{equation}
	A_{CP}^{\mathcal{J}} = \frac{\mathcal{J}-\bar{\mathcal{J}}}{2},\qquad \langle\mathcal{J}\rangle = \frac{\mathcal{J}+\bar{\mathcal{J}}}{2},
\end{equation}
whose predictions are listed in Table~\ref{tab:Lb2pV-observables}. Unfortunately, the CPVs $A_{CP}^{\mathcal{J}}$ in the $\Lambda_b\to p\rho^-,pK^{\ast -}$ decays are too tiny to search for experimentally.

\subsection{\texorpdfstring{Quasi-two-body $\Lambda_b \to pA$ decays}{Lb->pA}}

\begin{figure}[htbp]
	\begin{minipage}{0.45\linewidth}
		\includegraphics[width=\linewidth]{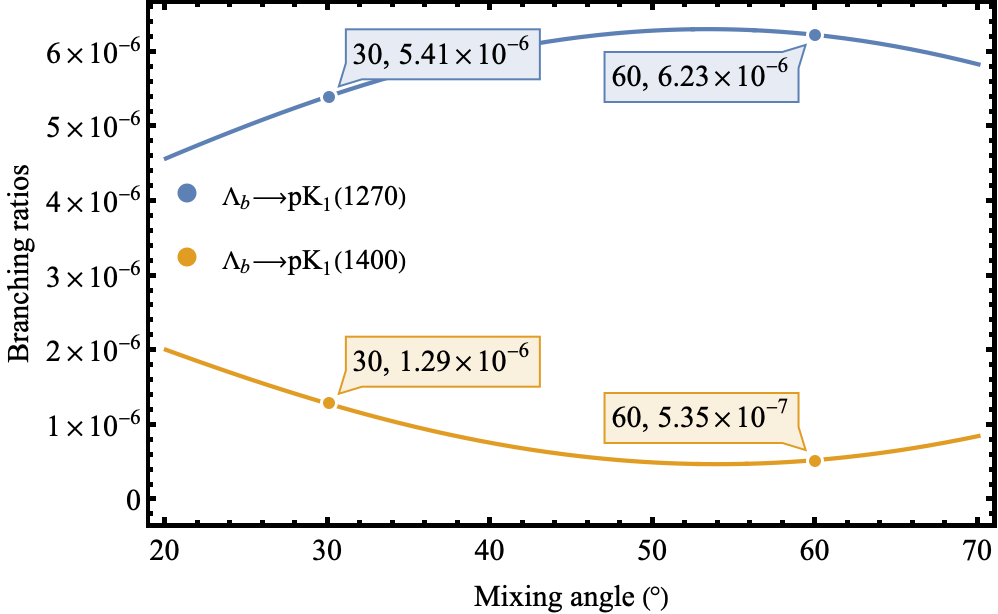}
	\end{minipage}
	\begin{minipage}{0.45\linewidth}
		\includegraphics[width=\linewidth]{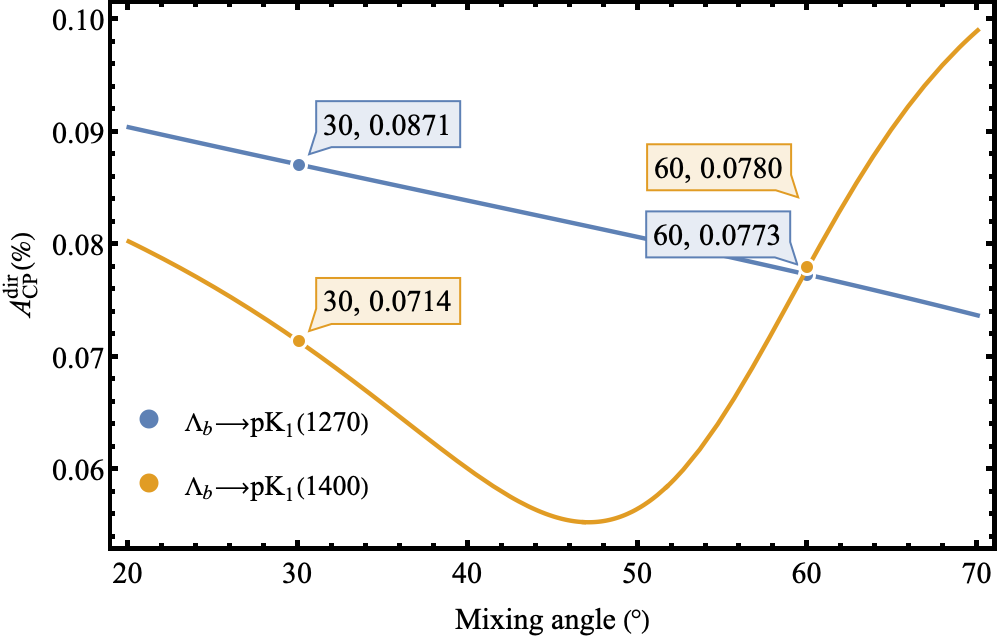}
	\end{minipage}
	
	\begin{minipage}{0.45\linewidth}
		\includegraphics[width=\linewidth]{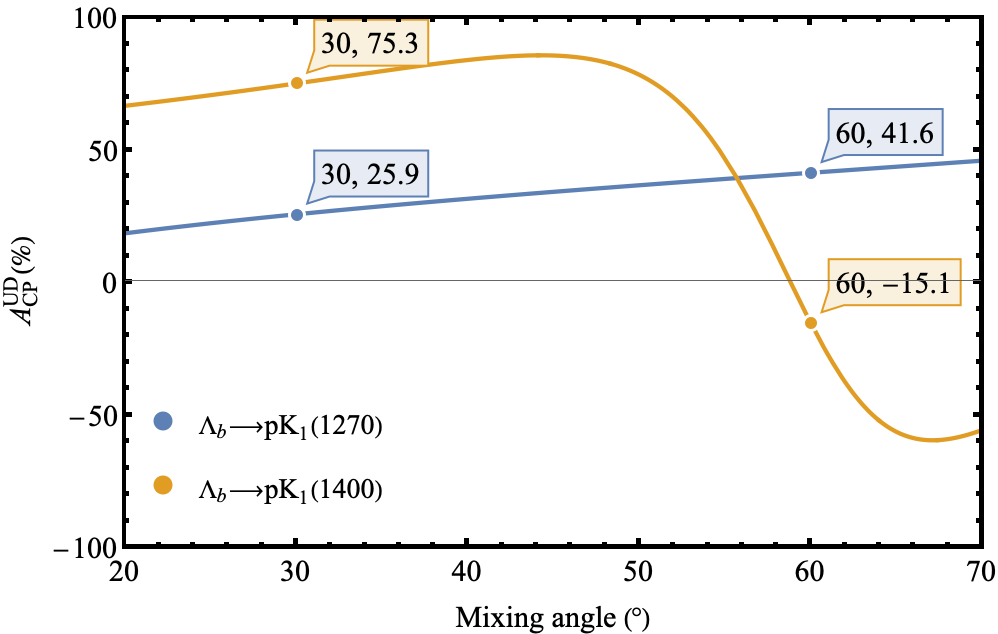}
	\end{minipage}
	\caption{(upper-left) Dependencies  of the $\Lambda_b\to pK_1$ branching ratios on the mixing angle $\theta_K$; (upper-right) same as the up-left panel but for the direct CPVs; (bottom) same as the up-left panel but for the CPVs in the up-down asymmetry.}
	\label{fig:depend-mixing-angle}
\end{figure}

The decays $\Lambda_b \to pA$, with $A$ denoting the axial-vector mesons $a^-_1(1260),K^-_1(1270),K^-_1(1400)$, share the same amplitude parametrization and  observables as  $\Lambda_b \to pV$ in the previous subsection. 
Because the $K_1(1270)$-$K_1(1400)$ mixing angle $\theta_K$ is not yet well determined \cite{Shi:2023kiy}, we take the typical values $\theta_K=30^\circ$ and $60^\circ$ for illustration. It is seen that most of our predictions in Table~\ref{tab:Lb2pA-observables} are insensitive to $\theta_K$. 
The predicted $\Lambda_b\to pa^-_1(1260)$ and $\Lambda_b \to pK^-_1(1270,1400)$ branching ratios are of the order $10^{-6}$, which can be verified experimentally.
The partial-wave CPVs of the $\Lambda_b \to pa^-_1(1260),pK^-_1(1270),pK^-_1(1400)$ decays are significant, exceeding $50\%$. 
The opposite signs between the partial-wave CPVs turn in the minor direct CPV of $\Lambda_b\to pa_1^-(1260)$. 
The direct CPVs of $\Lambda_b \to pK_1(1270)^-,pK_1(1400)^-$ can deviate from zero, 
both of which are above $+5\%$  in the range  $20^\circ\sim 70^\circ$ of the mixing angle as plotted in Fig.~\ref{fig:depend-mixing-angle}.

\begin{table}[htbp]
	\scriptsize
	\centering
	\renewcommand{\arraystretch}{1.5}
	\caption{Same as Table~\ref{tab:Lb2pP-observables} but for the $\Lambda_b\to pA$ decays.}
	\begin{tabular*}{160mm}{c@{\extracolsep{\fill}}ccc}
		\toprule[1pt]
		\toprule[0.7pt]
		& $Br(\times 10^{-6})$ & $A_{CP}^{dir}$ &$A_{CP}^{S^T}(\kappa_{S^T})$   \\
		\toprule[0.7pt]
		$\Lambda_b\to pa_1^-(1260)$ & $11.06^{+8.21+3.88+0.91+1.73}_{-4.30-3.32-0.46-0.06}$ & $-0.01^{+0.01+0.03+0.02+0.03}_{-0.00-0.01-0.02-0.00}$ &$-0.22^{+0.04+0.07+0.05+0.04}_{-0.03-0.07-0.07-0.01}(6\%)$    \\
		$\Lambda_b\to pK_1^-(1270)(30^\circ)$ & $5.48^{+3.63+1.94+0.27+2.49}_{-1.87-1.55-0.31-1.11}$ & $0.09^{+0.03+0.07+0.03+0.01}_{-0.04-0.02-0.02-0.00}$ &$0.34^{+0.00+0.01+0.01+0.00}_{-0.02-0.03-0.01-0.05}(8\%)$   \\
		$\Lambda_b\to pK_1^-(1400)(30^\circ)$ & $1.25^{+0.59+0.33+0.13+0.64}_{-0.39-0.19-0.19-0.31}$ & $0.06^{+0.03+0.05+0.03+0.00}_{-0.03-0.09-0.04-0.01}$ &$0.71^{+0.05+0.06+0.03+0.03}_{-0.02-0.16-0.04-0.13}(13\%)$   \\
		$\Lambda_b\to pK_1^-(1270)(60^\circ)$ & $6.28^{+3.97+1.93+0.18+2.79}_{-2.13-1.51-0.41-1.32}$ & $0.07^{+0.01+0.03+0.03+0.01}_{-0.04-0.04-0.03-0.00}$ & $0.46^{+0.00+0.00+0.02+0.01}_{-0.02-0.04-0.02-0.07}(9\%)$   \\
		$\Lambda_b\to pK_1^-(1400)(60^\circ)$ & $0.53^{+0.33+0.38+0.09+0.36}_{-0.16-0.19-0.22-0.13}$ & $0.08^{+0.11+0.09+0.12+0.00}_{-0.08-0.11-0.04-0.03}$ &$0.07^{+0.00+0.41+0.08+0.22}_{-0.12-0.09-0.15-0.10}(3\%)$   \\
		\toprule[0.7pt]
		& $A_{CP}^{S^L+D}(\kappa_{S^L+D})$ &  $A_{CP}^{P_1}(\kappa_{P_1})$ & $A_{CP}^{P_2}(\kappa_{P_2})$  \\
		\toprule[0.7pt]
		$\Lambda_b\to pa_1^-(1260)$ & $-0.11^{+0.02+0.01+0.02+0.02}_{-0.00-0.01-0.07-0.03}(46\%)$ & $0.18^{+0.03+0.02+0.04+0.09}_{-0.03-0.02-0.03-0.04}(40\%)$ & $-0.24^{+0.01+0.05+0.04+0.03}_{-0.02-0.09-0.06-0.06}(8\%)$  \\
		$\Lambda_b\to pK_1^-(1270)(30^\circ)$ & $-0.11^{+0.01+0.08+0.08+0.03}_{-0.04-0.06-0.03-0.00}(42\%)$ & $0.19^{+0.10+0.13+0.05+0.02}_{-0.06-0.09-0.11-0.01}(42\%)$ & $0.33^{+0.00+0.04+0.02+0.00}_{-0.02-0.03-0.02-0.03}(8\%)$ \\
		$\Lambda_b\to pK_1^-(1400)(30^\circ)$ & $0.81^{+0.09+0.17+0.07+0.04}_{-0.12-0.14-0.11-0.00}(17\%)$ & $-0.41^{+0.04+0.05+0.08+0.03}_{-0.07-0.05-0.11-0.04}(60\%)$ & $0.78^{+0.04+0.11+0.09+0.05}_{-0.06-0.20-0.04-0.10}(10\%)$ \\
		$\Lambda_b\to pK_1^-(1270)(60^\circ)$ & $0.06^{+0.01+0.08+0.07+0.03}_{-0.03-0.07-0.04-0.00}(37\%)$ & $-0.07^{+0.05+0.06+0.04+0.01}_{-0.06-0.05-0.05-0.02}(45\%)$ & $0.46^{+0.00+0.04+0.04+0.02}_{-0.01-0.03-0.02-0.06}(9\%)$ \\
		$\Lambda_b\to pK_1^-(1400)(60^\circ)$  & $-0.82^{+0.14+0.19+0.12+0.21}_{-0.07-0.09-0.07-0.02}(30\%)$ & $0.52^{+0.06+0.12+0.37+0.00}_{-0.01-0.14-0.03-0.07}(64\%)$ & $-0.28^{+0.27+0.04+0.03+0.03}_{-0.07-0.36-0.25-0.16}(3\%)$ \\
		\toprule[0.7pt]
		& $\alpha$ & $A_{CP}^\alpha$ & $\langle\alpha\rangle$  \\
		\toprule[0.7pt]
		$\Lambda_b\to pa_1^-(1260)$ & $-0.86^{+0.01+0.02+0.03+0.00}_{-0.03-0.05-0.05-0.03}$ & $0.02^{+0.00+0.01+0.01+0.00}_{-0.00-0.01-0.00-0.00}$ & $-0.88^{+0.01+0.02+0.04+0.00}_{-0.03-0.04-0.05-0.02}$\\
		$\Lambda_b\to pK_1^-(1270)(30^\circ)$ & $-1.00^{+0.00+0.00+0.00+0.00}_{-0.00-0.00-0.00-0.00}$ & $0.00^{+0.00+0.00+0.00+0.00}_{-0.00-0.00-0.00-0.00}$ & $-1.00^{+0.00+0.00+0.00+0.00}_{-0.00-0.00-0.00-0.00}$  \\
		$\Lambda_b\to pK_1^-(1400)(30^\circ)$ & $-1.00^{+0.00+0.01+0.01+0.00}_{-0.00-0.00-0.00-0.00}$ & $-0.01^{+0.01+0.01+0.01+0.00}_{-0.00-0.04-0.06-0.02}$ & $-0.98^{+0.01+0.05+0.07+0.02}_{-0.01-0.01-0.01-0.00}$\\
		$\Lambda_b\to pK_1^-(1270)(60^\circ)$ & $-1.00^{+0.00+0.00+0.00+0.00}_{-0.00-0.00-0.00-0.00}$ & $0.00^{+0.00+0.00+0.00+0.00}_{-0.00-0.00-0.00-0.00}$ & $-1.00^{+0.00+0.00+0.01+0.00}_{-0.00-0.00-0.00-0.00}$ \\
		$\Lambda_b\to pK_1^-(1400)(60^\circ)$ & $-0.96^{+0.01+0.10+0.10+0.04}_{-0.03-0.03-0.03-0.00}$ & $-0.00^{+0.01+0.01+0.02+0.00}_{-0.00-0.02-0.01-0.01}$ & $-0.96^{+0.00+0.09+0.08+0.05}_{-0.04-0.03-0.03-0.00}$  \\
		\toprule[0.7pt]
		& $\beta$ & $A_{CP}^\beta$ & $\langle\beta\rangle$  \\
		\toprule[0.7pt]
		$\Lambda_b\to pa_1^-(1260)$ & $-0.98^{+0.01+0.02+0.01+0.03}_{-0.01-0.02-0.01-0.00}$ & $-0.01^{+0.01+0.01+0.01+0.01}_{-0.01-0.01-0.01-0.00}$ & $-0.97^{+0.01+0.01+0.02+0.02}_{-0.01-0.01-0.02-0.01}$ \\
		$\Lambda_b\to pK_1^-(1270)(30^\circ)$ & $-0.99^{+0.00+0.03+0.03+0.00}_{-0.00-0.01-0.00-0.01}$ & $0.00^{+0.00+0.02+0.00+0.00}_{-0.01-0.02-0.00-0.01}$ & $-0.98^{+0.01+0.02+0.03+0.00}_{-0.00-0.00-0.00-0.00}$  \\
		$\Lambda_b\to pK_1^-(1400)(30^\circ)$ & $-0.15^{+0.28+0.17+0.13+0.25}_{-0.10-0.10-0.16-0.09}$ & $0.07^{+0.04+0.07+0.08+0.03}_{-0.01-0.02-0.04-0.03}$ & $-0.22^{+0.24+0.17+0.10+0.22}_{-0.10-0.11-0.15-0.06}$ \\
		$\Lambda_b\to pK_1^-(1270)(60^\circ)$ & $-0.95^{+0.04+0.06+0.04+0.01}_{-0.03-0.03-0.02-0.01}$ & $0.00^{+0.00+0.01+0.01+0.00}_{-0.01-0.01-0.01-0.00}$ & $-0.95^{+0.04+0.05+0.04+0.02}_{-0.02-0.03-0.01-0.01}$  \\
		$\Lambda_b\to pK_1^-(1400)(60^\circ)$ & $-0.01^{+0.01+0.07+0.10+0.09}_{-0.06-0.17-0.28-0.18}$ & $-0.26^{+0.09+0.09+0.03+0.03}_{-0.01-0.05-0.12-0.09}$ & $0.25^{+0.12+0.07+0.11+0.06}_{-0.15-0.16-0.18-0.09}$ \\
		\toprule[0.7pt]
		& $\gamma$ & $A_{CP}^\gamma$ & $\langle\gamma\rangle$  \\
		\toprule[0.7pt]
		$\Lambda_b\to pa_1^-(1260)$ & $-0.10^{+0.01+0.01+0.02+0.00}_{-0.01-0.02-0.02-0.01}$ & $0.04^{+0.01+0.02+0.01+0.01}_{-0.01-0.01-0.00-0.00}$ & $-0.14^{+0.01+0.01+0.02+0.00}_{-0.02-0.03-0.01-0.01}$  \\
		$\Lambda_b\to pK_1^-(1270)(30^\circ)$ & $-0.24^{+0.04+0.03+0.02+0.01}_{-0.03-0.03-0.02-0.04}$ & $-0.06^{+0.01+0.02+0.01+0.00}_{-0.01-0.01-0.01-0.00}$ & $-0.18^{+0.03+0.02+0.01+0.00}_{-0.02-0.02-0.01-0.04}$  \\
		$\Lambda_b\to pK_1^-(1400)(30^\circ)$ & $-0.61^{+0.15+0.23+0.13+0.00}_{-0.16-0.26-0.19-0.09}$ & $-0.27^{+0.06+0.09+0.06+0.00}_{-0.07-0.08-0.09-0.05}$ & $-0.34^{+0.08+0.14+0.07+0.00}_{-0.09-0.19-0.11-0.04}$  \\
		$\Lambda_b\to pK_1^-(1270)(60^\circ)$ & $-0.32^{+0.06+0.06+0.02+0.00}_{-0.04-0.06-0.03-0.06}$ & $-0.10^{+0.02+0.03+0.01+0.00}_{-0.02-0.01-0.01-0.01}$ & $-0.22^{+0.04+0.04+0.01+0.00}_{-0.02-0.05-0.02-0.05}$ \\
		$\Lambda_b\to pK_1^-(1400)(60^\circ)$ & $-0.05^{+0.02+0.02+0.01+0.02}_{-0.01-0.06-0.03-0.01}$ & $0.01^{+0.01+0.03+0.03+0.00}_{-0.01-0.01-0.00-0.01}$ & $-0.06^{+0.01+0.01+0.01+0.02}_{-0.01-0.08-0.06-0.01}$ \\
		\toprule[0.7pt]
		& $\Lambda$ & $A_{CP}^\Lambda$ & $\langle\Lambda\rangle$  \\
		\toprule[0.7pt]
		$\Lambda_b\to pa_1^-(1260)$ & $-0.97^{+0.00+0.02+0.01+0.03}_{-0.01-0.02-0.01-0.00}$ & $-0.01^{+0.01+0.01+0.01+0.01}_{-0.01-0.01-0.01-0.00}$ & $-0.96^{+0.01+0.01+0.02+0.02}_{-0.01-0.01-0.02-0.01}$ \\
		$\Lambda_b\to pK_1^-(1270)(30^\circ)$ & $-0.99^{+0.00+0.02+0.03+0.00}_{-0.00-0.01-0.00-0.01}$ & $0.00^{+0.00+0.02+0.00+0.00}_{-0.01-0.02-0.00-0.01}$ & $-0.99^{+0.01+0.01+0.02+0.00}_{-0.00-0.00-0.00-0.00}$  \\
		$\Lambda_b\to pK_1^-(1400)(30^\circ)$ & $-0.47^{+0.14+0.16+0.06+0.14}_{-0.02-0.11-0.14-0.08}$ & $-0.07^{+0.06+0.08+0.05+0.00}_{-0.02-0.04-0.04-0.03}$ & $-0.40^{+0.17+0.09+0.08+0.15}_{-0.06-0.11-0.13-0.05}$ \\
		$\Lambda_b\to pK_1^-(1270)(60^\circ)$ & $-0.96^{+0.03+0.05+0.03+0.01}_{-0.03-0.02-0.01-0.01}$ & $0.00^{+0.00+0.01+0.00+0.00}_{-0.01-0.00-0.01-0.00}$ & $-0.96^{+0.03+0.04+0.04+0.01}_{-0.02-0.02-0.01-0.01}$  \\
		$\Lambda_b\to pK_1^-(1400)(60^\circ)$ & $-0.06^{+0.10+0.08+0.10+0.11}_{-0.07-0.17-0.29-0.18}$ & $-0.23^{+0.09+0.08+0.03+0.02}_{-0.01-0.01-0.08-0.08}$ & $0.18^{+0.11+0.07+0.11+0.09}_{-0.14-0.17-0.22-0.10}$ \\
		\toprule[0.7pt]
		& $a_{UD}$ & $A_{CP}^{UD}$ & \\
		\toprule[0.7pt]
		$\Lambda_b\to pa_1^-(1260)$ & $-0.09^{+0.00+0.01+0.02+0.00}_{-0.01-0.01-0.01-0.01}$ & $-0.24^{+0.03+0.05+0.05+0.03}_{-0.03-0.09-0.06-0.06}$ &  \\
		$\Lambda_b\to pK_1^-(1270)(30^\circ)$ & $-0.19^{+0.03+0.02+0.01+0.01}_{-0.02-0.02-0.01-0.02}$ & $0.26^{+0.02+0.03+0.01+0.00}_{-0.03-0.08-0.04-0.04}$ &  \\
		$\Lambda_b\to pK_1^-(1400)(30^\circ)$ & $-0.38^{+0.06+0.10+0.05+0.00}_{-0.06-0.09-0.07-0.03}$ & $0.72^{+0.05+0.13+0.07+0.05}_{-0.05-0.23-0.03-0.12}$ &\\
		$\Lambda_b\to pK_1^-(1270)(60^\circ)$ & $-0.24^{+0.04+0.04+0.01+0.00}_{-0.02-0.03-0.02-0.03}$ & $0.40^{+0.02+0.03+0.02+0.01}_{-0.01-0.04-0.03-0.07}$ & \\
		$\Lambda_b\to pK_1^-(1400)(60^\circ)$ & $-0.04^{+0.02+0.02+0.01+0.02}_{-0.01-0.05-0.03-0.01}$ & $-0.19^{+0.12+0.14+0.00+0.06}_{-0.18-0.19-0.20-0.00}$ &  \\
		\toprule[0.7pt]
		\toprule[1pt]
	\end{tabular*}
	\label{tab:Lb2pA-observables}
\end{table}

\begin{figure}
	\centering
	\includegraphics[width=0.4\linewidth]{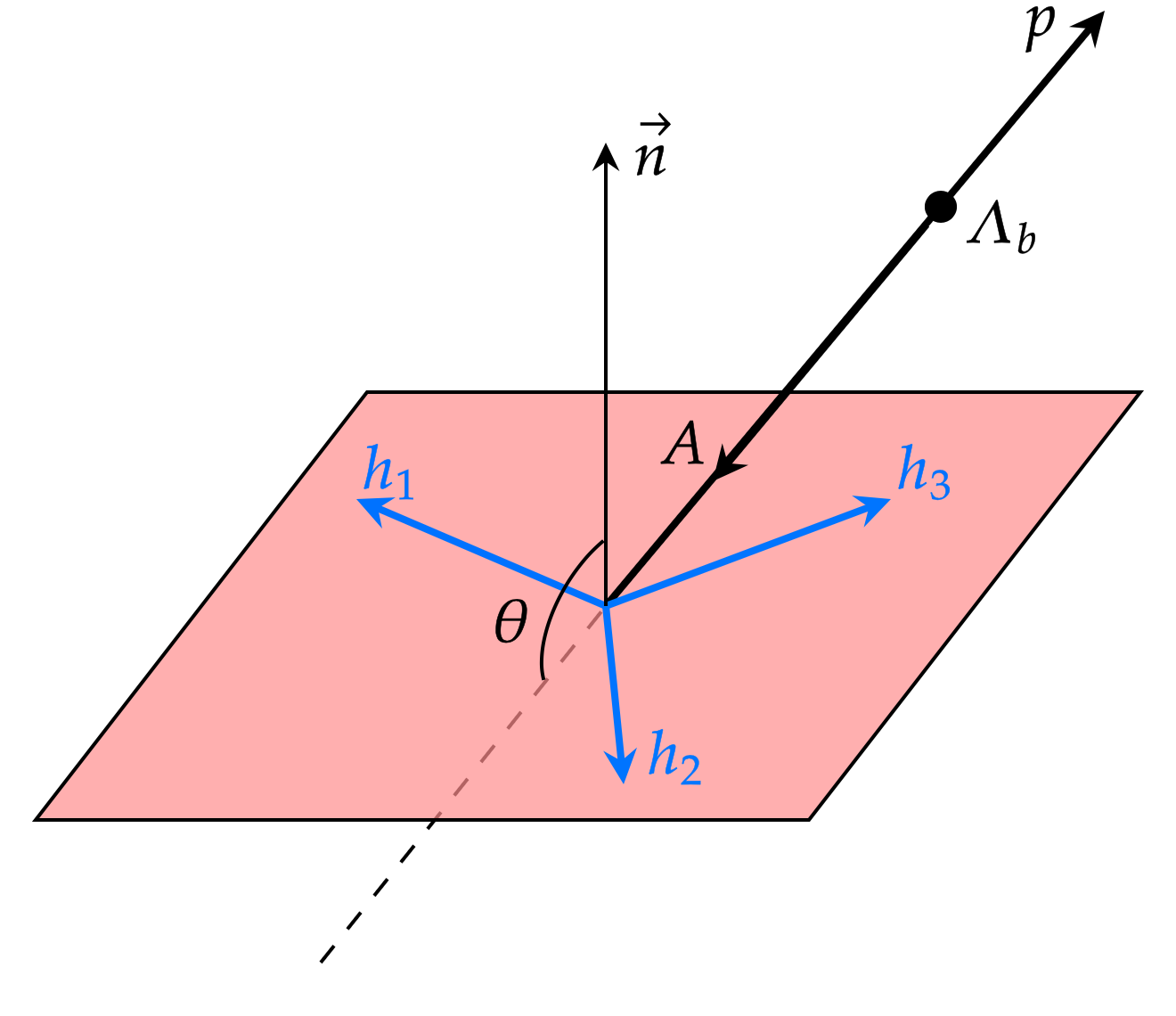}
	\caption{Configuration of the $\Lambda_b\to pA\to ph_1h_2h_3$ decay, where $\theta$ is the angle between the normal of the $A\to h_1h_2h_3$ decay plane and the $A$ meson momentum in the $\Lambda_b$ rest frame.}
	\label{fig:angle-distri-pa}
\end{figure}

The $\Lambda_b \to pa^-_1(1260),pK^-_1(1270),pK^-_1(1400)$ modes are actually the four-body decays, for $a^-_1(1260)$ and $K_1^-(1270,1400)$ decay into $\pi^-\pi^+\pi^-$ and $K^-\pi^+\pi^-$, respectively. 
We propose the promising observables associated with the angle distributions of the final states in these multi-body $\Lambda_b$ decays, which can be measured in the future.
The angular distribution for the $\Lambda_b\to pA\to ph_1h_2h_3$ decay  in Fig.~\ref{fig:angle-distri-pa} is given by~\cite{Wang:2024qff}
\begin{equation}
	\begin{split}
		\frac{d\Gamma}{d\cos\theta}\propto& \frac{ \left(|H_{1/2,1}|^2+|H_{-1/2,-1}|^2\right)\frac{1+\cos^2\theta}{2} + \left(|H_{1/2,0}|^2+|H_{-1/2,0}|^2\right)\sin^2\theta }{|H_{\frac{1}{2},0}|^2+|H_{-\frac{1}{2},0}|^2+|H_{-\frac{1}{2},-1}|^2+|H_{\frac{1}{2},1}|^2}\\
		&+\frac{ R \left(|H_{1/2,1}|^2-|H_{-1/2,-1}|^2\right)\cos\theta}{|H_{\frac{1}{2},0}|^2+|H_{-\frac{1}{2},0}|^2+|H_{-\frac{1}{2},-1}|^2+|H_{\frac{1}{2},1}|^2},
	\end{split}
\end{equation}
with $\theta$ being the angle between the normal of the $A \to h_1h_2h_3$ decay plane and the $A$ meson momentum in the $\Lambda_b$ rest frame, i.e., $\cos\theta=\vec n \cdot \vec p_A=(\vec p_1\times \vec p_2)\cdot \vec p_A$.
The nonperturbative factor $R$ parametrizes the effect and kinematics of the strong decay $A\to h_1h_2h_3$~\cite{Berman:1965gi}, which will be canceled in the proposed CPV observable below.
The up-down asymmetry is then constructed,
\begin{equation}
	\begin{split}
		A_{UD}=&\frac{\Gamma(\cos\theta>0) - \Gamma(\cos\theta<0)}{\Gamma(\cos\theta>0) + \Gamma(\cos\theta<0)}\\
		=&\frac{ R \left(|H_{1/2,1}|^2-|H_{-1/2,-1}|^2\right)}{\left(|H_{1/2,1}|^2+|H_{-1/2,-1}|^2+|H_{1/2,0}|^2+|H_{-1/2,0}|^2\right)}\\
		\equiv& R\cdot a_{UD},
		\label{eq:AUDR}
	\end{split}
\end{equation}
which defines the $R$-independent CPV
\begin{equation}
	A_{CP}^{UD}=\frac{A_{UD}+\bar{A}_{UD}}{A_{UD}-\bar{A}_{UD}},
	\label{eq:AUDCP}
\end{equation}
with $a_{UD}=Re(S^T P_2^\ast)$ and $\bar{A}_{UD}$ representing the charge conjugate of $A_{UD}$.

The numerical outcomes for $a_{UD}$ and $A_{CP}^{UD}$ are gathered in Table~\ref{tab:Lb2pA-observables} and their dependencies on the mixing angle are displayed in Fig.~\ref{fig:depend-mixing-angle}, where
$A_{CP}^{UD}$ of the $\Lambda_b\to pK_1(1270)$ decay is more stable against the variation of the mixing angle.
We highlight that  $A_{CP}^{UD}$ exceed $20\%$ in the $\Lambda_b \to pA$ decays owing to the large strong phase difference between the $S$- and $P$-wave amplitudes.
Therefore, $A_{CP}^{UD}$ with controllable uncertainties serve as  ideal observables for establishing CPVs in  $\Lambda_b$ baryon decays experimentally.
The measurements of $A_{CP}^{UD}$ are based on the four-body decays $\Lambda_b\to  p \pi^+ \pi^- \pi^-$ and $p K^- \pi^+ \pi^-$ for $pa_1(1260)$ and $pK_1(1270)$, respectively, both of which have abundant data at the LHCb.

The LHCb has announced the direct CPVs of the above modes from the Run 1 data corresponding to the integrated luminosity 3 fb$^{-1}$ at the center-of-mass energies $\sqrt{s}=$ 7 and 8 TeV \cite{LHCb:2019jyj}. The sum of particle and anti-particle channels gives the signal yields of $\Lambda_b\to pa_1$ and $pK_1$ around 800 and 1000, respectively. 
The yields would be enhanced by four times with the Run 2 data from twice the integrated luminosity of 6 fb$^{-1}$ and twice the cross section at a higher collision energy of $\sqrt{s}=13$ TeV.
More $b$ quarks and associated $b$-hadrons are produced at higher energies. As reported by the LHCb~\cite{LHCb:2016qpe}, the $b$-quark production cross sections are measured to be $72.0\pm0.3\pm6.8$ $\mu$b at $\sqrt{s}=7$ TeV and $144\pm1\pm21$ $\mu$b at $\sqrt{s}=13$ TeV, respectively, showing an increase by a factor of two from Run 1 to Run 2. 
The number of signal events in Run 2 is thus expected be four times greater than those in Run 1, with a factor of two from luminosities and another factor of two from cross sections. 
Using the combined data of Run 1 and Run 2, we would get the signal yields five times higher relative to Run 1, which amount to 4000 and 5000 events for $\Lambda_b\to p a_1$ and $p K_1$, respectively.

\begin{figure}[phtb]
	\centering
	\includegraphics[scale=0.3]{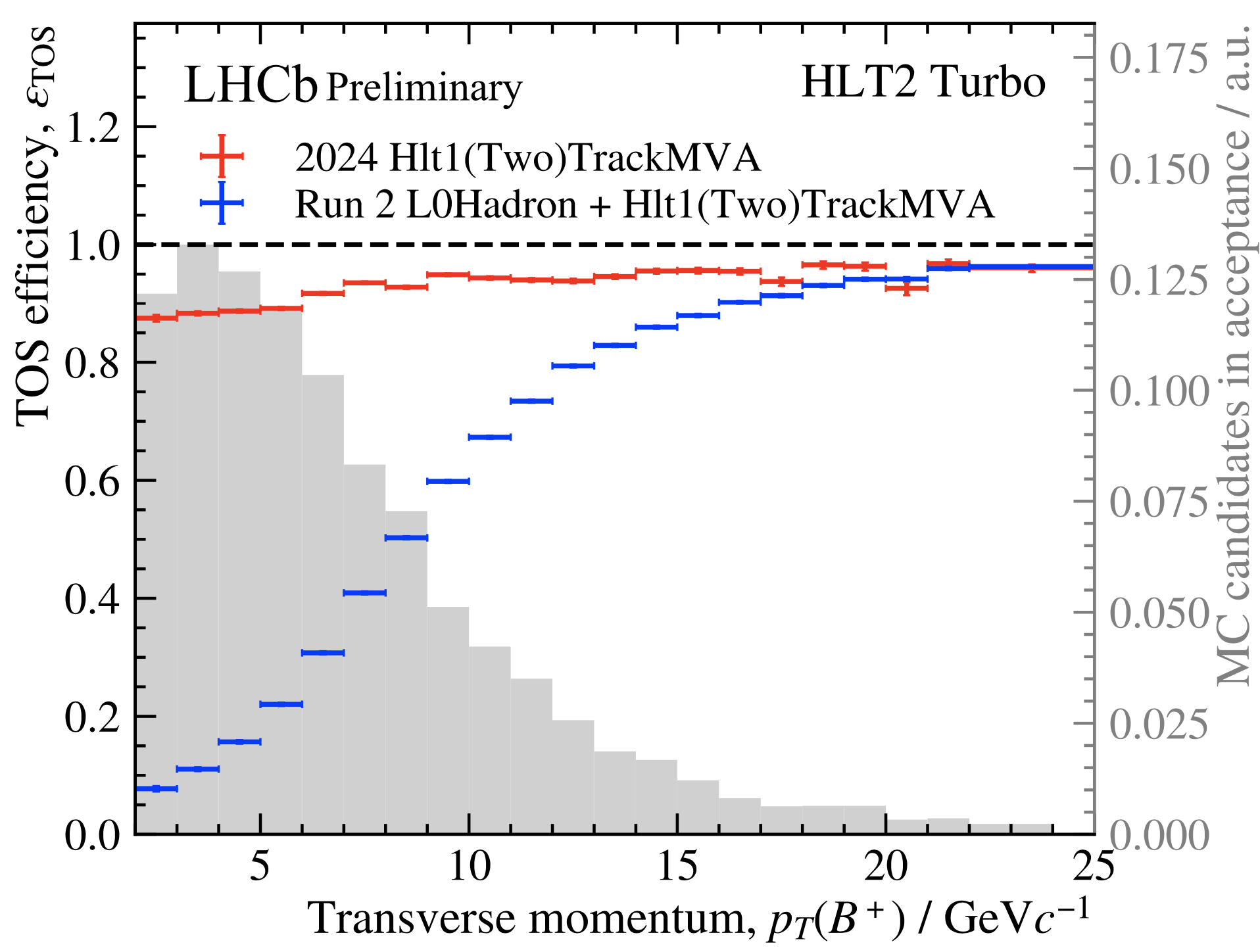}
	\caption{Comparison of trigger efficiencies for LHCb Run 2 and Run 3 taken from the LHCb public website: https://lbfence.cern.ch/alcm/public/figure/details/3837.   }\label{fig:trigger}
\end{figure}

The trigger efficiencies for Run 3 are higher than those for Run 2, especially in the low $p_T$ region, as indicated in Fig.~\ref{fig:trigger}.
The red points mark the efficiencies for Run 3 in 2024, and the blue points mark those for Run 2. 
The gray bars display the Monte-Carlo candidates in acceptance. 
The number of signal events can be deduced by convoluting the gray bars with the red or blue points.
The signal events in Run 3 are then found to be roughly twice as many as those in Run 2.
This figure depicts the trigger efficiencies for hadronic $B$ meson decays, which can be extrapolated to the case of hadronic $\Lambda_b$ baryon decays. 
According to the plan~\cite{Chen:2021ftn}, the LHCb will reach the
integrated luminosity of 50 fb$^{-1}$ at the end of Run 4.
Since LHCb Run 3 is currently ongoing and its final integrated luminosity remains uncertain, we will estimate the signal events for the combined dataset of Run 3 and Run 4 by assuming that the conditions and performance of Run 4 are identical to those of Run 3.
The ratio between the events from combined Run 3+4 and those from Run 2 is about $(50-9)$ fb$^{-1}/6$ fb$^{-1}\times 2\sim14$, where the factor of 2 comes from the enhancement of trigger efficiencies for Run 3 \cite{TriggerEfficiency} implied by Fig.~\ref{fig:trigger}. 
Taking into account the factor of 4 between the Run 2 and Run 1 data, the ratio of the events from Run 1+2+3+4 to those from Run 1 is nearly $1+4+4\times14\sim60$.
That is, the total signal events of $\Lambda_b\to p a_1$ and $p K_1$ from the full Run 1+2+3+4 data can attain $800\times 60=48000$ and $1000\times 60=60000$, respectively.

\begin{table}[phtb]
	\centering
	\renewcommand{\arraystretch}{1.3}
	\caption{Predictions for $A_{UD}$ and $A_{UD}^{CP}$ in the $\Lambda_b\to p a_1$, $p K_1$ decays.}
	\begin{tabular}{c|c|c|c}
		\hline
		\hline
		& $\Lambda_b\to p a_1$ & $\Lambda_b\to p K_1(\theta_K=30^\circ)$ & $\Lambda_b\to p K_1(\theta_K=60^\circ)$  \\
		\hline
		$A_{UD}$       & $(-9^{+2}_{-2})\%$ & $(-19^{+4}_{-4})\%$ & $(-24^{+6}_{-5})\%$ \\
		\hline
		$A_{UD}^{CP}$  & $(-24^{+8}_{-13})\%$ & $(+26^{+4}_{-10})\%$ & $(+40^{+4}_{-9})\%$ \\
		\hline
		\hline
	\end{tabular}
	\label{tab:AUD}
\end{table}

Both the $A_{UD}$ and $A_{UD}^{CP}$ values are essential for evaluating the experimental sensitivities, whose predictions are provided in Table~\ref{tab:AUD}.
Recall the definition of $A_{UD}$,
\begin{equation}
	\begin{split}
		A_{UD}=\frac{\Gamma(\cos\theta>0)-\Gamma(\cos\theta<0)}{\Gamma(\cos\theta>0)+\Gamma(\cos\theta<0)}
		=\frac{N_>-N_<}{N_>+N_<},
	\end{split}
\end{equation}
where $N$ are the numbers of signal events with $N_{>}=N(\cos\theta>0)$ and $N_{<}=N(\cos\theta<0)$. The statistical error of $A_{UD}$ follows the error propagation formula
\begin{align}
	\sigma_{A_{UD}}&=\sqrt{\left(\frac{\partial A_{UD}}{\partial N_>}\right)^2\sigma_{N_>}^2+\left(\frac{\partial A_{UD}}{\partial N_<}\right)^2\sigma_{N_<}^2}\nonumber\\
	&=\sqrt{\left(\frac{2N_<}{(N_>+N_<)^2}\right)^2\sigma_{N_>}^2+\left(\frac{-2N_>}{(N_>+N_<)^2}\right)^2\sigma_{N_<}^2},
	\label{eq:sigma_AUD}
\end{align}
with $\sigma_{N_{>,<}}$ being the statistical errors of $N_{>,<}$. 
The neglect of background events for a naive analysis leads to the approximate uncertainties $\sigma_{N_>}\approx\sqrt{N_>}$ and $\sigma_{N_<}\approx\sqrt{N_<}$.
As a simple estimate of the uncertainties and experimental sensitivities, we assume $N_>=N_<=\overline N_>=\overline N_<=N_{\rm tot}/4$, where $N_{\rm tot}=48000$ for $\Lambda_b\to p a_1$ and $N_{\rm tot}=60000$ for $\Lambda_b\to pK_1$. The statistical uncertainty of $A_{UD}$ is then $\sigma_{A_{UD}}\approx \sqrt{2/N_{\rm tot}}$,
inferring the experimental uncertainties $\sigma_{A_{UD}}(\Lambda_b\to p a_1)\approx 0.65\%$ and $\sigma_{A_{UD}}(\Lambda_b\to p K_1)\approx 0.58\%$ from the full Run 1+2+3+4 data. 

The background effects are expected to be less than $30\%$ on $\sigma_{A_{UD}}$, as reflected by the LHCb data; 
the signal yields of $\Lambda_b\to p\pi\pi\pi$ are $6646\pm105$ corresponding to the integrated luminosity of 3 fb$^{-1}$ in~\cite{LHCb:2016yco}, and $27600\pm200$ for 6.6 fb$^{-1}$ in~\cite{LHCb:2019oke}, where the errors include background contributions. 
The uncertainties of the pure signal events $\sqrt{6646}=81.5$ and $\sqrt{27600}=166.1$ are lifted by $29\%$ to 105 and by $20\%$ to 200, respectively. 
Hence, the uncertainties are found to be $\sigma_{A_{UD}}(\Lambda_b\to p a_1)\approx 0.85\%$ and $\sigma_{A_{UD}}(\Lambda_b\to p K_1)\approx 0.75\%$, amplified by a factor of 1.3. 

We then proceed to the statistical uncertainty of $A_{UD}^{CP}=(A_{UD}+\overline A_{UD})/(A_{UD}-\overline A_{UD})$. Note that the quantity $\overline A_{UD} $ acquires an opposite sign relative to $A_{UD}$ under the parity transformation, because $A_{UD}$ is parity-odd. 
Analogous to Eq.~(\ref{eq:sigma_AUD}), the uncertainty of $A_{UD}^{CP}$ is derived from 
\begin{align}
	\sigma_{A_{UD}^{CP}}=\sqrt{\left(\frac{\partial A_{UD}^{CP}}{\partial A_{UD}}\right)^2\sigma_{A_{UD}}^2 + \left(\frac{\partial A_{UD}^{CP}}{\partial \overline A_{UD}}\right)^2\sigma_{\overline A_{UD}}^2  }
	\approx \frac{1}{\sqrt{2}|A_{UD}|}\sigma_{A_{UD}},
\end{align}
where the last expression is justified by the assumptions $\overline A_{UD}=-A_{UD}$ and $\sigma_{\overline A_{UD}}=\sigma_{A_{UD}}$.
The magnitude of $A_{UD}$ is crucial for obtaining the uncertainty of $A_{UD}^{CP}$.
The insertion of the central values of $A_{UD}$ in Table~\ref{tab:AUD} yields the statistical uncertainties of $A_{UD}^{CP}$, 
$\sigma_{A_{UD}^{CP}}(\Lambda_b\to p a_1)\approx \frac{0.85\%}{\sqrt{2}\times 9\%}=6.7\%$,
$\sigma_{A_{UD}^{CP}}(\Lambda_b\to p K_1(\theta_K=30^\circ))\approx \frac{0.75\%}{\sqrt{2}\times 19\%}=2.8\%$
and $\sigma_{A_{UD}^{CP}}(\Lambda_b\to p K_1(\theta_K=60^\circ))\approx \frac{0.75\%}{\sqrt{2}\times 24\%}=2.2\%$.

Considering the differences of event numbers between particles and antiparticles caused by direct CPVs, between $N_>$ and $N_<$ caused by $A_{UD}$, and between $A_{UD}$ and $\overline A_{UD}$ arising from $A_{UD}^{CP}$, we refine the above crude estimates, arriving at the uncertainties $\sigma_{A_{UD}^{CP}}(\Lambda_b\to p a_1)=5.2\%$, $\sigma_{A_{UD}^{CP}}(\Lambda_b\to p K_1(\theta_K=30^\circ))=3.8\%$ and $\sigma_{A_{UD}^{CP}}(\Lambda_b\to p K_1(\theta_K=60^\circ))=3.5\%$.
The decrease of $\sigma_{A_{UD}^{CP}}(\Lambda_b\to p a_1)$ and the increase of $\sigma_{A_{UD}^{CP}}(\Lambda_b\to p K_1)$ stem from the negativity of ${A_{UD}^{CP}}(\Lambda_b\to p a_1)$ and the positivity of ${A_{UD}^{CP}}(\Lambda_b\to p K_1)$, respectively. 
Referring to the predicted $A_{UD}^{CP}$ in Table~\ref{tab:AUD}, we conclude that the CPV induced by the up-down asymmetry in $\Lambda_b\to p K_1$ could attain a statistic significance greater than $5\sigma$, while the CPV in $\Lambda_b\to p a_1$ is at the edge of $5\sigma$.

At last, we come to the issue of systematic uncertainties.
We can learn from the measurements of $\hat T$-odd triple-product-asymmetry CPVs of $\Lambda_b\to p\pi^-\pi^+\pi^-$ in~\cite{LHCb:2016yco} and ~\cite{LHCb:2019oke}, which
are defined as $C_{\hat T}=\vec p_p\cdot (\vec p_{\pi_{\rm fast}^-}\times \vec p_{\pi^+})$ for the $\hat T$-odd triple product of final-state particle momenta in the $\Lambda_b$ rest frame. Note that the quasi-time reversal transformation $\hat T$ reverses particle spins and momenta, while maintaining the initial and final states.
The definitions of the triple product asymmetries (TPAs)  
\begin{equation}
	A_{\hat T}=\frac{N(C_{\hat T}>0)-N(C_{\hat T}<0)}{N(C_{\hat T}>0)+N(C_{\hat T}<0)},~~~~\bar A_{\hat T}=\frac{\bar N(-\bar C_{\hat T}>0)-\bar N(-\bar C_{\hat T}<0)}{\bar N(-\bar C_{\hat T}>0)+\bar N(-\bar C_{\hat T}<0)},
\end{equation}
are similar to that for the up-down asymmetries in Eq.~(\ref{eq:AUDR}). 
The $\hat T$-odd $CP$- and $P$-violating asymmetries,
\begin{equation}
	a_{CP}^{\hat T\text{-odd}}=\frac{1}{2}(A_{\hat T}-\bar A_{\hat T}),~~~~a_{P}^{\hat T\text{-odd}}=\frac{1}{2}(A_{\hat T}+\bar A_{\hat T}), 
\end{equation}
correspond to the numerator and the denominator, respectively, of the definition for $A_{CP}^{UD}$ in Eq.~(\ref{eq:AUDCP}). 

It has been known that the main sources of systematic uncertainties in the TPA analysis include selection criteria, reconstruction and detector acceptance based on the control sample of $\Lambda_b^0\to \Lambda_c^+(\to pK^-\pi^+)\pi^-$.
The systematic uncertainties are evaluated as $0.32\%$ for both $a_{CP}^{\hat T\text{-odd}}$ and $a_{P}^{\hat T\text{-odd}}$ with the Run 1 data in~\cite{LHCb:2016yco}, and $0.2\%$ with more data in~\cite{LHCb:2019oke}. 
The systematic uncertainties of both $A_{UD}+\bar A_{UD}$ and $A_{UD}-\bar A_{UD}$ can thus be taken as $0.2\%$ under the assumption that it is unchanged with more data samples. 
Using the error propagation formula with the fully positive correlation between $A_{UD}+\bar A_{UD}$ and $A_{UD}-\bar A_{UD}$, we obtain the systematic uncertainty of $A_{CP}^{UD}$ as about $0.2\%/(2|A_{UD}|)$.
Namely, the systematic uncertainties of $A_{CP}^{UD}$ are $1.1\%$ for $\Lambda_b\to p a_1$, $0.5\%$ for $\Lambda_b\to p K_1(\theta_K=30^\circ)$ and $0.4\%$ for $\Lambda_b\to p K_1(\theta_K=60^\circ)$, which are less than the statistic uncertainties. 
Therefore, there is a great chance to identify the order-of-$20\%\sim 40\%$ up-down asymmetric CPVs in $\Lambda_b\to p a_1$ and $p K_1$ in the near future, and to establish CPVs in baryon decays.

\section{Summary}\label{sec:summary}

We have performed the first full QCD analysis on two-body hadronic $\Lambda_b\to ph$ decays in the PQCD approach. 
We calculated the various topological amplitudes, including their strong phases, by incorporating the higher-twist hadron DAs.
It is found that the subleading power corrections are in fact sizable in heavy baryon decays.

\begin{figure}[htbp]
	\centering
	\includegraphics[width=0.9\linewidth]{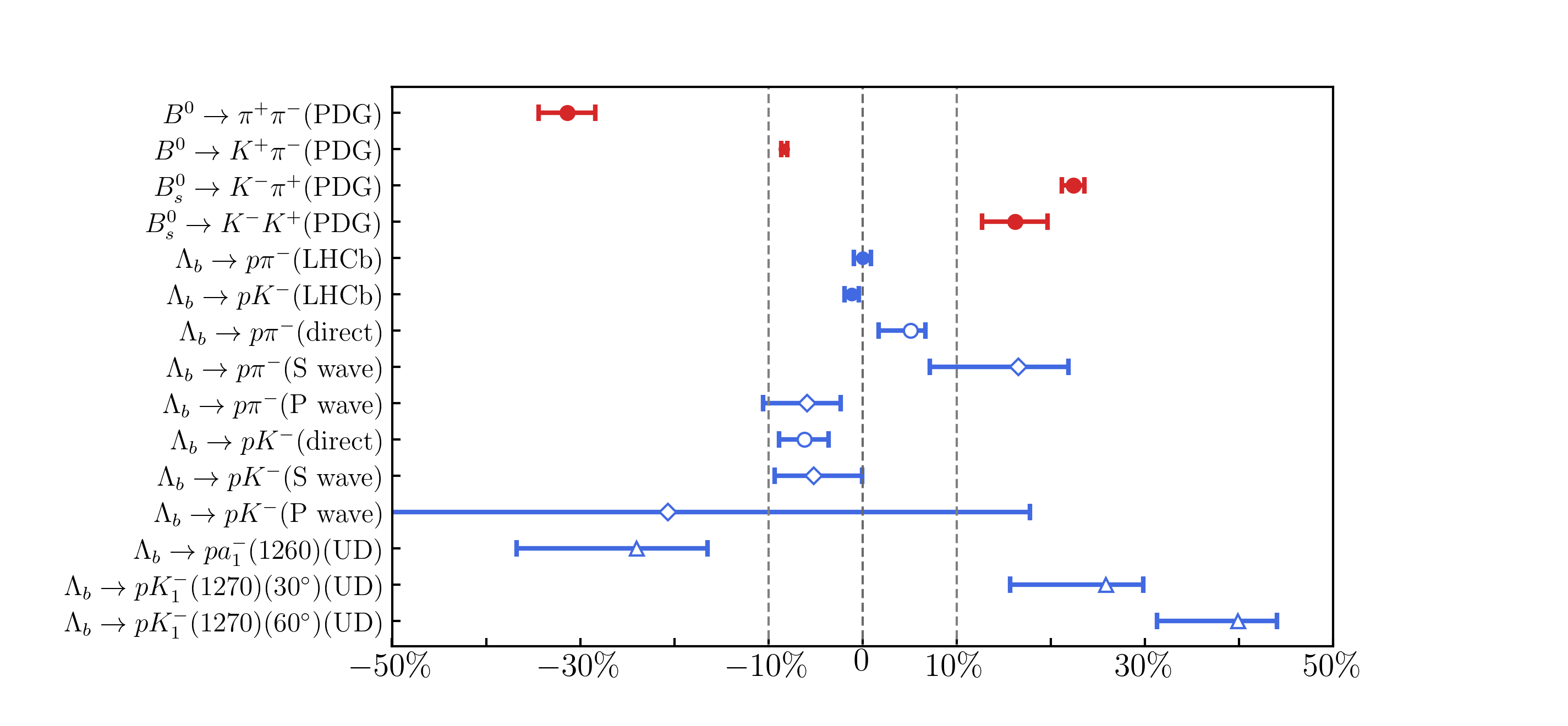}
	\caption{CPVs measured in $B$ meson and $\Lambda_b\to p\pi^-,pK^-$ decays, and our predictions.}
	\label{fig:expandpqcd}
\end{figure}

Our predictions for the total and partial-wave CPVs of the considered modes are summarized and compared with the available data for bottom hadron decays in Fig.~\ref{fig:expandpqcd}~\cite{Han:2024kgz}.
We elucidated the mechanism responsible for the measured small CPVs in $\Lambda_b\to p\pi^-,pK^-$, in contrast to the sizable CPVs in the similar $B$ meson decays. 
The partial-wave CPVs in $\Lambda_b\to p\pi^-$ attains $10\%$ potentially, but the destruction between them leads to the tiny CPV. 
The direct CPV of $\Lambda_b\to pK^-$ is primarily attributed to the modest $S$-wave CPV.

We have extended our investigation to the CPVs in the modes with vector and axial-vector final states. 
The partial-wave CPVs of the $\Lambda_b\to pV$, $pA$ decays also exceed $10\%$; for examples, the $P_2\text{-wave}$ CPVs of $\Lambda_b\to p\rho^-$ and $pa_1^-(1260)$ are $17\%$ and $-24\%$, respectively;
the $(D+S^L)\text{-wave}$ CPV of $\Lambda_b\to pK^{*-}$ is $27\%$;
the partial-wave CPVs of $\Lambda_b \to pK^-_1(1270)$ even amount to the order of $30\%$.
Nevertheless, the direct CPVs of the $\Lambda_b\to pK^{*-}$, $pa_1^-(1260)$ and $pK_1(1270)$ decays diminish as well, which trace back to the strong cancellation of the major $(D+S^L)$- and $P_1\text{-wave}$ CPVs, like the  $\Lambda_b\to p\pi^-$ case.

Our work suggests that certain partial-wave CPVs in bottom baryon decays, especially those related to angular distributions, can be sufficiently significant and probed to establish baryon CPVs. 
The decay asymmetry parameters of two-body $\Lambda_b$ modes are also predicted for future experimental confrontations.
It opens avenues for deeply understanding the dynamics in heavy baryon decays and their CPVs. 
The rich data samples and complicated dynamics in multi-body bottom baryon decays offer remarkable prospects for exploring baryon CPVs. 

\section*{Acknowledgements}
We acknowledge Jun Hua, Pei-Rong Li, Yan-Qing Ma, Ding-Yu Shao, Jian Wang and Yan-Xi Zhang for their valuable comments. J.J.H. and J.X.Y. wish to thank Fanrong Xu for the warm hospitality during their visit to Jinan University.
This work was supported in part by Natural Science Foundation of China under
Grant No. 12335003 and No. 12505113, by the Fundamental Research Funds for the Central Universities under No. lzujbky-2023-stlt01, lzujbky-2023-it12, lzujbky-2024-oy02 and lzujbky-2025-eyt01, and by the Super Computing Center at Lanzhou University.

\appendix

\section{Auxiliary functions}\label{app:aux-function}

The auxiliary functions $F_{1,2,3,4}$ as the results of the Fourier transformation of the hard decay kernels from the $k_T$ space to the $b$ space are collected below:
\begin{equation}
	\begin{split}
		&F_1(A,b)=\int d^2k_T\frac{e^{i\textbf{k}_T\cdot\textbf{b}}}{k^2+A} = 2\pi \left\{ K_0(\sqrt{Ab^2})\Theta(A) + \frac{i\pi}{2} H_0^{(1)}(\sqrt{-Ab^2})\Theta(-A)\right\},
	\end{split}
\end{equation}

\begin{equation}
	\begin{split}
		F_2(A,B,b)&=\int d^2k_T\frac{e^{i\textbf{k}_T\cdot\textbf{b}}}{(k^2+A)(k^2+B)}\\
		&=\int_{0}^{1}dz \frac{\pi b}{\sqrt{|Z_1|}}\left\{ K_1(\sqrt{Z_1b^2})\Theta(Z_1)
		-\frac{i\pi}{2}H_1^{(1)}(\sqrt{-Z_1b^2})\Theta(-Z_1) \right\},
	\end{split}
\end{equation}

\begin{equation}
	\begin{split}
		&F_3(A,B,C,b_1,b_2)=\int d^2k_{1T}\int d^2k_{2T}\frac{e^{i(\textbf{k}_{1T}\cdot\textbf{b}_1+\textbf{k}_{2T}\cdot\textbf{b}_2)}}{(k_1^2+A)(k_2^2+B)((k_1+k_2)^2+C)}\\
		&=\int_{0}^{1}\frac{dz_1dz_2}{z_1(1-z_1)} \frac{\pi^2\sqrt{X_2}}{\sqrt{|Z_2|}} \left\{  K_1(\sqrt{X_2Z_2})\Theta(Z_2) - \frac{i\pi}{2}H_1^{(1)}(\sqrt{-X_2Z_2})\Theta(-Z_2) \right\},
	\end{split}
\end{equation}

{\footnotesize
	\begin{equation}
		\begin{split}
			&F_4(A,B,C,D,b_1,b_2,b_3)=\int d^2k_{1T}\int d^2k_{2T}\int d^2k_{3T} \frac{e^{i(\textbf{k}_{1T}\cdot\textbf{b}_1+\textbf{k}_{2T}\cdot\textbf{b}_2+\textbf{k}_{3T}\cdot\textbf{b}_3)}}{(k_1^2+A)(k_2^2+B)(k_3^2+C)((k_1+k_2+k_3)^2+D)}\\
			&=\int_0^1 \frac{dz_1 dz_2 dz_3}{z_1z_2(1-z_1)(1-z_2)} \frac{\pi^3 \sqrt{X_3}}{\sqrt{|Z_3|}} \left\{  K_1(\sqrt{X_3Z_3})\Theta(Z_3)  -  \frac{i\pi}{2} H_1^{(1)}(\sqrt{-X_3Z_3} )\Theta(-Z_3)  \right\}.
		\end{split}
\end{equation}}
In the above expressions $H_v^{(1)}(z)$ is the  Hankel function of the first kind,
\begin{equation}
	H_v^{(1)}(z)=J_v(z)+iY_v(z),
\end{equation}
where $J_v(z)$ ($Y_v(z)$) is the Bessel function of the first (second) kind, and $K_v(z)$ is the modified Bessel function of second kind. The variables $Z_1$, $Z_2$, $Z_3$ and $X_2$, $X_3$ read
\begin{equation}
	\begin{split}
		Z_1=&Az+B(1-z),\\
		Z_2=&A(1-z_2)+\frac{z_2}{z_1(1-z_1)}\left[B(1-z_1)+Cz_1\right],\\
		Z_3=&A(1-z_3)+\frac{z_3}{z_2(1-z_2)} \left\{ B(1-z_2) + \frac{z_2}{z_1(1-z_1)}\left[C(1-z_1)+Dz_1 \right] \right\},\\
		X_2=&(b_1-z_1b_2)^2+\frac{z_1(1-z_1)}{z_2}b_2^2,\\
		X_3=&\left[ b_1-b_2z_2-b_3z_1(1-z_2) \right]^2 + \frac{z_2(1-z_2)}{z_3}(b_2-b_3z_1)^2 + \frac{z_1(1-z_1)z_2(1-z_2)}{z_2z_3}b_3^2,
	\end{split}
\end{equation}
with the Feynman parameters $z$'s. 
The above auxiliary functions appear in the PQCD factorization formulas for two-body hadronic $\Lambda_b$ baryon decays.

\section{Decay amplitudes}\label{app:decay-amplitudes}

\begin{figure}[htbp]
	\centering
	\includegraphics[width=0.7\linewidth]{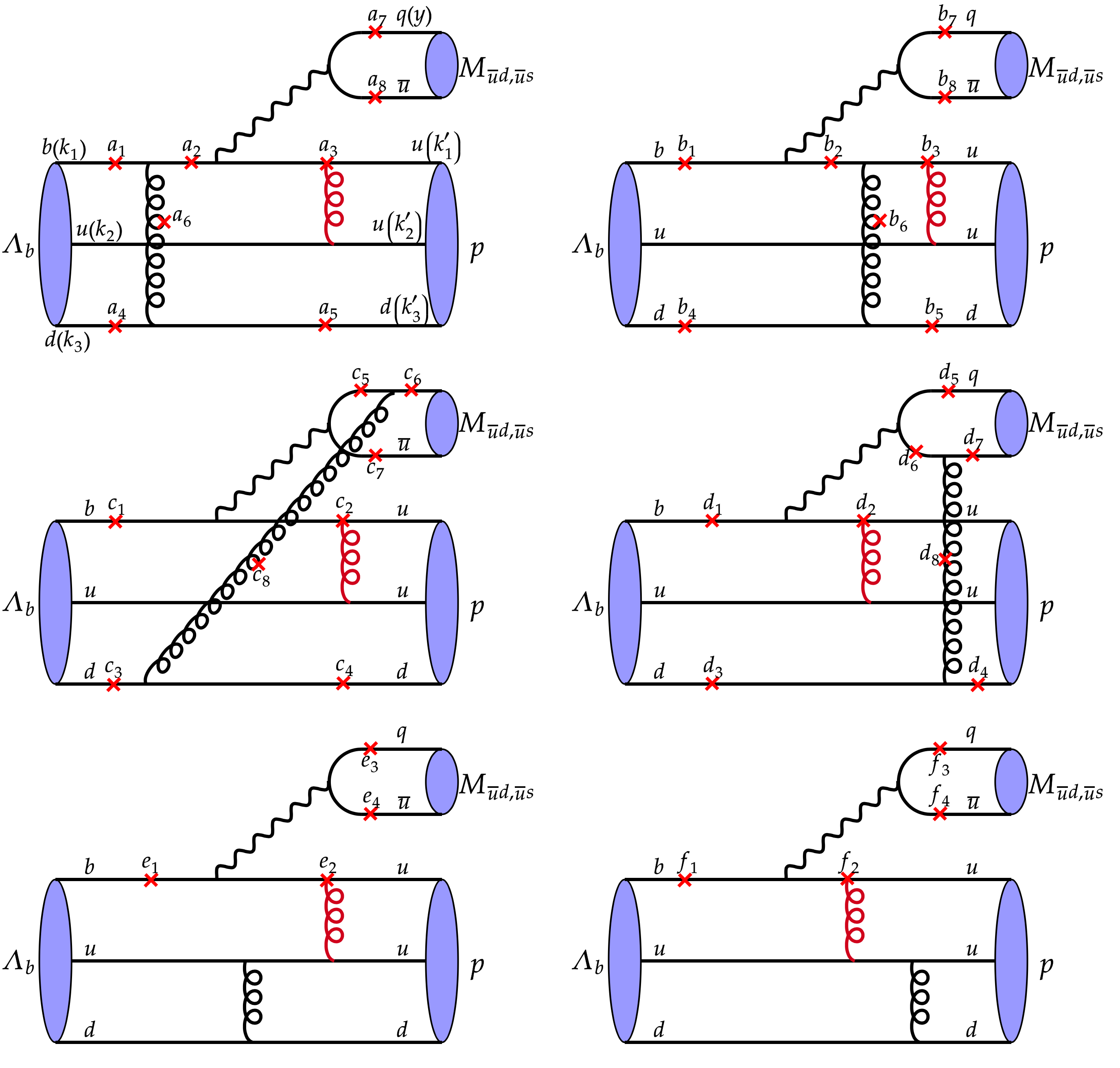}
	\caption{Feynman diagrams contributing to the external-$W$ emission topology $T$.}
	\label{fig:feynmanT}
\end{figure}

\begin{figure}[htbp]
	\centering
	\includegraphics[width=0.7\linewidth]{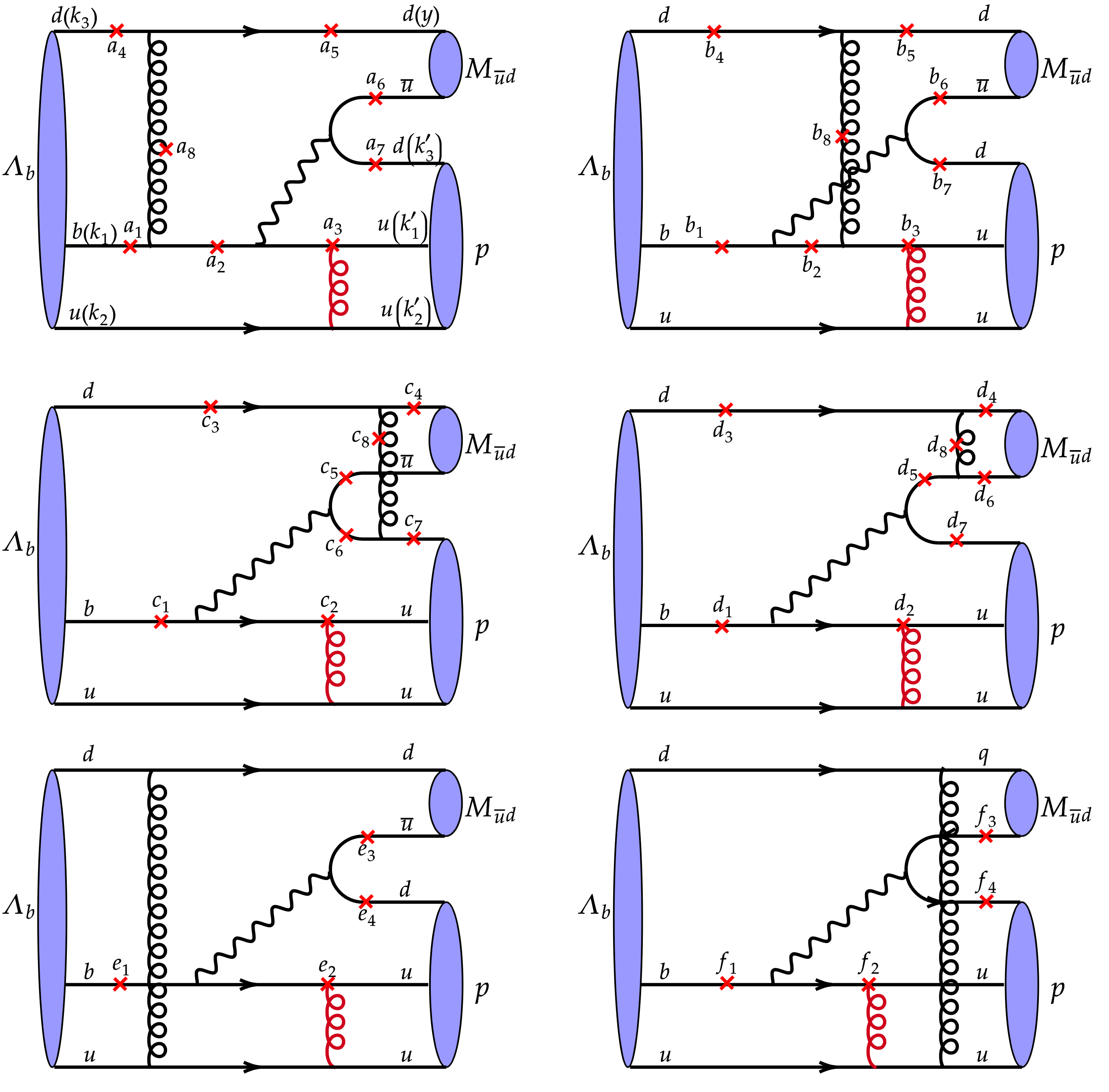}
	\caption{Feynman diagrams contributing to the $C_2$ topology.}
	\label{fig:feynmanCprime}
\end{figure}

\begin{figure}[htbp]
	\centering
	\includegraphics[width=0.7\linewidth]{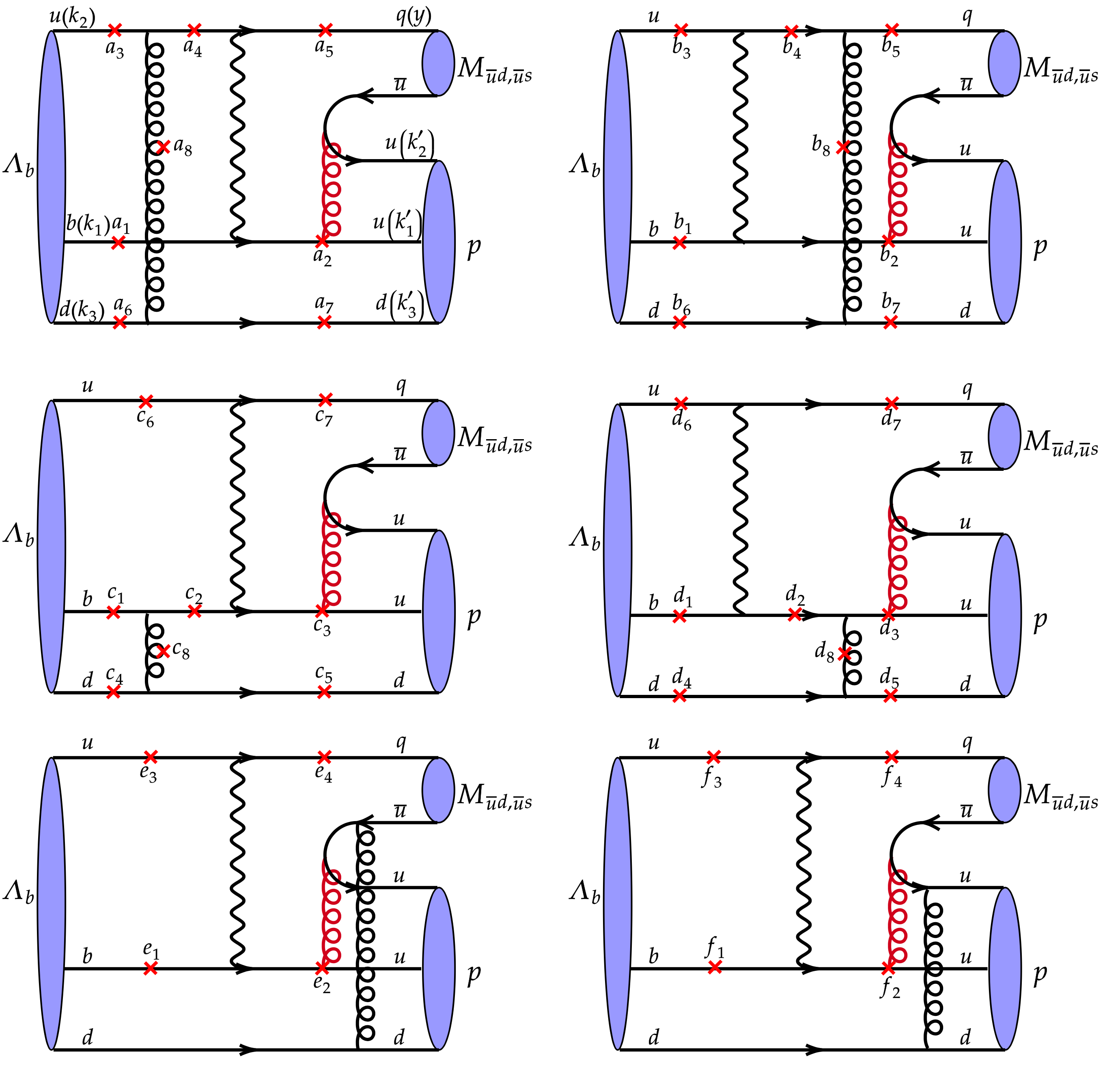}
	\caption{Feynman diagrams contributing to the $E_2$ topology.}
	\label{fig:feynmanE2}
\end{figure}

\begin{figure}[htbp]
	\centering
	\includegraphics[width=0.7\linewidth]{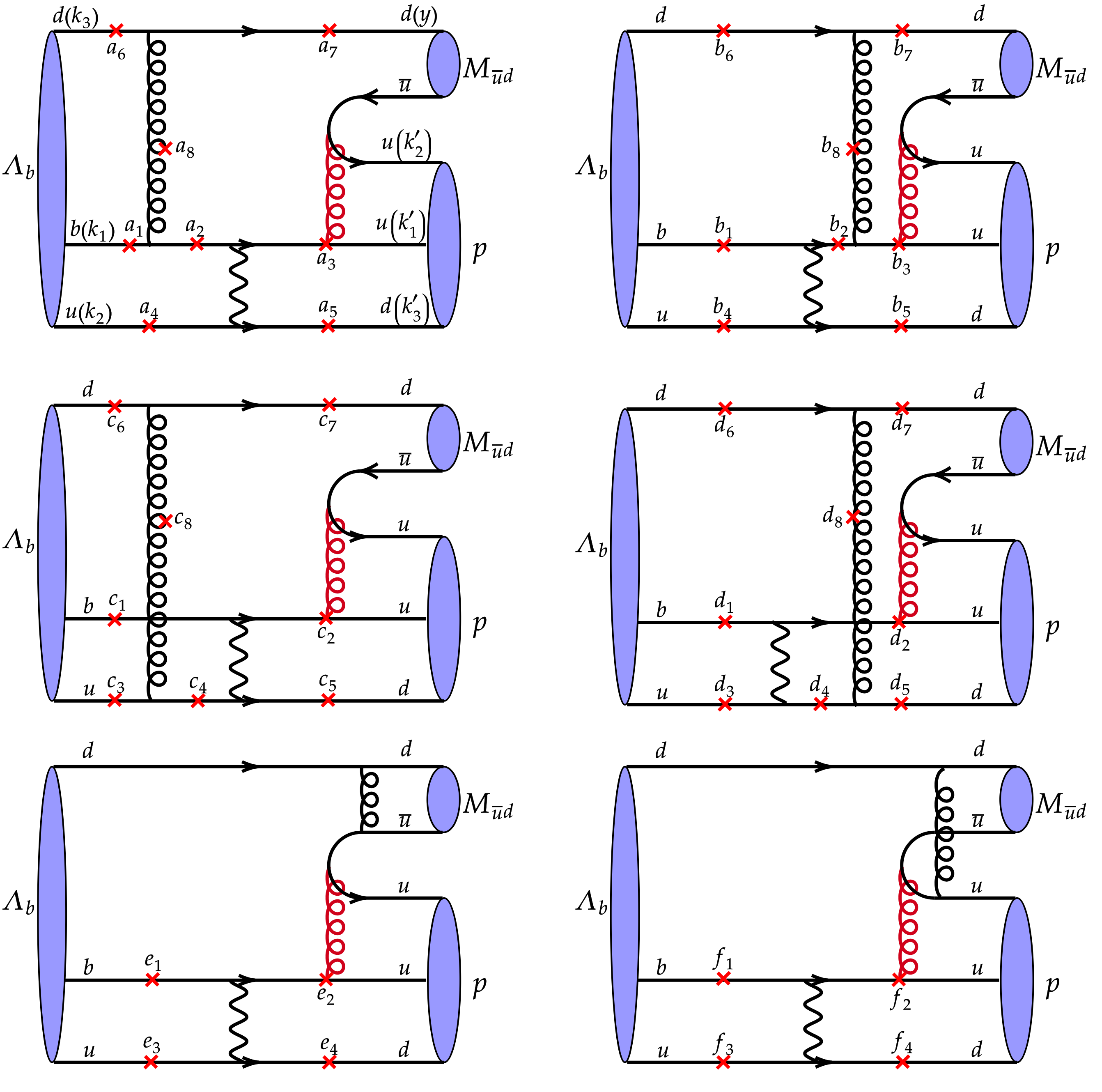}
	\caption{Feynman diagrams contributing to the $B$ topology.}
	\label{fig:feynmanB}
\end{figure}

\begin{figure}[htbp]
	\centering
	\includegraphics[width=0.7\linewidth]{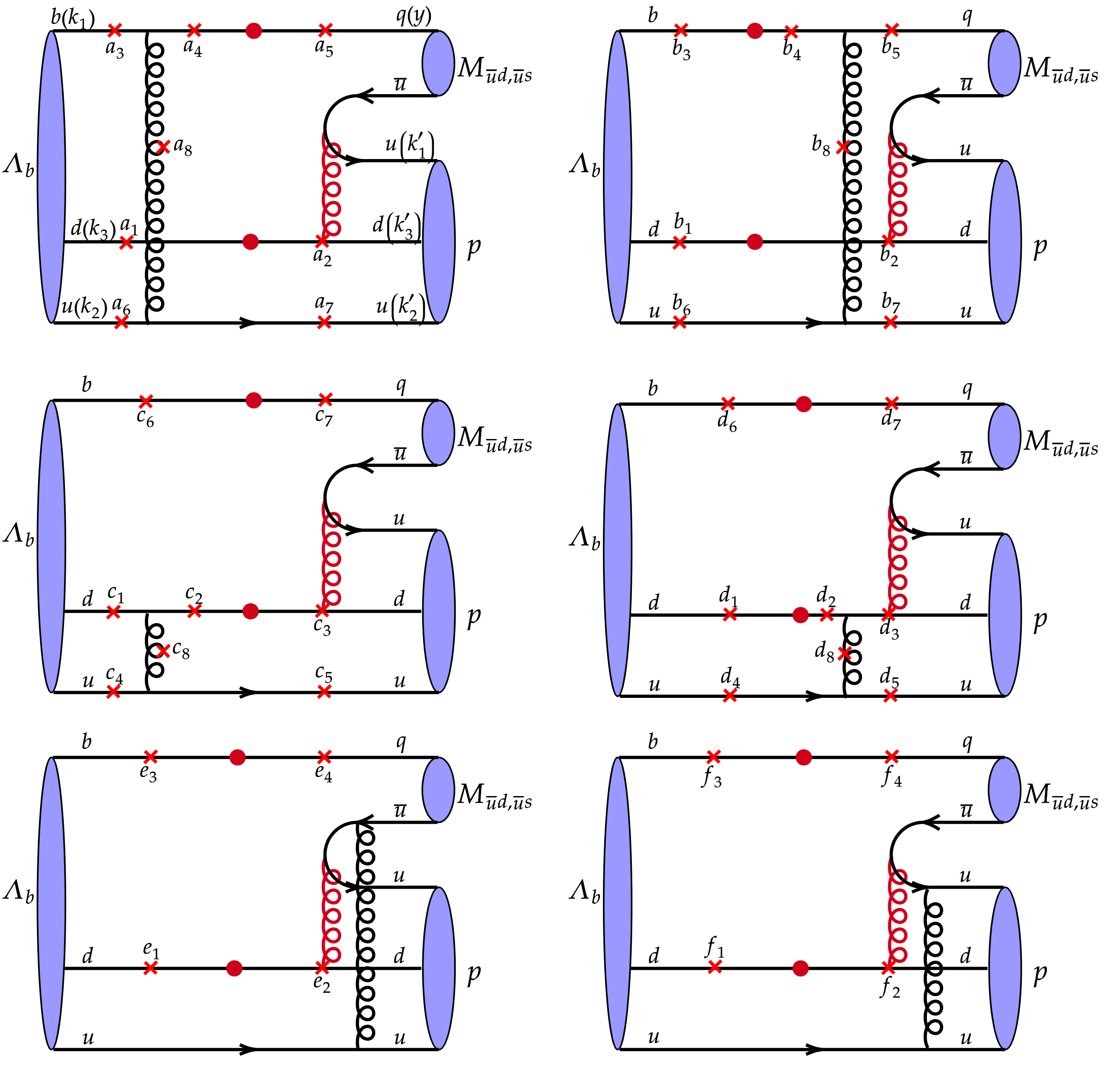}
	\caption{Feynman diagrams in the $PE_1^d$ topology, which also contribute to the $PE_2$ topology after the Feriz transformation.}
	\label{fig:feynmanPd}
\end{figure}

We present the PQCD factorization formula for each topological diagram, whose hard kernel involves two virtual gluons at the lowest order of the strong coupling $\alpha_s$. The Feynman diagrams to be evaluated are shown in Figs.~\ref{fig:feynmanT}, \ref{fig:feynmanE2}, \ref{fig:feynmanCprime}, \ref{fig:feynmanB} and \ref{fig:feynmanPd},  where the gluon in black stays fixed, while the gluon in red has one fixed end with another attaching to a cross at various locations. A Feynman diagram is  labeled by the crossed vertex. For instance, the upper-left diagram in Fig.~\ref{fig:feynmanT} is labeled as $a3$. We provide the formula for one typical diagram in each topology involved in the $\Lambda_b\to p\pi^-,pK^-$ decay. The amplitudes for the other modes $\Lambda_b\to p\rho^-,pK^{\ast -}, pa_1^-, pK_1^-$ from the same topology are similar, but with different meson DAs.


There are $40$ Feynman diagrams from the $T$ topology in Fig.~\ref{fig:feynmanT}, among which $16$ are factorizable without hard gluons attaching to the emitted meson, and $24$ are nonfactorizable. 
Applying the Feriz transformation to Fig.~\ref{fig:feynmanT} and replacing the tree operators by the penguin ones, we acquire the diagrams for the $PC_1$ topology.
The formula for the sum of the factorizable $T(b2)$ diagram and the corresponding $PC_1$ diagram is written as
\begin{equation}
	\begin{split}
		&\mathcal{A}_{T(b2)+PC_1(b2)}=\frac{8}{3} \frac{-i}{128N_c^2} \int[dx] \int[dx^\prime] \frac{1}{(2\pi)^5} \int b_1db_1 \int b_2^\prime db_2^\prime \int b_3^\prime db_3^\prime (4\pi\alpha_s(t^{T(b2)}))^2\\
		&(M_{T(b2)}+M_{PC_1(b2)})F_1(C,b_3^\prime)F_3(A,B,D,b_1,b_2^\prime) \text{exp}[-S^{T(b2)}([x],[x^\prime],y,[b],[b^\prime],b_q)].
	\end{split}
\end{equation}
The auxiliary functions $F_i$, with arguments 
\begin{equation}
	A=m_{\Lambda_b}^2 (x_2+x_3), \quad B=m_{\Lambda_b}^2 x_3(x_1^\prime+x_3^\prime), \quad C=m_{\Lambda_b}^2 x_3x_3^\prime, \quad D=m_{\Lambda_b}^2 x_2x_2^\prime,
\end{equation}
from the internal particle propagators, have been given in Appendix.~\ref{app:aux-function}. 
The functions 
\begin{equation}
	\begin{split}
		M_{T(b2)}=&\frac{G_F}{\sqrt{2}}\left\{V_{ub}V_{uq}^\ast\left[\frac{1}{3}C_1+C_2\right]\right\}M_{T(b2)}^{LL},\\
		M_{PC_1(b2)}=&-\frac{G_F}{\sqrt{2}}\left\{V_{tb}V_{tq}^\ast\left[\frac{1}{3}C_3+C_4+\frac{1}{3}C_9+C_{10}\right]\right\}M_{T(b2)}^{LL}\\
		&-\frac{G_F}{\sqrt{2}}\left\{V_{tb}V_{tq}^\ast\left[-\frac{2}{3}C_5-2C_6-\frac{2}{3}C_7-C_{8}\right]\right\}M_{T(b2)}^{SP}.
	\end{split}
\end{equation}
are associated with the tree and penguin contributions, respectively,
where the matrix element $M_{T(b2)}^{LL}$($M_{T(b2)}^{SP}$) corresponds to the insertion of the $(V-A)\otimes (V-A)$($(S-P)\otimes (S+P)$) operator.  
We display their explicit expressions as follows,
{\small
	\begin{align}
		&M_{T(b2)}^{LL}\equiv\langle h|(\bar{q}u)_{V-A}|0\rangle\langle p|(\bar{u}b)_{V-A}|\Lambda_b\rangle_{T(b2)}\nonumber\\
		=& \Big\{ 
		8 i f_P m_{\Lambda_b}^4 \psi_4 T_1 x_3 + m_p m_{\Lambda_b}^3 (-2 i f_P \psi_3^{-+}x_3 (P_1 + S_1 + 2 T_2  + T_3 - T_7 ) - 2 i f_P \psi_3^{+-} x_3(-A_3 + V_3)) \nonumber\\
		&+ 2 i f_P m_p^3 m_{\Lambda_b} \psi_3^{-+} (x_1^\prime + x_3^\prime)(A_6 + V_6) +  m_p^2 m_{\Lambda_b}^2 (-4 i f_P \psi_2 x_3 (-A_4 - A_5 - V_4 + V_5)\nonumber\\
		&+ 4 i f_P \psi_4 (x_1^\prime+x_3^\prime) (-A_4 - A_5 + P_2 - S_2 - T_4 - T_8 - V_4 + V_5))
		\Big\}\nonumber\\
		&+\Big\{
		8 i f_P m_{\Lambda_b}^4 \psi_4 T_1 x_3 + m_p m_{\Lambda_b}^3 (2 i f_P \psi_3^{-+}x_3 (-P_1 - S_1 - 2 T_2 - T_3 + T_7) + 2 i f_P \psi_3^{+-} x_3(A_3 - V_3))\nonumber\\
		&+ 2 i f_P m_p^3 m_{\Lambda_b} \psi_3^{-+}(x_1^\prime +x_3^\prime ) (A_6 + V_6 ) + m_p^2 m_{\Lambda_b}^2 (4 i f_P \psi_2x_3 (A_4 + A_5  + V_4  - V_5)\nonumber\\
		&- 4 i f_P \psi_4 (x_1^\prime+x_3^\prime)(A_4 + A_5  - P_2 + S_2  + T_4  + T_8  +  V_4  - V_5 ))
		\Big\}\gamma_5,
\end{align}}

{\small
	\begin{align}
		&M_{T(b2)}^{SP}\equiv\langle h|(\bar{q}u)_{S+P}|0\rangle\langle p|(\bar{u}b)_{S-P}|\Lambda_b\rangle_{T(b2)}\nonumber\\
		&= \Big\{
		-2 i f_P m_0^P (m_{\Lambda_b}^3 x_3 (4 \psi_4 T_1 + \psi_3^{+-} (A_1 + V_1) (x_2 + x_3)) + m_p^3 (A_6 \psi_3^{-+} + \psi_3^{-+} V_6 + 4 \psi_2 T_6 (x_2 + x_3))\nonumber\\
		& (x_1^\prime + x_3^\prime)- m_p m_{\Lambda_b}^2 (\psi_3^{-+} (P_1 + S_1 + 2 T_2 + T_3 - T_7) x_3 +  \psi_3^{+-} (-A_3 + V_3) x_3 + 2 \psi_2 (A_2 + A_3\nonumber\\
		& - P_1 + S_1 + T_3 + T_7 - V_2 + V_3) x_3 (x_2 + x_3) - 2 \psi_4 (A_2 + A_3 - V_2 + V_3) (x_2 + x_3) (x_1^\prime + x_3^\prime))\nonumber\\
		& +  m_p^2 m_{\Lambda_b} (2 \psi_2 (A_4 + A_5 + V_4 - V_5) x_3- 2 \psi_4 (A_4 + A_5 - P_2 + S_2 + T_4 + T_8 + V_4 - V_5) (x_1^\prime + x_3^\prime)\nonumber\\
		& - \psi_3^{+-} (P_2 + S_2 + T_4 + 2 T_5 - T_8) (x_2 + x_3) (x_1^\prime + x_3^\prime)+  \psi_3^{-+} (A_4 - V_4) (x_2 + x_3) (x_1^\prime + x_3^\prime)))
		\Big\}\nonumber\\
		&+\Big\{
		2 i f_P m_0^P (m_{\Lambda_b}^3 (4 \psi_4 T_1 x_3 -  \psi_3^{+-} (A_1 + V_1) x_3 (x_2 + x_3)) + m_p^3 (A_6 \psi_3^{-+} + \psi_3^{-+} V_6 - 4 \psi_2 T_6 (x_2 + x_3))\nonumber\\
		& (x_1^\prime + x_3^\prime)- m_p m_{\Lambda_b}^2 (\psi_3^{-+} (P_1 + S_1 + 2 T_2 + T_3 - T_7) x_3 +  \psi_3^{+-} (-A_3 + V_3) x_3- 2 \psi_2 (A_2 + A_3\nonumber\\
		& - P_1 + S_1 + T_3 + T_7 - V_2 + V_3) x_3 (x_2 + x_3) + 2 \psi_4 (A_2 + A_3 - V_2 + V_3) (x_2 + x_3) (x_1^\prime + x_3^\prime))\nonumber\\
		& + m_p^2 m_{\Lambda_b} (2 \psi_2 (A_4 + A_5 + V_4 - V_5) x_3- 2 \psi_4 (A_4 + A_5 - P_2 + S_2 + T_4 + T_8 + V_4 - V_5) (x_1^\prime + x_3^\prime)\nonumber\\
		& +  \psi_3^{+-} (P_2 + S_2 + T_4 + 2 T_5 - T_8) (x_2 + x_3) (x_1^\prime + x_3^\prime)-  \psi_3^{-+} (A_4 - V_4) (x_2 + x_3) (x_1^\prime + x_3^\prime)))
		\Big\}\gamma_5.
\end{align}}

The non-factorizable $T(c7)$ diagram yields
{\footnotesize
	\begin{equation}
		\begin{split}
			\mathcal{A}_{T(c7)}=&\frac{-1}{128N_c^2} \frac{1}{\sqrt{2N_c}}\int[dx] \int[dx^\prime] \int dy \frac{1}{(2\pi)^5} \int b_2db_2 \int b_3 db_3 \int b_q db_q (4\pi\alpha_s(t^{T(c7)}))^2\\
			&(M_{T(c7)}+M_{PC_1(c7)})F_4(D,C,B,A,b_2,b_3,b_q) \text{exp}[-S^{T(c7)}([x],[x^\prime],y,[b],[b^\prime],b_q)],
		\end{split}
\end{equation}}
with the arguments,
\begin{equation}
	A=m_{\Lambda_b}^2 x_3^\prime(x_3-y), \quad B=m_{\Lambda_b}^2 x_2^\prime(-1+x_2+y), \quad C=m_{\Lambda_b}^2 x_3x_3^\prime, \quad D=m_{\Lambda_b}^2 x_2x_2^\prime.
\end{equation}
The functions
\begin{equation}
	\begin{split}
		M_{T(c7)}=&\frac{G_F}{\sqrt{2}}\left\{V_{ub}V_{uq}^\ast\left[f_1 C_1+f_2 C_2\right]\right\}M_{T(c7)}^{LL},\\
		M_{PC_1(c7)}=&-\frac{G_F}{\sqrt{2}}\left\{V_{tb}V_{tq}^\ast\left[f_1 C_3+f_2C_4+f_1C_9+f_2C_{10}\right]\right\}M_{T(c7)}^{LL}\\
		&-\frac{G_F}{\sqrt{2}}\left\{V_{tb}V_{tq}^\ast\left[f_1C_5+f_2C_6+f_1C_7+f_2C_8\right]\right\}M_{T(c7)}^{LR},
	\end{split}
\end{equation}
where $f_1=8/3$ and $f_2=-2$ are the color factors, and the superscript $LR$ refers to the contribution from the $(V-A)\otimes(V+A)$ 
operator,
contain the matrix elements
{\small
	\begin{align}
		&M_{T(c7)}^{LL}\equiv\langle ph|(\bar{q}b)_{V-A}(\bar{u}u)_{V-A}|\Lambda_b\rangle_{T(c7)}\nonumber\\
		=&\Big\{
		8 m_{\Lambda_b} (4 m_p^3 \psi_2 T_6 x_2^\prime x_3^\prime + m_p^2 m_{\Lambda_b} x_3^\prime ((A_4 \psi_3^{-+} - 2 \psi_3^{+-} T_5 - \psi_3^{-+} V_4) x_2^\prime - 2 \psi_2 (P_2 - S_2 + T_4 + T_8) \nonumber\\
		&(-1 + x_2 + y)) + m_{\Lambda_b}^3 (x_3 - y) (A_1 \psi_3^{-+} x_2^\prime + \psi_3^{-+} V_1 x_2^\prime  - 2 A_1 \psi_2 (-1 + x_2 + y) + 2 \psi_2 (2 T_1 + V_1) \nonumber\\
		&(-1 + x_2 + y)) + m_p m_{\Lambda_b}^2 (2 \psi_4 (A_3 + V_3) x_2^\prime x_3^\prime - 2 \psi_2 (P_1 - S_1 + T_3 + T_7) x_2^\prime (x_3 - y)\nonumber\\
		& + \psi_3^{+-} (P_1 + S_1 + T_3 - T_7) x_3^\prime (-1 + x_2 + y))) \phi_m^A\Big\}\nonumber\\
		&+\Big\{
		8 m_{\Lambda_b} (-4 m_p^3 \psi_2 T_6 x_2^\prime x_3^\prime + m_p^2 m_{\Lambda_b} x_3^\prime ((-A_4 \psi_3^{-+} + 2 \psi_3^{+-} T_5 + \psi_3^{-+} V_4) x_2^\prime + 2 \psi_2 (-P_2 + S_2 - T_4 - T_8) \nonumber\\
		&(-1 + x_2 + y)) - m_{\Lambda_b}^3 (x_3 - y) (A_1 \psi_3^{-+} x_2^\prime + \psi_3^{-+} V_1 x_2^\prime + 2 A_1 \psi_2 (-1 + x_2 + y) - 2 \psi_2 (2 T_1 + V_1) \nonumber\\
		&(-1 + x_2 + y)) + m_p m_{\Lambda_b}^2 (-2 \psi_4 (A_3 + V_3) x_2^\prime x_3^\prime + 2 \psi_2 (P_1 - S_1 + T_3 + T_7) x_2^\prime (x_3 - y)\nonumber\\
		& + \psi_3^{+-} (P_1 + S_1 + T_3 - T_7) x_3^\prime (-1 + x_2 + y))) \phi_m^A
		\Big\}\gamma_5,
		\label{eq:hard-kernel-Tnf(c7)}
\end{align}}

{\footnotesize
	\begin{align}
		&M_{T(c7)}^{LR}\equiv\langle ph|(\bar{q}b)_{V-A}(\bar{u}u)_{V+A}|\Lambda_b\rangle_{T(c7)}\nonumber\\
		=&\Big\{
		8 m_0 (2 m_p^3 \psi_2 (-((A_6 - V_6) x_3^\prime (-1 + x_2 + y) (\phi_m^P - \phi_m^T)) + 2 T_6 x_2^\prime (x_3 - y) (\phi_m^P + \phi_m^T)) - 2 m_{\Lambda_b}^3 \psi_4 (-2 T_1 x_3^\prime \nonumber\\
		& (-1 + x_2 + y) (\phi_m^P - \phi_m^T) + A_1 x_2^\prime (x_3 - y) (\phi_m^P + \phi_m^T) - V_1 x_2^\prime (x_3 - y) (\phi_m^P + \phi_m^T)) + m_p^2 m_{\Lambda_b} (-2 \psi_3^{+-} T_5 x_2^\prime \nonumber\\
		& (x_3 - y) (\phi_m^P + \phi_m^T) + \psi_3^{+-} x_3^\prime (-1 + x_2 + y) ((A_4 + A_5 + P_2 + S_2 - T_4 + T_8 - V_4 + V_5) \phi_m^P \nonumber\\
		&+ (A_4 - A_5 + P_2 + S_2 - T_4 + T_8 - V_4 - V_5) \phi_m^T) + \psi_3^{-+} (((A_5 - 2 T_5 - V_4 + V_5) x_2^\prime x_3 + (P_2 + S_2 - T_4 + T_8) \nonumber\\
		& (-1 + x_2) x_3^\prime - (A_5 - 2 T_5 - V_4 + V_5) x_2^\prime y + (P_2 +S_2 - T_4 + T_8) x_3^\prime y) \phi_m^P - ((A_5 - 2 T_5 + V_4 + V_5) x_2^\prime x_3\nonumber\\
		& + (P_2 + S_2 - T_4 + T_8) (-1 + x_2) x_3^\prime - (A_5 - 2 T_5 + V_4 + V_5) x_2^\prime y + (P_2 + S_2 - T_4 + T_8) x_3^\prime y) \phi_m^T \nonumber\\
		& + A_4 x_2^\prime (x_3 - y) (\phi_m^P + \phi_m^T)) - 2 \psi_2 (-(((S_2 + T_4 + T_8) x_2^\prime x_3 + (A_4 + V_4) (-1 + x_2) x_3^\prime - (S_2 + T_4 + T_8) x_2^\prime y \nonumber\\
		& + (A_4 + V_4) x_3^\prime y) \phi_m^P) + (-((S_2 + T_4 + T_8) x_2^\prime x_3) + (A_4 + V_4) (-1 + x_2) x_3^\prime + (S_2 + T_4 + T_8) x_2^\prime y \nonumber\\
		& + (A_4 + V_4) x_3^\prime y) \phi_m^T + P_2 x_2^\prime (x_3 - y) (\phi_m^P + \phi_m^T))) + m_p m_{\Lambda_b}^2 (\psi_3^{-+} ((A_2 + A_3 + P_1 + S_1 - T_3 + T_7 + V_2 - V_3) x_2^\prime x_3 \nonumber\\
		& - 2 T_2 (-1 + x_2) x_3^\prime - ((A_2 + A_3 + P_1 + S_1 - T_3 + T_7 + V_2 - V_3) x_2^\prime + 2 T_2 x_3^\prime) y) \phi_m^P + \psi_3^{-+} ((A_2 - A_3 - P_1 \nonumber\\
		& - S_1 + T_3 - T_7 + V_2 + V_3) x_2^\prime x_3 + 2 T_2 (-1 + x_2) x_3^\prime + (-A_2 + A_3 + P_1 + S_1 - T_3 + T_7 - V_2 - V_3) x_2^\prime y + 2 T_2 x_3^\prime y) \phi_m^T \nonumber\\
		& + \psi_3^{+-} (((S_1 - T_3 + T_7) x_2^\prime x_3 + (A_2 + A_3 - 2 T_2 + V_2 - V_3) (-1 + x_2) x_3^\prime - (S_1 - T_3 + T_7) x_2^\prime y + (A_2 + A_3 \nonumber\\
		& - 2 T_2 + V_2 - V_3) x_3^\prime y) \phi_m^P + ((S_1 - T_3 + T_7) x_2^\prime x_3 + (A_2 - A_3 - 2 T_2 + V_2 + V_3) (-1 + x_2) x_3^\prime - (S_1 - T_3 + T_7) x_2^\prime y \nonumber\\
		& + (A_2 - A_3 - 2 T_2 + V_2 +V_3) x_3^\prime y) \phi_m^T + P_1 x_2^\prime (x_3 - y) (\phi_m^P + \phi_m^T)) + 2 \psi_4 (-((P_1 - S_1 - T_3 - T_7) x_3^\prime (-1 + x_2 + y) \nonumber\\
		& (\phi_m^P - \phi_m^T)) + A_3 x_2^\prime (x_3 - y) (\phi_m^P + \phi_m^T) + V_3 x_2^\prime (x_3 - y) (\phi_m^P + \phi_m^T))))
		\Big\} \nonumber\\
		&+\Big\{8 m_0 (2 m_p^3 \psi_2 (-((A_6 - V_6) x_3^\prime (-1 + x_2 + y) (\phi_m^P - \phi_m^T)) + 2 T_6 x_2^\prime (x_3 - y) (\phi_m^P + \phi_m^T)) + 2 m_{\Lambda_b}^3 \psi_4 (-2 T_1 x_3^\prime \nonumber\\
		& (-1 + x_2 + y) (\phi_m^P - \phi_m^T) + A_1 x_2^\prime (x_3 - y) (\phi_m^P + \phi_m^T) - V_1 x_2^\prime (x_3 - y) (\phi_m^P + \phi_m^T)) + m_p^2 m_{\Lambda_b} (-2 \psi_3^{+-} T_5 x_2^\prime \nonumber\\
		& (x_3 - y) (\phi_m^P + \phi_m^T) + \psi_3^{+-} x_3^\prime (-1 + x_2 + y) ((A_4 + A_5 + P_2 + S_2 - T_4 + T_8 - V_4 + V_5) \phi_m^P \nonumber\\
		& + (A_4 - A_5 + P_2 + S_2 - T_4 + T_8 - V_4 - V_5) \phi_m^T) + \psi_3^{-+} (((A_5 - 2 T_5 - V_4 + V_5) x_2^\prime x_3 + (P_2 + S_2 - T_4 + T_8) \nonumber\\
		& (-1 + x_2) x_3^\prime - (A_5 - 2 T_5 - V_4 + V_5) x_2^\prime y + (P_2 +S_2 - T_4 + T_8) x_3^\prime y) \phi_m^P - ((A_5 - 2 T_5 + V_4 + V_5) x_2^\prime x_3 \nonumber\\
		& + (P_2 + S_2 - T_4 + T_8) (-1 + x_2) x_3^\prime - (A_5 - 2 T_5 + V_4 + V_5) x_2^\prime y + (P_2 + S_2 - T_4 + T_8) x_3^\prime y) \phi_m^T \nonumber\\
		& + A_4 x_2^\prime (x_3 - y) (\phi_m^P + \phi_m^T)) + 2 \psi_2 (-(((S_2 + T_4 + T_8) x_2^\prime x_3 + (A_4 + V_4) (-1 + x_2) x_3^\prime - (S_2 + T_4 + T_8) x_2^\prime y \nonumber\\
		& + (A_4 + V_4) x_3^\prime y) \phi_m^P) + (-((S_2 + T_4 + T_8) x_2^\prime x_3) + (A_4 + V_4) (-1 + x_2) x_3^\prime + (S_2 + T_4 + T_8) x_2^\prime y \nonumber\\
		& + (A_4 + V_4) x_3^\prime y) \phi_m^T + P_2 x_2^\prime (x_3 - y) (\phi_m^P + \phi_m^T))) - m_p m_{\Lambda_b}^2 (\psi_3^{-+} ((A_2 + A_3 + P_1 + S_1 - T_3 + T_7 + V_2 - V_3) x_2^\prime x_3 \nonumber\\
		& - 2 T_2 (-1 + x_2) x_3^\prime - ((A_2 + A_3 + P_1 + S_1 - T_3 + T_7 + V_2 - V_3) x_2^\prime + 2 T_2 x_3^\prime) y) \phi_m^P + \psi_3^{-+} ((A_2 - A_3 - P_1 \nonumber\\
		& - S_1 + T_3 - T_7 + V_2 + V_3) x_2^\prime x_3 + 2 T_2 (-1 + x_2) x_3^\prime + (-A_2 + A_3 + P_1 + S_1 - T_3 + T_7 - V_2 - V_3) x_2^\prime y + 2 T_2 x_3^\prime y) \phi_m^T \nonumber\\
		& + \psi_3^{+-} (((S_1 - T_3 + T_7) x_2^\prime x_3 + (A_2 + A_3 - 2 T_2 + V_2 - V_3) (-1 + x_2) x_3^\prime - (S_1 - T_3 + T_7) x_2^\prime y + (A_2 + A_3 \nonumber\\
		& - 2 T_2 + V_2 - V_3) x_3^\prime y) \phi_m^P + ((S_1 - T_3 + T_7) x_2^\prime x_3 + (A_2 - A_3 - 2 T_2 + V_2 + V_3) (-1 + x_2) x_3^\prime - (S_1 - T_3 + T_7) x_2^\prime y \nonumber\\
		& + (A_2 - A_3 - 2 T_2 + V_2 +V_3) x_3^\prime y) \phi_m^T + P_1 x_2^\prime (x_3 - y) (\phi_m^P + \phi_m^T)) - 2 \psi_4 (-((P_1 - S_1 - T_3 - T_7) x_3^\prime (-1 + x_2 + y) \nonumber\\
		& (\phi_m^P - \phi_m^T)) + A_3 x_2^\prime (x_3 - y) (\phi_m^P + \phi_m^T) + V_3 x_2^\prime (x_3 - y) (\phi_m^P + \phi_m^T))))
		\Big\}\gamma_5.
\end{align}}

For the $C_2(c7)$ diagram in Fig.~\ref{fig:feynmanCprime}, we have
\begin{equation}
	\begin{split}
		\mathcal{A}_{C_2(c7)}=&\frac{-1}{128N_c^2} \frac{1}{\sqrt{2N_c}} \int[dx] \int[dx^\prime] \int dy \frac{1}{(2\pi)^5} \int b_2db_2 \int b_2^\prime db_2^\prime \int b_q db_q (4\pi\alpha_s(t^{C_2(c7)}))^2\\
		&(M_{C_2(c7)}+M_{PC_2(c7)})F_1(D,b_2^\prime)F_3(A,C,B,b_2,b_2^\prime,b_q) \text{exp}[-S^{C_2(c7)}([x],[x^\prime],y,[b],[b^\prime],b_q)],
	\end{split}
\end{equation}
with the arguments
\begin{equation}
	A=m_{\Lambda_b}^2 x_2(x_2^\prime + x_2^\prime), \quad B=m_{\Lambda_b}^2 (x_2^\prime -x_3 +x_3^\prime)(x_2-y), \quad C=m_{\Lambda_b}^2 x_3y, \quad D=m_{\Lambda_b}^2 x_2x_2^\prime.
\end{equation}

The functions
\begin{equation}
	\begin{split}
		M_{C_2(c7)}=&\frac{G_F}{\sqrt{2}}\left\{V_{ub}V_{ud}^\ast\left[f_1C_1+f_2C_2\right]\right\}M_{C_2(c7)}^{LL},\\
		M_{PC_2(c7)}=&-\frac{G_F}{\sqrt{2}}\left\{V_{tb}V_{td}^\ast\left[f_1C_3+f_2C_4+f_1C_9+f_2C_{10}\right]\right\}M_{C_2(c7)}^{LL}\\
		&-\frac{G_F}{\sqrt{2}}\left\{V_{tb}V_{td}^\ast\left[-2f_1C_5-2f_2C_6-2f_1C_7-2f_2C_{8}\right]\right\}M_{C_2(c7)}^{SP},
	\end{split}
\end{equation}
contain the color factors $f_1=-8/3$ and $f_2=-16/3$, and the matrix elements
{\small
	\begin{align}
		&M_{C_2(c7)}^{LL}\equiv\langle ph|(\bar{u}b)_{V-A}(\bar{d}u)_{V-A}|\Lambda_b\rangle_{C_2(c7)}\nonumber\\
		=&\Big\{
		8 m_0 m_p^3 \psi_3^{+-} (A_6 - V_6) (x_2^\prime + x_3^\prime) (x_2^\prime - x_3 + x_3^\prime) (\phi_m^P - \phi_m^T) + 8 m_0 m_{\Lambda_b}^3 \psi_3^{-+} (A_1 - V_1) x_2 (x_2 - y) \nonumber\\
		& (\phi_m^P + \phi_m^T)- 8 m_p m_{\Lambda_b}^2 (m_{\Lambda_b} \psi_3^{-+} (P_1 + S_1 - 2 T_2 - T_3 + T_7) (x_2^\prime + x_3^\prime) (x_2 - y) \phi_m^A - m_0 \psi_3^{-+} (A_3 + V_3) x_2\nonumber\\
		& (x_2^\prime - x_3 + x_3^\prime) (\phi_m^P - \phi_m^T) + 2 m_0 \psi_2 (P_1 + S_1 - 2 T_2 - T_3 + T_7) x_2 (x_2 - y) (\phi_m^P + \phi_m^T)  - 2 m_0 \psi_4 (P_1 + S_1\nonumber\\
		& - 2 T_2 - T_3 + T_7) (x_2^\prime + x_3^\prime) (x_2 - y) (\phi_m^P + \phi_m^T)) - 8 m_p^2 m_{\Lambda_b} (-2 \psi_4 (x_2^\prime + x_3^\prime) (x_2^\prime - x_3 + x_3^\prime)\nonumber\\
		& (A_5 m_{\Lambda_b} \phi_m^A - m_{\Lambda_b} V_5 \phi_m^A - m_0 (P_2 + S_2 - T_4 - 2 T_5 + T_8) (\phi_m^P - \phi_m^T)) + 2 \psi_2 (A_4 m_{\Lambda_b} (x_2^\prime + x_3^\prime) (x_2 - y) \phi_m^A\nonumber\\
		& + m_{\Lambda_b} V_4 (x_2^\prime + x_3^\prime) (x_2 - y) \phi_m^A - m_0 (P_2 + S_2 - T_4 - 2 T_5 + T_8) x_2 (x_2^\prime - x_3 + x_3^\prime) (\phi_m^P - \phi_m^T))\nonumber\\
		& - m_0 \psi_3^{+-} (A_4 + V_4) (x_2^\prime + x_3^\prime) (x_2 - y) (\phi_m^P + \phi_m^T))
		\Big\}\nonumber\\
		&+\Big\{
		-8 m_0 m_p^3 \psi_3^{+-} (A_6 - V_6) (x_2^\prime + x_3^\prime) (x_2^\prime - x_3 + x_3^\prime) (\phi_m^P - \phi_m^T) + 8 m_0 m_{\Lambda_b}^3 \psi_3^{-+} (A_1 - V_1) x_2 (x_2 - y) \nonumber\\
		& (\phi_m^P + \phi_m^T)- 8 m_p m_{\Lambda_b}^2 (m_{\Lambda_b} \psi_3^{-+} (P_1 + S_1 - 2 T_2 - T_3 + T_7) (x_2^\prime + x_3^\prime) (x_2 - y) \phi_m^A + m_0 \psi_3^{-+} (A_3 + V_3) x_2\nonumber\\
		& (x_2^\prime - x_3 + x_3^\prime) (\phi_m^P - \phi_m^T) + 2 m_0 \psi_2 (P_1 + S_1 - 2 T_2 - T_3 + T_7) x_2 (x_2 - y) (\phi_m^P + \phi_m^T)  - 2 m_0 \psi_4 (P_1 + S_1\nonumber\\
		& - 2 T_2 - T_3 + T_7) (x_2^\prime + x_3^\prime) (x_2 - y) (\phi_m^P + \phi_m^T)) - 8 m_p^2 m_{\Lambda_b} (-2 \psi_4 (x_2^\prime + x_3^\prime) (x_2^\prime - x_3 + x_3^\prime)\nonumber\\
		& (A_5 m_{\Lambda_b} \phi_m^A - m_{\Lambda_b} V_5 \phi_m^A + m_0 (P_2 + S_2 - T_4 - 2 T_5 + T_8) (\phi_m^P - \phi_m^T)) + 2 \psi_2 (A_4 m_{\Lambda_b} (x_2^\prime + x_3^\prime) (x_2 - y) \phi_m^A\nonumber\\
		& + m_{\Lambda_b} V_4 (x_2^\prime + x_3^\prime) (x_2 - y) \phi_m^A + m_0 (P_2 + S_2 - T_4 - 2 T_5 + T_8) x_2 (x_2^\prime - x_3 + x_3^\prime) (\phi_m^P - \phi_m^T))\nonumber\\
		& - m_0 \psi_3^{+-} (A_4 + V_4) (x_2^\prime + x_3^\prime) (x_2 - y) (\phi_m^P + \phi_m^T))
		\Big\}\gamma_5,
\end{align}}

{\small
	\begin{align}
		&M_{C_2(c7)}^{SP}\equiv\langle ph|(\bar{u}b)_{S-P}(\bar{d}u)_{S+P}|\Lambda_b\rangle_{C_2}\nonumber\\
		=&\Big\{
		8 m_p^3 T_6 (x_2^\prime + x_3^\prime) (m_{\Lambda_b} \psi_3^{+-} (x_2^\prime - x_3 + x_3^\prime) \phi_m^A + 2 m_0 \psi_2 (x_2 - y) (\phi_m^P - \phi_m^T)) - 8 m_{\Lambda_b}^3 T_1 x_2 (x_2^\prime - x_3 + x_3^\prime)\nonumber\\
		& (m_{\Lambda_b} \psi_3^{-+} \phi_m^A - 2 m_0 \psi_4 (\phi_m^P + \phi_m^T)) + 4 m_p^2 m_{\Lambda_b} (-m_{\Lambda_b} \psi_3^{+-} (P_2 - S_2 - T_4 -T_8) (x_2^\prime + x_3^\prime) (x_2^\prime - x_3 + x_3^\prime) \phi_m^A\nonumber\\
		& + 2 m_{\Lambda_b} \psi_4 (A_4 + A_5 - V_4 + V_5) (x_2^\prime + x_3^\prime) (x_2^\prime - x_3 + x_3^\prime) \phi_m^A + m_0 \psi_3^{-+} (A_4 + A_5 - V_4 + V_5) (x_2^\prime + x_3^\prime)\nonumber\\
		& (x_2 - y) (\phi_m^P - \phi_m^T) + m_0 \psi_3^{+-} (A_4 + A_5 - V_4 + V_5) x_2 (x_2^\prime - x_3 + x_3^\prime) (\phi_m^P + \phi_m^T)\nonumber\\
		& - 2 \psi_2 (m_{\Lambda_b}\phi_m^Ax_2(x_2^\prime-x_3+x_3^\prime)(A_4+A_5-V_4+V_5) + m_0(\phi_m^P-\phi_m^T)(x_2^\prime+x_3^\prime)(x_2-y)(P_2-S_2\nonumber\\
		& -T_4-T_8))) - 4 m_p m_{\Lambda_b}^2 (-m_{\Lambda_b} \psi_3^{-+} (-P_1 + S_1 + T_3 + T_7) x_2 (-x_2^\prime + x_3 - x_3^\prime) \phi_m^A + 2 m_{\Lambda_b} \psi_2 (A_2 + A_3\nonumber\\
		& + V_2 - V_3) x_2 (x_2^\prime - x_3 + x_3^\prime) \phi_m^A - m_0 \psi_3^{-+} (A_2 + A_3 + V_2 - V_3) (x_2^\prime + x_3^\prime) (x_2 - y) (\phi_m^P - \phi_m^T) \nonumber\\
		&- m_0 \psi_3^{+-} (A_2 + A_3 + V_2 - V_3) x_2 (x_2^\prime - x_3 + x_3^\prime) (\phi_m^P + \phi_m^T) - 2 \psi_4 (x_2^\prime - x_3 + x_3^\prime)\nonumber\\
		& (m_{\Lambda_b} (A_2 + A_3 + V_2 - V_3) (x_2^\prime + x_3^\prime) \phi_m^A - m_0 (P_1 - S_1 - T_3 - T_7) x_2 (\phi_m^P + \phi_m^T)))
		\Big\}\nonumber\\
		&+\Big\{
		8 m_p^3 T_6 (x_2^\prime + x_3^\prime) (m_{\Lambda_b} \psi_3^{+-} (x_2^\prime - x_3 + x_3^\prime) \phi_m^A + 2 m_0 \psi_2 (x_2 - y) (\phi_m^P - \phi_m^T)) + 8 m_{\Lambda_b}^3 T_1 x_2 (x_2^\prime - x_3 + x_3^\prime)\nonumber\\
		& (m_{\Lambda_b} \psi_3^{-+} \phi_m^A - 2 m_0 \psi_4 (\phi_m^P + \phi_m^T)) + 4 m_p^2 m_{\Lambda_b} (m_{\Lambda_b} \psi_3^{+-} (P_2 - S_2 - T_4 - T_8) (x_2^\prime + x_3^\prime) (x_2^\prime - x_3 + x_3^\prime) \phi_m^A\nonumber\\
		& + 2 m_{\Lambda_b} \psi_4 (A_4 + A_5 - V_4 + V_5) (x_2^\prime + x_3^\prime) (x_2^\prime - x_3 + x_3^\prime) \phi_m^A + m_0 \psi_3^{-+} (A_4 + A_5 - V_4 + V_5) (x_2^\prime + x_3^\prime)\nonumber\\
		& (x_2 - y) (\phi_m^P - \phi_m^T) + m_0 \psi_3^{+-} (A_4 + A_5 - V_4 + V_5) x_2 (x_2^\prime - x_3 + x_3^\prime) (\phi_m^P + \phi_m^T)\nonumber\\
		& - 2 \psi_2 (m_{\Lambda_b}\phi_m^Ax_2(x_2^\prime-x_3+x_3^\prime)(A_4+A_5-V_4+V_5) - m_0(\phi_m^P-\phi_m^T)(x_2^\prime+x_3^\prime)(x_2-y)(P_2-S_2\nonumber\\
		& -T_4-T_8))) + 4 m_p m_{\Lambda_b}^2 (m_{\Lambda_b} \psi_3^{-+} (P_1 - S_1 - T_3 - T_7) x_2 (x_2^\prime - x_3 + x_3^\prime) \phi_m^A + 2 m_{\Lambda_b} \psi_2 (A_2 + A_3\nonumber\\
		& + V_2 - V_3) x_2 (x_2^\prime - x_3 + x_3^\prime) \phi_m^A - m_0 \psi_3^{-+} (A_2 + A_3 + V_2 - V_3) (x_2^\prime + x_3^\prime) (x_2 - y) (\phi_m^P - \phi_m^T) \nonumber\\
		&- m_0 \psi_3^{+-} (A_2 + A_3 + V_2 - V_3) x_2 (x_2^\prime - x_3 + x_3^\prime) (\phi_m^P + \phi_m^T) - 2 \psi_4 (x_2^\prime - x_3 + x_3^\prime)\nonumber\\
		& (m_{\Lambda_b} (A_2 + A_3 + V_2 - V_3) (x_2^\prime + x_3^\prime) \phi_m^A + m_0 (P_1 - S_1 - T_3 - T_7) x_2 (\phi_m^P + \phi_m^T)))
		\Big\}\gamma_5.
\end{align}}

The $E_2(a4)$ diagram in Fig.~\ref{fig:feynmanE2} contributes
\begin{equation}
	\begin{split}
		\mathcal{A}_{E_2(a4)}=&\frac{-1}{128N_c^2} \frac{1}{\sqrt{2N_c}}\int[dx] \int[dx^\prime] \int dy \frac{1}{(2\pi)^5} \int b_1db_1 \int b_3 db_3 \int b_2^\prime db_2^\prime (4\pi\alpha_s(t^{E_2(a4)}))^2\\
		&(M_{E_2(a4)}+M_{PE_2(a4)})F_1(C,b_3)F_3(A,D,B,b_1,b_3,b_2^\prime) \text{exp}[-S^{E_2(a4)}([x],[x^\prime],y,[b],[b^\prime],b_q)],
	\end{split}
\end{equation}
with the arguments
\begin{equation}
	A=m_{\Lambda_b}^2 (x_2 + x_3)x_3^\prime, \quad B=m_{\Lambda_b}^2 (x_2^\prime +x_3^\prime)(-1+x_2+x_3+y), \quad C=m_{\Lambda_b}^2 x_3x_3^\prime, \quad D=m_{\Lambda_b}^2 x_2^\prime(-1+y).
\end{equation}
The functions
\begin{equation}
	\begin{split}
		M_{E_2(a4)}=&\frac{G_F}{\sqrt{2}}\left\{V_{ub}V_{uq}^\ast\left[f_1C_1+f_2C_2\right]\right\}M_{E_2(a4)}^{LL},\\
		M_{P^{E_1^u}(a4)}=&-\frac{G_F}{\sqrt{2}}\left\{V_{tb}V_{tq}^\ast\left[f_1C_3+f_2C_4+f_1C_9+f_2C_{10}\right]\right\}M_{E_2(a4)}^{LL}\\
		&-\frac{G_F}{\sqrt{2}}\left\{V_{tb}V_{tq}^\ast\left[-2f_1C_5-2f_2C_6-2f_1C_7-2f_2C_{8}\right]\right\}M_{E_2(a4)}^{SP},
	\end{split}
\end{equation}
involve the color factors $f_1=-8/3$ and $f_2=-16/3$, and  the matrix elements
{\small
	\begin{align}
		&M_{E_2(a4)}^{LL}\equiv\langle ph|(\bar{u}b)_{V-A}(\bar{q}u)_{V-A}|\Lambda_b\rangle_{E_2}\nonumber\\
		=&\Big\{8 (-4 m_{\Lambda_b}^4 \psi_4 T_1 x_3^\prime (-1 + x_2 + x_3 + y) \phi_m^A + m_{\Lambda_b}^3 (2 m_p x_3^\prime (\psi_4 (P_1 - S_1 + T_3 + T_7) (x_2^\prime + x_3^\prime)\nonumber\\
		& + (\psi_3^{-+} + \psi_3^{+-}) T_2 (-1 + x_2 + x_3 + y)) \phi_m^A +m_0 (\psi_3^{-+} + \psi_3^{+-}) (A_1 - V_1) (x_2 + x_3) (-1 + x_2 + x_3 + y)\nonumber\\
		& (\phi_m^P - \phi_m^T)) + m_0 m_p^3 (\psi_3^{-+} + \psi_3^{+-}) (A_6 - V_6) x_3^\prime (x_2^\prime + x_3^\prime) (\phi_m^P + \phi_m^T) + m_0 m_p^2 m_{\Lambda_b} (A_4 (\psi_3^{-+} + \psi_3^{+-}) x_3^\prime\nonumber\\
		& (-1 + x_2 + x_3 + y) (\phi_m^P - \phi_m^T) + \psi_3^{-+} V_4 x_3^\prime (-1 + x_2 + x_3 + y) (\phi_m^P - \phi_m^T) + \psi_3^{+-} V_4 x_3^\prime (-1 + x_2 + x_3 + y)\nonumber\\
		& (\phi_m^P - \phi_m^T) + 2 A_4 \psi_2 (x_2 + x_3) (x_2^\prime + x_3^\prime) (\phi_m^P + \phi_m^T) - 2 \psi_2 V_4 (x_2 + x_3) (x_2^\prime + x_3^\prime) (\phi_m^P + \phi_m^T)\nonumber\\
		& - 2 \psi_4 (A_5 + V_5) x_3^\prime (x_2^\prime + x_3^\prime) (\phi_m^P + \phi_m^T)) + m_p m_{\Lambda_b}^2 (-m_p (\psi_3^{-+} + \psi_3^{+-}) (P_2 + S_2 + T_4 - T_8) x_3^\prime (x_2^\prime + x_3^\prime) \phi_m^A\nonumber\\
		& - 2 m_0 \psi_2 (A_2 + V_2) (-1 + x_2) x_2 \phi_m^P + m_0 (A_3 (\psi_3^{-+} (x_2 + x_3) (x_2^\prime + x_3^\prime) + \psi_3^{+-} (x_2 + x_3) (x_2^\prime + x_3^\prime)\nonumber\\
		& + 2 \psi_4 x_3^\prime (-1 + x_2 + x_3 + y)) \phi_m^P + (2 \psi_2 (A_2 + V_2) x_3 + \psi_3^{+-} V_3 (x_2 + x_3) (x_2^\prime + x_3^\prime)\nonumber\\
		& - 2 (\psi_4 V_3 x_3^\prime (-1 + x_2 + x_3 + y) + \psi_2 V_2 (x_3 (2 x_2 + x_3) + (x_2 + x_3) y)\nonumber\\
		& + A_2 \psi_2 (x_3 (x_3 + y) + x_2 (2 x_3 + y)))) \phi_m^P + A_3 (\psi_3^{-+} (x_2 + x_3) (x_2^\prime + x_3^\prime) + \psi_3^{+-} (x_2 + x_3) (x_2^\prime + x_3^\prime)\nonumber\\
		& - 2 \psi_4 x_3^\prime (-1 + x_2 + x_3 + y)) \phi_m^T + (\psi_3^{+-} V_3 (x_2 +x_3) (x_2^\prime + x_3^\prime) + 2 A_2 \psi_2 (x_2 + x_3) (-1 + x_2 + x_3 + y)\nonumber\\
		& + 2 \psi_2 V_2 (x_2 + x_3) (-1 + x_2 + x_3 + y) + 2 \psi_4 V_3 x_3^\prime (-1 + x_2 + x_3 + y)) \phi_m^T\nonumber\\
		& + \psi_3^{-+} V_3 (x_2 + x_3) (x_2^\prime + x_3^\prime) (\phi_m^P + \phi_m^T))))\Big\}\nonumber\\
		&+\Big\{-8 (4 m_{\Lambda_b}^4 \psi_4 T_1 x_3^\prime (-1 + x_2 + x_3 + y) \phi_m^A + m_{\Lambda_b}^3 (2 m_p x_3^\prime (\psi_4 (P_1 - S_1 + T_3 + T_7) (x_2^\prime + x_3^\prime)\nonumber\\
		& - (\psi_3^{-+} + \psi_3^{+-}) T_2 (-1 + x_2 + x_3 + y)) \phi_m^A -m_0 (\psi_3^{-+} + \psi_3^{+-}) (A_1 - V_1) (x_2 + x_3) (-1 + x_2 + x_3 + y)\nonumber\\
		& (\phi_m^P - \phi_m^T)) + m_0 m_p^3 (\psi_3^{-+} + \psi_3^{+-}) (A_6 - V_6) x_3^\prime (x_2^\prime + x_3^\prime) (\phi_m^P + \phi_m^T) + m_0 m_p^2 m_{\Lambda_b} (-A_4 (\psi_3^{-+} + \psi_3^{+-}) x_3^\prime\nonumber\\
		& (-1 + x_2 + x_3 + y) (\phi_m^P - \phi_m^T) - \psi_3^{-+} V_4 x_3^\prime (-1 + x_2 + x_3 + y) (\phi_m^P - \phi_m^T) - \psi_3^{+-} V_4 x_3^\prime (-1 + x_2 + x_3 + y)\nonumber\\
		& (\phi_m^P - \phi_m^T) + 2 A_4 \psi_2 (x_2 + x_3) (x_2^\prime + x_3^\prime) (\phi_m^P + \phi_m^T) - 2 \psi_2 V_4 (x_2 + x_3) (x_2^\prime + x_3^\prime) (\phi_m^P + \phi_m^T)\nonumber\\
		& - 2 \psi_4 (A_5 + V_5) x_3^\prime (x_2^\prime + x_3^\prime) (\phi_m^P + \phi_m^T)) + m_p m_{\Lambda_b}^2 (-m_p (\psi_3^{-+} + \psi_3^{+-}) (P_2 + S_2 + T_4 - T_8) x_3^\prime (x_2^\prime + x_3^\prime) \phi_m^A\nonumber\\
		& + 2 m_0 \psi_2 (A_2 + V_2) (-1 + x_2) x_2 \phi_m^P + m_0 (A_3 (\psi_3^{-+} (x_2 + x_3) (x_2^\prime + x_3^\prime) + \psi_3^{+-} (x_2 + x_3) (x_2^\prime + x_3^\prime)\nonumber\\
		& - 2 \psi_4 x_3^\prime (-1 + x_2 + x_3 + y)) \phi_m^P + (-2 \psi_2 (A_2 + V_2) x_3 + \psi_3^{+-} V_3 (x_2 + x_3) (x_2^\prime + x_3^\prime)\nonumber\\
		& + 2 (\psi_4 V_3 x_3^\prime (-1 + x_2 + x_3 + y) + \psi_2 V_2 (x_3 (2 x_2 + x_3) + (x_2 + x_3) y)\nonumber\\
		& + A_2 \psi_2 (x_3 (x_3 + y) + x_2 (2 x_3 + y)))) \phi_m^P + A_3 (\psi_3^{-+} (x_2 + x_3) (x_2^\prime + x_3^\prime) + \psi_3^{+-} (x_2 + x_3) (x_2^\prime + x_3^\prime)\nonumber\\
		& + 2 \psi_4 x_3^\prime (-1 + x_2 + x_3 + y)) \phi_m^T + (\psi_3^{+-} V_3 (x_2 +x_3) (x_2^\prime + x_3^\prime) - 2 A_2 \psi_2 (x_2 + x_3) (-1 + x_2 + x_3 + y)\nonumber\\
		& - 2 \psi_2 V_2 (x_2 + x_3) (-1 + x_2 + x_3 + y) - 2 \psi_4 V_3 x_3^\prime (-1 + x_2 + x_3 + y)) \phi_m^T\nonumber\\
		& + \psi_3^{-+} V_3 (x_2 + x_3) (x_2^\prime + x_3^\prime) (\phi_m^P + \phi_m^T))))\Big\}\gamma_5,
\end{align}}

{\footnotesize
	\begin{align}
		&M_{E_2(a4)}^{SP}\equiv\langle ph|(\bar{u}b)_{S-P}(\bar{q}u)_{S+P}|\Lambda_b\rangle_{E_2}\nonumber\\
		=&\Big\{-4 m_p^3 (x_2^\prime + x_3^\prime) (m_{\Lambda_b} (\psi_3^{-+} + \psi_3^{+-}) (A_6 - V_6) x_3^\prime \phi_m^A - 4 m_0 \psi_2 T_6 (x_2 + x_3) (\phi_m^P - \phi_m^T))\nonumber\\
		& - 4 m_{\Lambda_b}^3 (-A_1 m_{\Lambda_b} (\psi_3^{-+} + \psi_3^{+-}) (x_2 + x_3) (x_2^\prime + x_3^\prime) \phi_m^A + m_{\Lambda_b} (\psi_3^{-+} + \psi_3^{+-}) V_1 (x_2 + x_3) (x_2^\prime + x_3^\prime) \phi_m^A\nonumber\\
		& - 4 m_0 \psi_4 T_1 x_3^\prime (-1 + x_2 + x_3 + y) (\phi_m^P + \phi_m^T)) - 4 m_p m_{\Lambda_b}^2 (2 m_{\Lambda_b} \psi_2 (A_2 + V_2) (x_2 + x_3) (x_2^\prime + x_3^\prime) \phi_m^A\nonumber\\
		& + \psi_3^{-+} (A_3 m_{\Lambda_b} (x_2 + x_3) (x_2^\prime + x_3^\prime) \phi_m^A + m_{\Lambda_b} V_3 (x_2 + x_3) (x_2^\prime + x_3^\prime) \phi_m^A - m_0 (2 T_2 x_3^\prime + (x_2 + x_3) \nonumber\\
		&((P_1 + S_1 - T_3 + T_7) x_2^\prime + (P_1 + S_1 - 2 T_2 - T_3 + T_7) x_3^\prime) - 2 T_2 x_3^\prime y) \phi_m^P + m_0 (-2 T_2 x_3^\prime + (x_2 + x_3)\nonumber\\
		&((P_1 + S_1 - T_3 + T_7) x_2^\prime + (P_1 + S_1 + 2 T_2 - T_3 + T_7) x_3^\prime) + 2 T_2 x_3^\prime y) \phi_m^T) + \psi_3^{+-} (A_3 m_{\Lambda_b} (x_2 + x_3) (x_2^\prime + x_3^\prime) \phi_m^A \nonumber\\
		&+ m_{\Lambda_b} V_3 (x_2 + x_3) (x_2^\prime + x_3^\prime) \phi_m^A - m_0 (2 T_2 x_3^\prime + (x_2 + x_3) ((P_1 + S_1 - T_3 + T_7) x_2^\prime + (P_1 + S_1 - 2 T_2 - T_3 + T_7) x_3^\prime)\nonumber\\
		&- 2 T_2 x_3^\prime y) \phi_m^P + m_0 (-2 T_2 x_3^\prime + (x_2 + x_3) ((P_1 + S_1 - T_3 + T_7) x_2^\prime + (P_1 + S_1 + 2 T_2 - T_3 + T_7) x_3^\prime) + 2 T_2 x_3^\prime y) \phi_m^T) \nonumber\\
		&+ 2 \psi_4 x_3^\prime (-m_{\Lambda_b} (A_3 - V_3) (x_2^\prime + x_3^\prime) \phi_m^A + m_0 (P_1 - S_1 - T_3 - T_7) (-1 + x_2 + x_3 + y) (\phi_m^P + \phi_m^T)))\nonumber\\
		&+ 4 m_p^2 m_{\Lambda_b} (2 m_{\Lambda_b} \psi_4 (A_5 + V_5) x_3^\prime (x_2^\prime + x_3^\prime) \phi_m^A + 2 \psi_2 (x_2 + x_3) (x_2^\prime + x_3^\prime) (m_{\Lambda_b} (-A_4 + V_4) \phi_m^A\nonumber\\
		&- m_0 (P_2 - S_2 - T_4 - T_8) (\phi_m^P - \phi_m^T)) + \psi_3^{-+} (A_4 m_{\Lambda_b} x_3^\prime (x_2^\prime + x_3^\prime) \phi_m^A + m_{\Lambda_b} V_4 x_3^\prime (x_2^\prime + x_3^\prime) \phi_m^A \nonumber\\
		&- 2 m_0 T_5 (x_2 + x_3) (x_2^\prime + x_3^\prime) (\phi_m^P - \phi_m^T) + m_0 (P_2 + S_2 - T_4 + T_8) x_3^\prime (-1 + x_2 + x_3 + y) (\phi_m^P + \phi_m^T)) \nonumber\\
		&+ \psi_3^{+-} (A_4 m_{\Lambda_b} x_3^\prime (x_2^\prime + x_3^\prime) \phi_m^A + m_{\Lambda_b} V_4 x_3^\prime (x_2^\prime + x_3^\prime) \phi_m^A - 2 m_0 T_5 (x_2 + x_3) (x_2^\prime + x_3^\prime) (\phi_m^P - \phi_m^T) \nonumber\\
		&+ m_0 (P_2 + S_2 - T_4 + T_8) x_3^\prime (-1 + x_2 + x_3 + y) (\phi_m^P + \phi_m^T)))\Big\}\nonumber\\
		&+\Big\{4 (-m_p^3 (x_2^\prime + x_3^\prime) (m_{\Lambda_b} (\psi_3^{-+} + \psi_3^{+-}) (A_6 - V_6) x_3^\prime \phi_m^A -  4 m_0 \psi_2 T_6 (x_2 + x_3) (\phi_m^P - \phi_m^T))\nonumber\\
		& +  m_{\Lambda_b}^3 (-A_1 m_{\Lambda_b} (\psi_3^{-+} + \psi_3^{+-}) (x_2 + x_3) (x_2^\prime + x_3^\prime) \phi_m^A + m_{\Lambda_b} (\psi_3^{-+} + \psi_3^{+-}) V_1 (x_2 + x_3) (x_2^\prime + x_3^\prime) \phi_m^A\nonumber\\
		& - 4 m_0 \psi_4 T_1 x_3^\prime (-1 + x_2 + x_3 + y) (\phi_m^P + \phi_m^T)) + m_p m_{\Lambda_b}^2 (2 m_{\Lambda_b} \psi_2 (A_2 + V_2) (x_2 + x_3) (x_2^\prime + x_3^\prime) \phi_m^A \nonumber\\
		&- \psi_3^{-+} (A_3 m_{\Lambda_b} (x_2 + x_3) (x_2^\prime + x_3^\prime) \phi_m^A + m_{\Lambda_b} V_3 (x_2 + x_3) (x_2^\prime + x_3^\prime) \phi_m^A + m_0 (2 T_2 x_3^\prime + (x_2 + x_3)\nonumber\\
		&((P_1 + S_1 - T_3 + T_7) x_2^\prime + (P_1 + S_1 - 2 T_2 - T_3 + T_7) x_3^\prime) - 2 T_2 x_3^\prime y) \phi_m^P - m_0 (-2 T_2 x_3^\prime + (x_2 + x_3) \nonumber\\
		&((P_1 + S_1 - T_3 + T_7) x_2^\prime + (P_1 + S_1 + 2 T_2 - T_3 + T_7) x_3^\prime) + 2 T_2 x_3^\prime y) \phi_m^T) - \psi_3^{+-} (A_3 m_{\Lambda_b} (x_2 + x_3) (x_2^\prime + x_3^\prime) \phi_m^A \nonumber\\
		&+ m_{\Lambda_b} V_3 (x_2 + x_3) (x_2^\prime + x_3^\prime) \phi_m^A + m_0 (2 T_2 x_3^\prime + (x_2 + x_3) ((P_1 + S_1 - T_3 + T_7) x_2^\prime + (P_1 + S_1 - 2 T_2 - T_3 + T_7) x_3^\prime)\nonumber\\
		& - 2 T_2 x_3^\prime y) \phi_m^P - m_0 (-2 T_2 x_3^\prime + (x_2 + x_3) ((P_1 + S_1 - T_3 + T_7) x_2^\prime + (P_1 + S_1 + 2 T_2 - T_3 + T_7) x_3^\prime) + 2 T_2 x_3^\prime y) \phi_m^T) \nonumber\\
		&+ 2 \psi_4 x_3^\prime (-m_{\Lambda_b} (A_3 - V_3) (x_2^\prime + x_3^\prime) \phi_m^A -m_0 (P_1 - S_1 - T_3 - T_7) (-1 + x_2 + x_3 + y) (\phi_m^P + \phi_m^T)))\nonumber\\
		&+ m_p^2 m_{\Lambda_b} (2 m_{\Lambda_b} \psi_4 (A_5 + V_5) x_3^\prime (x_2^\prime + x_3^\prime) \phi_m^A + 2 \psi_2 (x_2 + x_3) (x_2^\prime + x_3^\prime) (m_{\Lambda_b} (-A_4 + V_4) \phi_m^A \nonumber\\
		&+ m_0 (P_2 - S_2 - T_4 - T_8) (\phi_m^P - \phi_m^T)) - \psi_3^{-+} (A_4 m_{\Lambda_b} x_3^\prime (x_2^\prime + x_3^\prime) \phi_m^A +m_{\Lambda_b} V_4 x_3^\prime (x_2^\prime + x_3^\prime) \phi_m^A \nonumber\\
		&+ 2 m_0 T_5 (x_2 + x_3) (x_2^\prime + x_3^\prime) (\phi_m^P - \phi_m^T) - m_0 (P_2 + S_2 - T_4 + T_8) x_3^\prime (-1 + x_2 + x_3 + y) (\phi_m^P + \phi_m^T)) \nonumber\\
		&- \psi_3^{+-} (A_4 m_{\Lambda_b} x_3^\prime (x_2^\prime + x_3^\prime) \phi_m^A + m_{\Lambda_b} V_4 x_3^\prime (x_2^\prime + x_3^\prime) \phi_m^A + 2 m_0 T_5 (x_2 + x_3) (x_2^\prime + x_3^\prime) (\phi_m^P - \phi_m^T) \nonumber\\
		&- m_0 (P_2 + S_2 - T_4 + T_8) x_3^\prime (-1 + x_2 + x_3 + y) (\phi_m^P + \phi_m^T))))\Big\}\gamma_5.
\end{align}}

We have, for the $B(c3)$ diagram in Fig.~\ref{fig:feynmanB},
\begin{equation}
	\begin{split}
		\mathcal{A}_{B(c3)}=&\frac{-1}{128N_c^2} \frac{1}{\sqrt{2N_c}}\int[dx] \int[dx^\prime] \int dy \frac{1}{(2\pi)^5} \int b_1db_1 \int b_3 db_3 \int b_2^\prime db_2^\prime (4\pi\alpha_s(t^{B(c3)}))^2\\
		&(M_{B(c3)}+M_{PB(c3)})F_1(C,b_1-b_2^\prime)F_3(D,B,A,b_1,b_3,b_2^\prime) \text{exp}[-S^{B(c3)}([x],[x^\prime],y,[b],[b^\prime],b_q)],
	\end{split}
\end{equation}
with the arguments
\begin{equation}
	A=m_{\Lambda_b}^2 x_2^\prime(-1+x_2+y), \quad B=m_{\Lambda_b}^2 (-1+x_2)(x_2^\prime-x_3), \quad C=m_{\Lambda_b}^2 x_2^\prime(-1+y), \quad D=m_{\Lambda_b}^2 x_3y.
\end{equation}
The functions
\begin{equation}
	\begin{split}
		M_{B(c3)}=&\frac{G_F}{\sqrt{2}}\left\{V_{ub}V_{ud}^\ast\left[f_1C_1+f_2C_2\right]\right\}M_{B(c3)}^{LL},\\
		M_{PB(c3)}=&-\frac{G_F}{\sqrt{2}}\left\{V_{tb}V_{td}^\ast\left[f_1C_3+f_2C_4+f_2C_9+f_2C_{10}\right]\right\}M_{B(c3)}^{LL}\\
		&-\frac{G_F}{\sqrt{2}}\left\{V_{tb}V_{td}^\ast\left[-2f_1C_5-2f_2C_6-2f_1C_7-2f_2C_{8}\right]\right\}M_{B(c3)}^{SP},
	\end{split}
\end{equation}
contain the color factors $f_1=-10/3$ and $f_2=10/3$, and the matrix elements
{\small
	\begin{align}
		&M_{B(c3)}^{LL}\equiv\langle ph|(\bar{u}b)_{V-A}(\bar{d}u)_{V-A}|\Lambda_b\rangle_{B(c3)}\nonumber\\
		=&\Big\{8 m_p m_{\Lambda_b} (-2 m_{\Lambda_b}^2 \psi_4 (A_2 + A_3 - V_2 + V_3) x_2^\prime (x_2^\prime - x_3) \phi_m^A - m_0m_{\Lambda_b} \psi_3^{-+} (A_2 + A_3 - V_2 + V_3) (x_2^\prime - x_3) \nonumber\\
		&(-1 + x_2 + y) (\phi_m^P - \phi_m^T) - 2 m_0m_{\Lambda_b} \psi_2 (P_1 + S_1 + 2 T_2 + T_3 - T_7) (-1 + x_2) (-1 + x_2 + y) (\phi_m^P + \phi_m^T)\nonumber\\
		& -  m_p x_2^\prime (-2 m_{\Lambda_b} \psi_2 (A_4 + A_5 + V_4 - V_5) (-1 + x_2) \phi_m^A +  m_{\Lambda_b} \psi_3^{+-} (P_2 + S_2 + T_4 + 2 T_5 - T_8) \nonumber\\
		&(x_2^\prime - x_3) \phi_m^A + 2 m_0 \psi_4 (P_2 + S_2 + T_4 + 2 T_5 - T_8) (x_2^\prime - x_3) (\phi_m^P - \phi_m^T) \nonumber\\
		&+ m_0 \psi_3^{+-} (A_4 + A_5 + V_4 - V_5) (-1 + x_2) (\phi_m^P + \phi_m^T)))\Big\}\nonumber\\
		&+\Big\{8 m_p m_{\Lambda_b} (2 m_{\Lambda_b}^2 \psi_4 (A_2 + A_3 - V_2 + V_3) x_2^\prime (x_2^\prime - x_3) \phi_m^A + m_0m_{\Lambda_b} \psi_3^{-+} (A_2 + A_3 - V_2 + V_3) (x_2^\prime - x_3) \nonumber\\
		&(-1 + x_2 + y) (\phi_m^P - \phi_m^T) - 2 m_0m_{\Lambda_b} \psi_2 (P_1 + S_1 + 2 T_2 + T_3 - T_7) (-1 + x_2) (-1 + x_2 + y) (\phi_m^P + \phi_m^T)\nonumber\\
		& -  m_p x_2^\prime (-2 m_{\Lambda_b} \psi_2 (A_4 + A_5 + V_4 - V_5) (-1 + x_2) \phi_m^A -  m_{\Lambda_b} \psi_3^{+-} (P_2 + S_2 + T_4 + 2 T_5 - T_8) \nonumber\\
		&(x_2^\prime - x_3) \phi_m^A - 2 m_0 \psi_4 (P_2 + S_2 + T_4 + 2 T_5 - T_8) (x_2^\prime - x_3) (\phi_m^P - \phi_m^T)\nonumber\\
		&+ m_0 \psi_3^{+-} (A_4 + A_5 + V_4 - V_5) (-1 + x_2) (\phi_m^P + \phi_m^T)))\Big\}\gamma_5,
\end{align}}

{\small
	\begin{align}
		&M_{B(c3)}^{SP}\equiv\langle ph|(\bar{u}b)_{S-P}(\bar{d}u)_{S+P}|\Lambda_b\rangle_{B(c3)}\nonumber\\
		=&\Big\{8 m_p^3 T_6 x_2^\prime (x_2^\prime - x_3) (m_{\Lambda_b} \psi_3^{+-} \phi_m^A + 2 m_0 \psi_4 (\phi_m^P - \phi_m^T)) + 16 m_0 m_{\Lambda_b}^3 \psi_2 T_1 (-1 + x_2) (-1 + x_2 + y) \nonumber\\
		&(\phi_m^P + \phi_m^T) - 4 m_p^2 m_{\Lambda_b} (-2 m_{\Lambda_b} \psi_2 (A_5 + V_5) (-1 +x_2) x_2^\prime \phi_m^A + 2 \psi_4 x_2^\prime (x_2^\prime - x_3) (A_4 m_{\Lambda_b} \phi_m^A - m_{\Lambda_b} V_4 \phi_m^A \nonumber\\
		&+ m_0 (P_2 - S_2 - T_4 - T_8) (\phi_m^P - \phi_m^T)) + m_0 \psi_3^{-+} (A_4 - V_4) (x_2^\prime - x_3) (-1 + x_2 + y) (\phi_m^P - \phi_m^T)\nonumber\\
		& + \psi_3^{+-} x_2^\prime (m_{\Lambda_b} (P_2 - S_2 - T_4 - T_8) (x_2^\prime - x_3) \phi_m^A + m_0 (A_5 + V_5) (-1 + x_2) (\phi_m^P + \phi_m^T)))\nonumber\\
		& - 4 m_p m_{\Lambda_b}^2 (2 m_{\Lambda_b} \psi_4 (A_2 + V_2) x_2^\prime (x_2^\prime - x_3) \phi_m^A + m_0 \psi_3^{-+} (A_2 + V_2) (x_2^\prime - x_3) (-1 + x_2 + y) (\phi_m^P - \phi_m^T)\nonumber\\
		& + m_0 \psi_3^{+-} (A_3 - V_3) (-1 + x_2) x_2^\prime (\phi_m^P + \phi_m^T) + 2 \psi_2 (-1 + x_2) (-A_3 m_{\Lambda_b} x_2^\prime \phi_m^A + m_{\Lambda_b} V_3 x_2^\prime \phi_m^A\nonumber\\
		& + m_0 (P_1 - S_1 - T_3 - T_7) (-1 + x_2 + y) (\phi_m^P + \phi_m^T)))\Big\}\nonumber\\
		&+\Big\{8 m_p^3 T_6 x_2^\prime (x_2^\prime - x_3) (m_{\Lambda_b} \psi_3^{+-} \phi_m^A + 2 m_0 \psi_4 (\phi_m^P - \phi_m^T)) - 16 m_0 m_{\Lambda_b}^3 \psi_2 T_1 (-1 + x_2) (-1 + x_2 + y) \nonumber\\
		&(\phi_m^P + \phi_m^T) - 4 m_p^2 m_{\Lambda_b} (-2 m_{\Lambda_b} \psi_2 (A_5 + V_5) (-1 +x_2) x_2^\prime \phi_m^A + 2 \psi_4 x_2^\prime (x_2^\prime - x_3) (A_4 m_{\Lambda_b} \phi_m^A - m_{\Lambda_b} V_4 \phi_m^A\nonumber\\
		& - m_0 (P_2 - S_2 - T_4 - T_8) (\phi_m^P - \phi_m^T)) + m_0 \psi_3^{-+} (A_4 - V_4) (x_2^\prime - x_3) (-1 + x_2 + y) (\phi_m^P - \phi_m^T)\nonumber\\
		& +  \psi_3^{+-} x_2^\prime (m_{\Lambda_b} (-P_2 + S_2 + T_4 + T_8) (x_2^\prime - x_3) \phi_m^A + m_0 (A_5 + V_5) (-1 + x_2) (\phi_m^P + \phi_m^T))) \nonumber\\
		&+ 4 m_p m_{\Lambda_b}^2 (2 m_{\Lambda_b} \psi_4 (A_2 + V_2) x_2^\prime (x_2^\prime - x_3) \phi_m^A + m_0 \psi_3^{-+} (A_2 + V_2) (x_2^\prime - x_3) (-1 + x_2 + y) (\phi_m^P - \phi_m^T)\nonumber\\
		& + m_0 \psi_3^{+-} (A_3 - V_3) (-1 + x_2) x_2^\prime (\phi_m^P + \phi_m^T) - 2 \psi_2 (-1 + x_2) (A_3 m_{\Lambda_b} x_2^\prime \phi_m^A - m_{\Lambda_b} V_3 x_2^\prime \phi_m^A\nonumber\\
		& + m_0 (P_1 - S_1 - T_3 - T_7) (-1 + x_2 + y) (\phi_m^P + \phi_m^T)))\Big\}\gamma_5.
\end{align}}

The formula for the $PE_1^d(c2)$ diagram in Fig.~\ref{fig:feynmanPd} is written as
\begin{equation}
	\begin{split}
		\mathcal{A}_{PE_1^d(c2)}=&\frac{-1}{128N_c^2} \frac{1}{\sqrt{2N_c}}\int[dx] \int[dx^\prime] \int dy \frac{1}{(2\pi)^5} \int b_1db_1 \int b_2 db_2 \int b_q db_q (4\pi\alpha_s(t^{PE_1^d(c2)}))^2\\
		&M_{PE_1^d(c2)}F_1(D,b_3)F_3(A,B,C,b_1,b_2,b_q) \text{exp}[-S^{PE_1^d(c2)}([x],[x^\prime],y,[b],[b^\prime],b_q)],
	\end{split}
\end{equation}
with the arguments
\begin{equation}
	A=m_{\Lambda_b}^2 x_2^\prime(x_2 + x_3), \quad B=m_{\Lambda_b}^2 (x_1^\prime +x_2^\prime)(-1+x_2+x_3+y), \quad C=m_{\Lambda_b}^2 x_1^\prime(-1+y), \quad D=m_{\Lambda_b}^2 x_2x_2^\prime.
\end{equation}
The functions
\begin{equation}
	\begin{split}
		M_{PE_1^d(c2)}^{p\pi}=&-\frac{G_F}{\sqrt{2}}\left\{V_{tb}V_{td}^\ast\left[(f_1+f_2)(C_3+C_4-\frac{1}{2}C_9-\frac{1}{2}C_{10})\right]\right\}M_{PE_1^d(c2)}^{LL}\\
		&-\frac{G_F}{\sqrt{2}}\left\{V_{tb}V_{td}^\ast\left[f_1C_5+f_2C_6-\frac{1}{2}f_1C_7-\frac{1}{2}f_2C_{8}\right]\right\}M_{PE_1^d(c2)}^{LR}\\
		&-\frac{G_F}{\sqrt{2}}\left\{V_{tb}V_{td}^\ast\left[-2f_2C_5-2f_1C_6+f_2C_7+f_1C_{8}\right]\right\}M_{PE_1^d(c2)}^{SP},
	\end{split}
\end{equation}
\begin{equation}
	\begin{split}
		M_{PE_1^d(c2)}^{pK}=&-\frac{G_F}{\sqrt{2}}\left\{V_{tb}V_{ts}^\ast\left[(f_1C_3+f_2C_4-\frac{1}{2}f_1C_9-\frac{1}{2}f_2C_{10})\right]\right\}M_{PE_1^d(c2)}^{LL}\\
		&-\frac{G_F}{\sqrt{2}}\left\{V_{tb}V_{ts}^\ast\left[f_1C_5+f_2C_6-\frac{1}{2}f_1C_7-\frac{1}{2}f_2C_{8}\right]\right\}M_{PE_1^d(c2)}^{LR}.
	\end{split}
\end{equation}
contain the color factors $f_1=8/3$ and $f_2=16/3$, and the matrix elements
{\footnotesize
	\begin{align}
		&M_{PE_1^d(c2)}^{LL}\equiv\langle ph|(\bar{d}b)_{V-A}(\bar{d}d)_{V-A}|\Lambda_b\rangle_{PE_1^d(c2)}\nonumber\\
		=&\Big\{-8 m_{\Lambda_b}^3 (-1 + x_2 + x_3 + y) (2 A_1 m_{\Lambda_b} \psi_4 x_2^\prime \phi_m^A - 2 m_{\Lambda_b} \psi_4 V_1 x_2^\prime \phi_m^A + A_1 m_0 (\psi_3^{-+} + \psi_3^{+-}) (x_2 + x_3) (\phi_m^P - \phi_m^T)\nonumber\\
		&+ m_0 (\psi_3^{-+} + \psi_3^{+-}) V_1 (x_2 + x_3) (\phi_m^P - \phi_m^T)) - 8 m_0 m_p^3 (\psi_3^{-+} + \psi_3^{+-}) (A_6 + V_6) x_2^\prime (x_1^\prime + x_2^\prime) (\phi_m^P + \phi_m^T) \nonumber\\
		&+ 8 m_p m_{\Lambda_b}^2 (m_{\Lambda_b} x_2^\prime (-2 \psi_4 (A_3 + V_3) (x_1^\prime + x_2^\prime) + (\psi_3^{-+} + \psi_3^{+-}) (P_1 + S_1 - T_3 + T_7) (-1 + x_2 + x_3 + y)) \phi_m^A \nonumber\\
		&+ m_0 (2\psi_2(\phi_m^P-\phi_m^T)(P_1-S_1+T_3+T_7)(x_2+x_3)(-1+x_2+x_3+y) \nonumber\\
		&+ (\psi_3^{+-}+\psi_3^{-+})(\phi_m^P+\phi_m^T)(A_2+V_2)(x_2+x_3)(x_1^\prime+x_2^\prime)   -2\psi_4(\phi_m^P-\phi_m^T)(P_1-S_1-T_3-T_7)\nonumber\\
		&x_2^\prime(-1+x_2+x_3+y))) + 8 m_p^2 m_{\Lambda_b} (2 m_{\Lambda_b} (\psi_3^{-+} + \psi_3^{+-}) T_5 x_2^\prime (x_1^\prime + x_2^\prime) \phi_m^A \nonumber\\
		&+ m_0 (2 \psi_2 (S_2 + T_4 + T_8) (x_1^\prime + x_2^\prime) (x_2 + x_3) (\phi_m^P + \phi_m^T) - 2 P_2 (x_1^\prime + x_2^\prime) (-\psi_4 x_2^\prime + \psi_2 (x_2 + x_3)) (\phi_m^P + \phi_m^T) \nonumber\\
		&+ x_2^\prime (-2 \psi_4 (S_2 - T_4 - T_8) (x_1^\prime + x_2^\prime) \phi_m^P + (\psi_3^{-+} + \psi_3^{+-}) (A_5 + V_5) (-1 + x_2 + x_3 + y) \phi_m^P \nonumber\\
		&- 2 \psi_4 (S_2 - T_4 - T_8) (x_1^\prime + x_2^\prime) \phi_m^T - (\psi_3^{-+} + \psi_3^{+-}) (A_5 + V_5) (-1 + x_2 + x_3 + y) \phi_m^T)))\Big\}\nonumber\\
		&+\Big\{-8 m_{\Lambda_b}^3 (-1 + x_2 + x_3 + y) (2 A_1 m_{\Lambda_b} \psi_4 x_2^\prime \phi_m^A - 2 m_{\Lambda_b} \psi_4 V_1 x_2^\prime \phi_m^A + A_1 m_0 (\psi_3^{-+} + \psi_3^{+-}) (x_2 + x_3) (\phi_m^P - \phi_m^T) \nonumber\\
		&+ m_0 (\psi_3^{-+} + \psi_3^{+-}) V_1 (x_2 + x_3) (\phi_m^P - \phi_m^T)) + 8 m_0 m_p^3 (\psi_3^{-+} + \psi_3^{+-}) (A_6 + V_6) x_2^\prime (x_1^\prime + x_2^\prime) (\phi_m^P + \phi_m^T)\nonumber\\
		&- 8 m_p m_{\Lambda_b}^2 (-m_{\Lambda_b} x_2^\prime (2 \psi_4 (A_3 + V_3) (x_1^\prime + x_2^\prime) + (\psi_3^{-+} + \psi_3^{+-}) (P_1 + S_1 - T_3 + T_7) (-1 + x_2 + x_3 + y)) \phi_m^A \nonumber\\
		&+ m_0 (-2\psi_2(\phi_m^P-\phi_m^T)(P_1-S_1+T_3+T_7)(x_2+x_3)(-1+x_2+x_3+y) \nonumber\\
		&+ (\psi_3^{+-}+\psi_3^{-+})(\phi_m^P+\phi_m^T)(A_2+V_2)(x_2+x_3)(x_1^\prime+x_2^\prime)   +2\psi_4(\phi_m^P-\phi_m^T)(P_1-S_1-T_3-T_7)\nonumber\\
		&x_2^\prime(-1+x_2+x_3+y))) - 8 m_p^2 m_{\Lambda_b} (2 m_{\Lambda_b} (\psi_3^{-+} + \psi_3^{+-}) T_5 x_2^\prime (x_1^\prime + x_2^\prime) \phi_m^A \nonumber\\
		&+ m_0 (2 \psi_2 (S_2 + T_4 + T_8) (x_1^\prime + x_2^\prime) (x_2 + x_3) (\phi_m^P + \phi_m^T) - 2 P_2 (x_1^\prime + x_2^\prime) (-\psi_4 x_2^\prime + \psi_2 (x_2 + x_3)) (\phi_m^P + \phi_m^T) \nonumber\\
		&+ x_2^\prime (-2 \psi_4 (S_2 - T_4 - T_8) (x_1^\prime + x_2^\prime) \phi_m^P - (\psi_3^{-+} + \psi_3^{+-}) (A_5 + V_5) (-1 + x_2 + x_3 + y) \phi_m^P \nonumber\\
		&- 2 \psi_4 (S_2 - T_4 - T_8) (x_1^\prime + x_2^\prime) \phi_m^T + (\psi_3^{-+} + \psi_3^{+-}) (A_5 + V_5) (-1 + x_2 + x_3 + y) \phi_m^T)))\Big\}\gamma_5,
\end{align}}

{\small
	\begin{align}
		&M_{PE_1^d(c2)}^{LR}\equiv\langle ph|(\bar{d}b)_{V-A}(\bar{d}d)_{V-A}|\Lambda_b\rangle_{PE_1^d(c2)}\nonumber\\
		=&\Big\{-8 m_p^3 (x_1^\prime + x_2^\prime) (A_6 m_{\Lambda_b} (\psi_3^{-+} + \psi_3^{+-}) x_2^\prime \phi_m^A + m_{\Lambda_b} (\psi_3^{-+} + \psi_3^{+-}) V_6 x_2^\prime \phi_m^A + 2 A_6 m_0 \psi_2 (x_2 + x_3)  \nonumber\\
		&(\phi_m^P - \phi_m^T)- 2 m_0 \psi_2 V_6 (x_2 + x_3) (\phi_m^P - \phi_m^T)) + 8 m_{\Lambda_b}^3 (A_1 m_{\Lambda_b} (\psi_3^{-+} + \psi_3^{+-}) (x_1^\prime + x_2^\prime) (x_2 + x_3) \phi_m^A \nonumber\\
		&+ m_{\Lambda_b} (\psi_3^{-+} + \psi_3^{+-}) V_1 (x_1^\prime + x_2^\prime) (x_2 + x_3) \phi_m^A - 2 A_1 m_0 \psi_4 x_2^\prime (-1 + x_2 + x_3 + y) (\phi_m^P + \phi_m^T) \nonumber\\
		&+ 2 m_0 \psi_4 V_1 x_2^\prime (-1 + x_2 + x_3 + y) (\phi_m^P + \phi_m^T)) - 8 m_p^2 m_{\Lambda_b} (m_{\Lambda_b} (x_1^\prime + x_2^\prime) (-2 P_2 \psi_4 x_2^\prime + 2 \psi_4 (S_2\nonumber\\
		& - T_4 - T_8) x_2^\prime + (\psi_3^{-+} + \psi_3^{+-}) (A_5 + V_5) x_2^\prime + 2 P_2 \psi_2 (x_2 + x_3) - 2 \psi_2 (S_2 + T_4 + T_8) (x_2 + x_3)) \phi_m^A \nonumber\\
		&+ m_0 ( 2\psi_2(\phi_m^P-\phi_m^T)(A_5-V_5)(x_1^\prime+x_2^\prime)(x_2+x_3) - (\psi_3^{+-}+\psi_3^{-+})(P_2+S_2)(x_2^\prime(-1+2x_2+2x_3+y)\phi_m^P \nonumber\\
		&+ x_1^\prime(x_2+x_3)(\phi_m^P-\phi_m^T)+x_2^\prime(-1+y)\phi_m^T) + (\psi_3^{+-}+\psi_3^{-+})(T_4-T_8)(x_1^\prime(x_2+x_3)(\phi_m^P-\phi_m^T)\nonumber\\
		&+x_2^\prime(\phi_m^P-y\phi_m^P-(-1+2x_2+2x_3+y)\phi_m^T)) )) + 8 m_p m_{\Lambda_b}^2 (m_{\Lambda_b} (x_1^\prime + x_2^\prime) (2 \psi_2 S_1 x_2 - 2 \psi_2 T_3 x_2 - 2 \psi_2 T_7 x_2 \nonumber\\
		&+ \psi_3^{-+} V_2 x_2 + \psi_3^{+-} V_2 x_2 + 2 P_1 \psi_4 x_2^\prime - 2 \psi_4 S_1 x_2^\prime - 2 \psi_4 T_3 x_2^\prime - 2 \psi_4 T_7 x_2^\prime + 2 \psi_2 (S_1 - T_3 - T_7) x_3 \nonumber\\
		&+ (\psi_3^{-+} + \psi_3^{+-}) V_2 x_3 - 2 P_1 \psi_2 (x_2 + x_3) + A_2 (\psi_3^{-+} + \psi_3^{+-}) (x_2 + x_3)) \phi_m^A \nonumber\\
		&+ m_0 (  -2\psi_4(\phi_m^P+\phi_m^T)(A_2-V_2)x_2^\prime(-1+x_2+x_3+y) + (\psi_3^{+-}+\psi_3^{-+})(P_1+S_1)(x_2^\prime(-1+2x_2+2x_3+y)\phi_m^P \nonumber\\
		&+ x_1^\prime(x_2+x_3)(\phi_m^P-\phi_m^T)+x_2^\prime(-1+y)\phi_m^T) + (\psi_3^{+-}+\psi_3^{-+})(T_3-T_7)(x_1^\prime(x_2+x_3)(\phi_m^P-\phi_m^T)\nonumber\\
		&+x_2^\prime(\phi_m^P-y\phi_m^P-(-1+2x_2+2x_3+y)\phi_m^T))  ))\Big\}\nonumber\\
		&+\Big\{-8 m_p^3 (x_1^\prime + x_2^\prime) (A_6 m_{\Lambda_b} (\psi_3^{-+} + \psi_3^{+-}) x_2^\prime \phi_m^A + m_{\Lambda_b} (\psi_3^{-+} + \psi_3^{+-}) V_6 x_2^\prime \phi_m^A + 2 A_6 m_0 \psi_2 (x_2 + x_3)  \nonumber\\
		&(\phi_m^P - \phi_m^T)- 2 m_0 \psi_2 V_6 (x_2 + x_3) (\phi_m^P - \phi_m^T)) - 8 m_{\Lambda_b}^3 (A_1 m_{\Lambda_b} (\psi_3^{-+} + \psi_3^{+-}) (x_1^\prime + x_2^\prime) (x_2 + x_3) \phi_m^A\nonumber \\
		&+ m_{\Lambda_b} (\psi_3^{-+} + \psi_3^{+-}) V_1 (x_1^\prime + x_2^\prime) (x_2 + x_3) \phi_m^A - 2 A_1 m_0 \psi_4 x_2^\prime (-1 + x_2 + x_3 + y) (\phi_m^P + \phi_m^T) \nonumber\\
		&+ 2 m_0 \psi_4 V_1 x_2^\prime (-1 + x_2 + x_3 + y) (\phi_m^P + \phi_m^T)) + 8 m_p^2 m_{\Lambda_b} (m_{\Lambda_b} (x_1^\prime + x_2^\prime) (2 P_2 \psi_4 x_2^\prime + 2 \psi_4 (-S_2\nonumber\\
		& + T_4 + T_8) x_2^\prime + (\psi_3^{-+} + \psi_3^{+-}) (A_5 + V_5) x_2^\prime - 2 P_2 \psi_2 (x_2 + x_3) + 2 \psi_2 (S_2 + T_4 + T_8) (x_2 + x_3)) \phi_m^A \nonumber\\
		&+ m_0 ( 2\psi_2(\phi_m^P-\phi_m^T)(A_5-V_5)(x_1^\prime+x_2^\prime)(x_2+x_3) + (\psi_3^{+-}+\psi_3^{-+})(P_2+S_2)(x_2^\prime(-1+2x_2+2x_3+y)\phi_m^P \nonumber\\
		&+ x_1^\prime(x_2+x_3)(\phi_m^P-\phi_m^T)+x_2^\prime(-1+y)\phi_m^T) - (\psi_3^{+-}+\psi_3^{-+})(T_4-T_8)(x_1^\prime(x_2+x_3)(\phi_m^P-\phi_m^T)\nonumber\\
		&+x_2^\prime(\phi_m^P-y\phi_m^P-(-1+2x_2+2x_3+y)\phi_m^T)) )) + 8 m_p m_{\Lambda_b}^2 (m_{\Lambda_b} (x_1^\prime + x_2^\prime) (A_2 \psi_3^{-+} x_2 + A_2 \psi_3^{+-} x_2 - 2 \psi_2 S_1 x_2\nonumber\\
		& + 2 \psi_2 T_3 x_2 + 2 \psi_2 T_7 x_2 \nonumber\\
		&+ \psi_3^{-+} V_2 x_2 + \psi_3^{+-} V_2 x_2 - 2 P_1 \psi_4 x_2^\prime + 2 \psi_4 S_1 x_2^\prime + 2 \psi_4 T_3 x_2^\prime + 2 \psi_4 T_7 x_2^\prime + 2 \psi_2 (-S_1 + T_3 + T_7) x_3\nonumber \\
		&+ (\psi_3^{-+} + \psi_3^{+-}) (A_2 + V_2) x_3 + 2 P_1 \psi_2 (x_2 + x_3)) \phi_m^A\nonumber\\
		& - m_0 (  2\psi_4(\phi_m^P+\phi_m^T)(A_2-V_2)x_2^\prime(-1+x_2+x_3+y) + (\psi_3^{+-}+\psi_3^{-+})(P_1+S_1)(x_2^\prime(-1+2x_2+2x_3+y)\phi_m^P \nonumber\\
		&+ x_1^\prime(x_2+x_3)(\phi_m^P-\phi_m^T)+x_2^\prime(-1+y)\phi_m^T) + (\psi_3^{+-}+\psi_3^{-+})(T_3-T_7)(x_1^\prime(x_2+x_3)(\phi_m^P-\phi_m^T)\nonumber\\
		&+x_2^\prime(\phi_m^P-y\phi_m^P-(-1+2x_2+2x_3+y)\phi_m^T))  ))\Big\}\gamma_5,
\end{align}}

{\footnotesize
	\begin{align}
		&M_{PE_1^d(c2)}^{SP}\equiv\langle ph|(\bar{d}b)_{S-P}(\bar{d}d)_{S+P}|\Lambda_b\rangle_{PE_1^d(c2)},\nonumber\\
		=&\Big\{-4 m_{\Lambda_b}^3 (-1 + x_2 + x_3 + y) (2 A_1 m_{\Lambda_b} \psi_4 x_2^\prime \phi_m^A + 2 m_{\Lambda_b} \psi_4 V_1 x_2^\prime \phi_m^A + A_1 m_0 (\psi_3^{-+} + \psi_3^{+-}) (x_2 + x_3)  \nonumber\\
		&(\phi_m^P - \phi_m^T)+ m_0 (\psi_3^{-+} + \psi_3^{+-}) V_1 (x_2 + x_3) (-\phi_m^P + \phi_m^T)) - 4 m_p^3 (\psi_3^{-+} + \psi_3^{+-}) x_2^\prime (x_1^\prime + x_2^\prime) (2 m_{\Lambda_b} T_6 \phi_m^A \nonumber\\
		&+ m_0 (A_6 - V_6) (\phi_m^P + \phi_m^T)) + 4 m_p m_{\Lambda_b}^2 (m_{\Lambda_b} (\psi_3^{-+} + \psi_3^{+-}) (P_1 -S_1 + T_3 + T_7) x_2^\prime (-1 + x_2 + x_3 + y) \phi_m^A \nonumber\\
		&+ m_0 (  2\psi_2(P_1+S_1-T_3+T_7)(x_2+x_3)(-1+x_2+x_3+y)(\phi_m^P-\phi_m^T)+(\psi_3^{+-}+\psi_3^{-+})(\phi_m^P+\phi_m^T)\nonumber\\
		&(A_2-V_2)(x_1^\prime+x_2^\prime)(x_2+x_3)-2\psi_4(\phi_m^P-\phi_m^T)(P_1+S_1+T_3-T_7)x_2^\prime(-1+x_2+x_3+y)  ))\nonumber \\
		&- 4 m_p^2 m_{\Lambda_b} (2 A_4 m_{\Lambda_b} \psi_4 x_2^\prime (x_1^\prime + x_2^\prime) \phi_m^A - 2 m_{\Lambda_b} \psi_4 V_4 x_2^\prime (x_1^\prime + x_2^\prime) \phi_m^A \nonumber\\
		&+ m_0 (2 \psi_2 (S_2 + T_4 - T_8) (x_1^\prime + x_2^\prime) (x_2 + x_3) (\phi_m^P + \phi_m^T) + 2 P_2 (x_1^\prime + x_2^\prime) (-\psi_4 x_2^\prime + \psi_2 (x_2 + x_3)) (\phi_m^P + \phi_m^T) \nonumber\\
		&- x_2^\prime (2 \psi_4 (S_2 - T_4 + T_8) (x_1^\prime + x_2^\prime) \phi_m^P + (\psi_3^{-+} +  \psi_3^{+-}) (A_5 - V_5) (-1 + x_2 + x_3 + y) \phi_m^P \nonumber\\
		&+ 2 \psi_4 (S_2 - T_4 + T_8) (x_1^\prime + x_2^\prime) \phi_m^T - (\psi_3^{-+} + \psi_3^{+-}) (A_5 - V_5) (-1 + x_2 + x_3 + y) \phi_m^T)))\Big\}\nonumber\\
		&+\Big\{-4 m_{\Lambda_b}^3 (-1 + x_2 + x_3 + y) (2 A_1 m_{\Lambda_b} \psi_4 x_2^\prime \phi_m^A + 2 m_{\Lambda_b} \psi_4 V_1 x_2^\prime \phi_m^A + A_1 m_0 (\psi_3^{-+} + \psi_3^{+-}) (x_2 + x_3) \nonumber \\
		&(\phi_m^P - \phi_m^T)+ m_0 (\psi_3^{-+} + \psi_3^{+-}) V_1 (x_2 + x_3) (-\phi_m^P + \phi_m^T)) - 4 m_p^3 (\psi_3^{-+} + \psi_3^{+-}) x_2^\prime (x_1^\prime + x_2^\prime) (2 m_{\Lambda_b} T_6 \phi_m^A \nonumber\\
		&- m_0 (A_6 - V_6) (\phi_m^P + \phi_m^T)) + 4 m_p m_{\Lambda_b}^2 (m_{\Lambda_b} (\psi_3^{-+} + \psi_3^{+-}) (P_1 -S_1 + T_3 + T_7) x_2^\prime (-1 + x_2 + x_3 + y) \phi_m^A \nonumber\\
		&+ m_0 (  2\psi_2(P_1+S_1-T_3+T_7)(x_2+x_3)(-1+x_2+x_3+y)(\phi_m^P-\phi_m^T)-(\psi_3^{+-}+\psi_3^{-+})(\phi_m^P+\phi_m^T)\nonumber\\
		&(A_2-V_2)(x_1^\prime+x_2^\prime)(x_2+x_3)-2\psi_4(\phi_m^P-\phi_m^T)(P_1+S_1+T_3-T_7)x_2^\prime(-1+x_2+x_3+y)  )) \nonumber\\
		&- 4 m_p^2 m_{\Lambda_b} (2 A_4 m_{\Lambda_b} \psi_4 x_2^\prime (x_1^\prime + x_2^\prime) \phi_m^A - 2 m_{\Lambda_b} \psi_4 V_4 x_2^\prime (x_1^\prime + x_2^\prime) \phi_m^A\nonumber \\
		&+ m_0 (-2 \psi_2 (S_2 + T_4 - T_8) (x_1^\prime + x_2^\prime) (x_2 + x_3) (\phi_m^P + \phi_m^T) - 2 P_2 (x_1^\prime + x_2^\prime) (-\psi_4 x_2^\prime +  \psi_2 (x_2 + x_3)) (\phi_m^P + \phi_m^T) \nonumber\\
		&+ x_2^\prime (2 \psi_4 (S_2 - T_4 + T_8) (x_1^\prime + x_2^\prime) \phi_m^P - (\psi_3^{-+} + \psi_3^{+-}) (A_5 - V_5) (-1 + x_2 + x_3 + y) \phi_m^P \nonumber\\
		&+ 2 \psi_4 (S_2 - T_4 + T_8) (x_1^\prime + x_2^\prime) \phi_m^T + (\psi_3^{-+} + \psi_3^{+-}) (A_5 - V_5) (-1 + x_2 + x_3 + y) \phi_m^T)))\Big\}\gamma_5.
\end{align}}

\section{Numerical results for $\Lambda_b\to pV, pA$ decays}\label{app:appendix-numerical-results}

The numerical results for the $\Lambda_b\to p\rho^-,pK^{\ast -}, pa_1^-(1260),pK_1^-(1270), pK_1^-(1400)$ decay amplitudes are summarized here.
Several comments are in order:
\begin{itemize}
	\item
	Tables~\ref{tab:rho-invariant} and ~\ref{tab:Kstar-invariant} show the  invariant amplitudes for the $\Lambda_b\to p\rho^-$ and $\Lambda_b\to pK^{\ast -}$ modes, respectively.
	The $(S+P)\otimes (S-P)$ operators, coming from the Fierz transformation of the $(V-A)\otimes(V+A)$ ones, do not   contribute to the factorizable penguin diagrams, because a vector
	meson cannot be generated by the scalar or pseudoscalar current.
	There is thus no hierarchy between the invariant amplitudes in the $PC_1^f$ topology.
	The non-factorizable contributions are comparable to or even larger than the factorizable ones in all invariant amplitudes, contrary to the conventional power counting rules. 
	
	\item 
	The helicity amplitudes of the $\Lambda_b\to p\rho^-$ and $\Lambda_b\to pK^{\ast -}$ decays in Tables~\ref{tab:rho-helicity} and ~\ref{tab:Kstar-helicity}, respectively, reveal a hierarchical pattern for the factorizable topologies. The left-handed $(V-A)\otimes (V-A)$ current responsible for the factorizable topologies prefers the configuration with the proton spin being anti-parallel to its momentum and with a transversely polarized meson. As a consequence, the helicity amplitude $H_{-\frac{1}{2},0}$ dominates the $T_f$ and $PC_1^f$ topologies.
	
	\item Tables~\ref{tab:rho-partialwave} and ~\ref{tab:Kstar-partialwave} are for the partial-wave amplitudes in the  $\Lambda_b\to p\rho^-$ and $\Lambda_b\to pK^{\ast -}$ decays, respectively. The $S^L+D$ and $P_1$ waves are related to the $\lambda_V=0$ helicity state, and the $P_2-S^T$ and $P_2+S^T$ waves are related to the $\lambda_M=1$ and $\lambda_M=-1$ helicity states, respectively~\cite{Pakvasa:1990if}. Hence, the $T_f$ and $PC_1^f$ topologies dominate the $S^L+D$ and $P_1$ wave amplitudes, but are suppressed by helicity flip effects in the $P_2$ and $S^T$ waves.
	
\end{itemize}

\begin{sidewaystable}
	\centering
	\footnotesize
	\renewcommand{\arraystretch}{1.2}
\begin{tabular*}{230mm}{c@{\extracolsep{\fill}}cccccccccccc}
		\toprule[1pt]
		\toprule[0.7pt]
		& Real($A_1^L$) & Imag($A_1^L$) & Real($B_1^L$) & Imag($B_1^L$) & Real($A_2^L$) & Imag($A_2^L$) & Real($B_2^L$) & Imag($B_2^L$) & Real($A_1^T$) & Imag($A_1^T$) & Real($B_1^T$) & Imag($B_1^T$)\\
		\toprule[0.7pt]
		$T^f$ &  -162.20 &     0.00 &   222.31 &     0.00 &    58.98 &     0.00 &   -64.57 &     0.00 &  -162.20 &     0.00 &   222.31 &     0.00 \\
		$T^{nf}$ &     5.05 &     8.83 &     0.64 &   -17.28 &   -15.58 &  -172.86 &    -6.32 &   -25.32 &     1.52 &    -6.01 &     4.19 &     1.28 \\
		$T$ &  -157.15 &     8.83 &   222.95 &   -17.28 &    43.40 &  -172.86 &   -70.90 &   -25.32 &  -160.68 &    -6.01 &   226.51 &     1.28 \\
		$C_2$ &    91.63 &    58.69 &   118.67 &   115.65 &   206.61 &   163.16 &  -209.59 &  -180.04 &    25.15 &    73.11 &    19.06 &   128.59 \\
		$E_2$ &   -23.49 &   197.98 &    41.40 &  -284.37 &   -67.94 &   529.91 &   -74.14 &   486.93 &   -29.91 &   182.78 &    44.33 &  -251.50 \\
		$B$ &    -4.74 &    -2.65 &   -13.35 &     9.86 &   -12.43 &    -4.05 &    13.59 &   -21.22 &    -5.47 &    -8.45 &   -10.04 &    -4.43 \\
		\toprule[0.7pt]
		Tree &   -93.75 &   262.86 &   369.67 &  -176.14 &   169.64 &   516.17 &  -341.04 &   260.36 &  -170.92 &   241.42 &   279.86 &  -126.06 \\
\end{tabular*}
	\caption{Invariant amplitudes of the $\Lambda_b\to p\rho^-$ decay.}
	\label{tab:rho-invariant}
\end{sidewaystable}

\begin{sidewaystable}
	\centering
	\footnotesize
	\renewcommand{\arraystretch}{1.2}
	\begin{tabular*}{230mm}{c@{\extracolsep{\fill}}cccccccc}
		\toprule[1pt]
		\toprule[0.7pt]
		& Real($H_{1/2,1}$) & Imag($H_{1/2,1}$) & Real($H_{-1/2,-1}$) & Imag($H_{-1/2,-1}$) & Real($H_{1/2,0}$) & Imag($H_{1/2,0}$) & Real($H_{-1/2,0}$) & Imag($H_{-1/2,0}$) \\
		\toprule[0.7pt]
		$T^f$ &    29.39 &    -0.00 & -2084.58 &     0.00 &   -61.55 &     0.00 & 10376.39 &     0.00 \\
		$T^{nf}$ &   -29.27 &    33.29 &    -9.50 &   -45.09 &   235.30 &  1325.62 &  -402.27 & -3099.70 \\
		$T$ &     0.12 &    33.29 & -2094.08 &   -45.09 &   173.75 &  1325.62 &  9974.12 & -3099.70 \\
		$C_2$ &  -252.02 & -1070.82 &    75.78 &  -117.97 &   167.26 &   126.97 &  -274.43 &   544.89 \\
		$E_2$ &    -9.98 &   -28.57 &  -399.86 &  2353.61 &    78.59 &  -589.05 &  -148.59 &   354.39 \\
		$B$ &    82.07 &    75.56 &    10.74 &   -34.60 &  -145.74 &   -93.58 &  -162.44 &   -37.16 \\
		\toprule[0.7pt]
		Tree &  -179.81 &  -990.54 & -2407.42 &  2155.95 &   273.86 &   769.96 &  9388.65 & -2237.58 \\
\end{tabular*}
	\caption{Helicity amplitudes of the $\Lambda_b\to p\rho^-$ decay.}
	\label{tab:rho-helicity}
\end{sidewaystable}

\begin{sidewaystable}
	\centering
	\footnotesize
	\renewcommand{\arraystretch}{1.2}
	\begin{tabular*}{230mm}{c@{\extracolsep{\fill}}cccccccccccc}
		\toprule[1pt]
		\toprule[0.7pt]
		& Real($S^L$) & Imag($S^L$) & Real($S^T$) & Imag($S^T$) & Real($P_1$) & Imag($P_1$) & Real($P_2$) & Imag($P_2$) & Real($D$) & Imag($D$) & Real($D+S^L$) & Imag($D+S^L$)\\
		\toprule[0.7pt]
		$T^f$ &   162.20 &    -0.00 &   162.20 &    -0.00 &  -309.28 &     0.00 &   157.69 &     0.00 &   150.78 &    -0.00 &   312.97 &    -0.00  \\
		$T^{nf}$ &    -5.05 &    -8.83 &    -1.52 &     6.01 &     5.01 &    53.19 &     2.98 &     0.91 &   -14.07 &  -123.86 &   -19.12 &  -132.69  \\
		$T$ &   157.15 &    -8.83 &   160.68 &     6.01 &  -304.28 &    53.19 &   160.66 &     0.91 &   136.71 &  -123.86 &   293.86 &  -132.69  \\
		$C_2$ &   -91.63 &   -58.69 &   -25.15 &   -73.11 &     3.21 &   -20.15 &    13.52 &    91.21 &    78.38 &    71.22 &   -13.24 &    12.53  \\
		$E_2$ &    23.49 &  -197.98 &    29.91 &  -182.78 &     2.10 &     7.04 &    31.45 &  -178.39 &   -30.30 &   226.27 &    -6.81 &    28.29   \\
		$B$ &     4.74 &     2.65 &     5.47 &     8.45 &     9.24 &     3.92 &    -7.12 &    -3.14 &    -5.24 &    -0.96 &    -0.50 &     1.69   \\
		\toprule[0.7pt]
		Tree &    93.75 &  -262.86 &   170.92 &  -241.42 &  -289.72 &    44.01 &   198.51 &   -89.42 &   179.55 &   172.68 &   273.30 &   -90.18   \\
\end{tabular*}
	\caption{Partial wave amplitudes of the $\Lambda_b\to p\rho^-$ decay.}
	\label{tab:rho-partialwave}
\end{sidewaystable}


\begin{sidewaystable}
	\centering
	\footnotesize
	\renewcommand{\arraystretch}{1.2}
	\begin{tabular*}{230mm}{c@{\extracolsep{\fill}}cccccccccccc}
		\toprule[1pt]
		\toprule[0.7pt]
		& Real($A_1^L$) & Imag($A_1^L$) & Real($B_1^L$) & Imag($B_1^L$) & Real($A_2^L$) & Imag($A_2^L$) & Real($B_2^L$) & Imag($B_2^L$) & Real($A_1^T$) & Imag($A_1^T$) & Real($B_1^T$) & Imag($B_1^T$)\\
		\toprule[0.7pt]
		$T^f$ &  -194.45 &     0.00 &   266.80 &     0.00 &    71.23 &     0.00 &   -77.98 &     0.00 &  -194.45 &     0.00 &   266.80 &     0.00 \\
		$T^{nf}$ &     4.76 &    10.58 &     2.72 &   -20.50 &    -8.72 &  -203.69 &    -9.61 &   -29.00 &     2.22 &    -7.02 &     5.17 &     3.44 \\
		$T$ &  -189.69 &    10.58 &   269.53 &   -20.50 &    62.51 &  -203.69 &   -87.59 &   -29.00 &  -192.24 &    -7.02 &   271.98 &     3.44 \\
		$E_2$ &   -28.18 &   227.94 &    37.03 &  -316.34 &  -108.84 &   577.73 &   -82.87 &   549.87 &   -30.72 &   177.26 &    41.24 &  -260.86 \\
		\toprule[0.7pt]
		Tree &  -217.87 &   238.53 &   306.56 &  -336.84 &   -46.33 &   374.04 &  -170.46 &   520.87 &  -222.96 &   170.23 &   313.22 &  -257.41 \\
		\toprule[0.7pt]
		$PC_1^f$ &    -8.16 &     0.00 &    11.19 &     0.00 &     2.97 &     0.00 &    -3.23 &     0.00 &    -8.16 &     0.00 &    11.19 &     0.00 \\
		$PC_1^{nf}$ &     0.06 &     0.60 &     0.49 &    -1.65 &    -0.73 &    -6.16 &     0.01 &     0.88 &    -0.25 &    -3.90 &     0.84 &     6.11 \\
\end{tabular*}
	\caption{Invariant amplitudes of the $\Lambda_b\to pK^{\ast -}$ decay.}
	\label{tab:Kstar-invariant}
\end{sidewaystable}

\begin{sidewaystable}
	\centering
	\footnotesize
	\renewcommand{\arraystretch}{1.2}
	\begin{tabular*}{230mm}{c@{\extracolsep{\fill}}cccccccc}
		\toprule[1pt]
		\toprule[0.7pt]
		& Real($H_{1/2,1}$) & Imag($H_{1/2,1}$) & Real($H_{-1/2,-1}$) & Imag($H_{-1/2,-1}$) & Real($H_{1/2,0}$) & Imag($H_{1/2,0}$) & Real($H_{-1/2,0}$) & Imag($H_{-1/2,0}$) \\
		\toprule[0.7pt]
		$T^f$ &    36.78 &    -0.00 & -2491.55 &     0.00 &   -70.47 &     0.00 & 10702.76 &     0.00 \\
		$T^{nf}$ &   -38.20 &    29.82 &    -9.40 &   -61.50 &   133.06 &  1342.77 &  -266.10 & -3118.57 \\
		$T$ &    -1.42 &    29.82 & -2500.95 &   -61.50 &    62.59 &  1342.77 & 10436.65 & -3118.57 \\
		$E_2$ &     9.97 &    47.66 &  -389.44 &  2352.39 &   125.40 &  -132.47 &  -624.29 &    61.91 \\
		\toprule[0.7pt]
		Tree &     8.55 &    77.48 & -2890.39 &  2290.89 &   187.99 &  1210.31 &  9812.36 & -3056.66 \\
		\toprule[0.7pt]
		$PC_1^f$ &     1.55 &    -0.00 &  -104.51 &     0.00 &    -2.28 &     0.00 &   449.38 &     0.00 \\
		$PC_1^{nf}$ &    -2.21 &    -2.79 &    -5.48 &   -53.46 &    20.58 &    46.96 &     3.40 &  -101.76 \\
\end{tabular*}
	\caption{Helicity amplitudes of the $\Lambda_b\to pK^{\ast -}$ decay.}
	\label{tab:Kstar-helicity}
\end{sidewaystable}

\begin{sidewaystable}
	\centering
	\footnotesize
	\renewcommand{\arraystretch}{1.2}
	\begin{tabular*}{230mm}{c@{\extracolsep{\fill}}cccccccccccc}
		\toprule[1pt]
		\toprule[0.7pt]
		& Real($S^L$) & Imag($S^L$) & Real($S^T$) & Imag($S^T$) & Real($P_1$) & Imag($P_1$) & Real($P_2$) & Imag($P_2$) & Real($D$) & Imag($D$) & Real($D+S^L$) & Imag($D+S^L$)\\
		\toprule[0.7pt]
		$T^f$ &   194.45 &    -0.00 &   194.45 &    -0.00 &  -367.82 &     0.00 &   188.80 &     0.00 &   178.24 &    -0.00 &   372.69 &    -0.00 \\
		$T^{nf}$ &    -4.76 &   -10.58 &    -2.22 &     7.02 &     4.60 &    61.43 &     3.66 &     2.44 &    -9.04 &  -143.75 &   -13.81 &  -154.34 \\
		$T$ &   189.69 &   -10.58 &   192.24 &     7.02 &  -363.22 &    61.43 &   192.46 &     2.44 &   169.19 &  -143.75 &   358.88 &  -154.34 \\
		$E_2$ &    28.18 &  -227.94 &    30.72 &  -177.26 &    17.26 &     2.44 &    29.19 &  -184.59 &   -54.11 &   234.67 &   -25.94 &     6.72  \\
		\toprule[0.7pt]
		Tree &   217.87 &  -238.53 &   222.96 &  -170.23 &  -345.96 &    63.87 &   221.64 &  -182.15 &   115.08 &    90.91 &   332.95 &  -147.61  \\
		\toprule[0.7pt]
		$PC_1^f$ &     8.16 &    -0.00 &     8.16 &    -0.00 &   -15.47 &     0.00 &     7.92 &     0.00 &     7.47 &    -0.00 &    15.62 &    -0.00 \\
		$PC_1^{nf}$ &    -0.06 &    -0.60 &     0.25 &     3.90 &    -0.83 &     1.90 &     0.59 &     4.33 &    -0.53 &    -4.54 &    -0.59 &    -5.14  \\
\end{tabular*}
	\caption{Partial wave amplitudes of the $\Lambda_b\to pK^{\ast -}$ decay.}
	\label{tab:Kstar-partialwave}
\end{sidewaystable}


\begin{sidewaystable}
	\centering
	\footnotesize
	\renewcommand{\arraystretch}{1.2}
	\begin{tabular*}{230mm}{c@{\extracolsep{\fill}}cccccccccccc}
		\toprule[1pt]
		\toprule[0.7pt]
		& Real($A_1^L$) & Imag($A_1^L$) & Real($B_1^L$) & Imag($B_1^L$) & Real($A_2^L$) & Imag($A_2^L$) & Real($B_2^L$) & Imag($B_2^L$) & Real($A_1^T$) & Imag($A_1^T$) & Real($B_1^T$) & Imag($B_1^T$)\\
		\toprule[0.7pt]
		$T^f$ &     0.00 &   296.24 &    -0.00 &  -409.58 &    -0.00 &  -111.18 &     0.00 &   121.60 &     0.00 &   296.24 &    -0.00 &  -409.58 \\
		$T^{nf}$ &    12.02 &    -6.01 &   -26.71 &    -2.57 &  -253.52 &    15.21 &   -86.65 &    15.01 &   -11.02 &    -2.42 &     4.35 &    -7.12 \\
		$T$ &    12.02 &   290.23 &   -26.71 &  -412.15 &  -253.52 &   -95.97 &   -86.65 &   136.61 &   -11.02 &   293.83 &     4.35 &  -416.70 \\
		$C_2$ &     4.01 &  -109.12 &    75.01 &  -134.43 &    34.19 &  -256.32 &   -81.65 &   261.14 &    30.03 &   -66.00 &   107.05 &   -46.72 \\
		$E_2$ &   132.67 &    18.84 &  -194.56 &   -21.79 &   386.67 &    55.94 &   368.73 &    68.69 &   131.79 &    28.94 &  -191.35 &   -33.51 \\
		$B$ &     2.52 &   -14.87 &     1.25 &    -5.86 &     2.80 &   -34.90 &     8.93 &    16.22 &    13.09 &    -5.94 &    20.40 &    -3.95 \\
		\toprule[0.7pt]
		Tree &   151.22 &   185.08 &  -145.00 &  -574.24 &   170.14 &  -331.25 &   209.35 &   482.66 &   163.89 &   250.83 &   -59.55 &  -500.89 \\
\end{tabular*}
	\caption{Invariant amplitudes of the $\Lambda_b\to pa_1^-(1260)$ decay.}
	\label{tab:a1-invariant}
\end{sidewaystable}

\begin{sidewaystable}
	\centering
	\footnotesize
	\renewcommand{\arraystretch}{1.2}
	\begin{tabular*}{230mm}{c@{\extracolsep{\fill}}cccccccc}
		\toprule[1pt]
		\toprule[0.7pt]
		& Real($H_{1/2,1}$) & Imag($H_{1/2,1}$) & Real($H_{-1/2,-1}$) & Imag($H_{-1/2,-1}$) & Real($H_{1/2,0}$) & Imag($H_{1/2,0}$) & Real($H_{-1/2,0}$) & Imag($H_{-1/2,0}$) \\
		\toprule[0.7pt]
		$T^f$ &    -0.00 &   -59.16 &     0.00 &  3756.40 &     0.00 &    72.69 &     0.00 & -11416.70 \\
		$T^{nf}$ &    51.32 &    47.69 &   -90.62 &    16.58 &   609.70 &  -103.95 & -3116.18 &   298.85 \\
		$T$ &    51.32 &   -11.47 &   -90.62 &  3772.98 &   609.70 &   -31.26 & -3116.18 & -11117.84 \\
		$C_2$ &  -676.57 &   635.89 &  -289.78 &  -214.15 &   328.37 &    34.12 &   640.25 &   374.83 \\
		$E_2$ &    14.91 &   -35.12 &  1712.40 &   337.63 &   -74.15 &   232.15 &   493.87 &   326.34 \\
		$B$ &  -176.35 &    56.12 &    -7.80 &   -20.43 &   128.61 &    30.66 &    79.76 &    77.44 \\
		\toprule[0.7pt]
		Tree &  -786.69 &   645.43 &  1324.20 &  3876.03 &   992.53 &   265.67 & -1902.30 & -10339.24 \\
\end{tabular*}
	\caption{Helicity amplitudes of the $\Lambda_b\to pa_1^-(1260)$ decay.}
	\label{tab:a1-helicity}
\end{sidewaystable}

\begin{sidewaystable}
	\centering
	\footnotesize
	\renewcommand{\arraystretch}{1.2}
	\begin{tabular*}{230mm}{c@{\extracolsep{\fill}}cccccccccccc}
		\toprule[1pt]
		\toprule[0.7pt]
		& Real($S^L$) & Imag($S^L$) & Real($S^T$) & Imag($S^T$) & Real($P_1$) & Imag($P_1$) & Real($P_2$) & Imag($P_2$) & Real($D$) & Imag($D$) & Real($D+S^L$) & Imag($D+S^L$)\\
		\toprule[0.7pt]
		$T^f$ &    -0.00 &  -296.24 &    -0.00 &  -296.24 &     0.00 &   545.91 &     0.00 &  -287.06 &    -0.00 &  -256.66 &    -0.00 &  -552.90  \\
		$T^{nf}$ &   -12.02 &     6.01 &    11.02 &     2.42 &   120.62 &    -9.38 &     3.05 &    -4.99 &  -167.28 &    13.37 &  -179.30 &    19.38  \\
		$T$ &   -12.02 &  -290.23 &    11.02 &  -293.83 &   120.62 &   536.53 &     3.05 &  -292.05 &  -167.28 &  -243.29 &  -179.30 &  -533.52  \\
		$C_2$ &    -4.01 &   109.12 &   -30.03 &    66.00 &   -46.61 &   -19.68 &    75.03 &   -32.74 &    19.01 &   -92.73 &    15.01 &    16.40  \\
		$E_2$ &  -132.67 &   -18.84 &  -131.79 &   -28.94 &   -20.20 &   -26.88 &  -134.11 &   -23.49 &   160.01 &    23.37 &    27.33 &     4.53   \\
		$B$ &    -2.52 &    14.87 &   -13.09 &     5.94 &   -10.03 &    -5.20 &    14.30 &    -2.77 &     0.17 &   -12.62 &    -2.35 &     2.25   \\
		\toprule[0.7pt]
		Tree &  -151.22 &  -185.08 &  -163.89 &  -250.83 &    43.78 &   484.77 &   -41.73 &  -351.05 &    11.92 &  -325.26 &  -139.31 &  -510.34   \\
\end{tabular*}
	\caption{Partial wave amplitudes of the $\Lambda_b\to pa_1^-(1260)$ decay.}
	\label{tab:a1-partialwave}
\end{sidewaystable}


\begin{sidewaystable}
	\centering
	\footnotesize
	\renewcommand{\arraystretch}{1.2}
	\begin{tabular*}{230mm}{c@{\extracolsep{\fill}}cccccccccccc}
		\toprule[1pt]
		\toprule[0.7pt]
		& Real($A_1^L$) & Imag($A_1^L$) & Real($B_1^L$) & Imag($B_1^L$) & Real($A_2^L$) & Imag($A_2^L$) & Real($B_2^L$) & Imag($B_2^L$) & Real($A_1^T$) & Imag($A_1^T$) & Real($B_1^T$) & Imag($B_1^T$)\\
		\toprule[0.7pt]
		$T^f$ &     0.00 &   378.38 &    -0.00 &  -524.96 &    -0.00 &  -143.60 &     0.00 &   157.06 &     0.00 &   378.38 &    -0.00 &  -524.96 \\
		$T^{nf}$ &   -34.43 &     2.18 &    29.72 &    -4.78 &   -37.64 &     9.01 &  -362.56 &    37.73 &   -25.06 &     0.73 &    38.14 &   -10.20 \\
		$T$ &   -34.43 &   380.56 &    29.72 &  -529.74 &   -37.64 &  -134.59 &  -362.56 &   194.79 &   -25.06 &   379.10 &    38.14 &  -535.15 \\
		$E_2$ &  -144.36 &   -33.42 &   201.71 &    43.50 &  -450.06 &   -98.06 &  -382.73 &  -118.18 &   -57.35 &    -4.77 &    77.80 &    12.11 \\
		\toprule[0.7pt]
		Tree &  -178.79 &   347.14 &   231.43 &  -486.24 &  -487.70 &  -232.65 &  -745.29 &    76.62 &   -82.41 &   374.33 &   115.94 &  -523.04 \\
		\toprule[0.7pt]
		$PC_1^f$ &     0.00 &    15.86 &     0.00 &   -22.01 &     0.00 &    -5.99 &     0.00 &     6.50 &     0.00 &    15.86 &     0.00 &   -22.01 \\
		$PC_1^{nf}$ &    -9.57 &     0.85 &    12.46 &    -1.08 &   -21.33 &     1.48 &   -34.10 &     3.09 &   -11.29 &     0.32 &    16.50 &    -0.96 \\
\end{tabular*}
	\caption{Invariant amplitudes of the $\Lambda_b\to pK_1(1270)^-$ decay with mixing angle $\theta_K=30^\circ$.}
	\label{tab:K1270-invariant}
\end{sidewaystable}

\begin{sidewaystable}
	\centering
	\footnotesize
	\renewcommand{\arraystretch}{1.2}
	\begin{tabular*}{230mm}{c@{\extracolsep{\fill}}cccccccc}
		\toprule[1pt]
		\toprule[0.7pt]
		& Real($H_{1/2,1}$) & Imag($H_{1/2,1}$) & Real($H_{-1/2,-1}$) & Imag($H_{-1/2,-1}$) & Real($H_{1/2,0}$) & Imag($H_{1/2,0}$) & Real($H_{-1/2,0}$) & Imag($H_{-1/2,0}$) \\
		\toprule[0.7pt]
		$T^f$ &    -0.00 &   -68.10 &     0.00 &  4803.85 &     0.00 &    88.95 &     0.00 & -14484.42 \\
		$T^{nf}$ &   -10.66 &    41.32 &  -333.38 &    50.67 & -3199.56 &   248.60 & -2531.08 &   292.08 \\
		$T$ &   -10.66 &   -26.78 &  -333.38 &  4854.52 & -3199.56 &   337.54 & -2531.08 & -14192.34 \\
		$E_2$ &    18.28 &   -23.92 &  -720.14 &   -85.37 &   275.50 &  -302.29 &  -710.87 &  -451.08 \\
		\toprule[0.7pt]
		Tree &     7.62 &   -50.69 & -1053.51 &  4769.15 & -2924.06 &    35.25 & -3241.95 & -14643.42 \\
		\toprule[0.7pt]
		$PC_1^f$ &    -0.00 &    -2.85 &     0.00 &   201.41 &     0.00 &     2.65 &     0.00 &  -607.92 \\
		$PC_1^{nf}$ &    -1.77 &     2.28 &  -147.09 &     6.42 &  -132.76 &    15.41 &   -87.73 &     6.22 \\
\end{tabular*}
	\caption{Helicity amplitudes of the $\Lambda_b\to pK_1(1270)^-$ decay.}
	\label{tab:K1270-helicity}
\end{sidewaystable}

\begin{sidewaystable}
	\centering
	\footnotesize
	\renewcommand{\arraystretch}{1.2}
	\begin{tabular*}{230mm}{c@{\extracolsep{\fill}}cccccccccccc}
		\toprule[1pt]
		\toprule[0.7pt]
		& Real($S^L$) & Imag($S^L$) & Real($S^T$) & Imag($S^T$) & Real($P_1$) & Imag($P_1$) & Real($P_2$) & Imag($P_2$) & Real($D$) & Imag($D$)& Real($D+S^L$) & Imag($D+S^L$)\\
		\toprule[0.7pt]
		$T^f$ &    -0.00 &  -378.38 &    -0.00 &  -378.38 &     0.00 &   697.91 &     0.00 &  -367.80 &    -0.00 &  -328.16 &    -0.00 &  -706.54 \\
		$T^{nf}$ &    34.43 &    -2.18 &    25.06 &    -0.73 &   277.83 &   -26.21 &    26.72 &    -7.14 &    -2.02 &     4.29 &    32.41 &     2.11 \\
		$T$ &    34.43 &  -380.56 &    25.06 &  -379.10 &   277.83 &   671.70 &    26.72 &  -374.95 &    -2.02 &  -323.87 &    32.41 &  -704.43 \\
		$E_2$ &   144.36 &    33.42 &    57.35 &     4.77 &    21.11 &    36.52 &    54.51 &     8.49 &  -192.18 &   -40.64 &   -47.82 &    -7.21  \\
		\toprule[0.7pt]
		Tree &   178.79 &  -347.14 &    82.41 &  -374.33 &   298.94 &   708.23 &    81.23 &  -366.46 &  -194.20 &  -364.51 &   -15.41 &  -711.64  \\
		\toprule[0.7pt]
		$PC_1^f$ &    -0.00 &   -15.86 &    -0.00 &   -15.86 &     0.00 &    29.34 &     0.00 &   -15.42 &    -0.00 &   -13.74 &    -0.00 &   -29.60 \\
		$PC_1^{nf}$ &     9.57 &    -0.85 &    11.29 &    -0.32 &    10.69 &    -1.05 &    11.56 &    -0.68 &    -7.39 &     0.40 &     2.18 &    -0.45  \\
\end{tabular*}
	\caption{Partial wave of the $\Lambda_b\to pK_1(1270)^-$ decay.}
	\label{tab:K1270-partial}
\end{sidewaystable}


\begin{sidewaystable}
	\centering
	\footnotesize
	\renewcommand{\arraystretch}{1.2}
	\begin{tabular*}{230mm}{c@{\extracolsep{\fill}}cccccccccccc}
		\toprule[1pt]
		\toprule[0.7pt]
		& Real($A_1^L$) & Imag($A_1^L$) & Real($B_1^L$) & Imag($B_1^L$) & Real($A_2^L$) & Imag($A_2^L$) & Real($B_2^L$) & Imag($B_2^L$) & Real($A_1^T$) & Imag($A_1^T$) & Real($B_1^T$) & Imag($B_1^T$)\\
		\toprule[0.7pt]
		$T^f$ &     0.00 &   156.11 &     0.00 &  -216.25 &     0.00 &   -58.88 &     0.00 &    64.41 &     0.00 &   156.11 &     0.00 &  -216.25 \\
		$T^{nf}$ &    35.90 &    -9.89 &   -50.17 &     0.35 &  -320.19 &    26.95 &    88.01 &     2.81 &    -0.81 &    -4.76 &   -15.54 &    -2.98 \\
		$T$ &    35.90 &   146.23 &   -50.17 &  -215.90 &  -320.19 &   -31.92 &    88.01 &    67.22 &    -0.81 &   151.35 &   -15.54 &  -219.24 \\
		$E_2$ &   222.20 &    40.60 &  -323.31 &   -46.62 &   686.72 &   131.52 &   613.16 &   127.39 &   179.22 &    41.20 &  -257.22 &   -49.19 \\
		\toprule[0.7pt]
		Tree &   258.10 &   186.82 &  -373.48 &  -262.52 &   366.52 &    99.60 &   701.16 &   194.62 &   178.42 &   192.55 &  -272.76 &  -268.43 \\
		\toprule[0.7pt]
		$PC_1^f$ &     0.00 &     6.53 &     0.00 &    -9.06 &     0.00 &    -2.46 &     0.00 &     2.67 &     0.00 &     6.53 &     0.00 &    -9.06 \\
		$PC_1^{nf}$ &   -17.15 &     1.77 &    24.90 &    -3.08 &   -55.17 &     6.10 &   -47.81 &     5.59 &   -24.97 &     1.49 &    35.31 &    -2.61 \\
\end{tabular*}
	\caption{Invariant amplitudes of the $\Lambda_b\to pK_1(1400)^-$ decay with mixing abgle $\theta_K=30^\circ$.}
	\label{tab:K1400-invariant}
\end{sidewaystable}

\begin{sidewaystable}
	\centering
	\footnotesize
	\renewcommand{\arraystretch}{1.2}
	\begin{tabular*}{230mm}{c@{\extracolsep{\fill}}cccccccc}
		\toprule[1pt]
		\toprule[0.7pt]
		& Real($H_{1/2,1}$) & Imag($H_{1/2,1}$) & Real($H_{-1/2,-1}$) & Imag($H_{-1/2,-1}$) & Real($H_{1/2,0}$) & Imag($H_{1/2,0}$) & Real($H_{-1/2,0}$) & Imag($H_{-1/2,0}$) \\
		\toprule[0.7pt]
		$T^f$ &    -0.00 &   -33.73 &     0.00 &  1967.90 &     0.00 &    34.08 &     0.00 & -5381.29 \\
		$T^{nf}$ &    74.68 &    43.86 &    64.35 &   -17.17 &  2370.88 &  -276.73 & -2405.30 &   333.60 \\
		$T$ &    74.68 &    10.13 &    64.35 &  1950.73 &  2370.88 &  -242.65 & -2405.30 & -5047.68 \\
		$E_2$ &     1.30 &   -44.14 &  2299.26 &   484.10 &  -313.49 &   219.18 &   850.48 &   501.57 \\
		\toprule[0.7pt]
		Tree &    75.98 &   -34.01 &  2363.61 &  2434.83 &  2057.39 &   -23.47 & -1554.81 & -4546.11 \\
		\toprule[0.7pt]
		$PC_1^f$ &    -0.00 &    -1.35 &     0.00 &    82.39 &     0.00 &     0.97 &     0.00 &  -225.54 \\
		$PC_1^{nf}$ &     2.17 &     2.15 &  -317.96 &    21.19 &    31.88 &    -7.79 &   -83.06 &     8.73 \\
\end{tabular*}
	\caption{Helicity amplitudes of the $\Lambda_b\to pK_1(1400)^-$ decay.}
	\label{tab:K1400-helicity}
\end{sidewaystable}

\begin{sidewaystable}
	\centering
	\footnotesize
	\renewcommand{\arraystretch}{1.2}
	\begin{tabular*}{230mm}{c@{\extracolsep{\fill}}cccccccccccc}
		\toprule[1pt]
		\toprule[0.7pt]
		& Real($S^L$) & Imag($S^L$) & Real($S^T$) & Imag($S^T$) & Real($P_1$) & Imag($P_1$) & Real($P_2$) & Imag($P_2$) & Real($D$) & Imag($D$)& Real($D+S^L$) & Imag($D+S^L$)\\
		\toprule[0.7pt]
		$T^f$ &    -0.00 &  -156.11 &    -0.00 &  -156.11 &     0.00 &   283.94 &     0.00 &  -150.85 &    -0.00 &  -131.44 &    -0.00 &  -287.56 \\
		$T^{nf}$ &   -35.90 &     9.89 &     0.81 &     4.76 &     1.83 &    -3.02 &   -10.84 &    -2.08 &  -217.72 &    22.52 &  -253.61 &    32.41 \\
		$T$ &   -35.90 &  -146.23 &     0.81 &  -151.35 &     1.83 &   280.92 &   -10.84 &  -152.93 &  -217.72 &  -108.92 &  -253.61 &  -255.15 \\
		$E_2$ &  -222.20 &   -40.60 &  -179.22 &   -41.20 &   -28.51 &   -38.27 &  -179.43 &   -34.31 &   284.01 &    55.59 &    61.81 &    14.99  \\
		\toprule[0.7pt]
		Tree &  -258.10 &  -186.82 &  -178.42 &  -192.55 &   -26.69 &   242.64 &  -190.27 &  -187.24 &    66.29 &   -53.33 &  -191.81 &  -240.15  \\
		\toprule[0.7pt]
		$PC_1^f$ &    -0.00 &    -6.53 &    -0.00 &    -6.53 &     0.00 &    11.92 &     0.00 &    -6.32 &    -0.00 &    -5.50 &    -0.00 &   -12.03 \\
		$PC_1^{nf}$ &    17.15 &    -1.77 &    24.97 &    -1.49 &     2.72 &    -0.05 &    24.63 &    -1.82 &   -23.25 &     2.65 &    -6.10 &     0.88  \\
\end{tabular*}
	\caption{Partial wave amplitudes of the $\Lambda_b\to pK_1(1400)^-$ decay.}
	\label{tab:K1400-partial}
\end{sidewaystable}


\newpage


\begin{thebibliography}{200}
	\bibitem{Christenson:1964fg}
	J.~H.~Christenson, J.~W.~Cronin, V.~L.~Fitch and R.~Turlay,
	Evidence for the $2\pi$ Decay of the $K_2^0$ Meson, 
	Phys. Rev. Lett. \textbf{13}, 138-140 (1964)
	
	\bibitem{BaBar:2001ags}
	B.~Aubert \textit{et al.} (BaBar  Collaboration),
	Measurement of $CP$-violating asymmetries in $B^0$ decays to $CP$ eigenstates, 
	Phys. Rev. Lett. \textbf{86}, 2515-2522 (2001)
	
	\bibitem{Belle:2001zzw}
	K.~Abe \textit{et al.} (Belle  Collaboration),
	Observation of large $CP$ violation in the neutral $B$ meson system, 
	Phys. Rev. Lett. \textbf{87}, 091802 (2001)
	
	\bibitem{LHCb:2019hro}
	R.~Aaij \textit{et al.} (LHCb  Collaboration),
	Observation of $CP$ Violation in Charm Decays, 
	Phys. Rev. Lett. \textbf{122}, 211803 (2019)
	
	
	\bibitem{BESIII:2018cnd}
	M.~Ablikim \textit{et al.} (BESIII Collaboration ),
	Polarization and Entanglement in Baryon-Antibaryon Pair Production in Electron-Positron Annihilation, 
	Nature Phys. \textbf{15}, 631-634 (2019)
	
	\bibitem{BESIII:2021ypr}
	M.~Ablikim \textit{et al.} (BESIII  Collaboration),
	Probing CP symmetry and weak phases with entangled double-strange baryons, 
	Nature \textbf{606}, no.7912, 64-69 (2022)
	
	\bibitem{Belle:2022uod}
	L.~K.~Li \textit{et al.} (Belle  Collaboration),
	Search for $CP$ violation and measurement of branching fractions and decay asymmetry parameters for \ensuremath{\Lambda_c^+}
	\textrightarrow{}\ensuremath{\Lambda h^+} and \ensuremath{\Lambda_c^+}
	\textrightarrow{}\ensuremath{\Sigma^0 h^+} ($h=K,\pi$), 
	Sci. Bull. \textbf{68} (2023), 583-592
	
	\bibitem{LHCb:2016yco}
	R.~Aaij \textit{et al.} (LHCb  Collaboration),
	Measurement of matter-antimatter differences in beauty baryon decays, 
	Nature Phys. \textbf{13}, 391-396 (2017)
	
	
	\bibitem{LHCb:2017hwf}
	R.~Aaij \textit{et al.} (LHCb Collaboration),
	A measurement of the $CP$ asymmetry difference in $\varLambda_{c}^{+} \to pK^{-}K^{+}$ and $p\pi^{-}\pi^{+}$ decays, 
	JHEP \textbf{03}, 182 (2018)
	
	\bibitem{LHCb:2018fpt}
	R.~Aaij \textit{et al.} (LHCb Collaboration),
	Search for CP violation using triple product asymmetries in $\Lambda^{0}_{b}\to pK^{-}\pi^{+}\pi^{-}$, $\Lambda^{0}_{b}\to pK^{-}K^{+}K^{-}$ and $\Xi^{0}_{b}\to pK^{-}K^{-}\pi^{+}$ decays, 
	JHEP \textbf{08}, 039 (2018)
	
	
	\bibitem{LHCb:2018fly}
	R.~Aaij \textit{et al.} (LHCb Collaboration),
	Search for $C\!P$ violation in $\Lambda^0_b \to p K^-$ and $\Lambda^0_b \to p \pi^-$ decays, 
	Phys. Lett. B \textbf{787}, 124-133 (2018)
	
	
	\bibitem{LHCb:2019jyj}
	R.~Aaij \textit{et al.} (LHCb Collaboration),
	Measurements of $CP$ asymmetries in charmless four-body $\Lambda_b^0$ and $\Xi_b^0$ decays, 
	Eur. Phys. J. C \textbf{79}, 745 (2019)
	
	
	\bibitem{LHCb:2019oke}
	R.~Aaij \textit{et al.} (LHCb Collaboration),
	Search for $CP$ violation and observation of $P$ violation in $\Lambda_b^0 \to p \pi^- \pi^+ \pi^-$ decays, 
	Phys. Rev. D \textbf{102}, 051101 (2020)
	
	\bibitem{LHCb:2024iis}
	R.~Aaij \textit{et al.} (LHCb Collaboration),
	Measurement of $CP$ asymmetries in $\Lambda_b^0\to ph^{-}$ decays, 
	Phys. Rev. D \textbf{111}, 092004 (2025)
	
	
	\bibitem{LHCb:2024yzj}
	R.~Aaij \textit{et al.} (LHCb Collaboration),
	Study of $\Lambda_b^0$ and $\Xi_b^0$ Decays to $\Lambda h^+ h^{\prime-}$ and Evidence for $CP$ Violation in $\Lambda_b^0\to \Lambda K^+K^-$ Decays, 
	Phys. Rev. Lett. \textbf{134},  101802 (2025)
	
	\bibitem{LHCb:2025ray}
	R.~Aaij \textit{et al.} (LHCb Collaboration),
	Observation of charge-parity symmetry breaking in baryon decays, 
	[arXiv:2503.16954 [hep-ex]].
	
	\bibitem{Yu:2025ekh}
	F.~S.~Yu and C.~D.~L\"u,
	New horizon in particle physics: First observation of CP violation in baryon decays,
	[arXiv:2504.15008 [hep-ph]].
	
	\bibitem{Han:2024kgz}
	J.~J.~Han, J.~X.~Yu, Y.~Li, H.~n.~Li, J.~P.~Wang, Z.~J.~Xiao and F.~S.~Yu,
	Establishing $CP$ Violation in b-Baryon Decays,
	Phys. Rev. Lett. \textbf{134}, 221801 (2025) 
	
	
	
	
	
	\bibitem{Wang:2024qff}
	J.~P.~Wang, Q.~Qin and F.~S.~Yu,
	CP violation observables in baryon decays, 
	[arXiv:2411.18323 [hep-ph]].
	
	\bibitem{Korner:1992wi}
	J.~G.~Korner and M.~Kramer,
	Exclusive nonleptonic charm baryon decays, 
	Z. Phys. C \textbf{55}, 659-670 (1992)
	
	\bibitem{Zhao:2024ren}
	Y.~J.~Zhao, Z.~H.~Zhang and X.~H.~Guo,
	Decay-angular-distribution correlated CP violation in heavy hadron cascade decays, 
	Phys. Rev. D \textbf{110}, no.1, 013007 (2024)
	
	\bibitem{Geng:2021sxe}
	C.~Q.~Geng and C.~W.~Liu,
	Time-reversal asymmetries and angular distributions in \ensuremath{\Lambda}$_{b}$ \textrightarrow{} \ensuremath{\Lambda}V, 
	JHEP \textbf{11}, 104 (2021)
	
	\bibitem{Wang:2011uv}
	W.~Wang,
	Factorization of Heavy-to-Light Baryonic Transitions in SCET, 
	Phys. Lett. B \textbf{708}, 119-126 (2012)
	
	\bibitem{Han:2022srw}
	J.~J.~Han, Y.~Li, H.~n.~Li, Y.~L.~Shen, Z.~J.~Xiao and F.~S.~Yu,
	$\Lambda _b\rightarrow p$ transition form factors in perturbative QCD, 
	Eur. Phys. J. C \textbf{82}, no.8, 686 (2022)
	
	\bibitem{Beneke:1999br}
	M.~Beneke, G.~Buchalla, M.~Neubert and C.~T.~Sachrajda,
	QCD factorization for B ---\ensuremath{>} pi pi decays: Strong phases and CP violation in the heavy quark limit, 
	Phys. Rev. Lett. \textbf{83}, 1914-1917 (1999)
	
	\bibitem{Beneke:2000ry}
	M.~Beneke, G.~Buchalla, M.~Neubert and C.~T.~Sachrajda,
	QCD factorization for exclusive, nonleptonic B meson decays: General arguments and the case of heavy light final states, 
	Nucl. Phys. B \textbf{591}, 313-418 (2000)
	
	\bibitem{Bauer:2000yr}
	C.~W.~Bauer, S.~Fleming, D.~Pirjol and I.~W.~Stewart,
	An Effective field theory for collinear and soft gluons: Heavy to light decays, 
	Phys. Rev. D \textbf{63}, 114020 (2001)
	
	\bibitem{Bauer:2001yt}
	C.~W.~Bauer, D.~Pirjol and I.~W.~Stewart,
	Soft collinear factorization in effective field theory, 
	Phys. Rev. D \textbf{65}, 054022 (2002)
	
	\bibitem{Bauer:2002nz}
	C.~W.~Bauer, S.~Fleming, D.~Pirjol, I.~Z.~Rothstein and I.~W.~Stewart,
	Hard scattering factorization from effective field theory, 
	Phys. Rev. D \textbf{66}, 014017 (2002)
	
	\bibitem{Keum:2000wi}
	Y.~Y.~Keum, H.~N.~Li and A.~I.~Sanda,
	Penguin enhancement and $B \to K \pi$ decays in perturbative QCD, 
	Phys. Rev. D \textbf{63}, 054008 (2001)
	
	\bibitem{Lu:2000em}
	C.~D.~Lu, K.~Ukai and M.~Z.~Yang,
	Branching ratio and CP violation of B ---\ensuremath{>} pi pi decays in perturbative QCD approach, 
	Phys. Rev. D \textbf{63}, 074009 (2001)
	
	\bibitem{Keum:2000ph}
	Y.~Y.~Keum, H.~n.~Li and A.~I.~Sanda,
	Fat penguins and imaginary penguins in perturbative QCD, 
	Phys. Lett. B \textbf{504}, 6-14 (2001)
	
	\bibitem{Hsiao:2014mua}
	Y.~K.~Hsiao and C.~Q.~Geng,
	Direct CP violation in $\Lambda_b$ decays, 
	Phys. Rev. D \textbf{91}, no.11, 116007 (2015)
	
	\bibitem{Hsiao:2017tif}
	Y.~K.~Hsiao, Y.~Yao and C.~Q.~Geng,
	Charmless two-body anti-triplet $b$-baryon decays, 
	Phys. Rev. D \textbf{95}, no.9, 093001 (2017)
	
	\bibitem{Geng:2021nkl}
	C.~Q.~Geng, C.~W.~Liu and T.~H.~Tsai,
	Non-leptonic two-body decays of $\Lambda^0_b$ in light-front quark model, 
	Phys. Lett. B \textbf{815}, 136125 (2021)
	
	\bibitem{Wang:2024oyi}
	J.~P.~Wang and F.~S.~Yu,
	CP violation of baryon decays with N \ensuremath{\pi} rescatterings*,
	Chin. Phys. C \textbf{48}, no.10, 101002 (2024)
	
	\bibitem{Duan:2024zjv}
	Z.~D.~Duan, J.~P.~Wang, R.~H.~Li, C.~D.~Lv and F.~S.~Yu,
	Final-state rescattering mechanism of bottom-baryon decays,
	[arXiv:2412.20458 [hep-ph]].
	
	\bibitem{Zhu:2016bra}
	J.~Zhu, H.~W.~Ke and Z.~T.~Wei,
	The decay of $\Lambda _b\rightarrow p~K^-$ in QCD factorization approach, 
	Eur. Phys. J. C \textbf{76}, no.5, 284 (2016)
	\bibitem{Zhu:2018jet}
	J.~Zhu, Z.~T.~Wei and H.~W.~Ke,
	Semileptonic and nonleptonic weak decays of $\Lambda_b^0$, 
	Phys. Rev. D \textbf{99}, no.5, 054020 (2019)
	
	\bibitem{Lu:2009cm}
	C.~D.~Lu, Y.~M.~Wang, H.~Zou, A.~Ali and G.~Kramer,
	Anatomy of the perturbative QCD approach to the baryonic decays $\Lambda_b \to p\pi,pK$, 
	Phys. Rev. D \textbf{80}, 034011 (2009)
	
	\bibitem{Zhang:2022iun}
	C.~Q.~Zhang, J.~M.~Li, M.~K.~Jia and Z.~Rui,
	Nonleptonic two-body decays of $\Lambda_b \to \Lambda_c\pi,\Lambda_cK$ in the perturbative QCD approach,
	Phys. Rev. D \textbf{105} (2022) no.7, 073005
	
	\bibitem{Rui:2022sdc}
	Z.~Rui, C.~Q.~Zhang, J.~M.~Li and M.~K.~Jia,
	Investigating the color-suppressed decays $\Lambda_b \to \Lambda\psi$ in the perturbative QCD approach,
	Phys. Rev. D \textbf{106} (2022) no.5, 053005
	
	\bibitem{Rui:2023fiz}
	Z.~Rui and Z.~T.~Zou,
	Charmonium decays of beauty baryons in the perturbative QCD approach,
	Phys. Rev. D \textbf{109} (2024) no.3, 033013
	
	\bibitem{Buchalla:1995vs}
	G.~Buchalla, A.~J.~Buras and M.~E.~Lautenbacher,
	Weak decays beyond leading logarithms, 
	Rev. Mod. Phys. \textbf{68}, 1125-1144 (1996)
	
	\bibitem{Chau:1995gk}
	L.~L.~Chau, H.~Y.~Cheng and B.~Tseng,
	Analysis of two-body decays of charmed baryons using the quark diagram scheme, 
	Phys. Rev. D \textbf{54}, 2132-2160 (1996)
	
	\bibitem{Yu:2017zst}
	F.~S.~Yu, H.~Y.~Jiang, R.~H.~Li, C.~D.~L\"u, W.~Wang and Z.~X.~Zhao,
	Discovery Potentials of Doubly Charmed Baryons, 
	Chin. Phys. C \textbf{42}, no.5, 051001 (2018)
	
	\bibitem{Jia:2024pyb}
	C.~P.~Jia, H.~Y.~Jiang, J.~P.~Wang and F.~S.~Yu,
	Final-state rescattering mechanism of charmed baryon decays, 
	JHEP \textbf{11}, 072 (2024)
	
	\bibitem{Leibovich:2003tw}
	A.~K.~Leibovich, Z.~Ligeti, I.~W.~Stewart and M.~B.~Wise,
	Predictions for nonleptonic Lambda(b) and Theta(b) decays, 
	Phys. Lett. B \textbf{586}, 337-344 (2004)
	
	\bibitem{Ball:2008fw}
	P.~Ball, V.~M.~Braun and E.~Gardi,
	Distribution Amplitudes of the Lambda(b) Baryon in QCD, 
	Phys. Lett. B \textbf{665}, 197-204 (2008)
	
	\bibitem{Bell:2013tfa}
	G.~Bell, T.~Feldmann, Y.~M.~Wang and M.~W.~Y.~Yip,
	Light-Cone Distribution Amplitudes for Heavy-Quark Hadrons, 
	JHEP \textbf{11}, 191 (2013)
	
	\bibitem{Wang:2015ndk}
	Y.~M.~Wang and Y.~L.~Shen,
	Perturbative Corrections to $\Lambda_b \to \Lambda$ Form Factors from QCD Light-Cone Sum Rules, 
	JHEP \textbf{02}, 179 (2016)
	
	\bibitem{Ali:2012zza}
	A.~Ali, C.~Hambrock and A.~Y.~Parkhomenko,
	Light-cone wave functions of heavy baryons, 
	Theor. Math. Phys. \textbf{170}, 2-16 (2012)
	
	\bibitem{Groote:1997yr}
	S.~Groote, J.~G.~Korner and O.~I.~Yakovlev,
	An Analysis of diagonal and nondiagonal QCD sum rules for heavy baryons at next-to-leading order in alpha-s, 
	Phys. Rev. D \textbf{56}, 3943-3954 (1997)
	
	\bibitem{Huang:2022lfr}
	K.~S.~Huang, W.~Liu, Y.~L.~Shen and F.~S.~Yu,
	$\Lambda _b \rightarrow p, N^*(1535)$ form factors from QCD light-cone sum rules, 
	Eur. Phys. J. C \textbf{83}, no.4, 272 (2023)
	
	\bibitem{Braun:2000kw}
	V.~Braun, R.~J.~Fries, N.~Mahnke and E.~Stein,
	Higher twist distribution amplitudes of the nucleon in QCD, 
	Nucl. Phys. B \textbf{589}, 381-409 (2000)
	[erratum: Nucl. Phys. B \textbf{607}, 433-433 (2001)]
	
	\bibitem{Braun:2006hz}
	V.~M.~Braun, A.~Lenz and M.~Wittmann,
	Nucleon Form Factors in QCD, 
	Phys. Rev. D \textbf{73}, 094019 (2006)
	
	\bibitem{RQCD:2019hps}
	G.~S.~Bali \textit{et al.} [RQCD],
	Light-cone distribution amplitudes of octet baryons from lattice QCD, 
	Eur. Phys. J. A \textbf{55}, no.7, 116 (2019)
	
	\bibitem{Chernyak:1983ej}
	V.~L.~Chernyak and A.~R.~Zhitnitsky,
	Asymptotic Behavior of Exclusive Processes in QCD, 
	Phys. Rept. \textbf{112}, 173 (1984)
	
	\bibitem{Braun:1988qv}
	V.~M.~Braun and I.~E.~Filyanov,
	QCD Sum Rules in Exclusive Kinematics and Pion Wave Function, 
	Z. Phys. C \textbf{44}, 157 (1989)
	
	\bibitem{Braun:1989iv}
	V.~M.~Braun and I.~E.~Filyanov,
	Conformal Invariance and Pion Wave Functions of Nonleading Twist, 
	Z. Phys. C \textbf{48}, 239-248 (1990)
	
	\bibitem{Kurimoto:2001zj}
	T.~Kurimoto, H.~n.~Li and A.~I.~Sanda,
	Leading power contributions to B ---\ensuremath{>} pi, rho transition form-factors, 
	Phys. Rev. D \textbf{65}, 014007 (2002)
	
	\bibitem{Ball:2004ye}
	P.~Ball and R.~Zwicky,
	New results on $B \to \pi, K, \eta$ decay formfactors from light-cone sum rules, 
	Phys. Rev. D \textbf{71}, 014015 (2005)
	
	\bibitem{Ball:2006wn}
	P.~Ball, V.~M.~Braun and A.~Lenz,
	Higher-twist distribution amplitudes of the K meson in QCD, 
	JHEP \textbf{05}, 004 (2006)
	
	\bibitem{Li:2009tx}
	R.~H.~Li, C.~D.~Lu and W.~Wang,
	Transition form factors of B decays into p-wave axial-vector mesons in the perturbative QCD approach, 
	Phys. Rev. D \textbf{79}, 034014 (2009)
	
	\bibitem{Ali:2007ff}
	A.~Ali, G.~Kramer, Y.~Li, C.~D.~Lu, Y.~L.~Shen, W.~Wang and Y.~M.~Wang,
	Charmless non-leptonic $B_s$ decays to $PP$, $PV$ and $VV$ final states in the pQCD approach, 
	Phys. Rev. D \textbf{76}, 074018 (2007)
	
	
	\bibitem{Braun:2004vf}
	V.~M.~Braun and A.~Lenz,
	On the SU(3) symmetry-breaking corrections to meson distribution amplitudes, 
	Phys. Rev. D \textbf{70}, 074020 (2004)
	
	\bibitem{Ball:2005vx}
	P.~Ball and R.~Zwicky,
	SU(3) breaking of leading-twist K and K* distribution amplitudes: A Reprise, 
	Phys. Lett. B \textbf{633}, 289-297 (2006)
	
	
	\bibitem{Ball:2006nr}
	P.~Ball and R.~Zwicky,
	$|V_{td} / V_{ts}|$ from $B \to V \gamma$, 
	JHEP \textbf{04}, 046 (2006)
	
	\bibitem{Ball:2007rt}
	P.~Ball and G.~W.~Jones,
	Twist-3 distribution amplitudes of K* and phi mesons, 
	JHEP \textbf{03}, 069 (2007)
	
	
	
	\bibitem{Cheng:2013cwa}
	H.~Y.~Cheng,
	Mixing angle of $K_1$ axial vector mesons, 
	PoS \textbf{Hadron2013}, 090 (2013)
	
	\bibitem{Hatanaka:2008xj}
	H.~Hatanaka and K.~C.~Yang,
	B ---\ensuremath{>} K(1) gamma Decays in the Light-Cone QCD Sum Rules, 
	Phys. Rev. D \textbf{77}, 094023 (2008)
	[erratum: Phys. Rev. D \textbf{78}, 059902 (2008)]
	
	\bibitem{Cheng:2011pb}
	H.~Y.~Cheng,
	Revisiting Axial-Vector Meson Mixing, 
	Phys. Lett. B \textbf{707}, 116-120 (2012)
	
	\bibitem{Shi:2023kiy}
	Y.~J.~Shi, J.~Zeng and Z.~F.~Deng,
	Revisiting K1(1270)-K1(1400) mixing with QCD sum rules, 
	Phys. Rev. D \textbf{109}, no.1, 016027 (2024)
	
	\bibitem{Lepage:1977sw}
	G.~P.~Lepage,
	A New Algorithm for Adaptive Multidimensional Integration, 
	J. Comput. Phys. \textbf{27}, 192 (1978)
	
	\bibitem{Lepage:2020tgj}
	G.~P.~Lepage,
	Adaptive multidimensional integration: VEGAS enhanced, 
	J. Comput. Phys. \textbf{439}, 110386 (2021)
	
	\bibitem{Keum:2000ms}
	Y.~Y.~Keum and H.~n.~Li,
		Nonleptonic charmless B decays: Factorization versus perturbative QCD,
		Phys. Rev. D \textbf{63}, 074006 (2001)
	
	\bibitem{Cheng:1996cs}
	H.~Y.~Cheng,
	Nonleptonic weak decays of bottom baryons, 
	Phys. Rev. D \textbf{56}, 2799-2811 (1997)
	[erratum: Phys. Rev. D \textbf{99}, no.7, 079901 (2019)]
	
	\bibitem{Cheng:2011qh}
	H.~Y.~Cheng and S.~Oh,
	Flavor SU(3) symmetry and QCD factorization in $B \to PP$ and $PV$ decays, 
	JHEP \textbf{09}, 024 (2011)
	
	\bibitem{Geng:2020ofy}
	C.~Q.~Geng, C.~W.~Liu and T.~H.~Tsai,
	Nonleptonic two-body weak decays of $\Lambda_b$ in modified MIT bag model, 
	Phys. Rev. D \textbf{102}, no.3, 034033 (2020)
	
	\bibitem{Lee:1957qs}
	T.~D.~Lee and C.~N.~Yang,
	General Partial Wave Analysis of the Decay of a Hyperon of Spin 1/2, 
	Phys. Rev. \textbf{108}, 1645-1647 (1957)
	
	
	\bibitem{Berman:1965gi}
	S.~M.~Berman and M.~Jacob,
	SYSTEMATICS OF ANGULAR POLARIZATION DISTRIBUTIONS IN THREE-BODY DECAYS, 
	Phys. Rev. \textbf{139}, B1023 (1965)
	
	\bibitem{LHCb:2016qpe}
	R.~Aaij \textit{et al.} (LHCb Collaboration),
	Measurement of the $b$-quark production cross-section in 7 and 13 TeV $pp$ collisions, 
	Phys. Rev. Lett. \textbf{118} (2017) no.5, 052002
	[erratum: Phys. Rev. Lett. \textbf{119} (2017) no.16, 169901]
	
	\bibitem{Chen:2021ftn}
	S.~Chen, Y.~Li, W.~Qian, Z.~Shen, Y.~Xie, Z.~Yang, L.~Zhang and Y.~Zhang,
	Heavy Flavour Physics and CP Violation at LHCb: a Ten-Year Review, 
	Front. Phys. \textbf{18}, 44601 (2023)
	
	\bibitem{TriggerEfficiency}
	See https://lbfence.cern.ch/alcm/public/figure/details/3837
	
	\bibitem{Pakvasa:1990if}
	S.~Pakvasa, S.~P.~Rosen and S.~F.~Tuan,
	Parity Violation and Flavor Selection Rules in Charmed Baryon Decays, 
	Phys. Rev. D \textbf{42}, 3746-3754 (1990)
	
\end{thebibliography}
\end{document}